\documentclass[a4paper]{article}

\usepackage[
  top=1.5cm,
  bottom=2cm,
  left=2.5cm,
  right=2.5cm
]{geometry}
\usepackage{graphicx}
\graphicspath{{Figures/}}
\usepackage{cite}
\usepackage{comment}
\usepackage{authblk}
\usepackage{caption}
\usepackage{amsmath, lipsum} 
\numberwithin{equation}{section}
\usepackage{amssymb} 
\usepackage{bm} 
\usepackage{dsfont} 
\usepackage{mathrsfs} 
\usepackage{url}
\usepackage{float}
\usepackage[numbers,square,compress]{natbib}
\usepackage{siunitx}
\DeclareSIUnit\eV{e\kern-.05em V}
\RequirePackage{hyperref}
\hypersetup{%
  colorlinks=true,
  linkcolor=black,
  citecolor=blue,
  filecolor=blue,
  urlcolor=blue
}
\usepackage{mdframed} 
\usepackage{cancel}
\usepackage{xspace}
\usepackage[T1]{fontenc}
\usepackage{bold-extra}

\newcommand{\pot}[2]{#1 \times 10^{#2}}

\newcommand{\COBE}{{\it COBE}\xspace}
\newcommand{\Planck}{{\it Planck}\xspace}

\newcommand{\PIXIE}{\textit{PIXIE}\xspace}

\newcommand{\SphereX}{{\it SphereX}\xspace}

\newcommand{\CT}{\texttt{CosmoTherm}\xspace}
\newcommand{\COBEF}{\textit{COBE/FIRAS}\xspace}
\newcommand{\FOSSIL}{\textit{FOSSIL}\xspace}

\newcommand{\ymono}{\langle y\rangle\xspace}
\newcommand{\tmono}{\langle T_e\rangle\xspace}

\newcommand{\id}{{\,\rm d}} %

\newcommand{\cii}{[\textsc{C\,ii}]}

\usepackage{tikz}
\usetikzlibrary{arrows.meta,calc}

\newcommand{\cancelfix}[2][0]{%
  \sbox0{$\displaystyle #2$}%
  \dimen0=\dimexpr(\ht0+\dp0)/2\relax
  \begin{tikzpicture}[baseline=(X.base)]
    \node[inner sep=0pt] (X) {\usebox0};
    \pgfmathsetmacro\halfh{\the\dimen0/1pt}
    \pgfmathsetmacro\radA{\halfh/sin(38)}
    \pgfmathsetmacro\radB{\radA+6}
    \draw[semithick,-{Stealth[length=5pt,width=4pt]}]
      ($(X.center)-(38:\radA pt)$) --
      ($(X.center)+(38:\radB pt)$)
      node[inner sep=1pt,anchor=south west,font=\scriptsize]{$#1$};
  \end{tikzpicture}%
}

\AddToHook{cmd/section/before}{\vspace{2em}}

\usepackage{tikz}
\usepackage{tikz-feynman}
\usetikzlibrary{patterns}
\tikzfeynmanset{compat=1.1.0}
\tikzfeynmanset{every blob={/tikz/pattern=north east lines, /tikz/pattern color=black}}
\tikzfeynmanset{photon/.style={/tikz/draw=none,
    /tikz/postaction={/tikz/draw,
      /tikz/decoration={complete sines, amplitude=1.1mm, segment length=2.5mm},
      /tikz/decorate=true}}}

\title{{\bf Peering Beyond the Veil of Last Scattering: \\A View of the Universe with CMB Spectral Distortions}}
\author[1,*]{Bryce Cyr}
\affil[1]{Center for Theoretical Physics - A Leinweber Institute,
Massachusetts Institute of Technology, Cambridge, MA, USA}

\author[2]{Nabila Aghanim}
\affil[2]{Université Paris-Saclay, CNRS, Institut d'Astrophysique Spatiale, 91405 Orsay, France}

\author[3]{Ethan Baker}
\affil[3]{Physics Department, Boston University, Boston, MA 02215, USA}

\author[4,5]{Elia Stefano Battistelli}
\affil[4]{Dipartimento di Fisica, Sapienza Universit\'a di Roma, Roma, IT}
\affil[5]{INFN, Sezione di Roma, Roma, IT}

\author[6]{Richard~Battye}
\affil[6]{Jodrell Bank Centre for Astrophysics, University of Manchester, Manchester, UK}

\author[7]{José Luis Bernal}
\affil[7]{Instituto de F{\'i}sica de Cantabria (IFCA), CSIC-Univ. de Cantabria, Avda. de los Castros s/n, E-39005 Santander, Spain}

\author[6]{Jens Chluba}

\author[2]{Xavier Coulon}

\author[8]{William Coulton}
\affil[8]{Department of Physics, University of Oxford, Denys Wilkinson Building, Keble Road, Oxford
OX1 3RH, UK}

\author[4,5]{Paolo de Bernardis}

\author[9]{Eleonora Di Valentino}
\affil[9]{School of Mathematical and Physical Sciences, University of Sheffield, Hounsfield Road, Sheffield S3 7RH, United Kingdom}

\author[10,11]{Guillem Domènech}
\affil[10]{Institute for Theoretical Physics, Leibniz University Hannover, Appelstraße 2, 30167 Hannover, Germany}
\affil[11]{Max Planck Institute for Gravitational Physics,
Albert Einstein Institute, 30167 Hannover, Germany}

\author[6]{Sara~Evangelista}

\author[2]{Giulio Fabbian}

\author[12,13]{Fabio Finelli}
\affil[12]{INAF - Osservatorio di Astrofisica e Scienza dello Spazio, Via Gobetti 101, 40129 Bologna, Italy}
\affil[13]{INFN, Sezione di Bologna, Via Irnerio 46, 40126 Bologna, Italy}

\author[14]{J.~Colin Hill}
\affil[14]{Department of Physics, Columbia University, New York, NY 10027, USA}

\author[15]{Rishi Khatri}
\affil[15]{Department of Theoretical Physics, Tata Institute of Fundamental Research, Homi Bhabha Road, Mumbai 400005, India}

\author[16]{Alan Kogut}
\affil[16]{Observational Cosmology Laboratory, Goddard Space Flight Center, Greenbelt, USA}

\author[17]{Guilaine Lagache}
\affil[17]{Aix Marseille Univ, CNRS, CNES, LAM (Laboratoire d'Astrophysique de Marseille), Marseille, France}

\author[3]{Hongwan Liu}

\author[2]{Bruno Maffei}

\author[18]{Fereshteh Majidi}
\affil[18]{Department of Physics and Astronomy, University of British Columbia (UBC), Vancouver, Canada}

\author[19]{Abhishek~S.~Maniyar}
\affil[19]{Laboratoire d’Annecy de Physique Th\'eorique (LAPTh), CNRS/USMB, 9 Chemin de Bellevue BP110 - Annecy - F-74941 - ANNECY CEDEX - FRANCE}

\author[20,21]{C. J. A. P. Martins}
\affil[20]{Centro de Astrof\'{\i}sica da Universidade do Porto, Rua das Estrelas, 4150-762 Porto, Portugal}
\affil[21]{Instituto de Astrof\'{\i}sica e Ci\^encias do Espa\c co, Universidade do Porto, Rua das Estrelas, 4150-762 Porto, Portugal}

\author[4,5]{Silvia Masi}

\author[22]{Jeff McMahon}
\affil[22]{Dept. of Astronomy and Astrophysics, University of Chicago, Chicago, Il 60637}

\author[12,13]{Daniela~Paoletti}

\author[23]{Vivian Poulin}
\affil[23]{Laboratoire Univers \& Particules de Montpellier,
CNRS \& Université de Montpellier (UMR-5299), 34095 Montpellier, France}

\author[24]{Wenzer Qin}
\affil[24]{Center for Cosmology and Particle Physics, Department of Physics, New York University, New York, NY 10003, USA}

\author[7]{Mathieu Remazeilles}

\author[25,26]{Jos\'e~Alberto~Rubi\~no-Mart\'{\i}n}
\affil[25]{Instituto de Astrof\'{\i}sica de Canarias, E-38205 La Laguna, Tenerife, Spain}
\affil[26]{Departamento de Astrof\'{\i}sica, Universidad de La Laguna, E-38206 La Laguna,
Tenerife, Spain }

\author[27,28]{Alina Sabyr}
\affil[27]{Berkeley Center for Cosmological Physics, Department of Physics, University of California, Berkeley, CA 94720, USA}
\affil[28]{Lawrence Berkeley National Laboratory, One Cyclotron Road, Berkeley, CA 94720, USA}

\author[29]{Mayuri Sathyanarayana Rao}
\affil[29]{Astronomy \& Astrophysics, Raman Research Institute,
Bengaluru, India 560080}

\author[30]{Carlos~Sierra}
\affil[30]{Kavli Institute for Particle Astrophysics and Cosmology, Stanford University, Stanford, CA 94305, USA}

\author[31]{Jack Singal}
\affil[31]{Physics Department, University of Richmond, 138 UR Drive, Richmond, VA 23173 USA}

\author[1]{Tracy R.~Slatyer}

\author[23]{Elsa M. Teixeira}

\author[32,33]{Leander Thiele}
\affil[32]{Center for Data-Driven Discovery, Kavli IPMU (WPI), UTIAS, The University of Tokyo, Kashiwa, Chiba 277-8583, Japan}
\affil[33]{Kavli IPMU (WPI), UTIAS, The University of Tokyo, 5-1-5 Kashiwanoha, Kashiwa, Chiba 277-8583, Japan}

\author[34]{Léo Vacher}
\affil[34]{Laboratoire de Physique des 2 infinis Irène Joliot-Curie (IJCLab), Université Paris-Saclay, CNRS/IN2P3, Orsay, FR}

\author[35]{Duncan Watts}
\affil[35]{Institute of Theoretical Astrophysics, University of Oslo, Blindern, Oslo, Norway}

\makeatletter
\renewcommand{\@author}{\AB@authlist}

\newcommand{\printauthoraffiliations}{%
    \begingroup
    \setlength{\parindent}{0pt}%
    \AB@affillist
    \par
    \endgroup
}
\makeatother

\date{}

\begin{document}

\maketitle

\begin{center}
\small\textit{Author affiliations are listed in the Appendix.}
\end{center}

\begingroup
\renewcommand{\thefootnote}{*}
\footnotetext{Corresponding author: \texttt{brycecyr@mit.edu}}
\endgroup

\begin{abstract}
The frequency spectrum of the cosmic microwave background is the most precise blackbody ever measured in nature, with deviations constrained at the level of almost one part per million from the \COBEF satellite. Despite this fact, it is known that departures away from a perfect blackbody are present in standard $\Lambda$CDM cosmology, lurking just beneath the surface of our current observational bounds. These spectral distortions provide invaluable information relating to our thermal history in both the post- and pre-recombination epochs, allowing us to peer beyond the veil of last scattering and deep into the primordial Universe. Here, we present an overview of the underlying physics responsible for the generation of CMB spectral distortions at all epochs. As an illustration of this rich physics, we review a comprehensive set of mechanisms capable of generating distortions both within and beyond the standard $\Lambda$CDM paradigm. We also discuss the information that can be gleaned by going beyond the monopole (sky-averaged) spectrum and exploiting the spatial information present in anisotropic spectral distortions. To supplement our discussion of the diverse science of spectral distortions, we provide an overview of the upcoming and proposed experimental landscape. We highlight that the combination of the TMS, COSMO, and BISOU experiments will provide the first ever discovery of a monopole $y$-type distortion within the coming decade. From space, the proposed \FOSSIL experiment is forecasted to improve upon the original \COBEF measurement by roughly three orders of magnitude in sensitivity, bringing with it the detection of the $\Lambda$CDM $\mu$-type distortion sourced by the dissipation of small scale acoustic modes in the pre-recombination plasma. With transformational measurements on the horizon, CMB spectral distortions offer a uniquely sensitive probe of the thermal history of the Universe at redshifts $z \lesssim 2 \times 10^6$.
\end{abstract}

\newpage

\tableofcontents

\section{The Physics of CMB Spectral Distortions} \label{sec:Intro}
The cosmic microwave background (CMB) has proved to be a veritable treasure trove of high quality data over the past three decades. It can be argued that the current age of precision cosmology was ushered in by the launch of the \textit{COBE} satellite in the early 90s \citep{Mather:1990cobe, Boggess:1992cobe}, which had the \textit{DMR} (Differential Microwave Radiometer) \cite{Smoot:1992td, Bennett:1996ce, Kogut1996}, \textit{FIRAS} (Far Infrared Absolute Spectrophotometer) \cite{Mather1994, Wright:1994interp, Fixsen:1996nj, Fixsen2002, Fixsen:2009ug}, and \textit{DIRBE} (Diffuse Infrared Background Experiment) \cite{Puget1996, fixsen1998, Hauser:1998cib} instruments onboard. Each instrument provided pivotal information to cosmologists at the time, with \textit{DMR} detecting large scale temperature anisotropies at the $\Delta T/T \simeq 10^{-5}$ level, \textit{FIRAS} making the most precise measurement to date of the sky-averaged blackbody spectrum of the CMB, and \textit{DIRBE} providing the first ever measurement of the cosmic infrared background (CIB) and producing high quality maps of interplanetary and galactic dust emission.

\indent After the enormous success of the \textit{COBE} mission, the main aim of CMB experiments was to extend our understanding of the anisotropies, both in temperature and polarization. The amplitude and position of the first peak in the $TT$ power spectrum was first convincingly measured in 2000 by the BOOMERanG \cite{deBernardis2000} and MAXIMA instruments \cite{Hanany2000} (with TOCO \cite{Miller1999} providing important data regarding its position). Subsequently, a number of experiments \cite{Halverson2002, Kovac2002, Pearson2003, Dickinson2004, Kuo2004, Benoit2003} drilled further into the peak structure of the TT spectrum, and began performing $E$ and eventually $B$-mode \cite{Hanson2013, POLARBEAR2014, BICEPKeck2015, BICEPKeck2021, EssingerHileman2014} polarization measurements. Ultimately, the $\Lambda$CDM paradigm was established and then firmly consolidated by the high precision data taken by the WMAP \cite{Bennett2003, Spergel2003, Kogut2003, Bennett2013, Hinshaw2013} and Planck \cite{Planck:2018nkj, Planck:2018vyg} satellites, respectively. From the ground, the ACT \cite{ACT:2020gnv, AtacamaCosmologyTelescope:2025blo, Madhavacheril2024} and SPT \cite{Story2013, SPT-3G:2022hvq} observatories have also further deepened our understanding of the CMB on the very smallest scales (high $\ell$), probing deep into the so-called damping tail. With the information content of the temperature sky now largely exhausted, attention has shifted to polarization, gravitational lensing, and secondary anisotropies, a programme that will be driven by the Simons Observatory (SO) \cite{SimonsObs2019} and LiteBIRD \cite{LiteBIRD2023} in the coming decades.

Despite this boom in anisotropy science, observations of the CMB frequency spectrum have lagged behind, with the \COBEF data from more than three decades ago remaining the most precise measurement to date. On the one hand, this is a testament to the incredible capabilities of the \COBEF team, who managed to constrain deviations (distortions) away from a perfect blackbody spectrum at the $\Delta I/I \lesssim {\rm few} \times10^{-5}$ level across the main CMB band ($60-600$ GHz). On the other hand, it is well known that CMB spectral distortions exist\footnote{See \cite{Chluba2016} for an overview, with the different subsections below providing more complete reference lists.} within the $\Lambda$CDM model at the $\Delta I/I \lesssim 10^{-6}$ level, only an order of magnitude below the \COBEF sensitivity, providing numerous tantalizing science targets for CMB spectrometer missions. As we shall see, CMB spectral distortions provide a unique and highly sensitive probe of the thermal history of the Universe over a vast range of redshifts ($0 \lesssim z \lesssim 10^6$), and will yield important, complementary information to more traditional anisotropy measurements. 

At present, activities are underway which aim to measure different parts of the CMB spectrum from each observational domain (the ground, atmosphere, and space). From the northern and southern hemispheres respectively, the Tenerife Microwave Spectrometer (TMS) \cite{TMS} and the Cosmic Monopole Observer (COSMO) \cite{Masi:2021azs} will provide distortion measurements that will rival the sensitivity obtained by \COBEF. The balloon experiment BISOU \cite{Maffei:2021xur, Maffei:2024SPIE} will provide complementary measurements from the atmosphere with a targeted sensitivity of $\sigma(\Delta I) \simeq 10^{-6}$, bringing with it the first detection of a sky-averaged CMB spectral distortion. On top of the valuable data gathered by TMS, COSMO, and BISOU, in the next 5-10 years, each of these experiments will act as much-needed pathfinders for future satellite missions, such as \FOSSIL \cite{FOSSILPlaceholder}, \PIXIE \cite{Kogut2011, Kogut2016SPIE, Kogut:2024vbi}, or the SPECTER \cite{specter} concept. In addition, the ground-based proposal APSERa \cite{sathyanarayana2015detection} provides a unique path to measuring a part of the recombination line spectrum. More information on each of these missions can be found in Sec.~\ref{sec:missions}.

\subsection{CMB spectral distortion taxonomy}
Within the error bars of the \COBEF instrument, the frequency spectrum of the CMB appears to be consistent with that of a perfect blackbody. A blackbody spectrum is fully characterized by its temperature, measured to be $T_{\rm CMB,0} = 2.7255 \pm 0.0006$ K \cite{Fixsen:2009ug} in the case of the CMB. The specific intensity of a blackbody is given by
%
\begin{align}
    I^{\rm  bb}_{\nu} = \frac{2 h}{c^2} \frac{\nu^3}{{\rm e}^{h\nu/k_{\rm b}T_{\rm CMB}}-1},
\end{align}
%
where $h$ and $k_{\rm b}$ are the Planck and Boltzmann constants respectively, $c$ is the speed of light, $T_{\rm CMB}$ is the monopole temperature of the CMB, and $\nu$ the photon frequency. It is also convenient to introduce the dimensionless frequency
%
\begin{align}
    x = \frac{h\nu}{k_{\rm b} T_{\rm CMB}}.
\end{align}
%
In the absence of interactions that exchange energy between photons and matter, both $\nu$ and $T_{\rm CMB}$ redshift in the same way, implying that $x$ is redshift independent. For the purposes of studying spectral distortions, it is often more transparent to work with the photon occupation number $n(x)$, related to the specific intensity by
%
\begin{align}
    I_{\nu} &= \frac{2 k_{\rm b}^3}{c^2h^2} T^3_{\rm CMB} \,x^3 \, n(x),
\end{align}
%
where for a blackbody distribution the occupation number is simply given by
%
\begin{align}
    n_{\rm bb}(x) = \frac{1}{{\rm e}^x-1}.
\end{align}
%
A CMB spectral distortion is then defined as the non-blackbody part of the radiation field
%
\begin{align}
\Delta n(x) \equiv n(x) - n_{\rm bb}(x).
\end{align}
%
In addition to subtracting away the reference blackbody, one needs to further remove any pure temperature shifts that may have been introduced, which we will define shortly. The work of theorists is to determine the size and shape of $\Delta n(x)$ in response to any processes that perturb the cosmological plasma away from full thermal equilibrium.
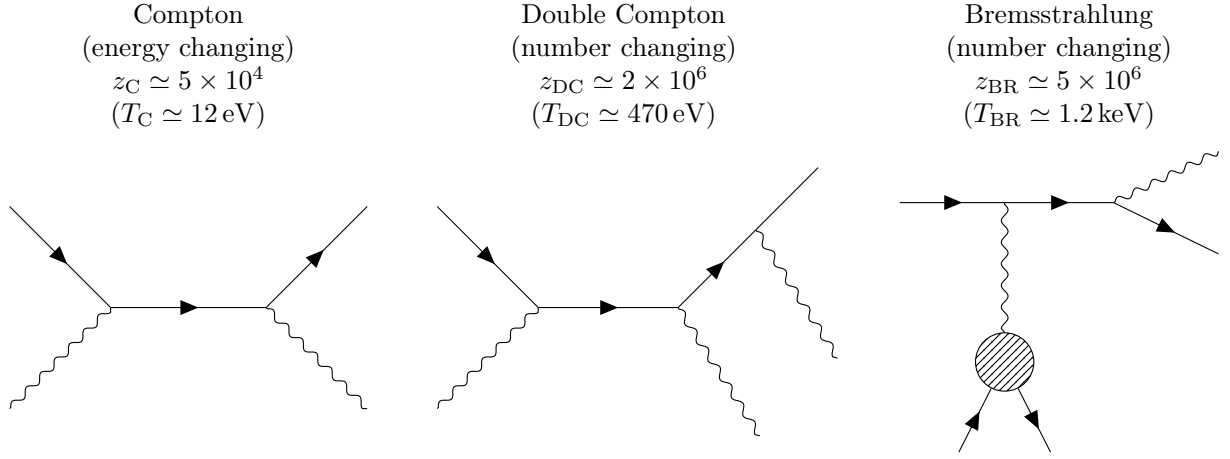
\begin{figure*}
  \centering
  \resizebox{\textwidth}{!}{%
  \begin{tikzpicture}
 
  \begin{scope}[shift={(0,0)}]
    \node[align=center] at (0,3.1) {%
      Compton\\
      (energy changing)\\
      $z_{\mathrm{C}} \simeq 5\times 10^{4}$\\
      $(T_{\mathrm{C}} \simeq 12\,\mathrm{eV})$};
    \begin{feynman}
      \vertex (v1) at (-1.0, 0.0);
      \vertex (v2) at ( 1.0, 0.0);
      \vertex (ei) at (-2.3, 1.3);
      \vertex (gi) at (-2.3,-1.3);
      \vertex (eo) at ( 2.3, 1.3);
      \vertex (go) at ( 2.3,-1.3);
      \diagram*{
        (ei) -- [fermion] (v1),
        (gi) -- [photon]  (v1),
        (v1) -- [fermion] (v2),
        (v2) -- [fermion] (eo),
        (v2) -- [photon]  (go),
      };
    \end{feynman}
  \end{scope}
 
  \begin{scope}[shift={(5.6,0)}]
    \node[align=center] at (0,3.1) {%
      Double Compton\\
      (number changing)\\
      $z_{\mathrm{DC}} \simeq 2\times 10^{6}$\\
      $(T_{\mathrm{DC}} \simeq 470\,\mathrm{eV})$};
    \begin{feynman}
      \vertex (v1) at (-1.1, 0.0);
      \vertex (v2) at ( 0.7, 0.0);
      \vertex (v3) at ( 1.7, 1.0);
      \vertex (ei) at (-2.4, 1.3);
      \vertex (gi) at (-2.4,-1.3);
      \vertex (g1) at ( 1.75,-1.65);
      \vertex (eo) at ( 2.5, 1.8);
      \vertex (g2) at ( 2.75,-0.65);
      \diagram*{
        (ei) -- [fermion] (v1),
        (gi) -- [photon]  (v1),
        (v1) -- [fermion] (v2),
        (v2) -- [photon]  (g1),
        (v2) -- [fermion] (v3),
        (v3) --           (eo),
        (v3) -- [photon]  (g2),
      };
    \end{feynman}
  \end{scope}
 
  \begin{scope}[shift={(11.2,0.35)}]
    \node[align=center] at (0,2.75) {%
      Bremsstrahlung\\
      (number changing)\\
      $z_{\mathrm{BR}} \simeq 5\times 10^{6}$\\
      $(T_{\mathrm{BR}} \simeq 1.2\,\mathrm{keV})$};
    \begin{feynman}
      \vertex (ei) at (-2.05, 1.0);
      \vertex (v1) at (-0.7,  1.0);
      \vertex (v2) at ( 0.7,  1.0);
      \vertex (gr) at ( 2.05, 1.66);
      \vertex (eo) at ( 2.05, 0.34);
      \vertex [blob] (b) at (-0.7,-1.05) {};
      \vertex (ni) at (-1.29,-2.21);
      \vertex (no) at (-0.11,-2.21);
      \diagram*{
        (ei) -- [fermion] (v1),
        (v1) -- [fermion] (v2),
        (v2) -- [fermion] (eo),
        (v2) -- [photon]  (gr),
        (v1) -- [photon]  (b),
        (ni) -- [fermion] (b),
        (b)  -- [fermion] (no),
      };
    \end{feynman}
  \end{scope}
 
  \end{tikzpicture}%
  }
  \caption{Lowest order processes that mediate thermalization in the early Universe. The redshift at which each process freezes out ($\Gamma_{\rm i}(z_{\rm i}) = H$) and becomes inefficient is also shown. The inefficiency of these processes at late times ($z \lesssim 2 \times 10^6$) is precisely what allows for distortions to form and persist until today. Straight and wavy lines denote electrons and photons respectively.}
  \label{fig:PP_processes}
\end{figure*}

The essential building blocks necessary to study the thermalization between photons and electrons was first introduced by Kompaneets \cite{Kompaneets:1957}, and applied to the microwave background radiation in a series of papers by Weymann \cite{Weymann:1966}, Zel'dovich, and Sunyaev \cite{Zeldovich1969, SunyaevZeldovich:1970b} in the late 60s. From here, further refinements where made to include external distortion heat sources \cite{ZeldovichIllarionovSunyaev:1972}, and the effects of photon number changing processes \cite{IllarionovSunyaev:1975, DaneseDeZotti:1982, Burigana:1991}. The now famous spectral distortion known as the Sunyaev-Zel'dovich effect (produced through the upscattering of CMB photons as they traverse through clusters of hot electrons) was described first in 1969 \cite{Zeldovich1969,SunyaevZeldovich:1970b,SunyaevZeldovich:1972} a little more than a decade before its first significant detection in 1984 \cite{Birkinshaw:1984}. Another type of distortion we will describe in detail below is sourced by the damping of small scale density perturbations in the pre-recombination plasma \cite{Silk1968, SunyaevZeldovich:1970a, Daly1991, Barrow1991, Chluba:2012gq}, and is a central target of future space missions. 

To understand how CMB spectral distortions can be produced, we first need to understand the main ingredients that govern the thermalization process in the early Universe. In order to re-establish full thermal equilibrium in the presence of a distortion, two types of processes need to be efficient in the plasma. First, one needs to be able to rapidly reprocess photon energy across the full spectrum, which at early times is handled primarily by the Compton scattering (CS) process. Second, one needs a way to efficiently create and destroy photons. At lowest order in electromagnetic coupling ($\alpha_{\rm em}$), this is mediated by both double Compton scattering (DC) and Bremsstrahlung (BR). At very early times ($z \gtrsim 2 \times 10^6$), all three of these processes are efficient (meaning that the rate of interactions $\Gamma_i > H$ exceeds the expansion rate of the Universe) and small departures from equilibrium are quickly thermalized. Figure~\ref{fig:PP_processes} shows the Feynman diagrams for these processes along with their approximate freeze-out redshifts. 

With all of this in mind, a clean taxonomy of distortion epochs begins to emerge \cite{Hu:1992dc, Hu1993}. A non-thermal injection of energy or entropy when CS, DC, and BR are all efficient is quickly thermalized, leading to an overall temperature shift (what will we shortly call a $G$-type spectrum, occurring in the $T$-era) in the monopole. As there is no \textit{a priori} expectation for what $T_{\rm CMB,0}$ should be, this temperature shift does not provide any constraining power, and the spectral distortion is effectively washed out\footnote{There are some exceptions, for example in the large distortion regime when the non-thermal injection is order $\Delta \rho/\rho\simeq \mathcal{O}(0.1)$, distortions can persist even if they are sourced at $z_{\rm inj} \simeq 10^{7}$ \cite{Chluba2020L,Acharya2021}.}. Following this, at $3\times 10^5 \lesssim z \lesssim 2\times 10^6$ number changing processes are completely frozen out, while Compton scattering remains fully efficient. Non-thermal processes active during this epoch typically induce a chemical potential in the photon field, giving rise to a very specific spectral shape known as a $\mu$-type distortion. At even later times ($z \lesssim 10^4$),  Compton scattering freezes out and we enter what is commonly known as the spectral free-streaming (SFS) or $y$-distortion epoch. The choice of language here depends on whether the non-thermal source is a direct photon injection, or something that heats the electrons. If the former, the distortion will typically retain its shape at sourcing, and if the latter, a characteristic $y$-type distortion is introduced into the photon field. 
\begin{figure*}[t]
    \centering
    \includegraphics[width=1\linewidth]{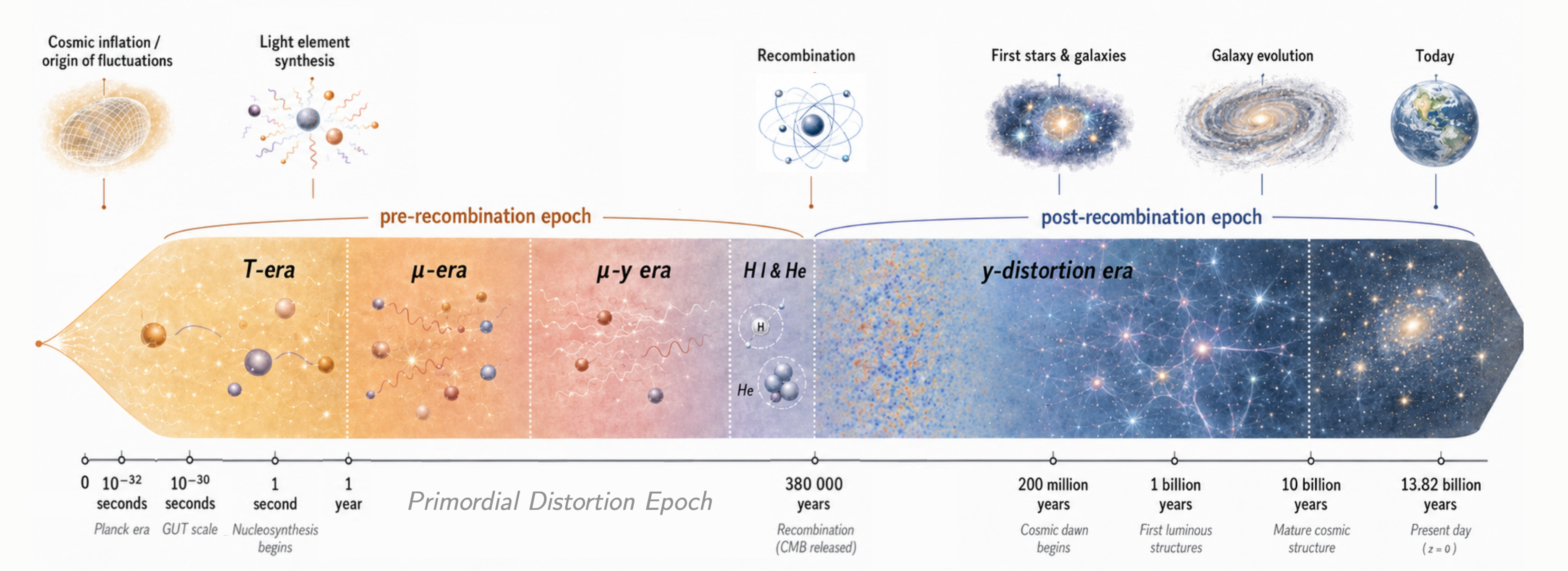}
    \caption{An illustration of the various spectral distortion epochs within the wider context of the thermal history of the Universe. This figure was initially generated by ChatGPT and then post-processed.}
    \label{fig:SD_epochs}
\end{figure*}

Beyond the three main distortion eras ($T,\mu,y$) there is another period in which non-trivial distortions can be sourced \cite{ChlubaSunyaev:2012, Khatri:2012tw, Chluba2013Green}. Referred to as the residual epoch, it lives at $10^4 \lesssim z \lesssim 3 \times 10^5$, a time in which number changing interactions are frozen out, and Compton scattering is becoming inefficient at reprocessing energy. In this regime, non-thermal energy release will produce a spectral distortion that is not simply a superposition of the $\mu$- and $y$-types, but will also contain important residual information that can be used to disentangle the exact time dependence of the non-thermal process. This information will be invaluable when performing model discrimination between different beyond the standard model (BSM) sources of energy and entropy injection \cite{Chluba2013fore}. An illustration of the different spectral distortion epochs is shown in Fig.~\ref{fig:SD_epochs}.

\begin{figure*}[t]
    \centering
    \includegraphics[width=0.48\linewidth]{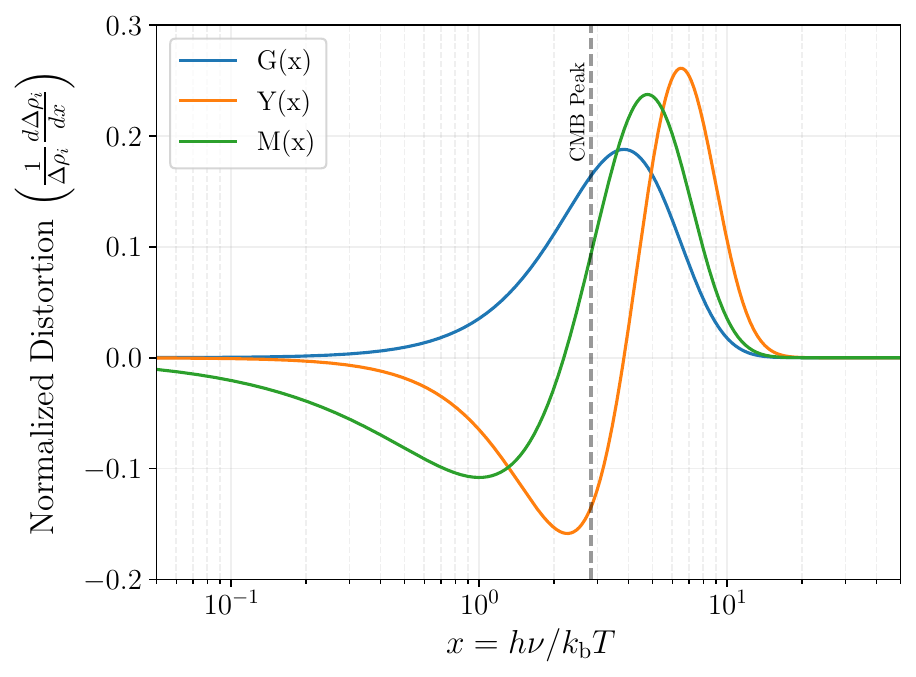}
    \includegraphics[width=0.48\linewidth]{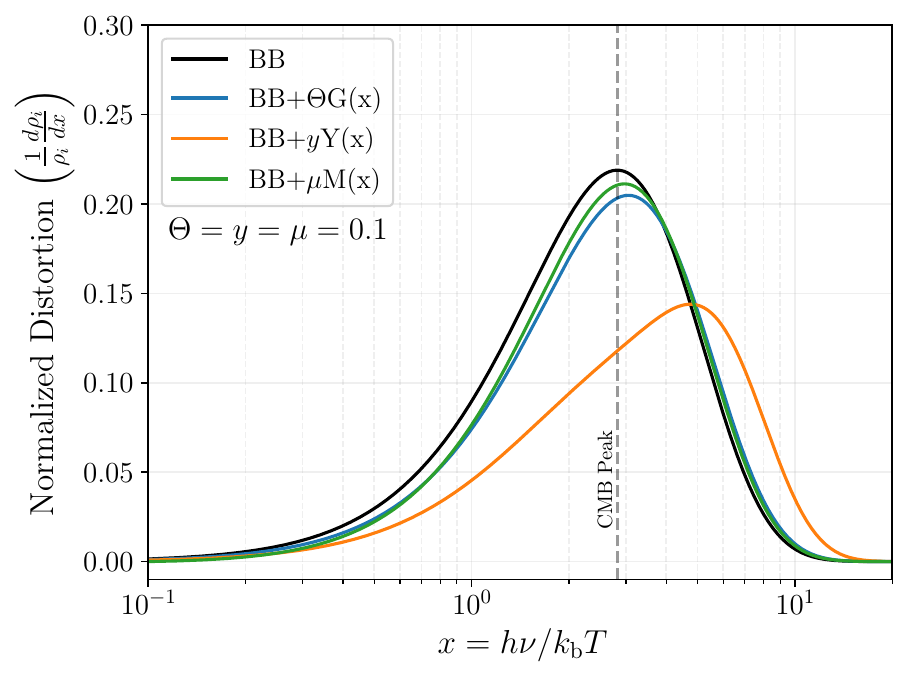}    
    \caption{Left: Individual distortion shapes normalized to their total energy density $\Delta \rho_{\rm i} = \int \id x \,\,(\id \Delta \rho_{\rm i}/\id x)$. Right: Full spectral distortion shapes with greatly exaggerated distortion amplitudes, for illustration. Here the normalization is over the full distorted spectrum.}
    \label{fig:dist_shapes}
\end{figure*}

Mathematically, the generation and subsequent evolution of CMB spectral distortions is governed by the Boltzmann equation. By treating the Thomson optical depth as a time parameter ($\id \tau = \sigma_{\rm T} N_{\rm e} c \, \id t$, where $\sigma_{\rm T}$ is the Thomson cross section and $N_{\rm e}$ is the number density of free electrons), the master equation for the photon occupation number can be written as \cite{ChlubaSunyaev:2012}
%
\begin{align} \label{eq:distortion_master}
    \frac{\partial n}{\partial \tau} =  \left. \frac{\partial n}{\partial \tau}\right|_{\rm CS} + \left. \frac{\partial n}{\partial \tau}\right|_{\rm DC} + \left. \frac{\partial n}{\partial \tau}\right|_{\rm BR} + S_{\rm heat}(\tau,x) + S_{\rm photons}(\tau,x).
\end{align}
%
As discussed above, the sky-averaged photon occupation number will evolve through Compton, double Compton, and Bremsstrahlung processes provided they are efficient at a given time $\tau(z)$. We have split the possible external source (or sink) terms up into two distinct pieces to further illustrate the physics. Here, $S_{\rm heat}$ refers to a source of energy injection that primarily heats the electrons, which will then transfer their energy to the photon bath through efficient or residual Compton interactions. The $S_{\rm photons}$ term refers to the direct addition or subtraction of photons in the primordial plasma. This distinction in source terms is less important in the $\mu$ epoch, but lead to qualitatively different distortion shapes at late times. The various distortion epochs can thus be mapped to different limits of Eq.~\eqref{eq:distortion_master},
%
\begin{align}
    \frac{\partial n}{\partial \tau} &=  \left. \frac{\partial n}{\partial \tau}\right|_{\rm CS} + \left. \frac{\partial n}{\partial \tau}\right|_{\rm DC} + \left. \frac{\partial n}{\partial \tau}\right|_{\rm BR} + S_{\rm heat}(\tau,x) + S_{\rm photons}(\tau,x) \hspace{30mm} (T-{\rm shift}),\\
    &=  \left. \frac{\partial n}{\partial \tau}\right|_{\rm CS} + \cancelfix{\left. \frac{\partial n}{\partial \tau}\right|_{\rm DC}} + \cancelfix{\left. \frac{\partial n}{\partial \tau}\right|_{\rm BR}} + S_{\rm heat}(\tau,x) + S_{\rm photons}(\tau,x) \hspace{14mm} (\mu-{\rm distortion}),\\
    &=  \cancelfix{\left. \frac{\partial n}{\partial \tau}\right|_{\rm CS}} + \cancelfix{\left. \frac{\partial n}{\partial \tau}\right|_{\rm DC}} + \cancelfix{\left. \frac{\partial n}{\partial \tau}\right|_{\rm BR}} + S_{\rm heat}(\tau,x) + \cancelfix{S_{\rm photons}}(\tau,x) \hspace{9.5mm} (y-{\rm distortion}),\\
    &=  \cancelfix{\left. \frac{\partial n}{\partial \tau}\right|_{\rm CS}} + \cancelfix{\left. \frac{\partial n}{\partial \tau}\right|_{\rm DC}} + \cancelfix{\left. \frac{\partial n}{\partial \tau}\right|_{\rm BR}} + \cancelfix{S_{\rm heat}}(\tau,x) + S_{\rm photons}(\tau,x) \hspace{6.5mm} ({\rm SFS \, \, distortion}).
\end{align}
%
As usual, the residual epoch lies between the $\mu$ and $y$ expressions as the $\partial n/\partial\tau|_{\rm CS}$ term starts to become irrelevant. An important concept in modern distortion literature is the notion of the boost operator (or generator) \cite{DaiChluba2014, Chluba:2022xsd, Chluba2025boost}, $\hat{\mathcal{O}}_x = -x\partial_x$, which allows us to build the main library of spectral distortion shapes. A typical decomposition of the distortion can be written as \citep{Chluba:2022xsd}
%
\begin{align} \label{eq:dist_decomp}
    \Delta n(x) = \Theta \, G(x) + \mu  \, M(x) + y \, Y(x) + \sum_{i\geq1} y_i \, Y_i(x),
\end{align}
%
where $\Theta$, $\mu$, $y$, and $y_i$ are the amplitudes of the temperature shift, $\mu$, $y$, and residual distortions respectively. The more terms one takes for the residual distortion, the more faithful the representation of the full distortion in this basis. The spectral shapes are given by operations of the boost operator on the blackbody spectrum,
%
\begin{subequations}
\label{eq:basis}
\begin{align} \label{eq:G-def}
    G(x) &= \hat{\mathcal{O}}_x n_{\rm bb}(x) = \frac{x \, {\rm e}^x}{({\rm e}^x-1)^2},\\ \label{eq:M-def}
    M(x) &= G(x) \left[\frac{1}{\beta_{\rm M}}- \frac{1}{x} \right],\\ \label{eq:Y-def}
    Y(x) &= \hat{\mathcal{O}}_x(\hat{\mathcal{O}}_x-3)n_{\rm bb}(x) = G(x) \left[x \frac{{\rm e}^x + 1}{{\rm e}^x-1}-4\right],\\ \label{eq:Yi-def}
    Y_{i}(x) &= \left(\frac{1}{4} \right)^i (\hat{\mathcal{O}}_x)^i Y(x).
\end{align}
\end{subequations}
%
Above, $\beta_{\rm M} \simeq 2.1923$ is a constant which arises when enforcing photon number conservation in the definition of the $\mu$-distortion. The temperature shift [$\Theta\, G(x)$] part of the distortion is intrinsically unobservable as it just relates a blackbody at one temperature to another through
%
\begin{align}
    n^{T+\Delta T}_{\rm bb}(x) \approx n^{T}_{\rm bb}(x) + \Theta \, G(x),
\end{align}
%
where we note that $\Theta = \Delta T/T\ll 1$. Another important point to mention is that the distortion basis written above is not orthogonal nor complete. As a result, such a simplified representation will typically capture most (but not all) of the distortion energy. Illustrative examples of the different distortion shapes are shown in Fig.~\ref{fig:dist_shapes}.

The standard $\Lambda$CDM paradigm provides a wealth of standard model distortion signatures to hunt for \cite{Chluba2016}. We reserve a detailed discussion of each source to their respective subsections, but briefly mention the perhaps most compelling cases. Chronologically, the dissipation of small scale modes in the pre-recombination plasma (described in Sec.~\ref{sec:small-scale-structure}) produce a $\mu$-type distortion with amplitude $|\mu| \simeq 2 \times 10^{-8}$ \cite{Silk1968, SunyaevZeldovich:1970a, Daly1991, Barrow1991, Chluba:2012gq}. Following this, the process of helium and hydrogen recombination produces a set of spectral lines with a very distinct distortion shape at the $|\Delta I/I| \simeq 10^{-9}$ \cite{Zeldovich68, Peebles1968, Dubrovich1975} (Sec.~\ref{CRR}). Post-recombination, the upscattering of the CMB as it traverses through hot clouds of electrons produce a global $y$-type distortion at the $y\simeq 10^{-6}$ level \cite{SunyaevZeldovich:1972, Refregier2000, daSilva2000, Hill2015}, just out of reach of the \COBEF satellite (Sec.~\ref{sec:SZ_effect}). Relativistic corrections to this global SZ distortion \cite{Wright1979, Sazonov1998, Itoh98, Challinor_1998, Hill2015} can also be observed with amplitude $y \simeq 10^{-7}$, similar in strength to what could be expected from the processes of cosmic reionization ($y \simeq 10^{-8}-10^{-7}$) \cite{Hu1994pert,
OhCoorayKamionkowski2003}, as discussed Sec.~\ref{sec:CD_and_reionization}.

\subsection{Detailed modeling of distortions}
Computing the various distortion amplitudes in Eq.~\eqref{eq:dist_decomp} requires the photon Boltzmann equation [Eq.~\eqref{eq:distortion_master}] to be solved. To do this, a variety of numerical and analytic tools have been developed. \texttt{CosmoTherm} \cite{ChlubaSunyaev:2012} is a C++ code purpose built to study the intricacies of the cosmological thermalization problem. Over the past 15 years, many upgrades and additional physics modules have been implemented into the codebase, allowing one to compute detailed distortion solutions from a wide range of scenarios, 
including a full treatments of the Compton scattering process in the large distortion regime \citep{Chluba2020L, Acharya2021}. \texttt{CosmoTherm} is capable of solving Eq.~\eqref{eq:distortion_master} at all redshifts throughout the distortion epoch, with a typical run beginning at $z = 10^7$, deep in the $T$-era.

Another codebase that has recently been upgraded to include a module capable of computing spectral distortions is \texttt{DarkHistory} \cite{LiuRidgwaySlatyer:2020, LiuQin:2023a, Liu2023}. This code, written in Python, was originally crafted to study the impacts of decaying or annihilating dark matter candidates on various cosmological observables. One of the main use cases for \texttt{DarkHistory} is its ability to treat high energy cascades, which occur when particles with $E \gg T_{\rm CMB}$ are injected into the background. These types of injections are commonplace when one considers generic decays of typical dark matter candidates such as the weakly interacting massive particle (WIMP). Unlike \texttt{CosmoTherm}, \texttt{DarkHistory} runs start at matter-radiation equality ($z_{\rm eq} \simeq 3400$), and therefore only cover a small part of the pre-recombination distortion epoch. Due to its cascade technology, however, \texttt{DarkHistory} is uniquely situated to compute distortions from high energy injections post-recombination. At the moment, \texttt{CosmoTherm} lacks such a capability, instead excelling at computing distortions when photons or electrons are injected into the plasma at or slightly above the characteristic temperature of the bath ($T_{\rm CMB}$), across a much larger range of redshifts.
However, distortions created by high-energy particle cascades in the pre-recombination era have been considered in \citep{Acharya:2018iwh}, although for a more limited range of models. Recently, an open-source spectral distortion codebase known as \texttt{spectroxide} \cite{Baker:2026xtc} was constructed with heavy human-AI interaction, and validated against a subset of published \texttt{CosmoTherm} results.

In addition to numerical solutions, various approximation for the $\mu$ and $y$ distortion amplitudes have been developed over the years. Roughly a decade ago, \texttt{CosmoTherm} was used to develop a Green's function formalism  that allowed for the determination of a simple, back-of-the-envelope approximation one can use to estimate the strength of a given $\mu$ or $y$ distortion. Practically these Green's function solutions were constructed by injecting $\delta$-functions of energy \cite{Chluba2013Green} and \cite{Chluba2015} entropy into \texttt{CosmoTherm} at different redshifts \footnote{See also \citep{ChlubaSunyaev:2012} and \citep{Khatri:2012tw} for earlier investigations.}, and computing the overlap of the final distortion with the typical $\mu$ and $y$ type parameters. The main results of this work can be neatly summarized,
%
\begin{align}
    \mu &\approx 1.401 \int_{z_{ \mu y}}^{\infty} \id z \left(\frac{1}{\rho_{\rm CMB}} \frac{\id \rho_{\rm inj}}{\id z} - \frac{4}{3N_{\rm CMB}}\frac{\id N_{\rm inj}}{\id z}\right) \mathcal{J}_{\rm bb}(z),\\
    y &\approx \frac{1}{4} \int_{z_{\rm rec}}^{z_{\mu y}} \id z \frac{1}{\rho_{\rm CMB}} \frac{\id \rho_{\rm inj}}{\id z},
\end{align}
%
where $z_{\mu y} = 5\times10^4$ is the redshift at which CS freezes out, $\rho_{\rm CMB}$ and $N_{\rm CMB}$ are the energy and number density of the CMB respectively, $\rho_{\rm inj}$ and $N_{\rm inj}$ are the non-thermal sources (or sinks) of photons or heat from some external process, and $\mathcal{J}_{\rm bb}(z)$ is the distortion visibility function, which models the efficiency of DC and BR processes at high redshifts. Explicitly, it is often written as
%
\begin{align}
    \mathcal{J}_{\rm bb}(z) \simeq {\rm e}^{-(z/z_{\rm th})^{5/2}},
\end{align}
%
where $z_{\rm th} \approx 2 \times 10^6$ is the redshift at which DC processes freeze-out \citep{Burigana:1991, Hu1993}. It should be stressed that the above expressions are approximations to the full problem, typically accurate at the $10\%$ level when comparing against the full numerical solution. Other related approximate forms exist, and their relative accuracies are compared in \citep{Chluba2016} and more recently in section 2.1 of \cite{Cyr:2023pgw}. The most recent frequency hierarchy decomposition introduced in \citep{Chluba:2022xsd} provides another efficient way of approximating energy release distortions using just a small number of distortion parameters.

\subsection{Current constraints}
To conclude this section, let us briefly mention the current constraints on distortion parameters. As mentioned above, the \COBEF instrument is to date the only space-based determination of the CMB frequency spectrum, and its raw data remains state-of-the-art. The original papers from the \textit{COBE} collaboration provided distortion limits of $|\mu| \leq 9\times 10^{-5}$ and $|y| \leq 1.5 \times 10^{-5}$, both at the $95\%$ confidence level \cite{Fixsen:1996nj}. A reanalysis of the \COBEF data was performed recently, where it was found that both limits could be improved by using modern foreground models and including pixel-to-pixel variability (as opposed to assuming the foregrounds are constant across the whole sky) \cite{Bianchini:2022dqh, Sabyr:2025hwd, Fabbian_2025}. Thus, the current best limits on the $\mu$ and $y$ distortion amplitudes stand at
%
\begin{equation}
\begin{aligned}
    |\langle \mu\rangle| &\leq 4.7 \times 10^{-5},\\
    \langle y \rangle    &\leq 5.2 \times 10^{-6}.
\end{aligned}
\qquad (95\%~\mathrm{C.L.})
\end{equation}
%
Despite the lack of a space-based spectrometer thus far, excitement for distortion science on the experimental front is beginning to bubble over, with various proposals, as well as \textit{bonafide} instruments and experimental roadmaps \citep{Chluba:2019kpb, Chluba2021}. A summary description and references can be found in Sec.~\ref{sec:missions}. All in all, the community is well-poised to begin unlocking the information hidden deep within the frequency spectrum of the cosmic microwave background. This data will be invaluable in sharpening our understanding of the precise thermal history of the Universe, both pre- and post-recombination.

\section{The Early Universe}
Observations of the CMB temperature and polarization anisotropies have established a remarkably successful picture of the early Universe, yet they sample only a relatively narrow slice of its history through their accessible angular scales ($\ell \lesssim 3000$). The frequency spectrum of the CMB retains a complementary record. Departures from a perfect blackbody, once generated at redshifts $z \lesssim {\rm few} \times 10^{6}$, can no longer be fully erased, and they preserve information about the primordial density field and the thermal history at epochs largely inaccessible to any other observable. Within $\Lambda$CDM, the dissipation of small-scale acoustic modes alone predicts a distortion of $\mu \simeq 2\times 10^{-8}$, setting a well-defined benchmark for future spectrometers such as \FOSSIL and \PIXIE. In this section, we survey the early Universe science accessible at this level of sensitivity, moving from the initial conditions laid down by inflation to precision tests of the recombination era.

In Sec.~\ref{sec:small-scale-structure}, we discuss the damping of primordial density fluctuations, which is capable of sourcing distortions and is sensitive to roughly ten $e$-folds of inflationary evolution beyond the reach of the anisotropies. Sec.~\ref{sec:BH-formation} discusses the constraining power of distortions to scenarios that require enhanced small scale power, for example the classic primordial black hole formation mechanisms. A detection or null result at the $\mu \sim 10^{-8}$ level would sharply narrow the space of viable black hole formation channels. Moreover, the same fluctuations inevitably generate a stochastic gravitational wave background, as discussed in Sec.~\ref{sec:GWBGs}. Spectral distortions probe such backgrounds both directly, through the dissipation of tensor modes in a picohertz window lying between CMB polarization searches and pulsar timing arrays, and indirectly, by constraining the density fluctuations that source them, informing the interpretation of their reported nanohertz background.

Exotic energy release from beyond the standard model scenarios is discussed in Sec.~\ref{sec:BSM}. Primordial magnetic fields, topological defect networks, evaporating and accreting primordial black holes, and proposed resolutions of the Hubble tension can all exchange energy or photons with the plasma, often imprinting spectral shapes that the standard $\mu$ and $y$ decomposition does not capture.

Finally, absolute spectroscopy strengthens the observational foundations of precision cosmology. The cosmological recombination radiation (discussed in Sec.~\ref{CRR}), emitted during the three recombination epochs of hydrogen and helium, offers a direct spectroscopic view of the Universe at and beyond last scattering, and responds sensitively to variations of fundamental constants, additional ionization or heating, and other non-standard recombination histories. Furthermore, an improved determination of the monopole temperature $T_0$ (Sec.~\ref{sec:T0}), still anchored by \COBEF after three decades, would sharpen both the predictions of Big Bang nucleosynthesis and the interpretation of CMB anisotropy data. Indeed, the current $T_0$ uncertainty is projected to degrade constraints on the baryon density by up to $50\%$ for upcoming high-resolution experiments. These targets demonstrate that a single measurement of the CMB frequency spectrum simultaneously tests inflation, gravitational wave backgrounds, exotic injection scenarios, and the thermal history underpinning $\Lambda$CDM itself, providing deep insights into the early Universe.

\subsection{A probe of inflationary small-scale structure}
\label{sec:small-scale-structure}
Measurements of the CMB temperature and polarisation anisotropies, together with large-scale structure and the Lyman-$\alpha$ forest, have established that a nearly scale-invariant spectrum of primordial scalar perturbations exists across comoving momentum scales of $10^{-4}\,\mathrm{Mpc}^{-1} \lesssim k \lesssim 1\,\mathrm{Mpc}^{-1}$ \cite{Bird:2010mp, Chabanier:2018rga, Ravoux:2023uvp, Karacayli:2023afs, Palanque-Delabrouille:2019iyz, Fernandez:2023ddy, DESI:2024jis, eBOSS:2020yzd, Philcox:2021kcw, Planck:2018vyg, AtacamaCosmologyTelescope:2025blo, SPT-3G:2025bzu}. The power spectrum of such fluctuations in the $\Lambda$CDM model are well-described on these scales by
%
\begin{align} \label{eq:scalar_pow_spec}
    \mathcal{P}_{\zeta}(k) = A_{\rm s} \left( \frac{k}{k_*}\right)^{n_{\rm s}-1},
\end{align}
%
where $\mathcal{P}_{\zeta}(k)$ is the dimensionless scalar power spectrum, $k_* = 0.05 \, {\rm Mpc^{-1}}$ is the pivot scale, $A_{\rm s}\simeq 2.1 \times 10^{-9}$ gives the amplitude of the perturbations, and $n_{\rm s} = 0.9679\pm 0.0033$ \cite{SPT-3G:2025bzu} describes the spectral tilt. The phase coherence (acoustic peak structure) of these perturbations is suggestive of the fact that these modes may stem from some common origin. This fact, and other cosmological evidence, has made Inflation the leading framework to describe the Universe in its infancy \cite{Guth80, Linde82, Starobinsky:1980te}. In the simplest inflationary models, the quantum fluctuations of a single (nearly stationary) scalar field ($\phi$) are driven super-horizon due to the rapid expansion of the background. These fluctuations then re-enter the Universe at later times, yielding this spectrum of nearly scale invariant perturbations.

Within the simplest single-field slow-roll models this power spectrum is expected to extend smoothly toward smaller scales, and current data shows no statistically significant running, ($n_{\rm run}\equiv\id n_{\rm s}/\id\ln k$), nor running of the running \cite{Planck:2018vyg, Cabass:2016ldu, AtacamaCosmologyTelescope:2025blo}. Even though one may expect the smooth continuation of Eq.~\eqref{eq:scalar_pow_spec} to small scales, the inflationary dynamics responsible for them have never been tested on scales below a few $\mathrm{Mpc}^{-1}$. Across much of the interval $1\,\mathrm{Mpc}^{-1}\lesssim k \lesssim 10^{4}\,\mathrm{Mpc}^{-1}$ the most robust constraint comes from \COBEF, which limits  enhancements of small-scale power to $\mathcal{P}_{\zeta}\lesssim10^{-5}$ \citep{Chluba:2012we}, still three to four orders of magnitude above the (extrapolated) amplitude measured on CMB scales. This leaves a vast and largely unexplored discovery space, precisely where a much richer inflationary phenomenology may be hiding, and where the primary anisotropies have been erased by photon diffusion (Silk damping).
 
The damping of primordial perturbations offers a clean and unavoidable route to rigorously testing this regime. The understanding that small-scale density perturbations must distort the CMB spectrum dates to the earliest studies of the relic radiation \cite{Silk1968, Sunyaev:1970er}, and the connection to the damping of acoustic waves was developed more quantitatively in the early 1990s \cite{Daly1991, Barrow1991, Hu:1994bz}. The mechanism itself is simple. As acoustic modes enter the horizon during radiation domination they set up standing sound waves in the tightly coupled photon-baryon plasma, with photons in different phases of a given wave carrying slightly different temperatures. Photon diffusion on the Silk damping scale then mixes these blackbodies, erasing the perturbations and depositing their energy into the radiation field. Because a superposition of blackbodies at different temperatures is not itself a blackbody \citep[e.g.,][]{Chluba2004}, the mixing leads to a distortion of the frequency spectrum \cite{Chluba:2012gq, Khatri2012short2x2}. A consistent second-order calculation of the effective heating rate through this process was given a little over a decade ago \cite{Chluba:2012gq}, which showed that roughly $1/3$ of the energy drawn from these dissipating modes produces a distortion, while the other $2/3$ led to a pure temperature shift ($\Theta G(x)$ see Sec.~\ref{sec:Intro} for details). The effective heating rate was also confirmed using a hydrodynamic formulation of the problem \citep{Pajer2012b, Inogamov2015AstL}.
 
Which part of this energy survives as an observable signal, and in what form, is fixed by the thermalisation history of the plasma. Upon horizon re-entry, a given mode oscillates until hits the damping scale, in the radiation era often approximated by \cite{Kaiser1983, Hu:1995en, Hu:1996mn, Chluba:2012gq}
%
\begin{align}
    k_{\rm D} \approx 4.0 \times 10^{-6} \, (1+z)^{3/2} \, {\rm Mpc^{-1}}.
\end{align}
%
At this point, the mode will rapidly dump its energy into the thermal plasma. Whether a $\mu$ or $y$ distortion is sourced depends on the exact redshift at which the damping occurs. In the $\mu$ distortion epoch ($5\times 10^4 \lesssim z \lesssim 2 \times 10^6$), this corresponds to a sensitivity to scales given by $50\,\mathrm{Mpc}^{-1}\lesssim k\lesssim 10^{4}\,\mathrm{Mpc}^{-1}$, with $y$-distortion information allowing one to probe down to $k\simeq 1 {\rm Mpc^{-1}}$. This corresponds to some ten $e$-folds of inflationary evolution beyond the six or seven probed by the CMB anisotropies. In the multipole language of anisotropies, the primary temperature damping tail terminates near $\ell\approx2000\text{--}3000$ ($k\approx0.2\,\mathrm{Mpc}^{-1}$). The dissipation signal instead carries information from modes that, projected onto the last-scattering surface, would populate $\ell\sim10^{4}\text{--}10^{8}$, far beyond the reach of any imaging experiment \citep{Chluba:2012gq, Chluba:2012we, Khatri:2013dha, Chluba2013PCA}.
 
In the tight coupling limit, it is possible to compute the non-thermal energy injection rate analytically. Following \cite{Chluba:2012gq, Chluba:2012we, Chluba:2013pya}, the effective rate of energy release during radiation domination is
%
\begin{equation}
    \frac{\id (Q_{\rm ac}/\rho_\gamma)}{\id \ln a} \approx
    4 \alpha_{\rm A}^{2}\,\frac{\id k_{\rm D}^{-2}}{\id \ln a}
    \int \id\ln k\; k^{2}\,\mathcal{P}_{\rm s}(k)\,e^{-2k^{2}/k_{\rm D}^{2}},
    \label{eq:heating_rate}
\end{equation}
%
where $\alpha_{\rm A}^{2}=(1+4R_\nu/15)^{-2}\approx0.81$ quantifies the suppression of the photon acoustic amplitude by free-streaming neutrino anisotropic stress ($R_\nu\approx0.41$). Integrating Eq.~\eqref{eq:heating_rate} over the $\mu$-era, weighted by the blackbody visibility $\mathcal{J}_{\rm bb}(z)\approx\exp[-(z/z_{\rm th})^{5/2}]$ with $z_{\rm th}\approx2\times10^{6}$, gives the compact and intuitive form
%
\begin{equation}
    \mu_{\rm ac} \simeq \int \id\ln k\; W_\mu(k)\,\mathcal{P}_{\rm s}(k),
    \label{eq:mu_window}
\end{equation}
%
where the dimensionless window $W_\mu(k)$ encodes the details of the thermalisation physics \citep[see][]{Chluba:2012we, Chluba:2013pya}. A recent overview of the exact shape of the window function can be found in \cite{Cyr:2023pgw}. Evaluating Eq.~\eqref{eq:mu_window} for the fiducial red-tilted $\Lambda$CDM spectrum ($A_{\rm s}\approx2.1\times10^{-9}$, $n_{\rm s}\approx0.965$) gives the predicted acoustic-damping signal\footnote{This number includes the smaller but inevitable adiabatic cooling distortion \citep{Chluba2005, ChlubaSunyaev:2012, Khatri2011BE} which is $\mu\simeq -\pot{0.3}{-8}$.} \cite{ChlubaSunyaev:2012, Chluba2016}
%
\begin{equation}
    \mu \simeq 2\times10^{-8}.
    \label{eq:mu_value}
\end{equation}
%
This signal is within reach of, and a central target for future space-based CMB spectrometer experiments such as \FOSSIL and \PIXIE.
 
The mapping between primordial power and the observed distortion depends on the type of initial perturbation. For adiabatic modes the heating efficiency is set by the single factor $\alpha_{\rm A}^{2}$ of Eq.~\eqref{eq:heating_rate}, but isocurvature perturbations stir the photon-baryon plasma with different amplitudes and phases. To capture the effects of isocurvature modes, $\alpha_{\rm A}^{2}$ is replaced by a mode-dependent efficiency $C^{2}(k)$ and the window function changes shape \cite{Chluba:2013pya}. Baryon and cold-dark-matter isocurvature behave almost identically to one another: in both cases the perturbed component is non-relativistic and gravitationally subdominant during radiation domination, so it drives the plasma only weakly until close to matter-radiation equality, weighting the dissipative heating toward later times and slightly favouring $y$ over $\mu$. The neutrino density and velocity modes instead sit in the relativistic sector and source metric perturbations efficiently well before matter radiation equality, depositing relatively more of their energy in the $\mu$-era. These shifts essentially redistribute the heating among the $\mu$, $y$, and residual components, albeit in a slightly less efficient way than for the adiabatic mode. From \COBEF alone, the dissipation of these modes already constrains the (ultra small-scale) baryon and CDM isocurvature amplitudes to roughly $A_{\rm b}\lesssim5\times10^{-2}$ and $A_{\rm c}\lesssim2\times10^{-3}$, and a \PIXIE-class measurement would tighten these limits by roughly three orders of magnitude \cite{Chluba:2013pya}. A detection or limit from \FOSSIL would therefore probe small-scale isocurvature in a regime that is challenging to access by other means.
%
\begin{figure}[ht]
    \centering
    \includegraphics[width=1\linewidth]{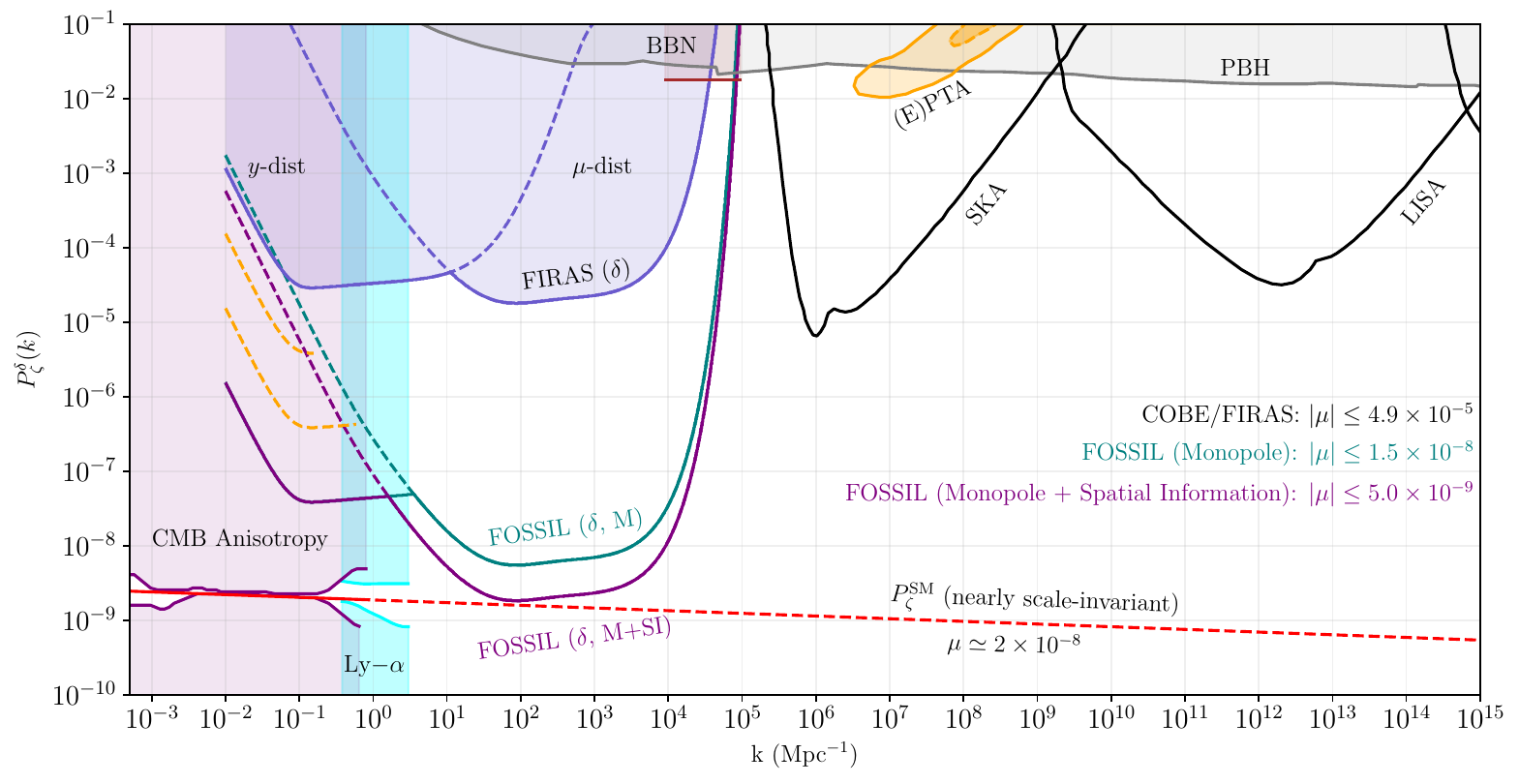}
    \caption{An extrapolation of the nearly scale-invariant spectrum of scalar perturbations to small scales. Forecasted constraints from \FOSSIL are given assuming no small-scale power except for a $\delta$-function spike at each wavenumber in order to be maximally conservative. Due to the integrated nature of the $\mu$-distortion, \FOSSIL/\PIXIE would have the sensitivity required to observe the standard model spectrum if it extends out to $k \simeq 10^4 \,\, {\rm Mpc}^{-1}$. Quantitative $\mu$ distortion forecasts are given on the right hand side of the plot to further stress the point that \FOSSIL has the reach to observe the diffusion damping signal.}
    \label{fig:PPS_general}
\end{figure}
%
 
A defining feature of the $\mu$-distortion is that it is an \emph{integrated} measure of the small-scale power, weighting $\mathcal{P}_{\zeta}(k)$ by $W_\mu(k)$ over several decades in wavenumber. Even modest deviations from the standard spectrum can therefore produce a large response in the resulting distortion \citep{Chluba:2012we, Clesse:2014pna}. A blue tilt or a localised excess of small-scale power raises $\mu$ above the $\Lambda$CDM value, whereas a strong negative running, a step-like suppression or an early end to inflation would lower it. 

An illustration of the extrapolated small scale power spectrum and the experimental reach of various proposed and upcoming experiments is shown in Fig.~\ref{fig:PPS_general}. We note that besides the CMB anisotropy and Lyman-$\alpha$ constraint curves, forecasts are given assuming $\delta$-function spikes at each wavenumber in the primordial spectrum. This is done to be maximally conservative, but has the side-effect of making it appear as though the \FOSSIL/\PIXIE missions would not reach the nearly scale-invariant line. We stress that because of the integrated nature of the $\mu$-distortion, \FOSSIL and \PIXIE would indeed have sensitivity to the extrapolated $P_{\zeta}^{\rm SM}$, despite what the figure appears to show. Constraints from gravitational wave observatories stem from the scalar induced gravitational wave (SIGW) process, detailed in Sec.~\ref{sec:GWBGs}.
 
Minimal, observationally allowed models of inflation make sharp predictions for this regime. Plateau models such as Starobinsky $R^{2}$ inflation \cite{Starobinsky:1980te} predict $n_{\rm s}\approx1-2/N\approx0.965$ for $N\approx50\text{--}60$ e-folds, with a small negative running $n_{\rm run}\approx-2/N^{2}\approx-7\times10^{-4}$ and negligible running of the running. Their power spectrum is essentially featureless down to the smallest dissipation scales, so they predict a $\mu$-distortion at the fiducial $\Lambda$CDM level; a \FOSSIL-class measurement could provide strong support for the simplest viable inflationary scenario across the entire distortion window \cite{Chluba:2015bqa,  Cabass:2016giw, Chluba2016}.

The situation is markedly different for scenarios that predict enhanced or strongly scale-dependent small-scale power. Many well-motivated mechanisms generically produce such power, including features or inflection points in the inflaton potential, bursts of particle production, waterfall transitions and axion inflation \cite{Chluba:2012we,Chluba:2015bqa}. For a localised, particle-production-type feature whose width is set by the production scale, a \PIXIE-class measurement would be sensitive to peak curvature amplitudes of $\mathcal{P}_{\rm s}\approx {\rm few}\times10^{-9}$ over $80\,\mathrm{Mpc}^{-1}\lesssim k\lesssim3000\,\mathrm{Mpc}^{-1}$ \cite{Chluba:2012we, Chluba:2015bqa}. This extreme sensitivity to small scale power underscores the quality of information that future distortion measurements could unlock. 
 
The dissipation of acoustic modes provides a calculable, linear and comparatively astrophysics-free mapping from the primordial perturbations to the CMB spectrum, providing the most robust test available of the small-scale power spectrum. The signal is guaranteed within $\Lambda$CDM, its amplitude is fixed by well-understood physics in the pre-recombination plasma, and its interpretation does not rely on the non-linear structure formation, galaxy bias or baryonic physics that limit late-time probes such as the Lyman-$\alpha$ forest or the inferred abundance of collapsed objects. A measurement at the level of $\mu \simeq 2 \times 10^{-8}$ would provide strong evidence for the simplest inflationary scenarios, while a non-detection would tell us where and how our current $\Lambda$CDM paradigm would need refinement. In either case the $\mu$-distortion opens a uniquely clean and otherwise inaccessible window onto the very early Universe \cite{Chluba:2019kpb, Lucca:2019rxf}.

\subsection{Black hole formation mechanisms} \label{sec:BH-formation}

At present, the origins of the largest supermassive black holes (SMBHs) are not well understood~\cite{Volonteri:2021sfo}. 
While the least exotic explanation is that SMBH seeds are the $\sim100\,M_\odot$ remnants of Population III stars, typical assumptions about the maximum rate of black hole accretion (i.e. no more than the Eddington limit) and mergers complicate this picture. 
Multiple ``heavy seed'' formation mechanisms with initial mass $M_{\rm BH} \gtrsim 10^4 M_\odot$ have therefore been proposed. 
Interest in heavy seed scenarios has recently been revived, given observations by the \textit{James Webb Space Telescope} (JWST) of so-called ``little red dots'' (LRDs)~\cite{2023ApJ...954L...4K,2023Natur.616..266L,2023ApJ...959...39H,2024A&A...691A.145M,2024ApJ...963..129M} and accreting black holes at $z>10$ ~\cite{2024Natur.627...59M,2024NatAs...8..126B,2024ApJ...960L...1N}, which may be indications of early and large SMBH seeds~\cite{2011BASI...39..145N,2024A&A...690A.182D,Zhang:2025asq,Matteri:2025vnv}.
These include primordial black holes (PBHs), which form from large initial density fluctuations and are already partly constrained by existing spectral distortion measurements, and variants of the ``direct collapse'' scenario, in which molecular cooling in collapsing baryons is suppressed, leading to the formation of a large compact object. 
The scenarios discussed below include not-quite-primordial black holes (NQPBHs), which form from moderate density fluctuations that collapse early enough for molecular cooling to be suppressed ($1+z > \mathcal{O}(100)$), and classic direct collapse black holes (DCBHs), which typically form when Population III star formation begins ($1+z \sim 30$) in halos with an additional source of radiation that destroys molecular hydrogen.

Spectral distortions of the cosmic microwave background (CMB) are highly complementary to the study of SMBH formation. 
Many of the scenarios above require large primordial density fluctuations, which produce sizable $\mu$-type spectral distortions (see Sec.~\ref{sec:small-scale-structure}). 
Moreover, because different formation mechanisms require different fluctuation amplitudes, a detection of the $\mu$-type spectral distortion would point toward a specific formation pathway, while a non-detection at the predicted sensitivity of proposed spectral distortion probes would greatly restrict the viable heavy seed mechanisms.

\subsubsection{Primordial black holes \label{sec:PBHs}}
Enhancements of the primordial power spectrum can give rise to overdense regions in the primordial plasma can collapse to form PBHs, as proposed by Zel’dovich and Novikov \cite{Zeldovich1967} and Hawking and Carr \cite{Hawking1971,Carr1974} in the 70s 
(see Refs.~\cite{Sasaki:2018dmp,Green:2020jor,Casanueva-Villarreal:2025kmd,Carr:2023tpt,Byrnes:2025tji} for recent reviews and Refs~\cite{Carr:2020gox,Carr:2026hot} for summaries of current constraints).
The formation of a PBH is qualitatively different from stellar collapse, where internal pressure fails to support a star. 
In PBH formation, the fate of the overdense region is determined before pressure can react. This is because in an expanding universe, regions that are overdense enough behave locally like a positively curved universe, which can recollapse \cite{Sasaki:2018dmp}. 
The time of recollapse determines the mass of the resulting PBH, while the number of regions that form PBHs depends on the statistics of the primordial density field. Below we provide rough estimates for the PBH masses probed by CMB spectral distortions, as well as the required amplitude of the primordial spectrum that yields a significant fraction of PBHs in the Universe. The derivation of accurate estimates is currently an ongoing active topic; we refer the interested reader to Refs.~\cite{Yoo:2018kvb,Yoo:2020dkz,Biagetti:2021eep,Ferrante:2022mui,Germani:2023ojx,Young:2024jsu,Fumagalli:2024kgg,Pi:2024ert}, and references therein, for recent advances. See also Ref.~\cite{Galoppo:2026eho} for recent claims that early collapse may also occur during the dark matter domination epoch within $\Lambda$CDM.

According to the standard cosmological model, the very early universe was filled with a thermal plasma of Standard Model particles with small random density fluctuations. The mean density of the plasma determines the expansion rate of the Universe (the Hubble parameter $H$) and the so-called Hubble radius, $R_H\sim c/H$ (a proxy for the causal size of the universe at a given time). The mass inside a Hubble volume during radiation domination is given by \cite{Sasaki:2018dmp,Green:2020jor}
\begin{align}
M_H\sim 5\times 10^{4}M_\odot\left(\frac{t}{1\,{\rm s}}\right)\sim 5\times 10^{4}M_\odot\left(\frac{k}{1.6\times 10^{4}\,{\rm Mpc^{-1}}}\right)^{-2}\,,
\end{align}
where we introduced the comoving wavenumber $k$ of a given fluctuation and used the relation $k/a=H$ at the time of collapse (i.e., when the physical size of the fluctuations equals the Hubble radius). 
Spectral distortions probe $1\,{\rm Mpc^{-1}}\lesssim k \lesssim 10^5 \,{\rm Mpc^{-1}}$, which corresponds to $10^{13}M_\odot\gtrsim M_H \gtrsim 1000 M_\odot$. Interestingly, this contains precisely the mass range where PBHs can be the seeds of SMBHs \cite{Chluba:2012we,Kawasaki:2012kn,Belotsky:2014kca,Nakama:2014vla,Clesse:2015wea,Nakama:2016kfq,Bernal:2017nec}.

The standard cosmological model also assumes that the random fluctuations in the density field follow Gaussian statistics, a prediction from the simplest models of cosmic inflation confirmed on the largest scales by CMB observations. 
Numerical simulations show that the threshold for PBH formation in the primordial plasma is about $\delta\rho/\rho\approx 0.4$ \cite{Escriva:2022duf}. 
Thus, the formation of a PBH is a rare process; it only occurs for fluctuations at the tail of the distribution. To have a substantial fraction of PBHs in the universe, the root-mean-square of the density fluctuations must be comparable to the collapse threshold, that is, $\sigma \sim \mathcal{O}(0.1)$. 
This estimate is robust even when including non-Gaussianity, which requires $\sigma\gtrsim \mathcal{O}(0.01)$ in optimistic cases \cite{Byrnes:2024vjt,Pritchard:2025yda,Sharma:2024img}. Such large fluctuations can be achieved in various inflationary models (see Ref.~\cite{Ozsoy:2023ryl} for a recent review), in which the inflaton field departs from standard slow-roll inflation after generating fluctuations on CMB scales. 
For a collection of models with large non-Gaussianities see, e.g., Refs.~\cite{Hooper:2023nnl,Allegrini:2026jqt}.

The large density fluctuations associated with PBHs can generate sizeable $\mu$-type distortions, as is discussed in Section~\ref{sec:small-scale-structure}.
Since \COBEF constrains $\mu<4.9 \times 10^{-5}$ \cite{Fixsen:1996nj, Bianchini:2022dqh}, we see that the variance must be $\sigma(k) \lesssim 10^{-2}$ for a sharply peaked primordial spectrum. 
This estimate naively excludes PBHs as seeds for SMBHs, unless primordial fluctuations are fairly non-Gaussian \cite{Nakama:2017xvq,Hooper:2023nnl,Allegrini:2026jqt}. 
Remarkably, non-Gaussianity opens a window for PBHs of mass $10^4 M_\odot \lesssim M_{\rm PBH} \lesssim 10^5 M_\odot$ to be the seeds of SMBHs, corresponding to $10^4 \,{\rm Mpc^{-1}}\lesssim k \lesssim 10^5 \,{\rm Mpc^{-1}}$. 
In terms of the fraction of PBHs as dark matter, the abundance would be $f_{\rm PBH} \equiv \rho_\mathrm{PBH} / \rho_\mathrm{DM}\sim 10^{-5}$ \cite{Kawasaki:2012kn,Unal:2020mts,Ziparo:2024nwh}, although the efficiency of baryonic accretion can perturb this value significantly. 
This fraction corresponds to $\sigma\sim 10^{-2}$ in the case of cubic non-Gaussianity \cite{Pritchard:2025yda}, hence allowing for PBHs as SMBH seeds \cite{Hooper:2023nnl,Allegrini:2026jqt,Dave:2026gsd}.
Note that smaller seeds with $M_{\rm PBH}\sim  10^3 M_\odot$ may also seed SMBHs at a higher initial abundance of $f_{\rm PBH} \sim 10^{-3}$ \cite{Prole:2025snf}.
%
\begin{figure}[ht]
    \centering
    \includegraphics[width=0.8\linewidth]{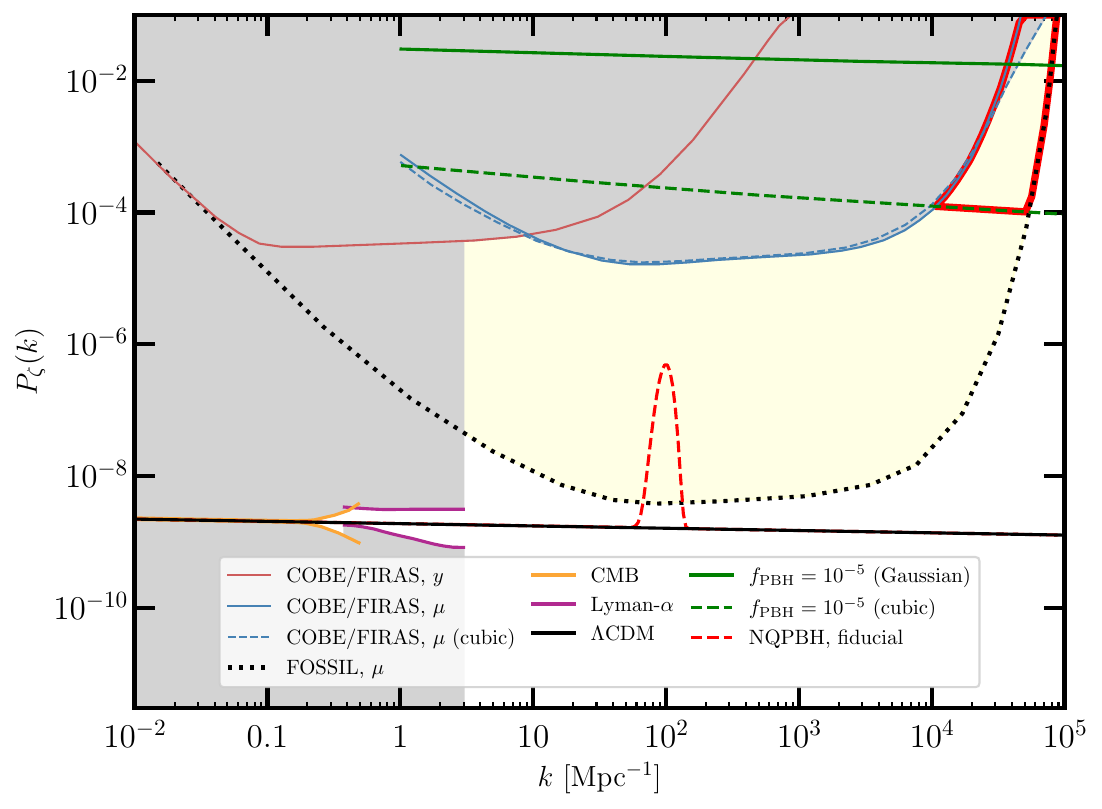}
    \caption{
        Benchmark models and limits on the primordial curvature power spectrum. 
        The forecasted sensitivity of a future probe such as \FOSSIL is shown in the black dotted line~\cite{Cyr:2023pgw}.
        Limits come from CMB anisotropies~\cite{Planck:2018jri}, Lyman-$\alpha$ forest~\cite{Bird:2010mp}, \COBEF $\mu$-type~\cite{Pritchard:2025yda} and $y$-type~\cite{Cyr:2023pgw} distortions.
        To showcase the parameter space for PBHs, we reproduce the curves from Ref.~\cite{Pritchard:2025yda} in green, which show the level of power necessary to produce $f_\mathrm{PBH} = 10^{-5}$ in the cases of Gaussian enhancements and including cubic non-Gaussianity.
        The interesting parameter space that \FOSSIL will probe for PBHs is highlighted in the thick red contour.
        We also reproduce a benchmark model for NQPBHs from Ref.~\cite{Qin:2025ymc} in red, which is a narrow peak in the power spectrum that produces a significant population of SMBH seeds. Both these theoretical models for PBHs predict a spectral distortion which will be detected at high statistical significance by \FOSSIL.
    }
    \label{fig:pspecs}
\end{figure}
%

Figure~\ref{fig:pspecs} shows some benchmarks for the PBH abundance in terms of the primordial curvature power spectrum, which are reproduced from Ref.~\cite{Pritchard:2025yda}.
Existing limits are shaded in grey, while the forecasted sensitivity for a future spectral-distortion probe such as \FOSSIL is shown as the black dotted line and highlighted by a yellow contour.
The solid green curve shows where Gaussian enhancements produce $f_\mathrm{PBH} = 10^{-5}$; the dashed green curve shows the same but with cubic non-Gaussianities.
The curve showing Gaussian enhancement is ruled out up to $k < 4 \times 10^4$~Mpc$^{-1}$, but when including non-Gaussianity, this constraint relaxes to $k < 10^4$~Mpc$^{-1}$.
\FOSSIL, however, will rule out even SMBHs produced from cubic non-Gaussianities up to $k = 6 \times 10^4$~Mpc$^{-1}$.

Future spectral distortions probes such as \PIXIE or \FOSSIL will provide the final test of PBHs as seeds of SMBHs \cite{Unal:2020mts} by searching for $\mu$-type spectral distortions down to $\mu\sim 10^{-8}$. 
Moreover, the implications for heavy-seed PBHs extend well beyond spectral distortions. 
First, heavy seeds in the early universe lead to earlier structure formation \cite{Carr:2018rid,Inman:2019wvr} and may be connected to the early galaxies seen by JWST \cite{Liu:2022bvr,Gouttenoire:2023nzr,Depta:2023uhy,Zhang:2025tgm,Zhang:2025oyl,DeLuca:2025nao}. 
Second, the existence of PBHs is accompanied by various gravitational wave (GW) signals. The merger of PBH-PBH binaries has a chirp frequency at the source frame of $f_{\rm merger}\approx 0.2 \,{\rm Hz} (10^4\,M_\odot/M_{\rm PBH})$. 
If PBHs are SMBH seeds, this predicts a signal in the LISA frequency range \cite{Bernal:2017nec}, either via a direct detection or the gravitational wave (GW) background from unresolved binaries. 
In addition, there is a GW background associated with PBH formation (see Ref.~\cite{Domenech:2021ztg} for a review) with a peak frequency of $f_{\rm formation}\sim 0.03 \,{\rm nHz} (10^4 M_\odot/M_{\rm PBH})^{1/2}$. While it apparently lies below the frequency accessible to PTAs, it may be accessible via pulsar parameter drifts \cite{10.1093/mnras/203.4.945,DeRocco:2022irl,DeRocco:2023qae,Zheng:2025tcm}. Furthermore, it is crucial to note that enhancements of fluctuations during inflation generally result in relatively broad peaks in the power spectrum \cite{Ozsoy:2023ryl}, leading to a broader mass distribution. 
Thus, the PBH-as-SMBH-seeds scenario is expected to contribute to the GW background at PTAs as well \cite{Unal:2020mts,Kite:2020uix,Cyr:2023pgw,Tagliazucchi:2023dai}. 
For the same reason, \FOSSIL can constrain realistic models leading to smaller SMBH seeds as in Ref.~\cite{Prole:2025snf}.

\subsubsection{Direct collapse and not-quite-primordial black holes}
``Direct collapse" refers to the monolithic collapse of baryons into a large central object in a halo, as opposed to the more standard fragmentation into several objects during collapse. 
This phenomenon is possible in the early universe under the correct conditions. 
Prior to onset of first star formation, the primary species responsible for the cooling and subsequent collapse of baryons in a halo are neutral atomic hydrogen (H) and molecular hydrogen (H$_2$).
Atomic hydrogen dominates the composition of baryons in our universe and is therefore readily abundant; the corresponding atomic cooling process is efficient at temperatures of $T_{\rm halo}\gtrsim 10^4$~K~\cite{Klessen:2023qmc}.
On the other hand, molecular hydrogen forms in much smaller abundances, but molecular cooling is efficient down to much lower temperatures of $\mathcal{O} (100)$~K~\cite{Klessen:2023qmc}.

The effect of these cooling mechanisms on the mass of forming objects can be understood through a heuristic argument based on the Jeans mass, which scales with temperature as $M_\mathrm{J} \propto T^{3/2}$.
Using the Jeans mass as a proxy for the fragment mass, higher temperatures lead to larger objects.
Hence, in environments where the abundance of H$_2$ is suppressed so that only atomic cooling is active, the gas temperature stays relatively high as it collapses, leading to the formation of a large compact object.
Additional requirements for direct collapse involve the gas having sufficiently low angular momentum~\cite{Koushiappas:2003zn,Lodato:2006hw} and low relative velocity to the dark matter halo~\cite{Qin:2025ymc}.

In the classic direct collapse black hole scenario, the first compact objects form at the expected redshifts of $1+z \sim 30$, at which point a significant amount of molecular hydrogen already exists cosmologically~\cite{Hirata:2006bt}.
However, if a halo is illuminated by sufficiently intense radiation at energies $11.2-13.6$~eV (known as the Lyman-Werner or LW band), this LW radiation may dissociate the H$_2$ and thus allow for direct collapse~\cite{Haiman:2012ic,Latif:2016qau}.
For example, this can be caused by feedback from Population III star formation: as the first galaxies form, they emit UV radiation that destroys H$_2$ in neighboring halos and causes direct collapse~\cite{2008MNRAS.391.1961D,2014MNRAS.445.1056V}.
In this case, there would be no non-standard contribution to CMB spectral distortions---the scenario does not require enhanced power and the necessary sources of radiation are also sources of the $y$-type distortion from reionization.
In other words, in the absence of a detected $\mu$-type distortion at the predicted \FOSSIL and \PIXIE sensitivities, this would likely be one of the few viable heavy seed scenarios remaining.
However, there are a number of examples in the literature on more exotic sources of LW radiation, which may indirectly source CMB spectral distortions.
Examples in the literature for exotic LW sources include dark matter decay and annihilation~\cite{Friedlander:2022ovf}, superconducting cosmic strings~\cite{Cyr:2022urs}, and accretion onto PBHs~\cite{Zhang:2025grn}.

Another way to realize the conditions for direct collapse is during the cosmic dark ages.
At $1+z \gtrsim 200$, the abundance of H$_2$ is naturally suppressed because the temperature of the cosmic microwave background is high enough to suppress the main chemical pathways by which the molecules form~\cite{Hirata:2006bt}.
Hence, in cosmologies where small halos begin to form by this redshift, it is possible to realize direct collapse.
As was shown in Ref.~\cite{Qin:2025ymc}, this requires an enhancement of about a few hundred in the early universe matter power spectrum at wavenumbers of $k \sim 100$~Mpc$^{-1}$ relative to the na\"{i}ve $\Lambda$CDM expectation.
If this enhancement is sourced primordially, then the necessary amount of power for this scenario is orders of magnitude less than what is needed to form PBHs---hence, such black holes are referred to as ``not-quite-primordial black holes" or NQPBHs.

NQPBHs provide a clear science target for future spectral distortion measurements; for the toy model power spectra studied in Ref.~\cite{Qin:2025ymc}, the level of enhancement needed to generate a significant population of SMBH seeds corresponds to $\mu \sim 10^{-6}$ for a narrow peak in the power spectrum.
This lies just outside of the \COBEF constraint but well within the expected sensitivity of e.g. \PIXIE or \FOSSIL, as shown in Figure~\ref{fig:pspecs}.
Hence, even if a future measurement is made of $\mu$-type spectral distortions, the experiment does not have to be as sensitive as \FOSSIL in order to severely constrain the NQPBH parameter space.
We remark that the enhancement in power that leads to NQPBHs would also create ultra-compact mini halos~\cite{Bringmann:2025cht}, which would therefore also be constrained by $\mu$-type distortions.

Were an experiment to find spectral distortions corresponding to small scale power around $10^4 \,{\rm Mpc^{-1}}\lesssim k \lesssim 10^5 \,{\rm Mpc^{-1}}$, it could have critical implications for the seeds of SMBHs, as well as early galaxy formation, GW signals at LISA, and the GW background at PTAs. 
The absence of power would strongly exclude both PBHs as SMBH seeds and NQPBHs from primordial overdensities, leaving direct collapse black holes as a viable heavy seed formation mechanism for SMBHs.

\subsection{Cosmological gravitational wave backgrounds} \label{sec:GWBGs}

Gravitational Waves offer a unique opportunity to test high-energy phenomena in the early universe. Unlike electromagnetic waves, they propagate virtually unimpeded through the hot primordial plasma, carrying pristine information from the earliest stages of cosmic history. Interestingly, the propagation of these relic GWs and the dynamics of their cosmic sources can leave sizeable deviations in the CMB blackbody spectrum. On the one hand, the propagation of GWs directly causes a local quadrupole anisotropy in the photon plasma that dissipates via Thomson scattering, which leads to spectral distortions \cite{Ota:2014hha,Chluba:2014qia, Kite:2020uix}. On the other hand, cosmic sources of GWs require a large degree of inhomogeneity \cite{Caprini:2018mtu}, implying in most cases significant density fluctuations, which later dissipate, leaving spectral distortions (see Sec.~\ref {sec:small-scale-structure}). Thus, CMB spectral distortions also provide indirect information about cosmic GW backgrounds. 

Probing GWs and their sources via CMB spectral distortions in an otherwise inaccessible regime is crucial now that we are exploring the Universe with GWs. The LIGO-VIRGO-KAGRA collaboration, operating at frequencies around $10-1000\,{\rm Hz}$, has detected more than 200 GW events \cite{LIGOScientific:2025slb,KAGRA:2021vkt,LIGOScientific:2026wfs}, most of which are binary black hole mergers. PTA collaborations recently reported on the possible detection of a ${\rm nHz}$ GW background \cite{EPTA:2023fyk,EPTA:2023sfo,EPTA:2023xxk,Zic:2023gta,Reardon:2023gzh,Reardon:2023zen,NANOGrav:2023hde,NANOGrav:2023gor,InternationalPulsarTimingArray:2023mzf,Xu:2023wog}, often attributed to the unresolved mergers of SMBHs (although with a GW background amplitude larger than expected \cite{NANOGrav:2023hvm,Sato-Polito:2023gym,Goncharov:2024htb,Sato-Polito:2025dqw}).  

Given the many detections and evidence, the future of GW observations looks bright. New photometric surveys like the Nancy Grace Roman Space Telescope may probe the $\mu{\rm Hz}$ window soon via relative astrometry \cite{Wang:2020pmf,Wang:2022sxn,Pardo:2023cag}, and may be complemented in the future by Laser Ranging Experiments \cite{Foster:2025nzf}. More GW detectors are on the way, such as LISA \cite{Barke:2014lsa,LISACosmologyWorkingGroup:2022jok},  Taiji \cite{Ruan:2018tsw}, TianQin \cite{Gong:2021gvw}, and ET \cite{ET:2019dnz}, and will cover a much wider GW frequency range, roughly from $10\,\mu \rm Hz$ to $\rm kHz$. At the other end, CMB B-mode polarization experiments, such as LiteBIRD \cite{LiteBIRD:2023zmo,LiteBIRD:2024wix} and the Simons Observatory \cite{SimonsObservatory:2025avm}, will look for signatures of very low frequency GWs, between $10^{-18}-10^{-15}\,\rm Hz$, expected to arise from quantum vacuum fluctuations during cosmic inflation \cite{Guzzetti:2016mkm}. CMB spectral distortions could directly probe GW frequencies from $10^{-15}\rm Hz$ to $10^{-10}\,\rm Hz$ \cite{Ota:2014hha,Chluba:2014qia}, bridging the gap between CMB polarization and PTAs \citep{ Kite:2020uix}.

It is plausible that some of the GWs that have been and will be detected have a cosmological origin \cite{Caprini:2018mtu,NANOGrav:2023hvm,Roshan:2024qnv}. For instance, a fraction of the binary black hole mergers detected by the LVK collaboration could be PBHs formed in the very early universe \cite{Wong:2020yig,Hutsi:2020sol} (see the discussion in Sec.~\ref{sec:PBHs}). Interestingly, the formation of PBHs is accompanied by the generation of a GW background \cite{Saito:2009jt,Bugaev:2009zh} (see Ref.~\cite{Domenech:2021ztg} for a review), which for LVK-PBHs is predicted to fall in the ${\rm nHz}$ window and may be related to the GW background reported by PTAs \cite{Inomata:2023zup,Franciolini:2023pbf,Domenech:2024rks,Iovino:2024tyg,Carr:2026hot}. Furthermore, LISA  may observe the merger of heavier PBH binaries with $M_{\rm PBH}\sim 10^4\,M_\odot-10^5\,M_\odot$ \cite{Bernal:2017nec}, a mass range of PBHs proposed as seeds of SMBHs \cite{Kawasaki:2012kn,Belotsky:2014kca,Nakama:2014vla,Clesse:2015wea,Nakama:2016kfq}. The companion GW background associated with such heavy PBH seed formation then falls well below ${\rm nHz}$, a range probed by CMB spectral distortions \cite{Cyr:2023pgw}. At such frequencies, there could also be cosmic GWs from inflation, phase transitions, and cosmic strings \cite{Guzzetti:2016mkm,Thorne:2017jft,Kite:2020uix}.  

CMB spectral distortions probes like \FOSSIL represent a promising, unique frontier for probing GW-related high-energy physics in an otherwise inaccessible regime, potentially providing new and complementary information to current and future GW detectors, such as CMB B-modes, PTAs, and LISA. We discuss in more detail below the direct and indirect probes of GWs using CMB spectral distortions, as well as their complementarity with other GW observations. 

\subsubsection{Direct gravitational wave probes}
GWs induce a local quadrupole anisotropy in the photon plasma that dissipates and yields spectral distortions \cite{Ota:2014hha,Chluba:2014qia}. While the underlying dissipative mechanism is qualitatively similar to the one discussed in Sec.~\ref{sec:small-scale-structure}, it is quantitatively described by a different window function than in Eq.~\eqref{eq:mu_window}, as derived in Ref.~\cite{Chluba:2014qia}. The main difference in the dissipation mechanism is that GWs themselves (more appropriately called tensor modes in this context) are not viscously damped; They can therefore source spectral distortions across a much wider range of scales, albeit with smaller amplitude than that generated by curvature fluctuations.

This effect can be used to probe very low-frequency GWs. For instance, the relation between a comoving scale and the GW frequency is given by \cite{Cyr:2023pgw}
\begin{align}\label{eq:fgws}
f_{\rm GW}\sim \frac{k}{2\pi}\approx 6\times 10^{-15}\,{\rm Hz}\left(\frac{k}{1\,{\rm Mpc}^{-1}}\right)
\end{align}
where $k$ is a given wavenumber, e.g., of a tensor mode. Spectral distortions probe $1\,{\rm Mpc^{-1}}\lesssim k \lesssim 10^5 \,{\rm Mpc^{-1}}$, which corresponds to $10^{-15}\,{\rm Hz}\lesssim f_{\rm GW} \lesssim 10^{-10}\,{\rm Hz}$. This is a frequency range not probed by any other experiment and bridges the frequency ranges probed by CMB B-mode polarization \cite{LiteBIRD:2023zmo,LiteBIRD:2024wix} and Pulsar Parameter Drifts \cite{DeRocco:2022irl,DeRocco:2023qae,Zheng:2025tcm}, which are  $10^{-18}\,{\rm Hz}\lesssim f_{\rm GW} \lesssim 10^{-15}\,{\rm Hz}$ and $10^{-11}\,{\rm Hz}\lesssim f_{\rm GW} < 10^{-9}\,{\rm Hz}$, respectively. 

Regarding the amplitude of the cosmic GW background, bounds from \COBEF, which found $\mu<4.9 \times 10^{-5}$ \cite{Fixsen:1996nj, Bianchini:2022dqh}, yield $\Omega_{\rm GW}<10^{-5}$ \cite{Kite:2020uix} around pHz frequencies. Currently, the best bound in a similar regime comes from CMB constraints on additional relativistic particles ($\Delta N_{\rm eff}$), which is given by $\Omega_{\rm GW}<10^{-6}$ for $f_{\rm GW}>10^{-15}\,{\rm Hz}$ \cite{Planck:2018vyg}. CMB spectral distortion experiments like \FOSSIL promise to directly probe the ${\rm pHz}$ GW background down to $\Omega_{\rm GW}\sim 10^{-8}$, improving current bounds by two orders of magnitude. 

Potential cosmic GW sources in the frequency range probed by \FOSSIL include cosmic strings, phase transitions \cite{Kite:2020uix} and GWs generated during inflation due to, e.g., spectator axion fields \cite{Thorne:2017jft,Putti:2024uyr}, primordial dark magnetic fields \cite{Kanno:2023fml} and modifications of gravity \cite{Fu:2020tlw,Addazi:2024gew}. It should be noted, however, that most of these GW sources also generate curvature/density fluctuations which contribute more strongly to spectral distortions, see, e.g., Refs.~\cite{Ramberg:2022irf,Putti:2024uyr,Xu:2025zsv} for a detailed discussion. However, there may be modified gravity models that only enhance the production of GWs as in Ref.~\cite{Fu:2020tlw,Addazi:2024gew}, without generating significant density fluctuations. 

\begin{figure}
    \centering
    \includegraphics[width=0.85\linewidth]{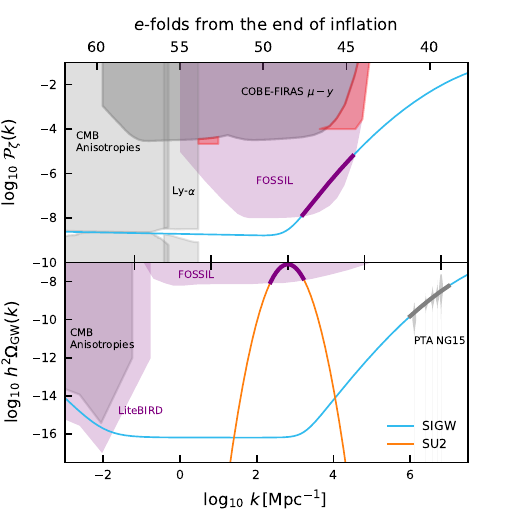}
    \caption{Indirect and direct probes of GWs with CMB spectral distortions, respectively, in the upper and lower panels. The bottom x-axis is the comoving scale, while the top x-axis shows the number of e-foldings from the end of inflation. The blue curve shows one example of primordial fluctuations that explain CMB anisotropies and the nHz GW background at PTAs, and which can be falsified by \FOSSIL \cite{Tagliazucchi:2023dai}. Upper panel: We show current constraints on the primordial power spectrum, in shaded gray regions, from CMB anisotropies \cite{Planck:2018vyg}, Lyman-$\alpha$ \cite{Bird:2010mp}, and \COBEF CMB spectral distortions. The purple-shaded region shows the many orders of magnitude by which \FOSSIL will improve, enabling it to probe models that connect CMB measurements with GWs at PTAs. The red shaded regions illustrate possible CMB B-mode contamination \cite{Cyr:2023pgw} (left red region) and possible SMBH seed formation \cite{Kawasaki:2012kn,Unal:2020mts,Hooper:2023nnl} (right red region). Lower panel: We show direct probes of the primordial GW background's amplitude. In shaded gray, we show current constraints from CMB anisotropies \cite{Planck:2018vyg}, and with gray violins, we show evidence for a nHz GW background at PTAs \cite{NANOGrav:2023gor}. The purple shaded regions show the expected GW sensitivity from the B-mode polarization survey LiteBIRD \cite{LiteBIRD2023,Cyr:2023pgw} and \FOSSIL. The solid orange line is a typical example of GWs generated during inflation, e.g., in spectator axion-U(1) \cite{Namba:2015gja} and axion-SU(2) inflation models \cite{Campeti:2020xwn} (see also Ref.~\cite{Putti:2024uyr}), which we parametrized using a broad log-normal bump. We thank Matteo Braglia for kindly providing this figure.}
    \label{fig:placeholder}
\end{figure}

We show the potential of \FOSSIL in directly probing GWs in the lower panel of Fig.~\ref{fig:placeholder}.
%
%
\subsubsection{Indirect GW probes and connection to GW observations}
Processes that generate detectable GWs require a large degree of inhomogeneity on a scale comparable to the GW frequency \cite{Caprini:2018mtu}. The same inhomogeneities in the density field that sourced the GWs can therefore inject energy into the plasma via dissipation, thereby generating CMB spectral distortions. This is how CMB spectral distortions indirectly probe GW generation, often yielding stronger bounds than direct searches \cite{Cyr:2023pgw}. 

For example, consider the GWs generated by the evolution of primordial density fluctuations with variance $\sigma(k)$. See, e.g., Refs.~\cite{Tomita,Matarrese:1992rp,Matarrese:1993zf,Matarrese:1997ay,Ananda:2006af,Baumann:2007zm,Kohri:2018awv,Espinosa:2018eve,Inomata:2018epa} for original works and Refs.~\cite{Domenech:2021ztg,Domenech:2025ior} for recent reviews. The amplitude of the resulting GW background is approximately given by \cite{Domenech:2021ztg}
\begin{align}\label{eq:estimate_OmegaIGWs}
\Omega_{\rm GW}(k)\sim 10^{-5}\sigma^4(k)\,.
\end{align}
Note that this rough estimate \eqref{eq:estimate_OmegaIGWs} is valid for any range of $k$ for $\sigma(k)\propto k^\alpha$ with $\alpha<3/4$ \cite{Atal:2021jyo}. For $\alpha>3/4$ and for a sharply peaked spectrum, the estimate \eqref{eq:estimate_OmegaIGWs} is only valid when evaluated at the peak scale, that is, at $\Omega^{\rm peak}_{\rm GW}\sim 10^{-5}\sigma^4(k_{\rm peak})$. Note that $\sigma(k)\propto k^\alpha$ with $\alpha>3/4$ must have an effective cut-off at $k=k_{\rm peak}$ \cite{Atal:2021jyo}. In these two cases, we can recover the frequency dependence of the GW spectrum from the universal scaling of low frequency tail, namely $\Omega_{\rm GW}\sim \Omega^{\rm peak}_{\rm GW} (f/f_{\rm peak})^3$ for $f<f_{\rm peak}$ \cite{Cai:2019cdl,Yuan:2019wwo}.

From now on, we focus on $\sigma(k)\propto k^\alpha$ with $\alpha>3/4$ and a peaked primordial spectrum for simplicity. We then use Eq.~\eqref{eq:estimate_OmegaIGWs} to get a rough idea of the constraining power of CMB spectral distortions. In particular, we use the fact that a sharp primordial spectrum that peaks on small scales will lead to CMB spectral distortions of the order of $\mu\sim \sigma^2(k_{\rm peak})$ (see Sec.~\ref{sec:small-scale-structure}). From Eq.~\eqref{eq:estimate_OmegaIGWs}, this implies that a detection of $\mu$-distortions roughly predicts an amplitude of the induced GW background given by $\Omega_{\rm GW}\sim10^{-5} \mu^2$. CMB spectral distortion probes such as \FOSSIL or \PIXIE aim to reach a precision of $\Delta\mu\sim 10^{-8}$, potentially constraining GW backgrounds from primordial fluctuations down to $\Omega_{\rm GW}\sim10^{-23}$. Note that this is a conservative estimate.  Constraints are expected to be tighter for a broad spectrum as CMB spectral distortions are an integrated effect, see Eq.~\eqref{eq:mu_window}.

More generally, processes that generate large inhomogeneities in the primordial Universe, such as cosmic inflation and phase transitions, generically spread power across different scales. For example, the spectrum of large density fluctuations generated during cosmic inflation often takes a peaked, broken power-law shape \cite{Byrnes:2018txb}. In this situation, spectral distortions can restrict the amount of GW energy at higher frequencies by constraining the large-scale tail of the curvature power spectrum. This type of analysis is crucial for inferring the amount of cosmic GWs that can contribute to the $\rm nHz$ GW background at PTAs \cite{Unal:2020mts,Cyr:2023pgw,Tagliazucchi:2023dai,Yi:2023tdk}.

To illustrate this point, consider a toy model where $\sigma(k)\propto k$ is in the range probed by spectral distortions until PTAs. Such a primordial spectrum can be motivated from single field inflation, concretely constant-roll inflation \cite{Motohashi:2014ppa,Motohashi:2019rhu,Ozsoy:2023ryl}, and is steep enough such that the low frequency tail of the induced GW background scales as $\Omega_{\rm GW}\sim f^3$ and could fit PTA data \cite{Franciolini:2023pbf,Cai:2023dls}. Using the fact that PTA GW background searches find $\Omega_{\rm GW}\sim 10^{-8}$ at $f_{\rm GW}\sim 10 \,\rm nHz$, which means from Eq.~\eqref{eq:estimate_OmegaIGWs} that $\sigma(k_{\rm peak})\sim 0.18$ at $k_{\rm peak}\sim 7\times 10^{6}\,{\rm Mpc}^{-1}$, the expected power of density fluctuations at smaller scales is then $\sigma(k)\sim 3\times 10^{-4}\times k/(10^{4}{\rm Mpc}^{-1})$. This leads to $\mu\sim \sigma^2(k_{\rm hi})\sim 10^{-7}$ (where we evaluated $\sigma$ at its maximum value inside the range probed by spectral distortions, roughly at $k_{\rm hi}\sim 10^{4}\,{\rm Mpc}^{-1}$). Such value of $\mu$ is far from the current \COBEF will be a great asset in understanding the nature of the GW background seen in PTAs \cite{Unal:2020mts,Cyr:2023pgw,Tagliazucchi:2023dai,Yi:2023tdk}.

We note that a similar logic applies to phase transitions, cosmic strings, and GWs sourced by spectator fields during inflation. For instance, the phase transition that sources GWs also generates inhomogeneities that dissipate, causing spectral distortions (see, e.g., Refs.~\cite{Kite:2020uix,Xu:2025zsv}). For a related discussion on cosmic string networks, see, Ref.~\cite{Ramberg:2022irf}, and for spectator axions during inflation, see Ref.~\cite{Putti:2024uyr}.

Another way in which CMB spectral distortions may inform other GW observations is by recalling that large density fluctuations may also collapse to form PBHs \cite{Sasaki:2018dmp,Green:2020jor,Carr:2020gox,Byrnes:2025tji}. These PBHs will form binaries, merge in the nearby Universe, and can be detected by ground and space-based GW interferometers \cite{Sasaki:2018dmp}. As discussed in Sec.~\ref{sec:PBHs}, spectral distortions probe $1\,{\rm Mpc^{-1}}\lesssim k \lesssim 10^5 \,{\rm Mpc^{-1}}$, which corresponds to $10^{13}M_\odot\gtrsim M_{\rm PBH} \gtrsim 1000 M_\odot$. The corresponding chirp frequency of such PBH binaries in the source frame is approximately given by the ISCO (innermost stable circular orbit) frequency, namely $f_{\rm merger}\approx 0.2 \,{\rm Hz} (10^4\,M_\odot/M_{\rm PBH})$ \cite{Maggiore:2007ulw}, which for the range of spectral distortions may be in the LISA and PTA range. Particularly interesting is the mass range $M_{\rm PBH}\sim 10^4M_\odot-10^5M_\odot$, in which PBHs can be SMBH seeds \cite{Kawasaki:2012kn,Unal:2020mts,Hooper:2023nnl,Ziparo:2024nwh} (see also Refs.~\cite{Ziparo:2024nwh,Prole:2025snf}) and be compatible with \COBEF bounds \cite{Pritchard:2025yda} when including non-Gaussianity of primordial fluctuations. In that case, we expect PBH binaries with such masses to be detectable in the future by LISA \cite{Bernal:2017nec} (see also Ref.~\cite{Bhowmick:2025qff} for a merger-driven growth of SMBHs with smaller seeds). For a study showing complications arising from the explanation of the GW background detected by the PTA array from supermassive PBH binaries, see Refs.~\cite{Gouttenoire:2023nzr,Depta:2023uhy}. In this way, \FOSSIL can predict the amount of expected PBH binaries in LISA.

This applies to any finite-time source, including GWs induced by primordial fluctuations and phase transitions. This means that a GW background generated by an inhomogeneity with wavenumber in the spectral distortion range may contaminate CMB B-mode polarization noticeably \cite{Cyr:2023pgw,Ireland:2025yqr,Greene:2024xgq,Greene:2026one,Zebrowski:2026pye}. For instance, consider inhomogeneities with a typical scale $k_*$. From Eq.~\eqref{eq:estimate_OmegaIGWs} the GW spectrum in the low frequency tail is approximately given by $\Omega_{\rm GW}\sim 10^{-5}\sigma^4(k_*) (f/f_*)^3$. CMB B-modes experiments, such as LiteBIRD aim to probe $\Omega_{\rm GW}\sim 10^{-17}$ at $f\sim 10^{-17}\,\rm Hz$ \cite{Cyr:2023pgw}. Inverting the relation, we find that the necessary variance to cause contaminated B-modes is $\sigma(k_*)\gtrsim 0.03 \left(k_*\times {\rm Mpc}\right)^{3/4}$. While this is slightly above the \COBEF bound ($\sigma<0.01$), it is close enough to require more precise estimations. Furthermore, the $f^3$ scaling may be valid only at very low frequencies. In fact, as shown in Refs.~\cite{Cyr:2023pgw,Ireland:2025yqr}, CMB B-mode contamination is possible with the current \COBEF bound for a sharply peaked primordial spectrum. A space-based distortion experiment such as \FOSSIL or \PIXIE will be crucial for understanding and excluding B-mode contamination from such GW sources. We summarize the capability of \FOSSIL in indirectly probing GWs in Fig.~\ref{fig:placeholder}.

\subsection{Beyond the Standard Model} \label{sec:BSM}
The distortion signals discussed thus far arise largely within the standard cosmological paradigm. However, the same inefficiency of thermalization that allows these signals to survive also renders the CMB spectrum an exquisitely sensitive calorimeter for new physics. Nearly any scenario that exchanges energy, entropy, or photon number with the photon bath at redshifts $z \lesssim {\rm few} \times 10^{6}$ will leave an imprint on the frequency spectrum, with the shape of the resulting distortion encoding both the magnitude and the epoch of the injection. Well-motivated extensions of the standard models of particle physics and cosmology provide a wealth of such sources, in some cases remaining active across the entire distortion window.

In this section, we highlight four representative classes of such models. Primordial magnetic fields dissipate their small-scale energy into the plasma both before and after recombination. Future spectrometers may probe their amplitude at the sub-nanoGauss level, shedding light on the origin of cosmic magnetism. Cosmic strings and other topological defects act as long-lived sources of energy and entropy, offering sensitivity to symmetry breaking scales as high as $\sim 10^{13}\,$GeV, a regime inaccessible to collider searches. Primordial black holes can inject energy through Hawking evaporation, gas accretion, and the annihilation of dark matter within the dense spikes that form around them. Evaporating black holes in particular provide a unique window onto the primordial power spectrum at wavenumbers as large as $k \sim 10^{17}\,\mathrm{Mpc}^{-1}$, probing the final e-folds of inflation. Finally, models invoked to resolve the Hubble tension generically modify either the recombination history or the pre-recombination expansion rate, leaving imprints on the cosmological recombination radiation and Silk damping signals that can break degeneracies present at the level of the anisotropies.

A recurring theme throughout this section is that new physics scenarios often source distortions with a variety of spectral shapes. Examples include the negative $\mu$-distortions produced by photon injection, as well as the non-thermal residuals imprinted by high-energy particle cascades. These features carry information beyond what is accessible through simple $\mu$ and $y$ estimates, enabling discrimination between injection scenarios. Moreover, several of these models exhibit synergies with CMB anisotropy measurements, gravitational wave observatories, and radio surveys. With the orders-of-magnitude improvement in sensitivity over \COBEF offered by concepts such as \FOSSIL and \PIXIE, each of these classes can be decisively tested, transforming the CMB frequency spectrum into a powerful discovery channel for fundamental physics.

\subsubsection{Primordial magnetic fields}
The origin of cosmic magnetism is an open question in modern cosmology. Magnetic fields on cosmological scales have been observed for almost a century. They have been detected in high redshift galaxies \cite{Kronberg:1993vk,Bernet:2008qp,Geach:2023enq}, posing a challenge for astrophysical magnetogenesis mechanisms, and seem to feature morphologies similar to that of the host galaxy, pointing towards a co-evolution of the magnetic fields and their hosts \citep{Beck:1995zs,1997FCPh...19....1V,Battaner:2000ef}. Cosmological magnetic fields are also observed in clusters of galaxies \citep{Carilli:2001hj,Govoni:2004as,Bonafede:2010wg,Botteon:2022umz} and recently also the bridges and regions connecting them \cite{Govoni981,Stuardi2021,10.1093/mnrasl/slaa142,Balboni2023}. 
Over the last decade and a half, gamma-ray observations have been teasing out evidence of magnetization in cosmic voids, providing lower limits on the magnetic field amplitude through the lack of secondary cascade photons from extreme blazars \cite{Neronov2010,Tavecchio2010,Taylor2011,MAGIC_IGMF_2023,Vovk2026} and recently also gamma ray bursts \cite{Vovk:2023qfk,Huang2023_BOAT,Burmeister2025_BOAT}.
Recent radio observations have also provided hints of their existence in cosmic filaments \cite{Vernstrom2021,Locatelli2021,Carretti2023,Carretti:2024bcf}, probing the vast range of magnetization scales in the Universe. In this regard the next decade will be transformative; with the advent of the Square Kilometre Array \cite{SKA_Magnetism_2026}, LOFAR-2 \cite{LOFAR2_WhitePaper}, and the Cherenkov Telescope Array \cite{Meyer2020_CTA_IGMF}, future observations will provide a high-definition description of cosmic magnetism in the Universe.

Despite our ever-increasing knowledge of cosmic magnetism, its ultimate origin continues to stand as one of the fundamental open questions in cosmology.
The most recent data on very high redshift galaxies and the magnetization of filaments and voids strongly supports the primordial hypothesis: cosmic magnetism is the relic of primordial magnetic fields (PMFs) generated in the early Universe and amplified by structure formation \cite{Subramanian:2018xlb,Vachaspati:2020blt}. Such primordial fields would not only seed the large scale cosmic magnetism observed today, but would be a smoking gun of non-standard processes in the early Universe, as PMFs require either first order phase transitions \cite{Quashnock:1988vs,Vachaspati:1991nm,Sigl:1996dm} or complex dynamics during inflation to be generated \cite{Turner:1987bw,Ratra:1991bn,Giovannini:2000dj,Demozzi:2009fu,Finelli:2000sh}. Thus, their precise characterization would provide an observational window onto the nature of fundamental physics \cite{Kandus:2010nw,Durrer:2013pga,Subramanian:2015lua}. Among the various energy injection mechanisms, PMFs leave a peculiar footprint on the CMB in both the anisotropies and spectral distortions. The latter are particularly interesting, as even though future instruments forecast strong constraining power on PMFs, the detailed modeling of the dissipation on the CMB frequency spectrum remains rather underexplored. Leveraging future spectral distortion data can help shed light on the small-scale behavior of PMFs, with important implications for other cosmological probes of primordial magnetism as we discuss below.

PMFs are typically modeled as a stochastic background acting as a massless, relativistic, part of the cosmological fluid. Unlike a perfect fluid, PMFs possess anisotropic pressure, and induce a Lorentz force on charged particles in the plasma, modifying their distribution and evolution. This implies that the presence of primordial $B$-fields will affect the Universe over a wide range of time-and-length scales, providing tantalizing signatures in several observational windows. For example, PMFs can impact Big Bang Nucleosynthesis, \cite{Grasso:1996kk,Kernan:1995bz,Cheng:1996yi,Kahniashvili:2010wm}, the clustering effects seen in large scale structure \cite{2013ApJ...770...47K,Kunze:2013hy,Kunze:2022mlr,Kunze:2021qxt}, very small scale baryonic effects \cite{Jedamzik:2018itu,Jedamzik:2020krr,Galli:2021mxk,Ralegankar:2023onp,Ralegankar:2023pyx}, the 21 cm signal \cite{Schleicher:2008hc,Shiraishi:2014fka,Minoda:2018gxj,Kunze:2019qpe}, the abundance of dwarf galaxies \cite{2020A&A...643A..54S} and the SZ effect \cite{Tashiro:2009hx}. Of these observables, the CMB provides the most comprehensive probe of PMFs, owing to the wide range of effects they can induce on the anisotropies. PMFs have a gravitational effect due to the magnetically-induced scalar, vector and tensor perturbations \cite{2004PhRvD..70d3011L,Finelli2008,Paoletti2009,Shaw:2009nf,Paoletti:2010rx,Shaw:2010ea,Paoletti:2012bb,Planck:2015zrl,Paoletti:2019pdi}. In addition, they induce a Faraday rotation generating a B-mode polarization from the E-mode as well as an additional effect similar to anisotropic birefringence \cite{1996ApJ...469....1K,Kosowsky:2004zh,2009PhRvD..80b3009K,2011PhRvD..84d3530P,Planck:2015zrl,POLARBEAR:2015pbf}. If modeled as a stochastic background, PMFs have a non-Gaussian impact on CMB anisotropies sourcing non-trivial bi- and tri-spectra \cite{Caprini:2009vk,Seshadri:2009sy,Trivedi:2010gi,Shiraishi:2011xvp,Shiraishi:2011fi,Shiraishi:2011dh,Trivedi:2011vt,Shiraishi:2012sn,Shiraishi:2012rm,Shiraishi:2013wua,Planck:2015zrl,Gao:2020gct,Rotti:2022gcl}. Finally the post-recombination dissipation via ambipolar diffusion and MHD decaying turbulence affects the thermal and ionization history modifying the CMB anisotropies in temperature and polarization \cite{Sethi:2004pe,Kunze2014,2015MNRAS.451.2244C,Planck:2015zrl,Paoletti:2018uic,Paoletti:2022gsn}. The constraints from these different effects depend on the characteristics of the PMFs and the parametrization choice. These bounds span from few picoGauss for blue tilted spectra to a few nanoGauss for the nearly scale invariant case on $1$ Mpc scales. Such constraints are expected to tighten significantly with upcoming data, in particular from future polarization power spectra \cite{LiteBIRD:2024twk}.

In addition to the anisotropies, PMFs can leave a strong impact on the CMB frequency spectrum. Magnetically induced perturbations survive Silk damping \cite{Brandenburg:1996fc}, and the presence of PMFs modifies the damping mechanism leading to much more complex dynamics driven by magnetosonic and Alfven waves. Nevertheless, PMFs are ultimately damped away on scales smaller than the usual Silk scale \cite{Subramanian1998, Jedamzik1998}. Indeed, this decay of the fields on small scales is a rather efficient mechanism to transfer damped magnetic energy into the plasma. This energy injection, which acts as a continuous source term after magnetogenesis, generating $\mu$, $y$, and residual distortions. The study of this damping regime is one of the most interesting open fields in the study of primordial magnetism. Current treatments rely either on approximations \cite{Subramanian1998, Jedamzik1998} or simulations which in turn depend on the assumptions for the initial conditions. The expected sensitivity of future experiments will require a detailed modeling of this damping \cite{Kunze2014}. 

On large scales in the early Universe, ideal magneto-hydrodynamic (MHD) evolution is ensured by the simple fact that the plasma is fully ionized prior to recombination. However, on small scales the interplay between the PMF pressure and fluid viscosity becomes relevant and the fields enter into a non-ideal MHD regime which provides a pathway for PMF decay~\cite{Jedamzik1998,Subramanian1998}. On non-linear (small) scales, this non-ideal MHD environment makes the exact estimation of the energy injection due to PMF decay rather complex, with many assumptions being made. Linear regimes were explored in \cite{Jedamzik:1999bm,Kunze2014}, who placed limits on the PMF amplitude of a few tens of nanoGauss using FIRAS data, while \cite{Subramanian1998} investigated the non-linear damping of Alfvén waves in order to derive a damping scale approximation. Non-linear effects have been also considered in \cite{Wagstaff:2015jaa}, where they limited their analysis to the case of PMFs generated through phase transitions, showing the importance of helicity in magnetogenesis. 

Helicity is the topological property describing the linkage, writhe and twisting of magnetic field lines and is a quantity which is conserved (with the exact form depending on the notation chosen). The conservation of helicity is one of the main drivers in the evolution of magnetic fields in the early Universe and it is a desirable condition for the development of the inverse cascade, which provides a transfer of energy from the small to the larger scales. This is a necessary condition for post-inflationary PMFs to reach cosmological coherence lengths \cite{PhysRevD.62.103008,Hindmarsh:2002tc}. Helicity impacts the evolution of the fields on small scales \footnote{Not only small scales, as primary large and intermediate CMB anisotropies are affected \cite{Ballardini:2014jta}.}, and as a consequence, the amplitude of magnetically-induced spectral distortions for helical fields. The most recent estimation of the spectral distortion (accounting for non-linear MHD), is based on treatments involving reconnection-driven turbulence and Hosking integral conservation which leads to energy transfers even with a net zero helicity by quasi-conserving the local helicity on very small scales \cite{Hosking:2020pzi,Hosking:2023jpp,Zhou:2024mnras,Verma:2011pre}. Current constraints from the updated FIRAS bounds on the (maximal helicity) combination of PMF amplitude and correlation length are $B^{2, \rm initial}_{[10^{-7} {\rm Gauss}]} l^{\rm initial}_{[{\rm pc}]}<0.6$. The next generation of space-based spectrometers (e.g. \FOSSIL or \PIXIE), with their projected sensitivity can improve this constraint by up to five orders of magnitude in $\mu$, allowing for possible limits at the sub-nanoGauss level on the amplitude of the PMFs \cite{Uchida:2024iog}.

While the damping on small scales is active from the generation time of magnetic fields, the ambient conditions of the cosmological plasma during and after recombination allow for the development of a new kind of damping for PMFs. The rapid decrease of the free electron fraction causes a drop in plasma viscosity, thereby driving a sharp increase in the magnetic Reynolds number and triggering the development of MHD decaying turbulence. At the same time the residual ionized component is affected by the Lorentz force induced by the PMFs whereas the neutral component is not. The interaction among the two components dissipates the magnetic fields in a process known as ambipolar diffusion \cite{Sethi:2004pe,Sethi2009,Kunze2014, 2015MNRAS.451.2244C,Planck:2015zrl,Kunze:2015oaa,Paoletti:2018uic,Paoletti:2022gsn}.
Both MHD decaying turbulence and ambipolar diffusion inject energy in the plasma heating it, with the ambipolar diffusion and the MHD decaying turbulence energy injection rates dependent on the Lorentz force and the magnetic energy density respectively \citep{Sethi:2004pe,Sethi2009, Schleicher2008b}.
These effects modify the CMB anisotropies and provide strong constraints \cite{Paoletti:2018uic,Paoletti:2022gsn}, but a stronger impact should be expected on $y$-spectral distortions due to the post-recombination energy injection.

The level of sensitivity that future spectrometers (especially \FOSSIL and \PIXIE) are expected to reach will enable the detection (or strongly constrain) PMFs on the sub-nanoGauss level from both $\mu$ and $y$ distortions induced by MHD decay from both pre- and post-recombination dissipation processes. Rather than just an additional probe, spectral distortions offer an observational window on the nature of primordial magnetism. The behaviour of PMFs on small scales is a key ingredient to understanding their impact on structure formation, and also their implication for the Hubble tension \cite{Jedamzik:2020krr}, a regime where the underlying theory remains poorly understood and subject to strong assumptions. Future spectrometers like \FOSSIL will open an observational window onto primordial magnetism on scales not reachable by CMB anisotropies, providing crucial insights into the role of PMFs in the evolution of Universe.

\subsubsection{Cosmic strings and other topological defects}
\label{CSTD}
Topological defects are a generic prediction of phase transitions in the early Universe, forming whenever the underlying particle physics model admits degenerate vacuum states with the appropriate symmetry breaking pattern \cite{Kibble1976, Vilenkin2000, Hindmarsh1994}. If such transitions occur at high energy scales, the resulting defects (cosmic strings, domain walls, textures, or monopoles) persist into the late Universe, injecting energy and entropy into the primordial plasma across a broad range of redshifts.

Domain walls and monopoles form from the breaking of discrete symmetries and local symmetries respectively, and their production must be suppressed during the evolution of the Universe since they would quickly come to dominate, and many mechanisms have been discussed to achieve this.
Cosmic strings and global monopoles/textures can last until the present epoch. Constraints on topological defects are usually expressed in terms of the symmetry breaking scale, $\eta$. In the fiducial case of cosmic strings this leads to a constraint on the dimensionless mass per unit length $G\mu$. Observations of the anisotropy and polarization of the CMB impose a model independent limit of $G\mu<2\times 10^{-7}$ and there are equivalent limits on other scaling defect scenarios~\cite{Planck2013topo}. Even stronger limits may exist from the stochastic gravitational wave background they produce, but this is less certain since it depends on the modeling of small scale physics.

Modeling the evolution of topological defects is uncertain since it involves tracking a wide range of scales, from the microscopic string core to the macroscopic size of observable Universe, over a significant timescale (i.e. the age of Universe). Hence, various approaches have been developed to perform this task, with independent methods failing to converge thus far on a clear picture. This leads to some level of uncertainty on constraints from the production of a gravitational wave background from cosmic strings, as discussed above  Simulations based on the line-like Nambu-Goto action have the largest dynamic range and suggest a distribution of sub-horizon loops will form, oscillating and radiating throughout all of cosmic time. Predictions of spectral distortion signatures have been made on the basis of this, with some details below. Field theory simulations appear to suggest that the scaling regime is somewhat different and that loops decay very quickly.

Regardless of the precise details of the modeling, the  spectral sensitivity of future spectral distortion experiments provide a unique opportunity to probe these scenarios in regions of parameter space that are inaccessible to anisotropy experiments, gravitational wave observatories, and more traditional collider searches.

\begin{figure*}[t]
    \centering
    \includegraphics[width=0.7\linewidth]{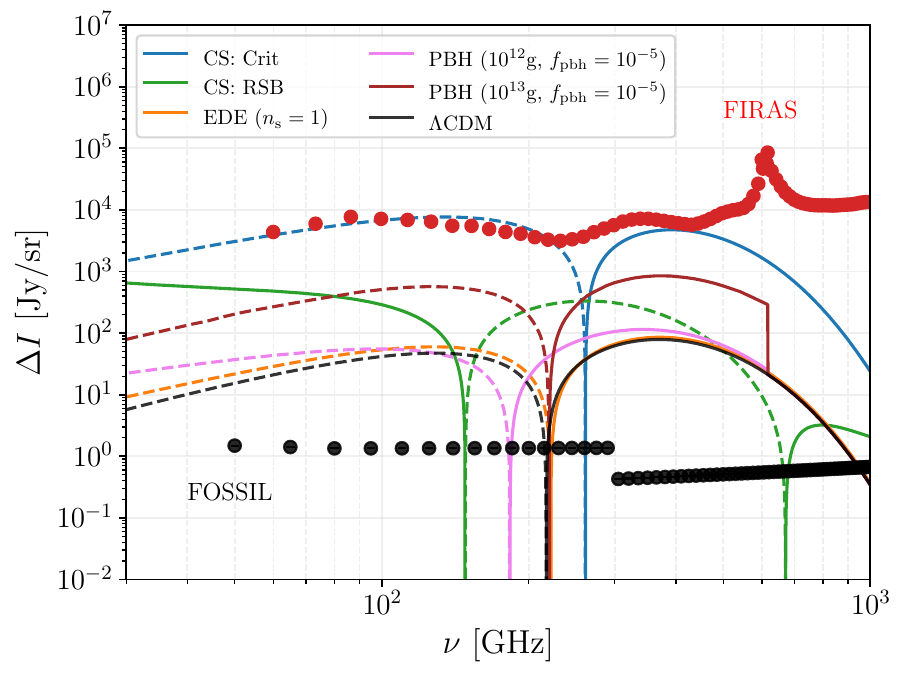}
    \caption{Total primordial (pre-recombination) distortion signatures in new physics scenarios. Solid curves correspond to photon excess relative to a blackbody, while dashed curves are photon decrements. CS: Crit are superconducting string scenarios with equal amounts of energy injection into electromagnetic species and gravitational waves, while CS: RSB is the best-fit value of $I$ and $G\mu$ to the observed radio synchrotron background (RSB) \cite{Cyr:2023yvj}. Each of these scenarios provide distinct avenues for detection and model discrimination through their spectral variation away from $\Lambda$CDM.}
    \label{fig:BSM)_distortion}
\end{figure*}

\paragraph{Superconducting cosmic strings.}
In rather general classes of cosmological phase transitions (see, for example, \cite{Davis:1995kk, Battye:2024dvw}), cosmic strings can acquire electromagnetic currents, leading to copious photon emission from the decay of transient structures (cusps) on oscillating string loops \cite{Witten1984, Ostriker1986, Cai2011, Acharya2019}. These models are typically characterized by two parameters: the dimensionless string tension $G\mu$ (not to be confused with the $\mu$-distortion) and the current $I$, and will generically act as a continuous source of both non-thermal energy and photon number into the plasma across the full distortion window. In \cite{Cyr2023}, it was shown that the direct injection of photons by such a network can produce a strong \emph{negative} $\mu$-distortion in a significant region of parameter space, an effect that had been overlooked in prior treatments \cite{Tashiro2012, Miyamoto2013}. This arises because the photon number injection can outpace the energy injection, driving the chemical potential negative. Using the thermalization code \texttt{CosmoTherm} \cite{ChlubaSunyaev:2012}, it was demonstrated that a full spectral shape analysis of the \COBEF residuals yields considerably stronger constraints than simple $\mu$ and $y$ estimates alone, underscoring the importance of non-standard distortion shapes. Some specific distortion shapes for superconducting cosmic strings can be seen in Fig.~\ref{fig:BSM)_distortion}.

The \COBEF limits on string parameters have just recently been superseded by constraints from CMB anisotropy measurements. Future spectral distortion experiments will probe an important and currently unexplored wedge of the $G\mu$--$I$ parameter space through detection of this negative $\mu$-distortion \cite{Cyr2023}. This represents a qualitatively new observable with no counterpart in anisotropy data, encoding direct information about the microphysics of the string network. Moreover, the low-frequency photon background generated by superconducting strings may contribute to the anomalous radio excess observed by ARCADE-2 \cite{Fixsen:2009xn} and the LWA \cite{Dowell2018} (see \S \ref{sec:RSB}), providing synergies between spectral distortions and existing radio surveys. Importantly, the time-dependence of the string injection, which is active across both the $\mu$ and $y$ eras, produces a characteristic spectral shape that can be distinguished from other injection scenarios with the high sensitivity measurements of the CMB spectrum, allowing this model to be definitively probed with spectral distortion data.

\paragraph{Ordinary cosmic strings.}
Even without electromagnetic currents, cusp annihilations on cosmic string loops can still source spectral distortions. In this case, the energy is released in hadronic jets (as opposed to direct photon emission) which perturb the plasma in a way that is significantly non-thermal. It was found in \cite{Anthonisen2015} that the fractional energy density released peaks for string tensions of $G\mu \sim 10^{-13}$, which corresponds to an effective probe of cosmological phase transitions at temperatures of $T \sim 10^{11}$--$10^{13}$\,GeV (redshifts of $10^{23} \lesssim z \lesssim 10^{25}$). The signal computed in \cite{Anthonisen2015} fell below the \COBEF sensitivity, but with future measurements one would constrain tensions in the range $10^{-15} \lesssim G\mu \lesssim 10^{-12}$, probing a class of intermediate-scale particle physics models that are otherwise difficult to reach. In addition to cusp annihilations, distortions can also be induced by the damping of acoustic waves sourced by the motion of cosmic strings in the primordial plasma \cite{Tashiro2012b}. The amplitude of this signal is somewhat uncertain, but is thought to sit roughly around $\mu \simeq 10^{-8}$.

\paragraph{Global textures.}
Textures arise in models where a global symmetry is spontaneously broken in a way that permits extended, unstable field configurations. These collapse at close to the speed of light and unwind, releasing energy into photons and neutrinos. If they existed, the expected distribution of textures would have produced both $\mu$- and $y$-type distortions with an amplitude that scales with the temperature of the phase transition in which they are formed \cite{Brandenberger2020}. Current \COBEF bounds on the texture energy scale are slightly weaker than those from the CMB angular power spectrum, but future measurements will improve these constraints by several orders of magnitude, potentially providing the leading bound on the general classes of texture-forming phase transitions in the early Universe.

The spectral signatures of defect networks, negative $\mu$-distortions from entropy injection, non-standard spectral shapes from redshift-dependent energy release, and low-frequency photon excesses, are qualitatively distinct from what can be learned through any other cosmological observable, providing a direct link between precision spectroscopy of the CMB and fundamental physics at energy scales far beyond the reach of terrestrial experiments.

\subsubsection{Evaporating and accreting PBHs}
Primordial black holes (PBHs) are a compelling dark matter candidate and a powerful probe of primordial density fluctuations on scales far smaller than those accessible through CMB anisotropies or large-scale structure, as discussed in Sec.~\ref{sec:PBHs}. Formed from the gravitational collapse of overdense regions in the radiation-dominated era~\cite{Zeldovich1967,Hawking1971,Carr1974}, PBHs can span an enormous range of masses, and their abundance at any mass directly constrains the amplitude of primordial curvature perturbations on the corresponding scale~\cite{Cyr:2023pgw,SatoPolito2019,Josan2009}\footnote{Note that there are non-standard formation mechanisms that require a more detailed distortion analysis \cite{Hawking:1982bubbles, Kodama:1982fopt, Crawford:1982qcd, Hawking:1989strings, Polnarev:1991strings, Caldwell:1996loops, Rubin:2000walls, Deng:2017walls, Deng:2017bubbles, Deng:2018distortions, Cotner:2017susy, Cotner:2019frag, Liu:2021fopt, Kawana:2021fermiballs, Baker:2021trapping, Gouttenoire:2023supercooled, Flores:2020drq, Dvali:2021confinement}.}. Through Hawking evaporation, gas accretion, and the enhanced annihilation of particle dark matter in their vicinity, PBHs can also inject electromagnetic energy into the primordial plasma across a wide range of cosmic epochs, producing spectral distortions whose amplitude and shape encode both the PBH abundance and the underlying BSM particle content.

\paragraph{Hawking evaporation.}
A black hole of mass $M$ radiates a modified greybody spectrum at the Hawking temperature $T_\mathrm{BH} \propto M^{-1}$, emitting all kinematically accessible Standard Model degrees of freedom~\cite{Hawking1974,Hawking1975}. For PBHs in the mass range $M \sim (10^{11}$--$10^{13})\,$g, the lifetimes are sufficiently short that they fully evaporate within the distortion epoch, injecting a significant fraction of their energy directly into the baryon-photon plasma. This violent evaporation process can thus produce $\mu$-type, $y$-type, and intermediate $i$-type spectral distortions depending on the precise injection redshift~\cite{Tashiro2008,Acharya2019,Acharya2020,Auffinger2022}.

A critical recent advance has been the proper treatment of the full Hawking emission spectrum, including quarks, gluons, and the subsequent hadronic and electromagnetic cascades. It has been shown~\cite{Acharya2019} that accounting for all Standard Model channels and evolving the resulting particle cascades through the expanding universe strengthens the distortion constraints on PBH abundance by more than an order of magnitude relative to earlier analyses that considered only primary photon emission~\cite{Tashiro2008} (e.g. neglecting the subsequent energy cascade). Crucially, the resulting distortion shapes differ from the standard $\mu$ and $y$ templates, and the non-thermal character of the particle cascade carries additional information that can be searched for in the residuals of a distortion experiment~\cite{Acharya2019}. The understanding of these features have been further refined~\cite{Acharya2021} by noting that for extreme scenarios with globally large energy injections ($\Delta \rho/\rho \lesssim 0.1$) the linear thermalization approximation breaks down, introducing corrections to the derived PBH constraints.

\COBEF limits constrain the PBH dark matter fraction $f_\mathrm{PBH}$ across the mass range discussed above, and provide a bound below $M\sim 10^{13}\,$g~\cite{Acharya2020,Lucca:2019rxf, Chluba2020L}. The roughly $10^3$-fold improvement in sensitivity from \FOSSIL or \PIXIE (when compared to \COBEF) will strengthen these limits by several orders of magnitude~\cite{Acharya2019,Acharya2020, Chluba2020L}, pushing deep into the parameter space relevant for constraining the primordial power spectrum on ultra-small scales ($10^{14}\,\mathrm{Mpc^{-1}} \lesssim k \lesssim 10^{17}\,\mathrm{Mpc^{-1}}$). This measurement will open an observational window onto a fundamentally distinct part of the inflationary history, providing powerful constraints on the final $\sim 40$ e-folds before the end of inflation~\cite{Acharya2020}. \FOSSIL or \PIXIE's broad spectral coverage will further enable the extraction of non-thermal distortion features unique to PBH evaporation, allowing it to be distinguished from other energy injection scenarios active in the pre-recombination plasma~\cite{Acharya2019}.

\paragraph{Gas accretion onto massive PBHs.}
PBHs with masses $M \gtrsim 1\,M_\odot$ have a negligibly low Hawking temperature, but can accrete baryonic gas, converting a fraction of the accreted rest-mass energy into radiation that heats and ionizes the surrounding medium~\cite{Ricotti2008,Ali-Haimoud2016,Poulin2017,Agius2024}. CMB anisotropy constraints currently dominate in this mass range (conservatively scaling with $f_\mathrm{PBH} \lesssim 10^{-1}(10 \, \, M_{\odot}/M)$ for $M \gtrsim 10\,M_\odot$~\cite{Ali-Haimoud2016, Agius2024}), with the induced spectral distortion signals remaining below \COBEF sensitivity~\cite{Ricotti2008,Ali-Haimoud2016,Aloni2017}. \FOSSIL will enable the first meaningful spectral distortion search for accreting PBHs, providing an independent and complementary probe, particularly in the $M \sim (10^3$--$10^4)\,M_\odot$ regime where significant theoretical uncertainties persist in the modeling of radiative feedback and accretion geometry~\cite{Poulin2017,Agius2024}. 

\paragraph{Annihilation of dark matter spikes around PBHs.}
Even when $f_{\rm PBH} \ll 1$, the spatial distribution of the dominant dark matter component can be qualitatively different than the case of $f_{\rm PBH} = 0$. Prior to matter-radiation equality, PBHs accrete particle dark matter, generating steep density spikes ($\rho \propto r^{-9/4}$) and forming objects known as ultra-compact minihalos~\cite{Eroshenko2016,Boudaud2021}. If the particle dark matter can annihilate, 
these density spikes serve as optimal regions to probe the dark matter cross-section due to the enhanced energy injection compared to what one would expect in a smooth background~\cite{UtrillaMena2022}. For velocity-independent ($s$-wave) annihilation at the thermal cross section, the energy injected between matter-radiation equality and recombination produces $y$-type distortions already constrainable by \COBEF. For dark matter masses $1 \, {\rm GeV} \leq m_\chi \leq 10^3\,$GeV, the resulting bounds are $\Omega_\mathrm{PBH} \lesssim 5\times 10^{-3}(m_\chi/\mathrm{GeV})^{1/3}$, independent of PBH mass since the per-halo annihilation rate $\Gamma_\mathrm{ann} \propto M_\mathrm{PBH}$ exactly compensates the decreasing number density $n_\mathrm{PBH} \propto M_\mathrm{PBH}^{-1}$~\cite{Yang2022,UtrillaMena2022,Si2026}. For velocity-suppressed ($p$-wave) annihilation, the constraints are relaxed and acquire an explicit mass dependence ($f_\mathrm{PBH} \propto M_\mathrm{PBH}^{-2/13}$), reflecting the interplay between the cross section $\langle \sigma v \rangle_p \propto v^2$ and the radial velocity profile $v(r) \propto r^{-1/2}$ within the spike~\cite{Si2026}. With e.g. \FOSSIL-level sensitivity ($y \sim 10^{-8}$), it will be possible to probe $f_\mathrm{PBH}$ as low as $\sim 10^{-5}$ across the asteroid-to-solar-mass range for GeV-scale WIMPs~\cite{Si2026}, providing a unique probe of mixed dark matter scenarios sensitive to both PBH abundance and particle dark sector physics.

\subsubsection{The Hubble tension} \label{H0tension}
Another avenue opened by a CMB spectral distortion experiment is the ability to test scenarios proposed to resolve the ``Hubble tension''. The Hubble tension refers to the discrepancy between ``direct'' late-Universe measurements of the Hubble constant $H_0$ and ``indirect'' values inferred from early-universe observations under the assumption of the $\Lambda$CDM cosmological model.
More precisely, the most statistically significant tension arises between measurements calibrated using the sound horizon -- the imprint of acoustic waves propagating in the primordial plasma until recombination -- and those relying on the calibration of the supernova (SNe) distance ladder. This discrepancy has now risen above $7\sigma$ between CMB-inferred~\cite{Planck:2018vyg, Planck:2018nkj,ACT:2020gnv,AtacamaCosmologyTelescope:2025blo,SPT-3G:2025bzu} and direct late-Universe measurements of $H_0$, with the most recent combined \emph{Planck}/ACT/SPT analysis predicting $H_0 = (67.24 \pm 0.35)$ km/s/Mpc in $\Lambda$CDM~\cite{SPT-3G:2025bzu}, while direct measurements~\cite{Freedman:2020dne,Birrer:2020tax,Riess:2021jrx,Anderson:2023aga,Scolnic:2023mrv,Jones:2022mvo,Anand:2021sum,Freedman:2021ahq,Uddin:2023iob,Huang:2023frr,Li:2024yoe,Pesce:2020xfe,Kourkchi:2020iyz,Schombert:2020pxm,Blakeslee:2021rqi,deJaeger:2022lit,Murakami:2023xuy,Breuval:2024lsv,Freedman:2024eph,Riess:2024vfa,Vogl:2024bum,Scolnic:2024hbh,Said:2024pwm,Boubel:2024cqw,Scolnic:2024oth,Li:2025ife,Jensen:2025aai,Riess:2025chq,Newman:2025gwg,H0DN:2025lyy} converge on the value measured by the ``$H_0$ distance network'' (H0DN), providing the most precise local determination, $H_0 = 73.5 \pm 0.81$ km/s/Mpc~\cite{H0DN:2025lyy}. This result combines different distance-ladder measurements while consistently accounting for their correlations. The direct determination incorporates multiple independently calibrated anchors and explicitly propagates correlations between rungs of the distance ladder, finding no evidence that known systematic effects can reconcile the result with the value inferred from early-Universe measurements in $\Lambda$CDM.
Alongside these direct and indirect determinations, there is a broader community-wide effort to search for possible systematic origins of the tension in either (or both) of the observational methodologies~\cite{Rigault:2014kaa,NearbySupernovaFactory:2018qkd,Addison:2017fdm,CSP:2018rag,Jones:2018vbn,Efstathiou:2020wxn,Brout:2020msh,Mortsell:2021nzg,Mortsell:2021tcx,Freedman:2021ahq,Garnavich:2022hef,Kenworthy:2022jdh,Riess:2022mme,Feeney:2017sgx,Breuval:2020trd,Javanmardi:2021viq,Wojtak:2022bct,Riess:2021jrx,Sharon:2023ioz,Murakami:2023xuy,Riess:2023bfx,Bhardwaj:2023mau,Brout:2023wol,Dwomoh:2023bro,Uddin:2023iob,Riess:2024ohe,Freedman:2024eph,Riess:2024vfa}.

This discrepancy motivates the search for new physics solutions to this tension (see Refs.~\cite{DiValentino:2021izs,Abdalla:2022yfr,CosmoVerseNetwork:2025alb} for reviews). Among early-Universe solutions, it is useful to distinguish between models that directly modify recombination microphysics and the photon spectrum, producing distinctive spectral distortion signatures, and those that affect the inferred value of $H_0$ indirectly through changes in the expansion history or primordial perturbations. Models in the first class generically imprint observable features in CMB spectral distortions, allowing decisive tests by high-accuracy measurements, such as those planned with future CMB spectral distortion experiments. In contrast, models in the second class modify the sound horizon primarily by altering the background evolution, typically leaving only secondary or degenerate effects in the distortion signal. This distinction is particularly important because models belonging to the two classes can produce similar shifts in the sound horizon while leaving qualitatively different signatures in the spectral distortion spectrum. Thus, a sensitive measurement of spectral distortions would allow us to break the degeneracy between these two classes of models.

\paragraph{Non-standard recombination.} 
Among early-Universe solutions to the Hubble tension, a particularly well-motivated class consists of models that operate during or prior to recombination by modifying the ionization and photon-baryon plasma dynamics, thereby producing primary spectral distortion signatures. These scenarios either alter the atomic and scattering processes governing the formation of neutral hydrogen and helium or inject energy and photons into the primordial plasma. In specific realizations, such modifications can shift the timing or duration of recombination, reduce the sound horizon, and thus impact the inferred value of $H_0$.
Such effects can arise from changes to fundamental constants, which directly modify the microphysics of recombination. In particular, variations of the fine-structure constant $\alpha_{\rm em}$ and the electron mass $m_e$ alter the energy levels of the hydrogen atom and the Thomson scattering rate~\cite{Hart:2017ndk,Hart:2019dxi,Sekiguchi:2020teg,Hart:2021kad,Lee:2022gzh,Chluba:2023xqj,Greene:2023cro,Greene:2024qis,Baryakhtar:2024rky,Seto:2024cgo,Mirpoorian:2024fka,Lynch:2024hzh,Toda:2024ncp,Schoneberg:2024ynd,Smith:2025uaq,GarciaEscudero:2025lef,Toda:2025kcq}. 
These effects can also come about from additional baryon clumping produced, for instance, by primordial magnetic fields~\cite{Jedamzik:2020krr,Thiele:2021okz,Rashkovetskyi:2021rwg,Galli:2021mxk}, which modify the local recombination dynamics and thus the effective ionization history. More generally, non-standard energy or photon injection around recombination -- such as from decaying, coupled, or annihilating dark matter~\cite{Anchordoqui:2020djl,Bjaelde:2012wi,Poulin:2016nat,Bringmann:2018jpr,Pandey:2019plg,Nygaard:2020sow,Blinov:2020uvz,Kumar:2018yhh,Yadav:2019jio,Becker:2020hzj,Archidiacono:2019wdp,Fung:2021wbz,Binder:2017lkj}, or from primordial black holes~\cite{Nesseris:2019fwr,Flores:2020drq} -- will also modify the free-electron fraction through additional ionization and heating, thereby potentially impacting the inferred sound horizon (see also~\cite{CosmoVerseNetwork:2025alb} and references therein). Photon injection or photon-number non-conservation mechanisms provide additional realizations of this class of models.
While several of these scenarios can be partially degenerate at the level of CMB anisotropies, they generically produce distinctive signatures in the CMB spectral distortions, in particular by modifying the spectrum of cosmological recombination radiation and potentially inducing additional $\mu$- and $y$-type distortions. In particular, modifications of the recombination history leave characteristic signatures in the Cosmological Recombination Radiation (CRR), discussed in Sec.~\ref{CRR}, providing an independent observational probe complementary to CMB anisotropies. These signatures constitute a primary target for future high-sensitivity CMB spectrometers. As such, future space-based spectral distortion missions represents a unique opportunity to detect those models and provide a decisive test of early-universe solutions that modify the recombination history.

\paragraph{Modified pre-recombination expansion.}
A second class of early-Universe solutions to the Hubble tension consists of models that primarily modify the background expansion and therefore induce only secondary or parameter-degenerate effects in the CMB spectrum. More generally, this class encompasses any early-time modification of the pre-recombination expansion rate -- including additional relativistic species~\cite{Vagnozzi:2019ezj}, dark radiation~\cite{Blinov:2020hmc}, early modified gravity~\cite{Braglia:2020auw} scenarios, or other fields contributing non-negligibly to the cosmic energy budget prior to recombination (see also~\cite{CosmoVerseNetwork:2025alb} and references therein) -- all of which reduce the sound horizon while inducing only indirect or parameter-degenerate effects in the spectral distortion signal. A prominent example involves a frozen scalar field acting as ``Early Dark Energy'' (EDE) at redshifts close to matter-radiation equality and rapidly diluting thereafter. EDE boosts the expansion history prior to recombination so as to reduce the sound horizon, leaving a specific signature on the CRR spectrum~\cite{Hart:2022agu} (see Sec.~\ref{CRR}). Moreover, to leave the CMB power spectra unaffected, such models generically predict an increase in the value of the primordial tilt $n_s$ that boosts the $\mu$-distortion signal from the dissipation of small-scale acoustic waves \cite{Lucca:2020fgp}. EDE models tend to prefer $n_s \simeq 1$ (as opposed to in $\Lambda$CDM where $n_s = 0.96$), which can lead to an increase in the Silk damping $\mu$-distortion by roughly a factor of $1.5$, as can be seen in Fig.~\ref{fig:BSM)_distortion}. As with the exotic recombination models, future CMB spectral distortion experiments will thus provide a complementary and independent test of EDE models and the opportunity to confirm -- or exclude -- these scenarios.

\begin{figure*}[t]
    \centering
    \includegraphics[width=0.8\columnwidth]{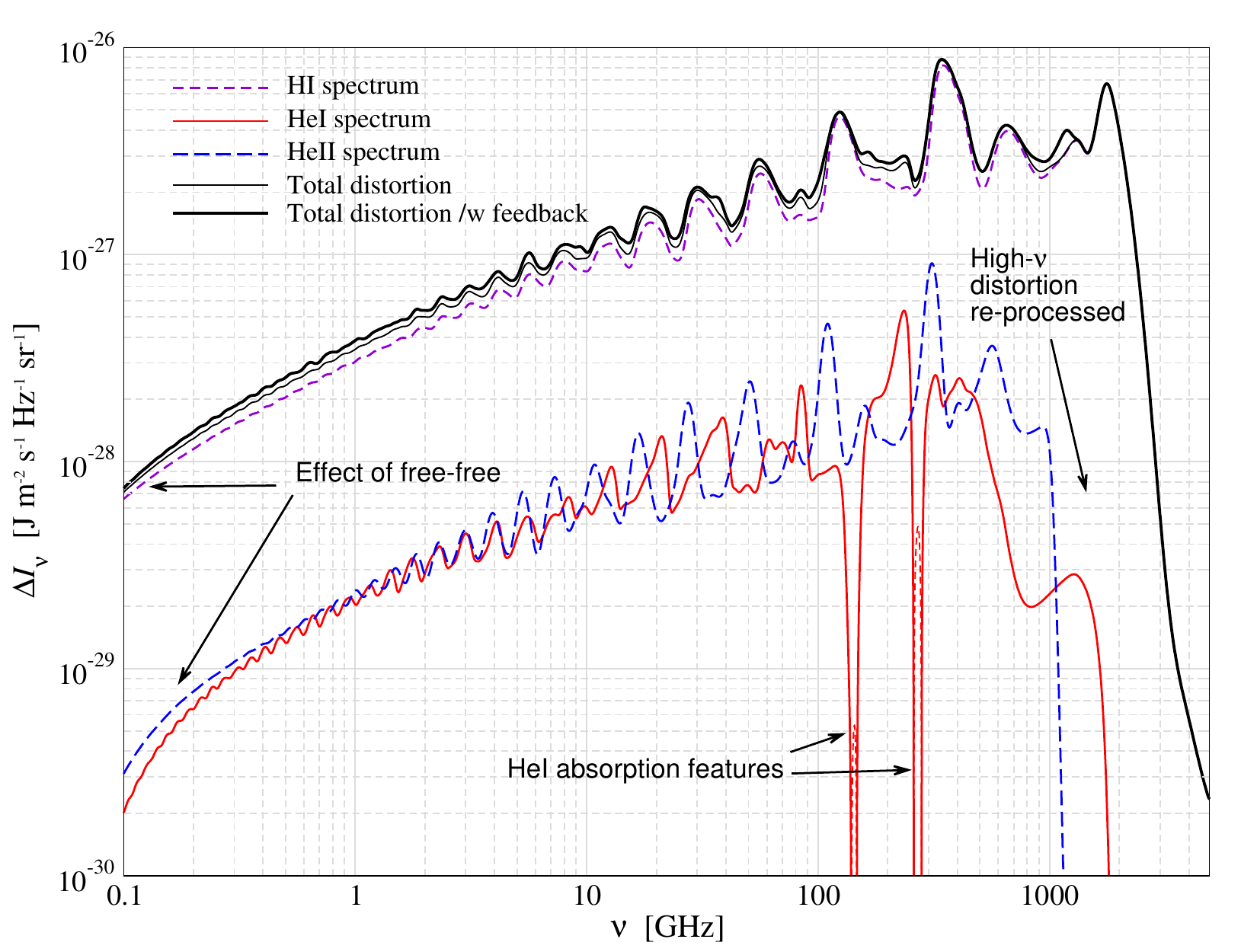}
    \caption{The cosmological recombination radiation (CRR) as obtained with {\tt CosmoSpec}. This {\it fingerprint} from the cosmological recombination process encodes valuable information about the time and duration of the {\it three} recombination epochs. The characteristic frequency pattern, extending over decades in frequency, is hard to mimic by foregrounds or instrumental effects and thus can principally be extracted using sensitive CMB spectrometer approaches \citep[e.g.,][]{Hart2020CRR}. Figure is taken from \citep{Chluba2016CosmoSpec}.}
    \label{fig:CRR}
\end{figure*}

\subsection{Cosmological Recombination Radiation} 
\label{CRR}
The process of cosmological recombination leaves a subtle but unavoidable distortion on the CMB due to the line emission from hydrogen and helium atoms. After redshifting from $z \approx 10^3$, this signal appears today as a faint but characteristic pattern of spectral features across the microwave bands \citep[e.g.,][]{Dubrovich1975,RybickiDell94,Burgin2003, Kholu2005,
Jose2006,Chluba2006,Jose2008,Yacine2013RecSpec,Chluba2016CosmoSpec}.  
This signal (see Fig.~\ref{fig:CRR}), known as the Cosmological Recombination Radiation (CRR), encodes direct information about the physical conditions and dynamics of the recombination era. 
Since it includes contributions from hydrogen as well as two distinct helium recombination epochs, the CRR probes periods of cosmic history that extend beyond the last-scattering surface observed through CMB anisotropies \cite{sunyaev2009signals, Chluba:2007zz, Hart:2022agu}. 

Understanding the recombination process is essential for the correct interpretation of CMB temperature and polarisation anisotropies \cite{Zeldovich68, Sunyaev:1970er, Peebles1970}. Testing the underlying physical assumptions during this epoch is a key objective of precision cosmology \cite{Hu1995, Lewis2006, Fendt2009, Jose2010, Shaw2011}. 
This makes the CRR a particularly powerful probe of new physics affecting recombination, including several proposed solutions to the Hubble tension (see Sec.~\ref{H0tension}).
In particular, the \FOSSIL satellite forecasts a sensitivity on $A_{\rm CRR}$, the CRR amplitude [$\sigma(A_{\rm CRR}) \simeq 0.8$], to detect for the first time the Balmer-$\alpha$ recombination line from $z\approx 1300$, providing a unique observational test of our theoretical description of the early Universe. Broader perspectives on the observational challenges can be found here \citep{Vince2015,
sathyanarayana2015detection, Hart2020CRR}

\subsubsection{Non-standard recombination}
Modifications to recombination could be envisaged in several cases. If fundamental constants vary during the recombination epoch, this alters the energy levels of the hydrogen atom and induces small changes in the scattering rates~\cite{Hart:2017ndk,Sekiguchi:2020teg,Lee:2022gzh,Greene:2023cro,Baryakhtar:2024rky,Lynch:2024hzh,Schoneberg:2024ynd,Mirpoorian:2024fka,Toda:2025kcq}. If PMFs are present during recombination, additional baryonic clumping may occur, which accelerates the recombination process~\cite{Jedamzik:2020krr,Thiele:2021okz,Rashkovetskyi:2021rwg,Galli:2021mxk}. Dark matter annihilations and interactions with the plasma can also impact the recombination process leaving potentially detectable spectral imprints on the CRR \citep[see e.g.,][]{CosmoVerseNetwork:2025alb, Chluba2016CosmoSpec}. In addition to modifications during recombination, a transient period of early ($z \sim z_{\rm eq}$) dark energy (EDE), in which the background expansion of the Universe is briefly accelerated, would leave specific imprints on the CRR~\cite{Hart:2022agu}, and can also boost the Silk damping $\mu$ distortion by a factor of 1.5 (see Fig. \ref{fig:BSM)_distortion}), as the model requires more power on small scale modes in order to match other observational constraints~\cite{Lucca:2020fgp}. Several of these scenarios have been proposed as early-Universe solutions to the Hubble tension (see Sec.~\ref{H0tension} and Ref.~\cite{DiValentino:2024yew} for a recent review). Although many of them can produce similar reductions of the sound horizon, they generally leave distinct signatures in the CRR, making CRR observations a powerful complementary probe for distinguishing between competing models. 

\subsubsection{Resonant scattering signals}
Interactions between CMB photons and matter can introduce frequency-dependent signatures through resonant line emission \cite{Loeb2001,Zaldarriaga2002, Jose2005,Carlos2005,Carlos2006,Carlos2007,Carlos2017}, Rayleigh scattering effects \cite{Yu2001,Lewis2013}, and collisional emission processes \cite{Carlos2017,Righi2008b,Hernandezetal2007b,Gongetal2012}. These signals provide independent probes of the recombination era, the dark ages, and reionization. Detecting such frequency-dependent features, even on large angular scales, is challenging due to the accuracy required for inter-channel calibration. Because of this reason, they are clear targets for spectrometers. 

In particular, some of these signals associated to the cosmological hydrogen recombination epoch could be detected with spectrometers providing moderate angular resolution ($\ell \lesssim 300$). For instance, the resonant scattering of CMB photons through the H$\alpha$ line during cosmological recombination \cite{Jose2005,Carlos2007} could be measured if the accuracy of the relative calibration between bands is below $0.1$\,$\mu$K.
Such a detection would provide a proof of concept for the methodology, which could then be extended to other spectral lines (e.g. P$\alpha$) and to new polarization observables (e.g. TE and EE signals associated with H$\alpha$) in future missions combining CMB spectroscopy and high-resolution imaging capabilities.

\subsection{Improved determination of $T_0$}
\label{sec:T0}

\begin{figure}
    \centering
    \includegraphics[width=0.55\linewidth]{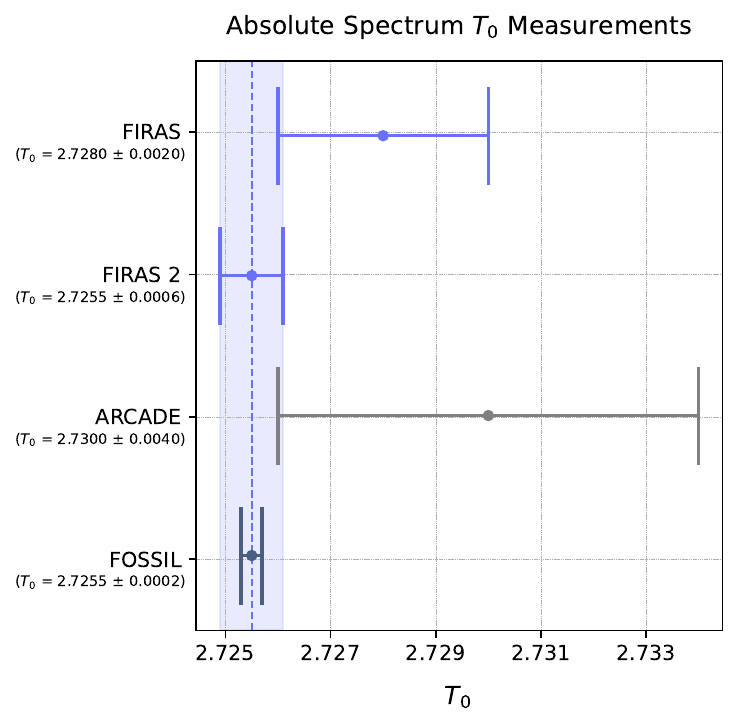}
    \caption{Whisker plot of absolute CMB temperature measurements, showing FIRAS (1996, 2009) \cite{Fixsen:1996nj,Fixsen:2009ug} in blue, ARCADE (2011) \cite{Fixsen:2009xn, Seiffert:2011} in gray, and the \FOSSIL forecast in navy, with an improvement in the error of about one order of magnitude. Points denote monopole $T_0$ values with $1\sigma$ uncertainties. The blue band marks the FIRAS 2009 re-analysis point for reference.}
    \label{fig:T0}
\end{figure}

The present-day temperature, $T_0$, is nominally the most precisely determined cosmological parameter. Still, despite the extraordinary progress in CMB anisotropy measurements over the past three decades, the most precise cosmological determination of $T_0$ still comes from the FIRAS instrument aboard \COBE, $T_0 = 2.72548 \pm 0.00057$ K \citep{Fixsen:1996nj,Fixsen:2009ug}. Direct local (but extrasolar) values of $T_0$ can be obtained from observations of interstellar molecular clouds in the Milky Way \citep{RothMeyer} and the Magellanic Clouds, from the low rotational level transitions of cyanogen (CN). These extrasolar and extragalactic values are in good agreement with the FIRAS estimation, but have significantly larger uncertainties. Thus, cosmological models predicting at $z=0$ non-Planckian spectra of the background radiation can be ruled out from local data (and without model assumptions) if they deviate by percent level from the blackbody spectrum. However, these local observations do not allow to distinguish between the standard model and cosmological models with a blackbody spectrum but different $T(z)$ dependencies.

A fundamental prediction of the hot Big Bang theory is that after electron-positron annihilation, the temperature of the cosmic microwave background (CMB) scales with redshift as
\begin{equation} 
T_{\rm CMB}(z) = T_0(1+z) \, ,
\label{eq:tz_scal} 
\end{equation}
under the assumptions of an initial blackbody spectrum and following photon number conservation and adiabatic cooling in an expanding Universe~\cite{Peebles:1994xt}. This simple relation links the present-day CMB monopole temperature $T_0$ to the thermal history of the Universe at much earlier times. In this sense, the CMB acts as a cosmic chronometer, primarily determined by the monopole calibration. For the purpose of this relation, measurements of $T_0$ complement measurements of the temperature at non-zero redshifts, for which two main observational windows are currently available. In the lower range $z\leq1$, $T_{\rm CMB} (z)$ can be estimated through measurements of the Sunyaev-Zel'dovich (SZ) effect in galaxy clusters~\cite{Sunyaev:1970er,SunyaevZeldovich:1972,2002ApJ...580L.101B,Luzzi:2009ae,Gelo:2022vsa,SPT:2013gam, Hurier:2013ona,Planck:2018nkj,ACT:2020gnv,SPT-3G:2022hvq,2021ApJ...922..136L} while high-resolution spectroscopy of molecular or atomic excitation lines in distant astrophysical environments enable measurement in the typical range $1\leq z\leq3.5$ \cite{2009IEEEP..97.1463W,NOEMA,Gelo:2022vsa,Srianand:2008fz,Noterdaeme:2010gv,Noterdaeme:2010tm,Noterdaeme:2016ykw,Muller:2012kv,Cui:2005qw,2001ApJ...547L...1G,Molaro:2001jv, Klimenko:2021wti,Srianand:2000wu,Riechers:2022tcj}. These methods probe different redshift ranges and also different astrophysical environments, possibly highlighting the presence of new physics at different cosmological epochs. 

Although this linear scaling is robust within standard $\Lambda$CDM cosmology, deviations could arise in scenarios where photon number is not conserved or where adiabatic evolution is violated, such as models with energy injection/removal processes, photon interactions with light particles, late-time entropy production, vacuum energy decay, or even more general modifications of gravity (see, \textit{e.g.}, Refs.~\cite{Ellis:2013cu,Bassett:2003vu,Santana:2017zvy,Azevedo:2021npm,Cai:2015emx,Cipriano:2024jng,Avgoustidis:2010ju,Csaki:2001yk,Bassett:2003zw,Chen:1994ch,Cillis:1996qy,Raffelt:1987im,Menard:2009yb,Khoury:2003aq,Khoury:2003rn,Burrage:2007ew,Barrow:1999is,Lee:2021xwh,Goncalves:2019xtc}). Therefore, testing the temperature–redshift relation provides a clean and largely model-independent probe of fundamental physics. However, several common astrophysical processes can also be erroneously interpreted as non-standard physics. Moreover, in models where this relation is violated the physical mechanisms behind the deviations from the standard evolution will, in general, create spectral distortions. This can be used as a consistency test for new physics.

\subsubsection{Measuring $T_0$: from FIRAS to future facilities} 

The FIRAS instrument aboard \COBE measured the CMB monopole temperature to be $T_0 = 2.72548 \pm 0.00057\,\mathrm{K}$ \cite{Fixsen:1996nj,Fixsen:2009ug}, a value that remains the reference for absolute CMB thermometry. The uncertainty of this measurement was dominated by systematic effects associated with the external blackbody calibrator, rather than by statistical noise. In particular, calibrator thermometry ($\sim 0.5-0.7\,$mK), calibrator emissivity ($\sim 0.2-0.3\,$mK), and statistical uncertainty ($< 0.1\,$mK) limited the measurement to the mK level.

Future facilities, under development or being planned, will play a pivotal role by providing the first major update of the CMB monopole temperature since FIRAS. This will lead to significant improvements on other cosmological sectors, examples of which are discussed in Sect. \ref{casestudy1}. It will also mildly impact temperature-redshift tests \cite{Ruchika:2025sbb,Luzzi:2009ae, Avgoustidis:2011aa,Lima:2000ay, LoSecco:2001zz, Hofmann:2023wyk, Euclid:2020ojp, SPT:2013gam, Bengaly:2020vly,Avgoustidis:2015xhk,de_Martino_2012,DeBernardis:2006ii,Chluba:2014wda}, although evidently these are not bottlenecked by the $T_0$ uncertainty. Furthermore, the foreseen broad frequency ranges enable control of low-frequency foregrounds, particularly relevant for interpreting claims such as the ARCADE radio excess \cite{Fixsen:2009xn, Seiffert:2011}. Clarifying whether this excess results from astrophysical foregrounds or points to new physics affecting the thermal history of the Universe requires a precise absolute spectral measurement \cite{Fornengo:2011cn,Singal:2017jlh}. 

Facilities being planned benefit from three decades of technological progress in cryogenic metrology, detector stability, and calibrator design. In particular, thermometer systems with long-term stability and absolute accuracy better than $10\,\mu\mathrm{K}$ at temperatures near $2.7\,\mathrm{K}$ are required. At the same time, external calibrators must achieve emissivities exceeding $1-10^{-6}$ across the relevant frequency range, with precisely characterised spectral and angular dependence. The control of internal thermal gradients at the few-microkelvin level and the minimization of stray light and parasitic reflections are similarly critical \cite{Kogut2011}. Meeting these requirements would enable a new generation of absolute CMB measurements with an order-of-magnitude improvement in precision (see Fig. \ref{fig:T0}), substantially reducing systematic uncertainties and providing a robust metrological foundation for future studies of early-Universe physics.

\subsubsection{Case study: Big Bang nucleosynthesis and recombination history}\label{casestudy1}

Big Bang Nucleosynthesis (BBN) describes the production of the light elements (${}^4$He, D, ${}^3$He, ${}^7$Li) during the first few minutes of the Universe, when the expansion rate was governed by the radiation energy density and nuclear reaction rates operated at $T \sim \mathrm{MeV}$ \cite{Pitrou:2018cgg}. An accurate measurement of the CMB temperature is crucial to BBN physics in several ways. The present-day CMB temperature determines the photon number density today, $n_\gamma = \big(2\zeta(3)/\pi^2\big)T_0^3$, while BBN is very sensitive to the baryon-to-photon ratio, $\eta=n_b/n_\gamma$. A measurement of $T_0$ thus allows to convert the determination of $\eta$ into a measurement of $\omega_b= m_pn_b$ (with $m_p$ the proton mass). When compared with the CMB determination of $\omega_b$, it allows for a stringent test of the standard model of particle physics and cosmology. Second, it sets the radiation density $\rho_{\rm rad} = \rho_\gamma(1+0.2271 N_{\rm eff})$ that dominates the expansion rate $H\propto \sqrt{\rho_{\rm rad}}$ at BBN times. Thus, a precise measurement of $T_0$ anchors the expansion history normalization. Finally, it also determines the neutrino temperature, $T_{\nu,0}=\big(4/11\big)^{1/3}T_0$, which alters both the neutrino energy density (and the expansion rate) and the weak interaction rates that dictate the proton-neutron ratio, a key quantity in setting the primordial helium abundance $Y_p$. A novel, more accurate, measurement of $T_0$ would thus allow to further refine BBN prediction, and combined with CMB constraints, provide us with a stringent test of exotic physics such as unstable heavy relics leading to late energy and entropy production \cite{Kawasaki:1994sc,Poulin:2015opa, Bolliet:2020ofj}, photon–dark sector interactions \cite{Fradette:2014sza,Li:2020roy,Chluba2024,Caputo:2025avc}, violations of adiabatic evolution \cite{Evans:2019jcs,Sobotka:2022vrr} and non-standard expansion history \cite{Allahverdi:2020bys,Sobotka:2023bzr}.

\begin{figure}
    \centering
    \includegraphics[width=0.55\linewidth]{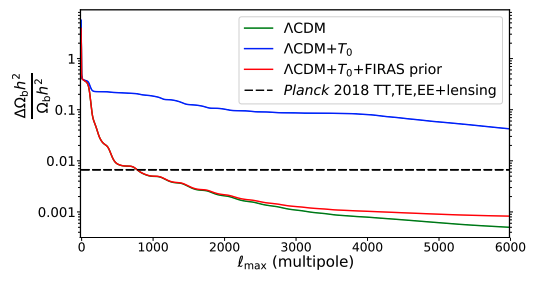}
    \includegraphics[width=0.4\linewidth]{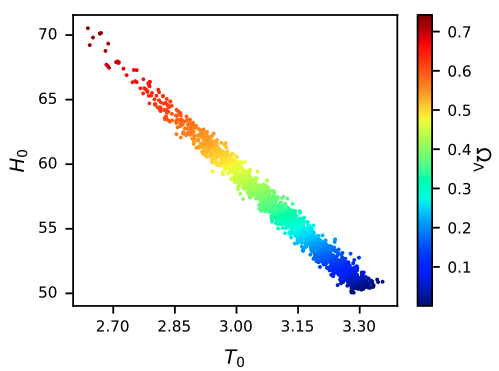}
    \caption{left panel: Estimated Fisher-matrix uncertainties for $\Omega_bh^2$ as a function of the maximum $\ell$, using lensed TT, TE, and EE power spectra under the assumption of the cosmic-variance limit. Errors on $T_0$ from FIRAS degrade the precision on $\Omega_bh^2$ by $50\%$ for future experiments. Right panel: correlation between $H_0$ and $T_0$ under $\Lambda$CDM when analyzing {\it Planck} data. Taken from Ref.~\cite{Wen:2020txi}.}
    \label{fig2}
\end{figure}
    
Another important role played by $T_0$ is in setting the physics of CMB anisotropies. The determination of all cosmological parameters is intrinsically tied to an accurate determination of $T_0$ \cite{Chluba:2007zz,Hamann:2007sk,Ivanov:2020mfr,Wen:2020txi,Bose:2020cjb}.  First, the era of recombination is sensitive to physical energy scales  at fixed baryon and dark matter densities per photon , $\bar{\omega}\equiv \omega_i/T_0^3$. Just like for BBN studies, an accurate measurement of the temperature fixes the absolute photon energy and number densities ($\rho_\gamma \propto T_0^4$, $n_\gamma \propto T_0^3$). This is key to translating relative constraints on baryon and CDM energy densities into absolute ones, and therefore to guiding our DM model-building efforts. In fact, current uncertainties in $T_0$ will impact the uncertainties on $\omega_b$ by up to $50\%$ for future CMB experiments, which plan to measure temperature anisotropies up to $\ell \sim 5000$ with high-accuracy. This is illustrated in Fig.~\ref{fig2}, left panel. Second, the normalization of the transfer function of CMB anisotropies, which are relative measurements with respect to $T_0$, explicitly depends on the product $A_sT_0^{n_s-1}$, assuming a power-law dependent primordial power spectrum.  Hence, accurate measurement of $T_0$ is relevant to inflationary physics. Third, the mapping between a physical sound horizon at last scattering and the observed angular acoustic scale depends on the distance to the last-scattering surface, $d_{\rm LSS}$. As $T_0$ sets the absolute energy scale entering the expansion rate and $d_{\rm LSS}$, it leads to a tight correlation between $H_0-T_0$ only weakly broken by late integrated Sachs-Wolfe effect and gravitational lensing, suggesting that $T_0$ may play a role in the ``Hubble tension'' between early- and late-time determinations of the Hubble constant \cite{Ivanov:2020mfr,Bose:2020cjb,ACT:2025tim}. This is illustrated in Fig.~\ref{fig2}, right panel. A novel, accurate measurement of $T_0$ would thus allow us to firmly rule out a mis-determination of $T_0$ as the origin of the tension, while sharpening constraints on exotic recombination scenarios, cosmological and inflationary parameters, and exotic post-BBN/post-CMB thermal histories.

Beyond anisotropies, $T_0$ plays a central role in CMB spectral distortions. Since distortions are defined as deviations from a perfect blackbody spectrum, their interpretation requires precise knowledge of the monopole temperature that sets the spectral baseline. In particular, the cosmological recombination radiation (CRR) spectrum depends explicitly on $T_0$, as the frequency and amplitude of the predicted line features scale with the absolute photon temperature, implying that uncertainties in $T_0$ propagate directly into predicted distortion signals \cite{Chluba:2007zz}. More generally, any $\mu$- or $y$-type distortion generated by energy release in the early Universe is measured relative to the blackbody defined by $T_0$. An accurate determination of $T_0$ is essential for disentangling primordial energy injection, late-time heating, or exotic thermal histories from instrumental calibration effects. Just as for recombination and CMB anisotropies, a precise measurement of $T_0$ anchors the absolute thermodynamic scale of the CMB and strengthens the constraining power of future spectral distortion experiments.
\section{The Late Universe}
Following recombination, the Universe transitions from a nearly uniform plasma into the rich hierarchy of structures we observe today, and the CMB spectrum evolves alongside it. While the primordial signals of the previous section may sometimes require sensitivities orders of magnitude beyond current limits, the late-time distortions are comparatively bright: the thermal Sunyaev-Zel'dovich effect guarantee a Compton-$y$ distortion at the level of $\langle y \rangle \simeq 10^{-6}$, with relativistic corrections hiding just a bit deeper \citep{Hill2015}. The challenge accordingly shifts from raw detection to decomposition. The science to be done in this epoch typically shifts constraining exotic physics to performing a census of the thermal and radiative content of the Universe. In this section, we describe the late-time science accessible to an absolute spectrometer, moving from the thermal history of cosmic gas to the broadband backgrounds it will map.

In Sec.~\ref{sec:SZ_effect}, we discuss the thermal Sunyaev-Zel'dovich effect, whose sky-averaged amplitude directly probes the thermal energy stored in cosmic gas. A high-significance measurement of $\langle y \rangle$ and its relativistic correction (yielding the mean electron temperature, $\langle T_{\rm e} \rangle \sim {\rm keV}$) would deliver percent-level constraints on the supernova and AGN feedback processes that increasingly limit weak lensing cosmology. Sec.~\ref{sec:CD_and_reionization} extends this picture to the diffuse intergalactic medium. There, reionization contributes a modest ($y_{\rm reion} \sim 10^{-7}$) yet unavoidable sub-component, while at Cosmic Dawn, low-frequency distortions modify the global 21cm signal both directly and through soft photon heating of the intergalactic gas. Moreover, the unexplained radio synchrotron background, several times brighter than known source populations can accommodate, is the subject of Sec.~\ref{sec:RSB}. An absolute measurement at frequencies below $\sim 50\,$GHz would cleanly discriminate among its proposed origins.

The same instrument that measures these distortions simultaneously functions as a full-sky, absolutely calibrated spectral survey. As discussed in Sec.~\ref{sec:LIM}, broad frequency coverage captures the CO rotational ladder and the \cii{} line across most of cosmic history, enabling line-intensity mapping of large-scale structure from the Epoch of Reionization to the present. Sec.~\ref{sec:cosmic_noon} focuses this capability on cosmic noon ($1 \lesssim z \lesssim 3$), where a volume-averaged accounting of molecular gas and star formation would help resolve long-standing questions about the cosmic baryon cycle. Finally, Sec.~\ref{Sec:CIB} discusses the cosmic infrared background, for which absolute measurements of the monopole and dipole at $\nu \lesssim 400\,$GHz would anchor the history of dust-obscured star formation out to $z \gtrsim 5$. These targets demonstrate that the reach of CMB spectroscopy extends well beyond the primordial Universe, providing a detailed thermal and radiative inventory of cosmic structure formation.

\subsection{The Sunyaev-Zel'dovich effect}
\label{sec:SZ_effect}
\begin{figure}
    \centering
\includegraphics[width=.65\textwidth]{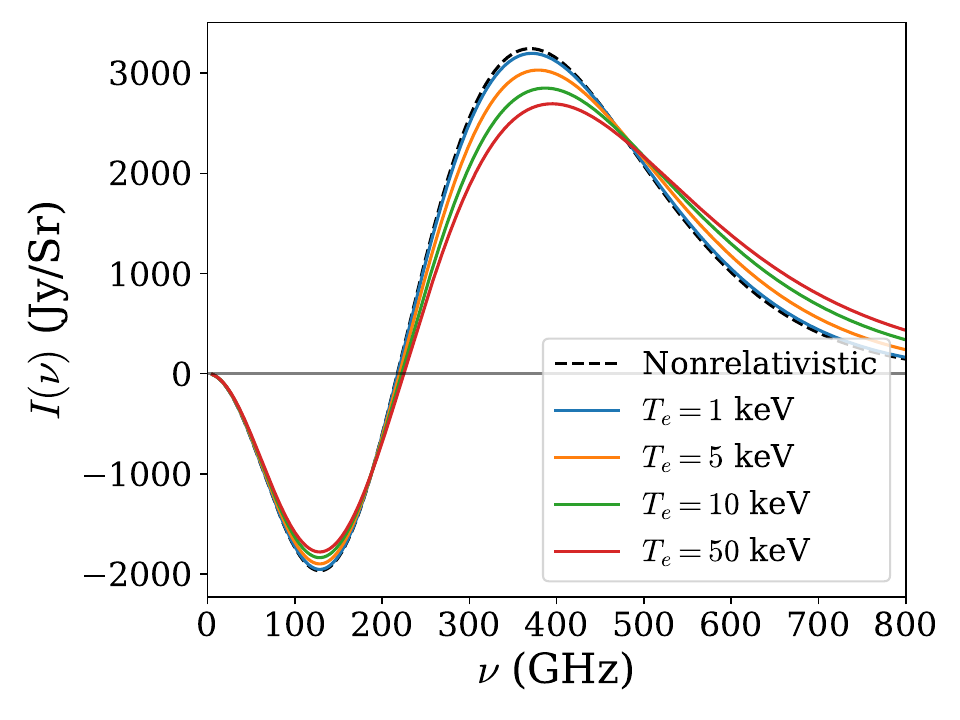}
    \caption{The frequency dependence of the relativistic thermal SZ signal (in Jy/sr) for a range of different electron temperatures.}
    \label{fig:responses} 
\end{figure}

In the late-time Universe Compton scattering of CMB blackbody photons off hot ionized electrons produces a spectral distortion~\cite{Zeldovich1969,
SunyaevZeldovich:1972}, known as the Sunyaev Zel'dovich effect (SZ). The SZ effect comes in two main flavors, the kinematic SZ effect, caused by the clusters bulk motion with respect to the CMB, and the thermal SZ effect (tSZ) due to the random thermal motion \citep[see][for broad review]{Mroczkowski2012}.
The tSZ signal is typically decomposed into two contributions: a non-relativistic Compton-$y$ signal and a relativistic Compton-$y$ signal \cite{Challinor_1998,Itoh98,Nozawa_1998, Chluba2012SZpack}. The non-relativistic signal is sourced as the line-of-sight integrated electron pressure,
\begin{align}\label{eq:compton-y}
 y^\text{nonrel}(\mathbf{n}, \nu) = g(\nu)\frac{\sigma_T}{m_e c^2}\int\mathrm{d}l\, P_e(\mathbf{n},l)=g(\nu)y(\mathbf{n})\,,
\end{align}
whilst the relativistic component carries an additional temperature-dependent weight,
\begin{align}\label{eq:T_e-y}
    y^\text{rel}(\mathbf{n}, \nu) = \frac{\sigma_T}{m_e c^2}\int\mathrm{d}l\, \tilde{g}(\nu,T_e(\mathbf{n},l)) P_e(\mathbf{n},l)\,,
\end{align}
where $\sigma_T$ is the Thomson scattering cross section, $m_e$ is the electron mass, $T_e$ is the electron temperature, and $P_e$ is the electron pressure. The spectral signature of the two components is distinct. For the non-relativistic signal the frequency function is 
\begin{align}
    g(\nu)=\frac{2(k_BT_\mathrm{CMB})^3}{(hc)^2}\frac{x^4e^x}{(e^x-1)^2}\left[x\coth(x/2)-4\right]
\end{align}
where $x=h\nu/k_BT_\mathrm{CMB}$. The relativistic response $\tilde{g}(\nu,T_e(\mathbf{n},l))$ can be computed numerically using packages like \textsc{szpack} \citep{Chluba2012SZpack, Chluba2012moments}. The spectral responses are shown in Fig.~\ref{fig:responses}.
\FOSSIL will measure the monopoles of the tSZ signals,
\begin{equation}
    \langle y^\text{nonrel, rel}(\nu) \rangle \equiv \int \frac{d^2\mathbf{n}}{4\pi} y^\text{nonrel, rel}(\mathbf{n}, \nu)\,.
\end{equation}
The mean Compton-$y$ signal is expected to be $\langle y \rangle\simeq 1\times 10^{-6}$ with percent level relativistic corrections. \FOSSIL will have a 200 $\sigma$ measurement of the non-relativistic signal and detect the relativistic correction at 20 $\sigma$. Measuring the relativistic effect through anisotropies is very difficult with the highest-SNR measurements obtained from stacks of bright clusters at $\sim 3-4 \sigma$ \cite{Remazeilles_2024,n7p5-pc66}. \FOSSIL's measurement would represent an 5 factor improvement. Measurement of the relativistic effect gives us access to the Compton-$y$ weighted electron temperature, defined as 
\begin{align}
    \langle T_e\rangle = \frac{1}{\langle y \rangle}\int \frac{\mathrm{d}^2\mathbf{n}}{4\pi}\frac{\sigma_T}{m_e c^2}\int\mathrm{d}l\, T_e(\mathbf{n},l)P_e(\mathbf{n},l)\,.
\end{align}
The theoretical expectation for this quantity is roughly $\langle T_e \rangle \simeq \text{keV}$ with substantial uncertainty.

\subsubsection{Scientific background}
The average Compton-y is a direct probe of the thermal energy content of the Universe, $E^{th}$ - this arises as it is simply the integral of the electron pressure over the volume of the Universe \cite{Hill2015}. The total electron thermal energy can be decomposed into different contributions
\begin{align}
E^{th} = E^\mathrm{collapse}+E^\mathrm{injected}-E^\mathrm{cooling}.
\end{align}
The first term captures the heating of cosmic gas as it falls into gravitational potentials and coverts that energy into thermal energy through collisions, turbulent dissipation and shocks. The second term captures energy injected through a diverse range of other processes; the key cosmological ones are heating of gas through the radiative and mechanical energy released from supernovae and from quasar- and radio- mode active galactic nuclei (AGN) - known as supernova  and AGN feedback. The final term accounts for cooling of the cosmic gas, through expansion and radiative processes. Whilst the first and last term are well understood, the amount of energy injected by feedback is poorly constrained. It has long been known that without significant feedback energy clusters and groups would experience strong cooling flows, which are not observed; however, the strength, efficiency and coupling mechanisms of these feedback processes is poorly characterized \cite[e.g.][]{Valentini_2025}. 

Gaining new insights into how feedback heats and displaces cosmic gas is vital for understanding the formation and evolution of galaxy groups and clusters; these processes shut off star formation in galaxies, heating up and expelling gas into the intergalactic medium.  In recent years, there has been great interest in this question from cosmologists as uncertainties in the gas properties are becoming a limiting systematic for cosmological probes such as weak lensing \cite[e.g.][]{Bigwood_2024,Pranjal_2025}. 

Accurately modeling supernova and AGN feedback processes in cosmological simulations is very challenging to the vast dynamical scales between the scales of supernova shocks and AGN jets, and cosmological scales. Typically these processes are included via subgrid models, prescriptions to include the impact of the unresolved processes on the cosmic scales; however, the lack of ab initio modeling means that the subgrid models need calibrating. 

Observationally measuring the density and temperature of diffuse gas is difficult, especially in low density and lower temperature regions. Recent advances, such as measurements of gas properties via the eROSITA and ACT \cite{Siegel_2025,Hadzhiyska_2025}, have suggested that the strength of feedback could be much larger than suggested by existing cosmological simulations. 
\subsubsection{Extracting information from the thermal SZ monopole}
\begin{figure}
    \centering
\includegraphics[width=.45\textwidth]{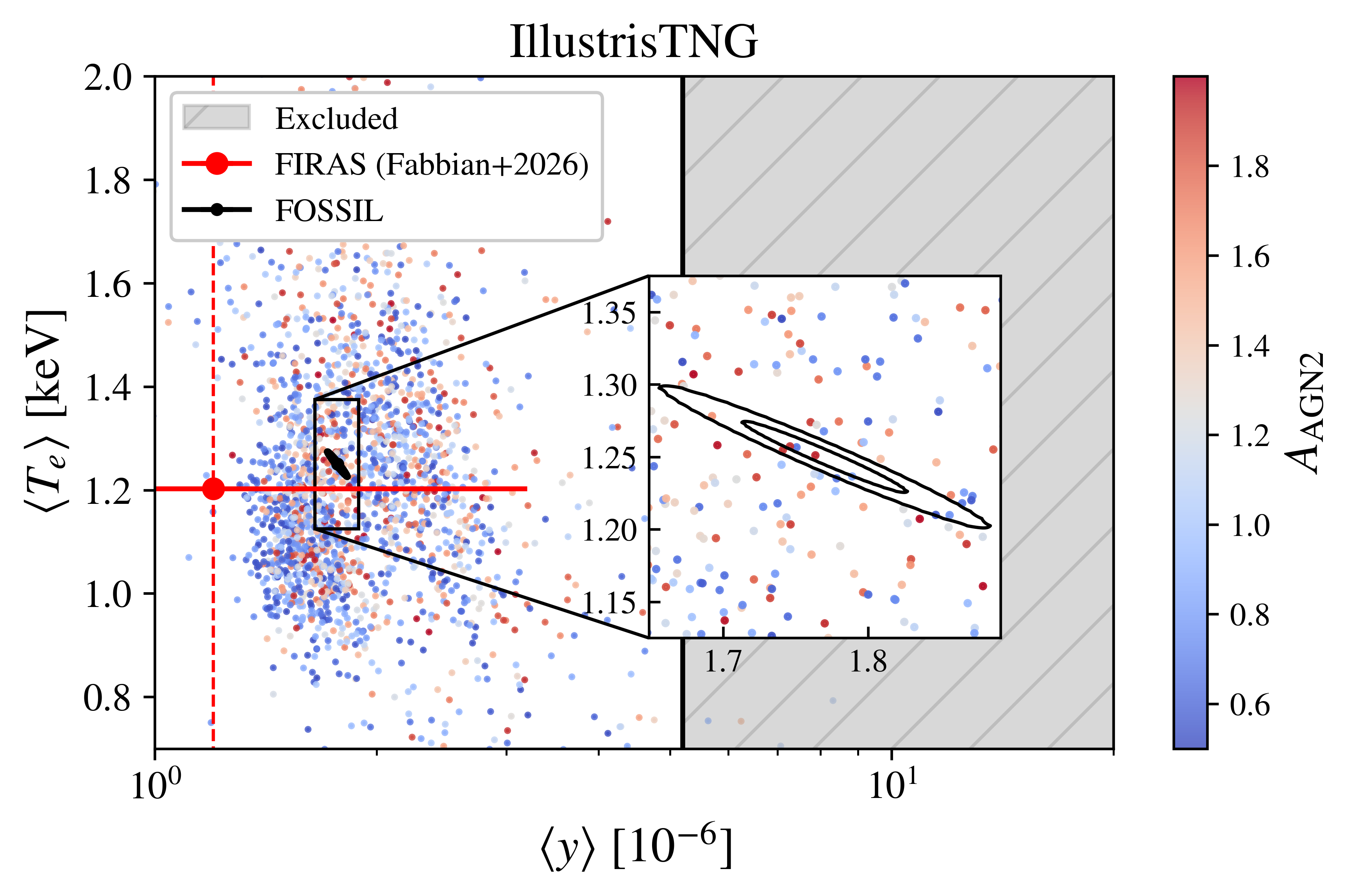}
\includegraphics[width=.45\textwidth]
{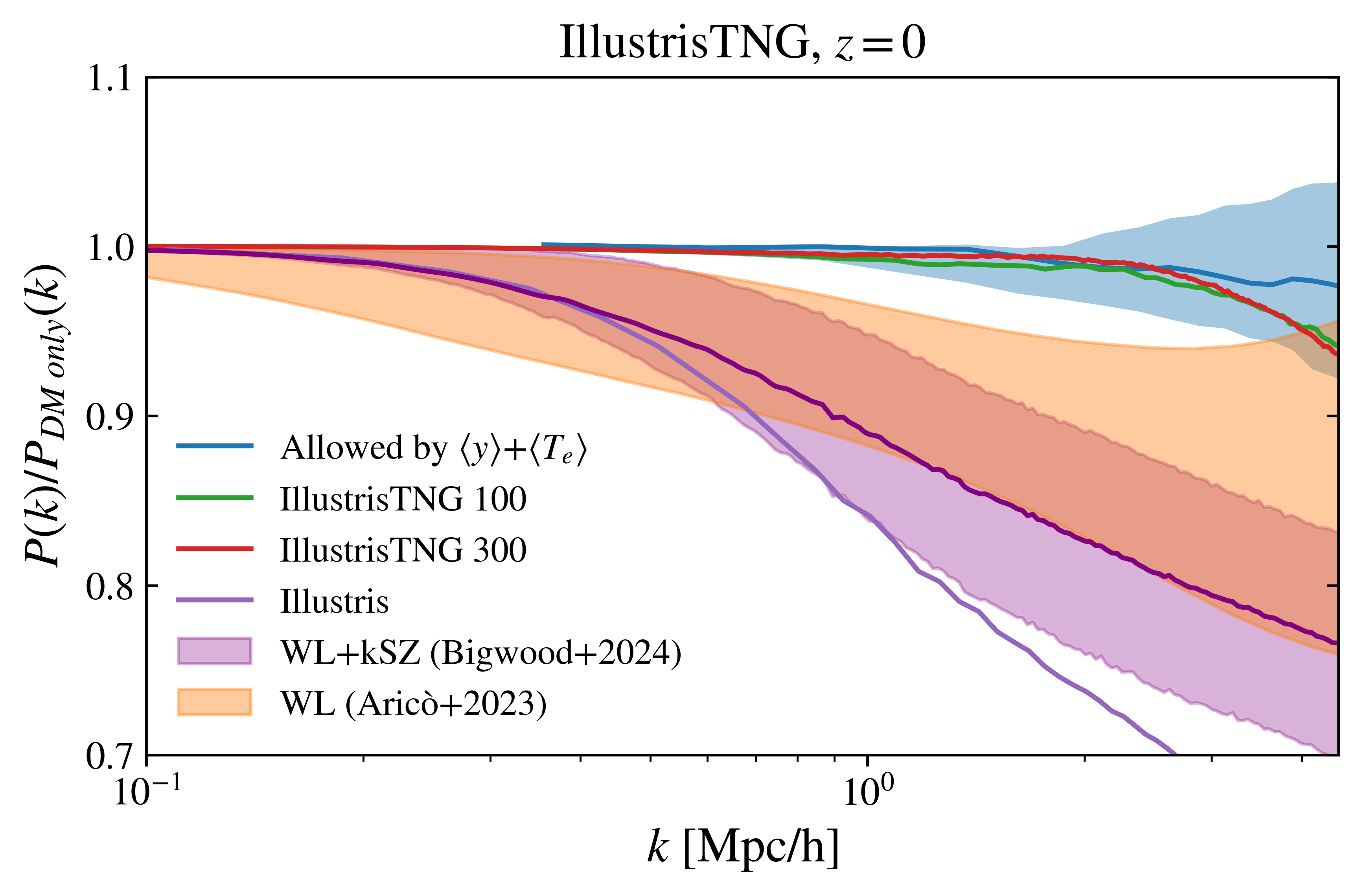}
    \caption{Top left: Predictions of $\tmono$, $\ymono$ based on the SIMBA galaxy formation model. Each points show results for a single simulation of the CAMELS suite color coded by values of the $A_{AGN2}$ parameter describing the velocity of AGN jets. The state of the art measurements from \cite{Fabbian_2025} are shown in red. As $\tmono$ is unconstrained by the data, we fix it to the median of the predictions. The excluded area at 95\% C.L. is shown in grey. Top right: Same as the left panel for the IllustrisTNG feedback model, which predicts weaker feedback effects compared to SIMBA. In both SIMBA and IllustrisTNG  models, \FOSSIL will be able to pinpoint the physical mechanisms underlying feedback with high precision. Bottom: forecast for the matter power spectrum suppression as constrained by \FOSSIL measurements of $\ymono$ and $\tmono$ at 68\% C.L.  (blue). We used the CAMELS IllustrisTNG as simulated data and show the allowed values currently constrained from DES cosmic shear alone (orange) and with kSZ data (purple). The fiducial suppression of the Illustris , IllustrisTNG-100 and IllustrisTNG- 300 simulations in solid lilac, green and red respectively. \FOSSIL will allow to set high-precision constrain on the baryon-induced suppression at high precision even for low-feedback model such as those implemented in IllustrisTNG.}
    \label{fig:AGN2-ASN2} 
\end{figure}
Extensive suites of simulations, with different strength and implementations of feedback processes, has been revealed the astrophysical information accessible in mean Compton-$y$ and $T_e$. Fig.~\ref{fig:AGN2-ASN2} demonstrates how varying the speed of supernova winds (denoted $A_\text{ASN2}$) and jet-speeds (denoted $A_\text{AGN2}$) leads to significant shifts in the mean Compton-$y$ and electron temperatures \cite{Villaescusa-Navarro-2021,Fabbian_2025}. \FOSSIL constraints will provide precision, i.e. percent level, constraints on the subgrid models characterizing these feedback processes. Equivalently, these constraints on feedback strengths can mapped into the space of cosmological observations - namely the suppression on the matter-power spectrum. These constraints are also shown in Fig.~\ref{fig:AGN2-ASN2} and show that the matter power spectrum suppression can be measured at the few percent level - competitive with upcoming approaches but providing access to lower mass and higher redshift objects. 
We can decompose the Compton-$y$ signal, Eq.~\ref{eq:compton-y}, into the different mass and redshift contributions utilizing the halo model formalism as
\begin{align}
    \langle y \rangle = \frac{\sigma_T}{m_e c^2}\int{\mathrm{d}z} \frac{(1+z)^2}{4\pi H(z)}\int\mathrm{d}m \frac{\mathrm{d}n}{\mathrm{d}m} P_e(m,z)
\end{align}
and similarly for the mean electron temperature, as 
\begin{align}
    \langle T_e \rangle = \frac{\sigma_T}{m_e c^2 \langle y \rangle}\int{\mathrm{d}z} \frac{(1+z)^2}{4\pi H(z)}\int\mathrm{d}m \frac{\mathrm{d}n}{\mathrm{d}m} P_e(m,z) T_e(m,z).
\end{align}
\begin{figure}[h]
    \centering
\includegraphics[width=.5\textwidth]{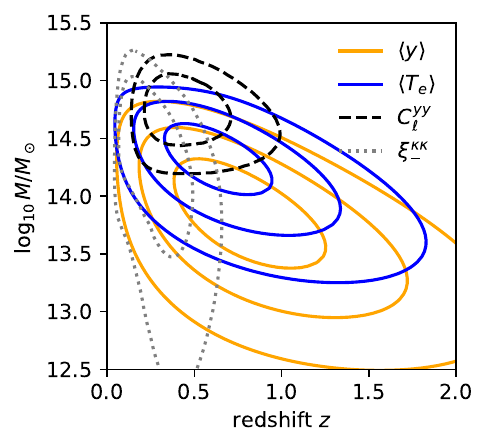}
    \caption{Contours of the relative differential contributions to the mean Compton-$y$ and mean electron temperature, $T_e$, as a function of halo mass and redshift. We compare the sensitivity of the ACT/SPT tSZ power spectrum $C_\ell^{yy}$ (at $\ell \sim 3000$) and DES weak lensing correlation function $\xi_-^{\kappa\kappa}$ as computed by \cite{Lucie-Smith-2025}.}
    \label{fig:differential_contributions} 
\end{figure}

These relations highlight that, in the case where the pressure-halo and temperature-halo relations are well understood, the mean Compton-$y$ and $T_e$ encode cosmological information through the halo mass function. For the studies of feedback, the halo mass function helps dictate which masses and redshift contribute to the two signals. The mass and redshift dependence is shown in Fig.~\ref{fig:differential_contributions}. Interestingly the Compton-$y$ signal is more sensitive to lower mass and higher redshift objects than other probes \cite{Thiele_2022}. This highlights the importance of \FOSSIL's gas thermodynamics measurements: it will give access to heating processes on otherwise inaccessible mass and redshift ranges.  The mass dependence of gas temperatures, approximately $T_e\propto M^{\frac{2}{3}}$, explains why the gas temperature is more sensitive to high masses (and low redshifts where such objects are more abundant). 

\subsection{Probing Cosmic Dawn and the Epoch of Reionization} \label{sec:CD_and_reionization}
The thermal Sunyaev-Zel'dovich (SZ) effect discussed in Sec.~\ref{sec:SZ_effect} is the dominant source of the late-time $y$-distortion, produced primarily by the hot ($k_{\rm b} T_{\rm e} \simeq$ few keV) gas bound in collapsed structures such as galaxy groups and clusters. A smaller but guaranteed contribution to the sky-averaged signal arises from the diffuse intergalactic medium (IGM), which is photoheated to $T_{\rm e} \sim 10^4$ K during the epoch of reionization (EoR) and kept ionized thereafter. This diffuse component is of particular interest because it carries information about the thermal state of the IGM during the first billion years, an epoch that remains only loosely constrained by other observables. In what follows we describe how the CMB frequency spectrum probes both reionization and the preceding era of Cosmic Dawn,  emphasizing the synergies between spectral distortions and the redshifted 21cm line.
 
As CMB photons traverse the ionized plasma, inverse Compton scattering transfers a small amount of energy from the warm free electrons to the radiation field, imprinting a Compton-$y$ distortion with the characteristic spectral shape $Y(x)$ introduced in Sec.~\ref{sec:Intro} \cite{Zeldovich1969, SunyaevZeldovich:1972}. The sky-averaged amplitude is set by the mean electron pressure integrated along the line of sight, 
%
\begin{equation}
\langle y \rangle = \frac{\sigma_{\rm T}}{m_{\rm e} c^2} \int \frac{c \, \id z}{(1+z) H(z)} \, \left\langle n_{\rm e}(\mathbf{x}, z) \, k_{\rm b} \left[ T_{\rm e}(\mathbf{x}, z) - T_\gamma(z) \right] \right\rangle ,
\label{eq:y_los}
\end{equation}
%
where $\sigma_{\rm T}$ is the Thomson cross section, $T_\gamma(z) = T_0 (1+z)$ is the CMB blackbody temperature (denoted $T_{\rm CMB}$ elsewhere), $n_{\rm e}$ is the free-electron density including the contribution of ionized helium, and the angle brackets denote a spatial average over the electron distribution. Because density, ionization fraction and temperature are correlated, the integrand cannot in general be replaced by the product of separate means. The factorization into a mean density and temperature is permissible only if the latter is the electron-density-weighted temperature at each redshift,
%
\begin{equation}
\bar{T}_{\rm e}(z) \equiv \frac{\langle n_{\rm e} T_{\rm e} \rangle}{\langle n_{\rm e} \rangle} .
\label{eq:Te_z}
\end{equation}
%
The monopole distortion measures the volume-averaged electron pressure, or equivalently the optical-depth-weighted thermal energy, and is linear in $n_{\rm e}$. It therefore does not receive the direct $n_{\rm e}^2$ clumping enhancement characteristic of the free-free processes discussed below \cite{TrombettiBurigana2014}, although inhomogeneity can still affect $\langle y \rangle$ through correlations between density, temperature and ionization state.
 
A systematic second-order treatment of the Compton scattering during reionization, including the $y$-type contribution generated by velocity-squared terms and its relation to the secondary CMB anisotropies, was developed in \cite{Hu1994pert}. Separately, early studies used limits on the thermal Compton distortion to constrain how early, hot and complete reionization could have been \cite{Bartlett1991, TegmarkSilk1994, Tegmark1994b}. More contemporary treatments tend to focus instead on its use as a quantitative probe of the IGM thermal history.
 
For an approximately uniform reionized medium, Eq.~(\ref{eq:y_los}) reduces to
%
\begin{equation}
y_{\rm reion} = \frac{k_{\rm b} \tau_{\rm reion}}{m_{\rm e} c^2} \, \langle T_{\rm e} - T_\gamma \rangle_\tau \simeq 8.4 \times 10^{-8} \left( \frac{\langle T_{\rm e} \rangle_\tau}{10^4 \, {\rm K}} \right) \left( \frac{\tau_{\rm reion}}{0.05} \right) ,
\label{eq:y_reion_estimate}
\end{equation}
%
where the optical-depth-weighted temperature is the electron-column average over the reionized interval,
%
\begin{equation}
\langle T_{\rm e} - T_\gamma \rangle_\tau \equiv \frac{1}{\tau_{\rm reion}} \int_{\rm reion} \left[ \bar{T}_{\rm e}(z) - T_\gamma(z) \right] \id\tau , \qquad \id\tau = \sigma_{\rm T} \langle n_{\rm e} \rangle \, \frac{c \, \id z}{(1+z) H(z)} ,
\label{eq:Te_tau_int}
\end{equation}
%
and we have used $T_{\rm e} \gg T_\gamma$ in the numerical estimate. For a medium near $2 \times 10^4$ K the amplitude is roughly $y_{\rm reion}\sim 2 \times 10^{-7}$.
 
The practical obstacle to this program is that the measured monopole $y$ distortion is not specific to reionization. Instead, it is comprised of contributions from many sources, which can be decomposed schematically as
%
\begin{equation}
y_{\rm tot} = y_{\rm reion} + y_{\rm groups/clusters} + y_{\rm WHIM} + y_{\rm feedback} + \cdots ,
\label{eq:y_decomposition}
\end{equation}
%
where the listed contributions are not necessarily mutually exclusive. The reionization signal lies roughly a factor of a few below the more dominant global SZ cluster contribution as described in Sec.~\ref{sec:SZ_effect}. Current models and tomographic measurements place the total mean signal at $y_{\rm tot} \simeq 1$--$2 \times 10^{-6}$, roughly an order of magnitude below the original FIRAS limit, with more than half of the integrated mean $y$-signal originating at $z < 1$ \cite{Hill2015, Chiang2020}. The reionization term in Eq.~(\ref{eq:y_reion_estimate}) is thus a $\sim 10^{-7}$ sub-component beneath the low-redshift signal, and the central measurement challenge is whether one will be able to dig it out of the data.
 
Several observables help to disentangle the contributions. Likely the most promising is the ability to cross-correlate spatially varying $y$ maps with tracers of known redshift to empirically account for the low-redshift contribution \cite{Chiang2020}. Present analyses reconstruct the SZ pressure out to $z \sim 1$, but more sensitive $y$-maps and deeper galaxy catalogs could allow for this approach to be extended to much higher redshifts. Another approach is to use relativistic temperature corrections to the SZ spectrum to further separate the hot, late time component \cite{Hill2015} (Sec.~\ref{sec:SZ_effect}), from the generally cooler reionization IGM. 

The kinetic SZ effect produced by the bulk motion of electrons during patchy reionization will also imprint a distinct small-scale anisotropy that can be used to constrain the timeline and morphology of the process \cite{ZhangPenTrac2004, Battaglia2013}, while the thermal SZ power spectrum provides a further handle on the unresolved pressure field \cite{Refregier2000}. Once the resolved and statistically reconstructed low-redshift contributions have been constrained, the residual monopole can be interpreted jointly with models and external measurements of the unresolved ionized gas to bound the high-redshift contribution. The total signal is comfortably within the sensitivity forecast for the spectrometers described in Sec.~\ref{sec:missions}, which would measure $y_{\rm tot}$ at high significance. That reach, however, applies to the total $y$-template. Isolating a separately fitted reionization component is substantially harder, and it is this decomposition, rather than the raw detection of $y_{\rm tot}$, that limits what can be learned about the EoR.

\subsubsection{Spectral distortions at Cosmic Dawn}
At redshifts preceding reionization ($z \sim 15$--$30$), an era typically referred to as Cosmic Dawn, the formation of the first stars produce a characteristic absorption trough in the redshifted 21cm (spin-flip) line of neutral hydrogen \cite{Field:1958rri, PritchardLoeb2012, FurlanettoOhBriggs2006}. The sky-averaged observable is the differential brightness temperature of the line against the background radiation field, which in the homogeneous limit may be written
%
\begin{equation}
\overline{\Delta T}_{\rm b} \approx 27 \, \bar{x}_{\rm HI} \left( \frac{\Omega_{\rm b} h^2}{0.023} \right) \left( \frac{0.15}{\Omega_{\rm m} h^2} \, \frac{1+z}{10} \right)^{1/2} \left( 1 - \frac{T_{\rm R}}{T_{\rm S}} \right) \, {\rm mK} ,
\label{eq:dTb}
\end{equation}
%
where $\bar{x}_{\rm HI}$ is the mean neutral fraction, $T_{\rm S}$ is the spin temperature of the hyperfine transition, and $T_{\rm R}$ is the brightness temperature of the background radiation at the rest-frame frequency $\nu_{21} = 1420$ MHz. Under standard assumptions the background is given by the CMB alone, so that $T_{\rm R}(z) = T_\gamma(z) = T_0 (1+z)$ in, with the redshifted $\nu_{21}$ lying in the Rayleigh-Jeans tail. This means that deviations away from the CMB (i.e. a spectral distortion) in the Rayleigh-Jeans tail can have strong impacts on the cosmic dawn signal.
 
The first claimed detection of the global 21cm cosmic dawn signal was reported in 2018 by the EDGES collaboration, who saw an absorption feature centered at 78 MHz ($z_{\rm CD} \simeq 17$) \cite{Bowman2018}. The amplitude of this absorption signal was roughly twice as strong as what one expected under standard astrophysical assumptions, which led to a flurry of activity in both the experimental and theoretical communities. On the experimental side, its cosmological interpretation is currently under dispute, after a follow-up experiment (SARAS 3) with comparable sensitivity to EDGES was not able to recover the signal. The SARAS 3 collaboration have thus rejected the EDGES best-fit template at $2\sigma$ \cite{Singh2022}.

On the theoretical side, two possible explanations for a deeper absorption trough were explored, both exploiting the $(1-T_{\rm R}/T_{\rm S})$ scaling. One possible resolution was to posit a way to cool the gas more efficiently, thus lowering $T_{\rm S}$. The most explicit realization of this was to introduce a millicharged dark matter candidate \cite{Barkana:2018lgd, Munoz:2018pzp, Fialkov:2018xre, Falkowski:2018qdj, Liu:2019knx} whose interactions with the electrons will slowly sap their energy. Alongside these optimistic papers, constraints on the model from other observables \cite{Berlin:2018sjs, Barkana:2018qrx, Kovetz:2018zan, Slatyer:2018aqg, Creque-Sarbinowski:2019mcm} drastically reduced the viable parameter space, though in principle the millicharged dark matter explanation is not fully excluded \cite{Liu:2019knx}.

The second avenue of study was to amplify the signal by allowing there to be an excess radio background above the CMB present at Cosmic dawn \cite{Feng:2018rje, Ewall-Wice:2018bzf, Mirocha:2018cih, 2019MNRAS.486.1763F, Ewall-Wice:2019may, Reis:2020arr, Fraser:2018acy, Pospelov:2018kdh, Moroi:2018vci, AristizabalSierra:2018emu, Chianese:2018luo, Brandenberger:2019lfm, Choi:2019jwx}. Indeed, such a resolution appeared at first glance reasonable, when taking into account data from ARCADE-2 and the LWA, who report to this day a radio monopole brighter than the CMB below $\sim 1$ GHz (Sec.~\ref{sec:RSB}) \cite{Fixsen:2009xn}. Shortly after the EDGES detection, however, it was shown in \cite{Feng:2018rje} that more than $\simeq 10\%$ of the ARCADE-2 excess would cause an absorption trough deeper than what EDGES had detected. This was, however, not the full story \cite{AcharyaCyrChluba2023, CyrAcharyaChluba2024}.
\begin{figure}
    \centering
    \includegraphics[width=.48\textwidth]{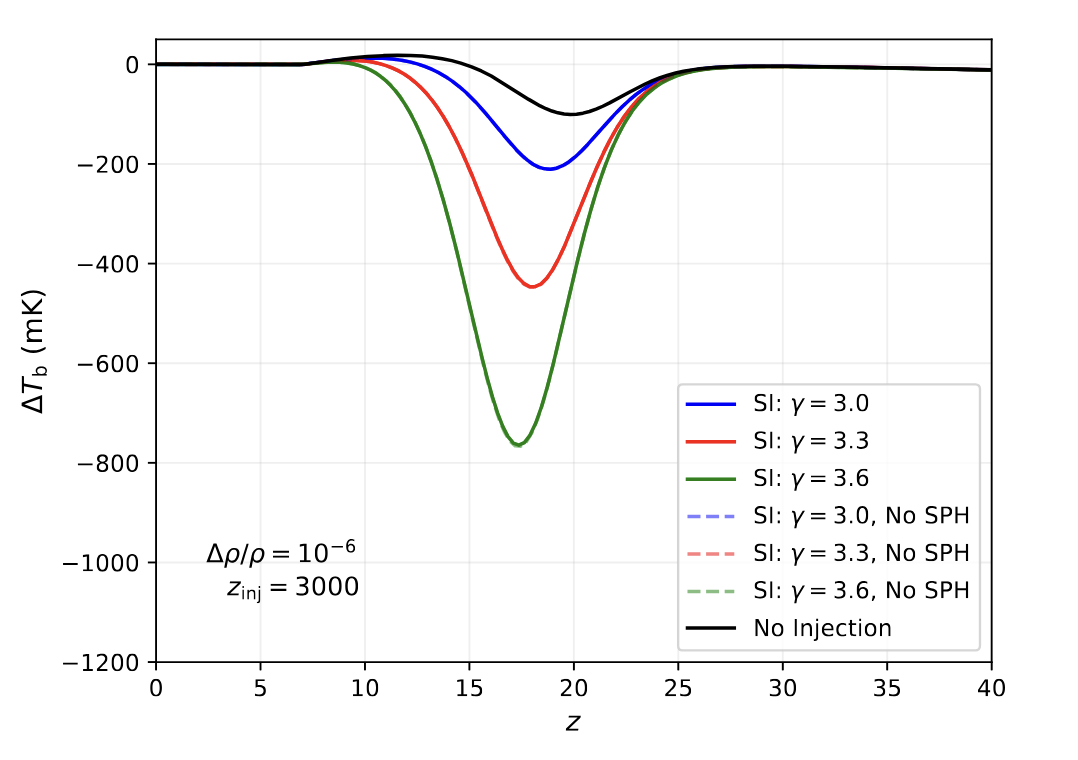}
    \includegraphics[width=.48\textwidth]{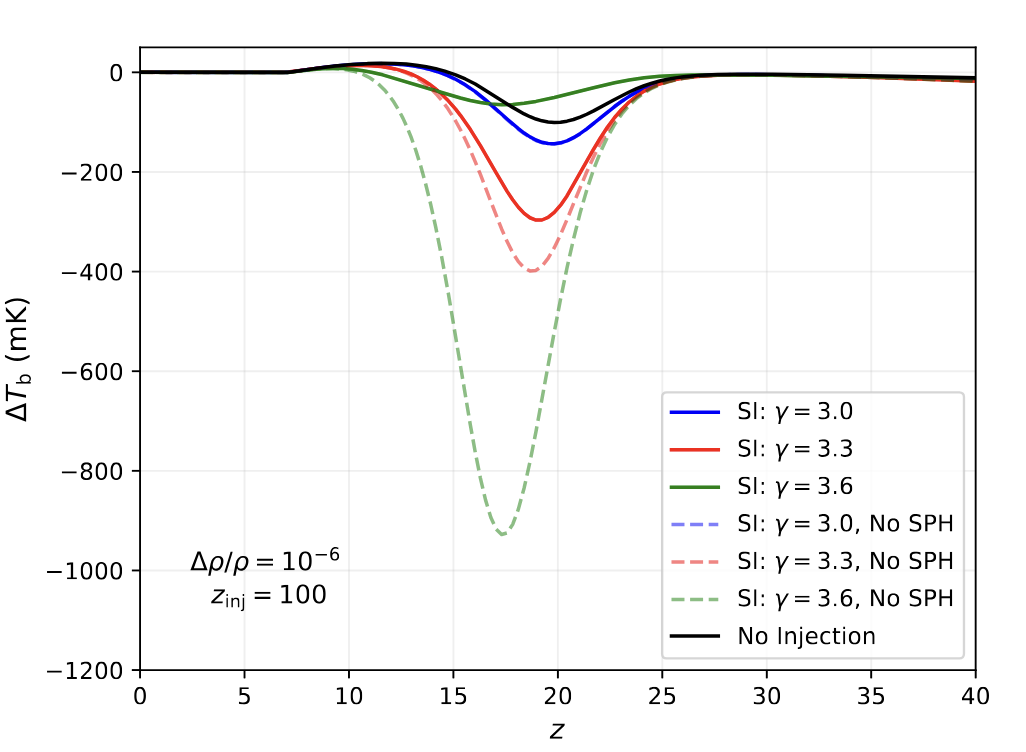}
    \caption{The response of the global 21cm brightness temperature in the presence of soft photon backgrounds. Case studies are shown with and without soft photon heating for free-free ($\gamma = 3.0$), synchrotron ($\gamma = 3.6$), and an intermediate type spectral tilt ($\gamma = 3.3$). Left: the response when the background is sourced pre-recombination ($z_{\rm inj} \simeq z_{\rm eq}$). Right: same, but for a post-recombination ($z_{\rm inj} = 100$) scenario, which showcases the need to propery account for this free-free heating effect. Figure adapted from \cite{CyrAcharyaChluba2024}.}
    \label{fig:SPH} 
\end{figure}

The link between low-frequency distortions and the 21cm signal is in fact richer than a shift of the background temperature $T_{\rm R}$. Recent work \cite{AcharyaCyrChluba2023, CyrAcharyaChluba2024} has shown that the lowest-frequency portion of an excess radio background is efficiently absorbed by the IGM through free-free absorption (inverse Bremsstrahlung), depositing energy in the gas and raising the matter temperature. This in turn alters the spin temperature through the usual coupling channels, and leads to a qualitatively different phenomenology than was previously understood \cite{Feng:2018rje}. This effect, termed soft photon heating (SPH), becomes important for backgrounds with a sufficiently steep low-frequency spectrum and produces a shallower absorption feature than would otherwise be expected. One can describe radio backgrounds in terms of the photon occupation number $\Delta n(x) \propto x^{-\gamma}$, or a radio brightness temperature $T_{\rm R} \propto \nu^{-\beta}$ where the two indices are related by $\gamma = \beta + 1$ in the Rayleigh-Jeans tail. A a synchrotron-like spectrum corresponds to $\beta \simeq 2.6$ ($\gamma \simeq 3.6$), while a free-free emission spectrum is given by $\beta \simeq 2$ ($\gamma \simeq 3$). A notable consequence is that a significant radio excess can be partially hidden from 21cm observations, since the associated heating compensates for the deepening of the absorption trough \cite{AcharyaCyrChluba2023, CyrAcharyaChluba2024}, as can be seen in Fig.~\ref{fig:SPH}. Soft photon heating therefore couples the amplitude of a Rayleigh-Jeans distortion to the thermal state of the Cosmic Dawn IGM, and must be accounted for when using either probe to constrain exotic sources of low-frequency photons. 

A spectrometer targeting the Rayleigh-Jeans tail, such as TMS in the $10$--$20$ GHz band, measures the present-day absolute sky spectrum and can constrain the amplitude and shape of any such background. The connection to the Cosmic Dawn 21cm signal is, however, indirect and model dependent. Relating a present-day monopole at $10$--$20$ GHz to the radiation field at $z \simeq 17$ requires both a large extrapolation in frequency, down to the much lower bands relevant to the reported excess, and an assumption about when the background was produced. A component generated at $z < 17$ would not contribute to $T_{\rm R}$ during Cosmic Dawn, whereas a background produced early enough would be related to its present-day spectrum primarily by cosmological redshifting, provided that subsequent absorption, emission and source evolution are negligible or consistently modelled. A low-frequency spectral measurement therefore constrains the viable source models and their extrapolations, rather than determining $T_{\rm R}$ at Cosmic Dawn directly.

\subsection{Radio surveys and the synchrotron background} \label{sec:RSB}

Investigations of the total emission received from the Universe at different wavelengths, known as the photon or electromagnetic {\it background radiation}, have been important in astrophysics and cosmology.  Throughout most of the electromagnetic spectrum, from microwave to gamma rays, the observed level of surface brightness and anisotropy of the background radiation is roughly consistent with that expected from known emission mechanisms from astrophysical and cosmological sources  \cite{2001ARA&A..39..249H,Planck:2018nkj,2009MNRAS.399.1694G,2017ApJ...837...19C,2016PhRvL.116o1105A}.  However, in the radio region of the electromagnetic spectrum, both the observed surface brightness level of the photon background and the level of its anisotropy angular power are seemingly much larger than that which can be produced by known source classes, resulting in an unexplained anomaly of contemporary cosmology.

As recently summarized in \cite{2023PASP..135c6001S} the surface brightness of the level of diffuse background radiation on the sky, from at least $\sim$20~MHz to 3~GHz, as a function of frequency $\nu$ is, in radiometric temperature units, \begin{equation}
T_\mathrm{BGND}(\nu) = 30.4 \pm 2.6 \mathrm{K} \, \left(
\frac{\nu}{310\mathrm{~MHz}}
\right)^{-2.66 \pm 0.04} \, + \, T_{\rm CMB}
\label{T_B}
\end{equation}
where $T_{\rm CMB}$ is the frequency-independent contribution of 2.725~K due to the CMB. \cite{Fixsen:2009xn,2011ApJ...730..138S,Dowell2018}.  Because the spectral index of $-2.6$ is consistent with synchrotron radiation, this extra term has been referred to as the radio synchrotron background (RSB).  

This level of surface brightness is several times higher than can be produced by known classes of discrete radio sources in the universe, particularly the AGN and star-forming galaxies \cite{2012ApJ...758...23C,2021A&A...648A..10H,2023MNRAS.520.2668H,10.1093/mnras/stad116}.  Producing the surface brightness of the RSB with discrete extragalactic sources would require an enormous number of a new class of low-flux sources \cite{2012ApJ...758...23C}.  Many lines of evidence disfavor a more local origin, such as in a heretofore undiscovered radio halo surrounding our Galaxy (e.g. \cite{2011ApJ...734....4K,2023PASP..135c6001S,2015ApJ...799L..10S}) or the local bubble \cite{2021MNRAS.502.2807K}.

In a parallel situation to the surface brightness, the anisotropy level of the RSB is also higher than that which could result from models of known source classes \cite{2022MNRAS.509..114O,2023MNRAS.523.5034C}.  The measured angular power spectrum in the range of angular scales where the RSB dominates the power indicates that either very numerous but highly clustered point sources or extended diffuse sources produce the RSB \cite{2023MNRAS.523.5034C}.  
There are other observational constraints on possible origin scenarios for the RSB.  Any possible source class and emission mechanism must: i) involve sources which do not follow the known correlation between radio and far-infrared emission in galaxies so as not to overproduce the background radiation at far-infrared wavelengths \cite{2011MNRAS.418..691P,2010MNRAS.409.1172S,2024MNRAS.530.2994T}, ii) if from synchrotron emission, involve sources with magnetic fields of around 1$\mu$G at redshift $z=0$ and increasing with redshift, so as not to  overproduce the observed level of the X-ray background through inverse-Compton upscattering of the CMB and/or optical/UV background \cite{2010MNRAS.409.1172S}, iii) not overproduce the observed limits on the 21cm absorption trough, due to the possible presence of the RSB at redshifts $z\sim 10$ increasing the temperature of the background relative to the temperature of the 21~cm transition \cite{2019MNRAS.486.1763F} -- however if the RSB is produced by energy injections at even higher ($z \gtrsim 20$) redshifts this effect can be suppressed and consistent with observed limits on the 21~cm trough \cite{CyrAcharyaChluba2024} -- and iv) not overproduce the observed cross-correlation angular power spectrum between the RSB and optical source catalogs, indicating that the vast majority of its sources, if discrete, must be at $z \gtrsim 0.5$ \cite{2024MNRAS.530.2994T}.

Some RSB origin scenarios that have been suggested include supernovae of massive population~III stars \cite{2014MNRAS.441.1147B}, emission from Alfv\'{e}n reacceleration in merging galaxy clusters \cite{2016JCAP...10..004F}, annihilating dark matter (DM) in halos or filaments \cite{Fornengo:2011cn,2012PhRvD..86j3003H,2015PhRvD..91h3501F,2019JCAP...11..047F} or ultracompact halos \cite{2013PhRvD..87h3519Y}, ``dark'' stars in the early universe \cite{2009ApJ...705.1031S}, dense nuggets of quarks which form the dark matter and in which primordial antimatter persists \cite{2013PhLB..724...17L}, accretion onto primordial black holes \cite{2023AAS...24141905C,2022MNRAS.510.4992M,2022MNRAS.517.2454A}, decays to dark photons \cite{Pospelov:2018kdh}, dark photon decays \cite{2023PhRvD.107l3033C,2023MNRAS.521.3939A}, and other injections of photons in the early universe from processes \cite{2024MNRAS.527.9450A} including superconducting cosmic strings discussed in Sec.~\ref{CSTD} \cite{Cyr:2023yvj}.  

Determining the spectral shape of the RSB at higher (and lower) frequencies than currently characterized can distinguish between different proposed origin scenarios, as different scenarios predict different spectral behavior at frequencies above and below those where the RSB is currently characterized.  Thus having a precise measurement of the absolute sky spectrum at around 50~GHz or lower, as would be possible with a dedicated spectral distortions space mission, would be an enormously powerful lever. According to Equation \ref{T_B}, if the synchrotron spectrum of the RSB continues unbroken it would have a radiometric temperature of $55~\mu\mathrm K$ at 50~GHz, which should be distinguishable from higher or lower scenarios involving spectral breaks given the anticipated absolute temperature sensitivity expected with missions such as TMS or \FOSSIL.  Also, there are spectral signatures of some of the photon injection scenarios discussed above beyond just the RSB background zero-level, such as emission and absorption lines, which could be accessible to spectral distortion measurements, or more likely which spectral distortion measurements could constrain.

\subsubsection{Observational synergies with radio surveys}
The diffuse microwave and radio sky contains multiple components -- the CMB, Galactic synchrotron emission, Galactic free-free emission, the RSB, Galactic spinning dust emission, and the AME discussed Sec.~\ref{AME} -- which all have different spatial structures and (often spatially varying) frequency dependence \cite[e.g.][]{2011ApJ...734....4K}.  Depending on the frequency range and science goal of interest, various of these components must be separated to study any given one, and this component separation, which relies on modeling the multiple components, is a major analysis challenge  \cite[e.g.][]{2016A&A...594A..10P}.  

Observations from spectral distortions probes, even at the level of sensitivity of of a balloon- or ground-based project, can aid in constraining the Galactic synchrotron and other foregrounds to isolate the CMB temperature and polarization signals and potentially other radio and microwave signals of interest such as the AME.  Such observations would be powerful in combination with the anticipated new absolutely calibrated sky map at 310~MHz \cite{2023PASP..135c6001S}, and in combination with with other GHz measurements such as the 5~GHz C-Band All-Sky Survey (C-BASS \cite{2018MNRAS.480.3224J}) and the 10--40~GHz Q-U-I JOint TEnerife (QUIJOTE \cite{2023MNRAS.519.3383R}) experiment.  It has been noted for example by the Planck collaboration \cite{2016A&A...594A..10P} that {\it all} CMB foreground analyses, from the 1980s to the present, ultimately rely on the 408~MHz Haslam map \cite{1982A&AS...47....1H}, which itself relies for its calibration on highly uncertain, blocked aperture, bare dipole measurements from the 1960s \cite{1962MNRAS.124...61P}.  Therefore additional sensitive absolutely calibrated microwave and radio measurements are needed for more precise foreground modeling in many investigations.

For understanding the RSB in particular, a measurement of the level of the RSB from a spectral distortions space mission would be additionally powerful because it would be in combination with those of the upcoming absolutely calibrated Tenerife Microwave Spectromoeter (TMS \cite{TMS}) at 10--20~GHZ for high frequency constraints, and the upcoming Lunar Surface Electromagnetic Experiment (LuSEE-Night) at 10 MHz for low frequency constraints \cite{2023arXiv230110345B}, allowing a determination of the RSB spectral shape over almost three orders of magnitude in frequency.

\subsection{Line Intensity Mapping} \label{sec:LIM}
Line-intensity mapping (LIM) has emerged as one of the most promising techniques for studying the large-scale structure of the Universe across most of cosmic history
\cite{Kovetz:2017status,Kovetz:2019jzq,Bernal:2022jap,Chang:2026ake}.
Rather than detecting individual galaxies, LIM measures the cumulative emission from all sources in a given spectral line as a function of sky position and frequency.
The spectral information provides a direct mapping between observed frequency and redshift, allowing tomographic reconstruction of the three-dimensional matter distribution over enormous cosmological volumes.

The key advantage of LIM is that it remains sensitive even when individual galaxies are too faint to detect.
Traditional galaxy surveys become increasingly sparse toward high redshift, eventually becoming shot-noise dominated because only the brightest objects are observable.
Intensity mapping instead measures the integrated emission from the entire galaxy population, including sources many orders of magnitude below the individual detection threshold.
As a result, LIM provides efficient access to the large-scale structure throughout the Epoch of Reionization, Cosmic Noon, and well into the low-redshift Universe.

Unlike continuum surveys, LIM simultaneously probes both cosmology and astrophysics.
Different emission lines trace different phases of the interstellar medium, molecular gas, ionized gas, star formation, and metal enrichment.
At the same time, fluctuations in these lines trace the underlying dark matter density field, enabling measurements of the expansion history, the growth of structure, primordial fluctuations, neutrino masses, and other extensions of the standard cosmological model.


\subsubsection{Why a space-based spectroscopic mission?}

A space mission designed to measure the absolute sky spectrum over a broad frequency range naturally provides an exceptionally powerful platform for line-intensity mapping
\cite{Kogut2011,Chluba2021}.
While spectral distortions are measured from the sky-averaged signal, the same observations simultaneously produce spectrally resolved intensity maps over the full sky.
This combination is unique to an absolute spectroscopic survey and enables a wealth of large-scale structure science with essentially the same data set.
These two measurements carry complementary information: the distortions record energy injection and departures from a thermal spectrum across cosmic history, while the line maps locate the galaxies and intergalactic gas behind much of the late-time signal.

Broad frequency coverage allows multiple emission lines to be observed over a wide range of redshifts.
In particular, the \cii{} $158\,\mu$m fine-structure line and the ladder of CO rotational transitions together provide nearly continuous tomographic coverage from the nearby Universe to the Epoch of Reionization.
Additional far-infrared cooling lines become available at the highest frequencies, providing complementary probes of galaxy evolution and the interstellar medium
\cite{Silva:2019jbe}.

Space observations also provide several decisive advantages over ground-based LIM surveys from an observational point of view. First, bypassing the atmosphere enables more stable measurements with well-controlled calibration and provides access to frequency bands that are inaccessible or severely attenuated from the ground due to the opacity of the atmosphere. On top of that, it circumvents the challenge that water lines in the atmosphere entails for LIM in these frequencies from the ground. Second, space observations allow for coherent full-sky mapping which opens the possibility to measurements on the largest observable scales and grants direct overlap with essentially every current and future galaxy, CMB and LIM experiment.

These capabilities make a future space-based spectroscopic mission an ideal anchor dataset for multi-tracer cosmology.


\subsubsection{Scientific landscape}

\begin{figure}[t]
\centering
\includegraphics[width=0.6\textwidth]{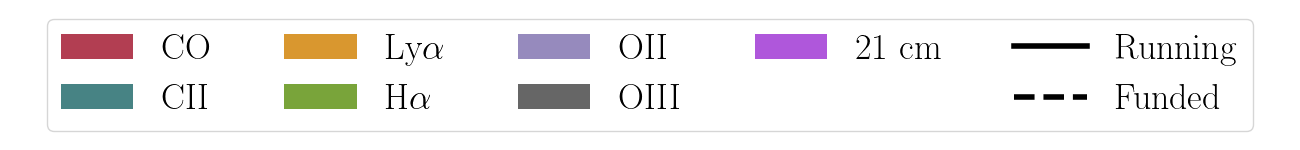}
\includegraphics[width=0.85\textwidth]{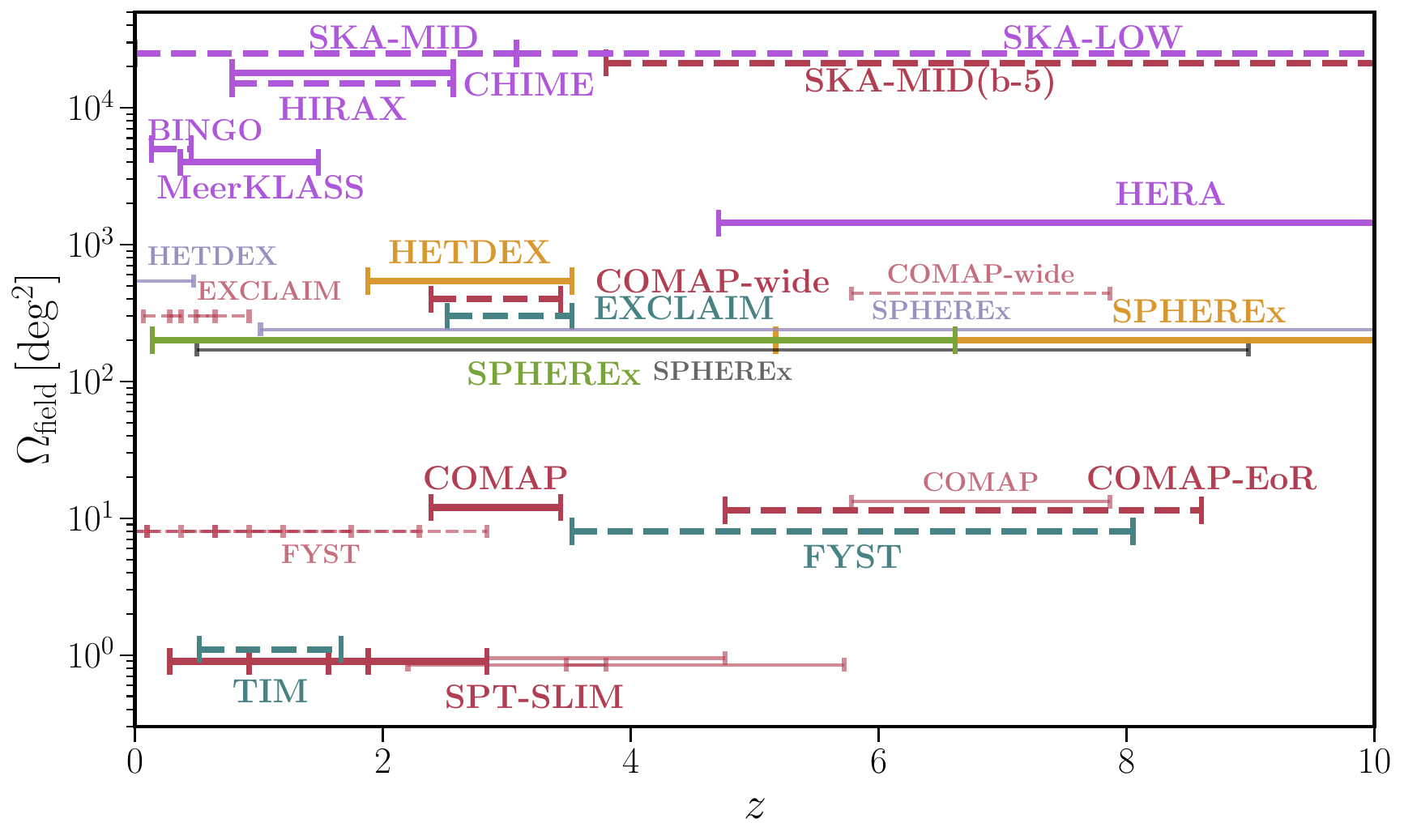}
\caption{
Ongoing and future LIM experiments, indicating the redshift coverage for each of the targeted (thick) and interloper (thin) spectral lines, and the survey sky area. A space-based spectroscopic mission could do a full-sky survey of millimeter and sub-millimeter lines across all these redshifts. Adapted and updated from Ref.~\cite{Bernal:2022jap}.}
\label{fig:landscape}
\end{figure}

\begin{figure}[h]
\centering
\includegraphics[width=0.85\textwidth]{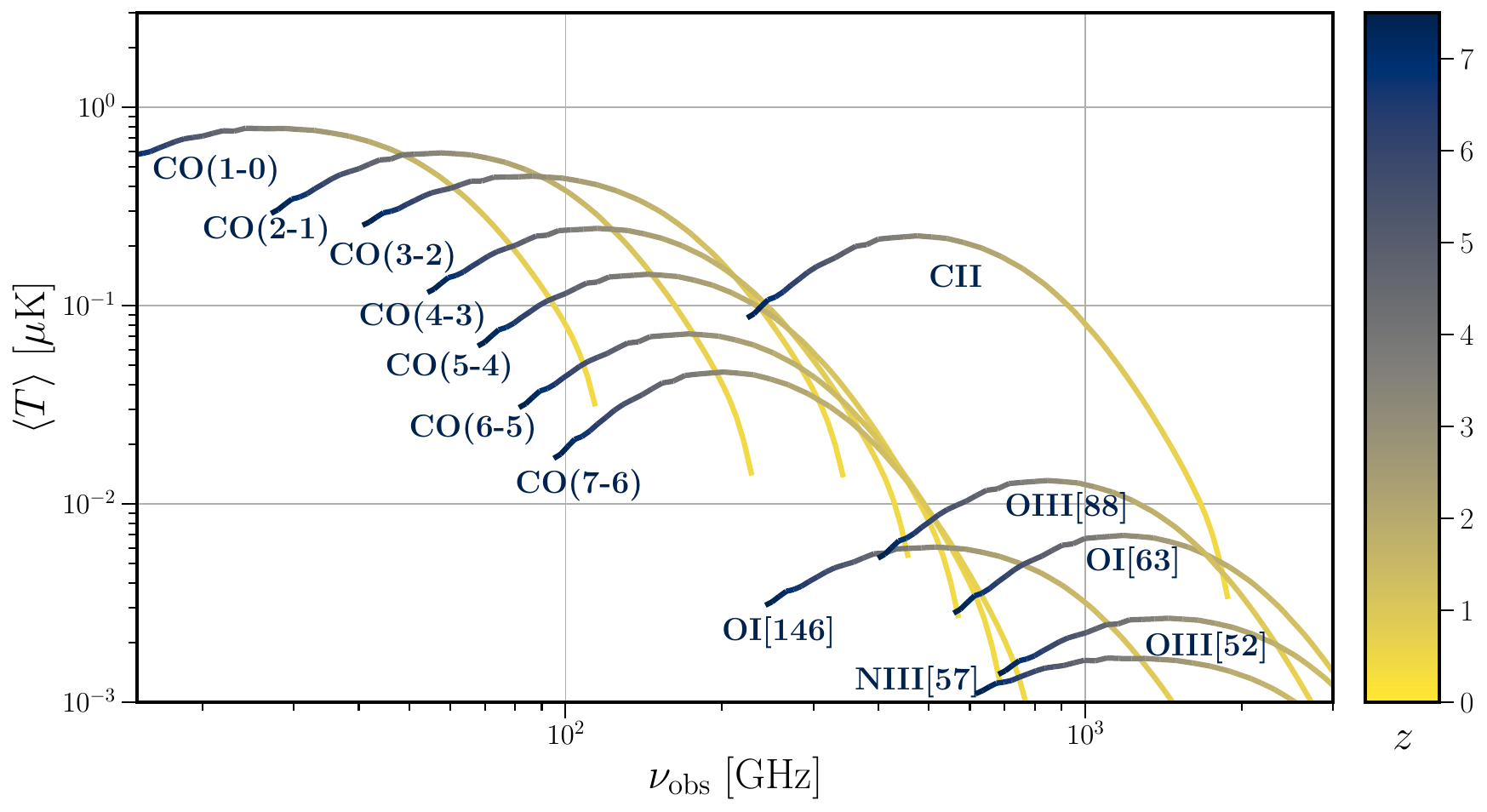}
\caption{ Main detectable spectral lines by a broad-band space-based spectrometer targeting absolute CMB spectral distortions. Inspired by Ref.~\cite{Silva:2019jbe} and updated using the models employed in  Ref.~\cite{Bernal:2022jap}.
}
\label{fig:lines}
\end{figure}

Comparing to current and future LIM experiments (see Fig.~\ref{fig:landscape}), a space-based broadband spectroscopic survey would have unprecedented potential to map the large-scale structure using LIM techniques. Being space-based, it could carry out a full-sky survey which could be complemented with smaller surveys with better angular resolution. Furthermore, the broad frequency coverage, paired with a good spectral resolution, can collect the contributions from many spectral lines, as shown in Fig.~\ref{fig:lines}. The accessible redshift ranges of the principal emission lines span nearly the entire history of galaxy formation while overlapping numerous existing and planned line-intensity mapping experiments
\cite{Dore:2014cca,Concerto:2020apex,Aravena:2021ccat,Vieira:2020tim}. Furthermore, the large number of narrow frequency channels can enable multi-line correlations, which can significantly increase the robustness against line interlopers and the sensitivity to the astrophysical processes triggering the emission of the lines (see e.g., Ref.~\cite{Cheng:2024nfy}).

The combination of broad frequency coverage and nearly full-sky observations produces an unprecedented three-dimensional map of cosmic structure.
The survey naturally overlaps essentially all major cosmological datasets, including galaxy surveys, CMB experiments, weak lensing surveys, and 21-cm observations, enabling powerful cross-correlation analyses and multi-tracer techniques.


\subsubsection{Cosmological opportunities}

The enormous volume accessible to future LIM surveys opens several unique cosmological opportunities.

\paragraph{Expansion history.}

Baryon Acoustic Oscillations can be measured well beyond the redshift reach of conventional galaxy surveys.
By extending distance measurements into the era of peak star formation, LIM offers an important new probe of dark energy, early dark energy, and the expansion history
\cite{Bernal:2019gfq,Karkare:2018sar,Bernal:2019usersguide}. Although BAO measurements can be challenging due to the large telescope beam of a spectrometer of this kind, the good spectral resolution would enable Alcock-Paczynski tests along the line of sight. 

\paragraph{Growth of structure.}

Tomographic maps of line emission measure the evolution of cosmic structure over most of the observable Universe.
Combined with galaxy surveys and CMB lensing, LIM enables improved measurements of the growth rate, bias evolution, and the suppression of power produced by massive neutrinos and other light relics. 

\paragraph{Primordial physics.}

The combination of enormous survey volume and full-sky coverage provides access to ultra-large scales approaching the horizon.
These scales contain information about primordial non-Gaussianity
\cite{MoradinezhadDizgah:2018fng},
relativistic projection effects, and possible departures from General Relativity.
The availability of multiple emission lines further enables multi-tracer analyses capable of substantially reducing cosmic variance
\cite{MoradinezhadDizgah:2018fng,Schaan:2021hhy,Sato-Polito:2022wiq}.

\paragraph{Cross-correlation science.}

Perhaps the greatest strength of LIM lies in cross correlations.
Joint analyses with galaxy surveys improve calibration and bias constraints, while correlations with CMB lensing directly connect luminous tracers to the underlying matter distribution.
Cross correlations with 21-cm observations provide a unique window into the morphology and evolution of cosmic reionization.


\subsubsection{Astrophysical opportunities}

The same observations simultaneously address major questions in galaxy evolution. The \cii{} and CO intensity fields trace different phases of the interstellar medium and therefore probe the evolution of molecular gas, star formation, and metal enrichment over cosmic time.
Additional far-infrared cooling lines provide diagnostics of gas density, metallicity, and ionization conditions within galaxies.

Because LIM measures the integrated emission from all galaxies, including sources too faint to detect individually, it provides an essentially unbiased census of the cosmic radiation budget.
This enables measurements of the cosmic star-formation history, the build-up of metals, the evolution of molecular gas and dust, and the accompanying growth of the cosmic infrared background, as well as the connection between galaxies and the surrounding intergalactic medium.

Finally, broad frequency coverage naturally allows searches for unexpected spectral features, including radiative decays of dark matter candidates and other exotic physics that would appear as anomalous emission lines
\cite{Bernal:2020dmdecay,Bernal:2021cnb}.

\subsection{Probing star formation at Cosmic Noon}
\label{sec:cosmic_noon}

Establishing a coherent history of cosmic star formation, and of the baryon cycle that drives it, is a central goal of extragalactic astrophysics. The epoch known as ``cosmic noon'' ($1 \lesssim z \lesssim 3$) marks the peak of the cosmic star-formation history, during which galaxies assembled the majority of their present-day stellar mass \cite{MadauDickinson2014}. Optical and near-infrared surveys have mapped the unobscured star formation of this epoch in detail, but a substantial fraction of the activity takes place in dust-enshrouded environments, where the ultraviolet light of young, massive stars is absorbed by dust grains and re-emitted in the far-infrared and submillimeter bands, giving rise to the cosmic infrared background (CIB; \cite{Puget1996}, Sec.~\ref{Sec:CIB}).

Understanding what regulates this activity requires going beyond star-formation-rate (SFR) tracers to the phases of the interstellar medium (ISM) that fuel it: the molecular gas reservoir, traced by the rotational lines of carbon monoxide (CO), and the photodissociation regions surrounding star-forming clouds, traced by the \cii{} 158\,$\mu$m fine-structure line. Ground-based interferometers such as ALMA and NOEMA have delivered exquisite observations of these tracers in individual high-redshift galaxies \cite{LeFevre2020,Tacconi2020}, but such observations are time intensive, cover small fields of view, and are often biased toward the most luminous systems. Line-intensity mapping (Sec.~\ref{sec:LIM}) is the natural complement: by measuring the aggregate emission of these lines on large angular scales, it provides a complete, volume-averaged census of the line-emitting gas, including the faint galaxy majority that is individually inaccessible to current instruments, linking gas content to the star-formation history in a way that resolved detections cannot \cite{Kovetz:2017status,Bernal:2022jap,Chang:2026ake}.

The absolutely calibrated spectrometers described in Sec.~\ref{sec:missions} are well suited to this measurement. Their broad frequency coverage, from tens of GHz up to ${\sim}2$\,THz, captures the redshifted CO and \cii{} emission from cosmic noon, and a spectral resolution of order $15$\,GHz, while coarse compared to dedicated spectroscopic receivers, is well matched to intensity mapping, where the goal is to recover the statistical properties of the line emission rather than the kinematics of individual galaxies. The sensitivity of such instruments, driven by the requirement to measure $\mu$-type distortions, suffices to detect the faint, aggregate glow of CO and \cii{} emission from galaxies well below the detection thresholds of current surveys. Moreover, absolute calibration against an internal blackbody reference gives access not only to the line fluctuations but to the \textit{absolute monopole} intensity of these line backgrounds, a fundamental cosmological observable that has remained essentially unconstrained since \COBEF.

\subsubsection{Tracing molecular gas and star formation with CO and \cii}
\label{LIM_CO_rhoH2}
Molecular hydrogen, the raw fuel of star formation, lacks a permanent dipole moment and is very difficult to observe directly in cold environments; CO serves as its primary proxy. The rotational transitions of CO form a ladder of lines at rest-frame frequencies $\nu_{J\rightarrow J-1} = J \times 115.27$\,GHz, whose relative strengths depend on the density and temperature of the gas, providing a diagnostic of the excitation state of the molecular reservoir. At cosmic noon several low- and mid-$J$ transitions redshift into the observing bands; for example, the CO(3--2) transition, with rest frequency $345.8$\,GHz, is observed at ${\sim}115$\,GHz at $z=2$. For any given redshift slice at cosmic noon, multiple transitions are therefore measured simultaneously. The aggregate CO intensity constrains the evolution of the cosmic molecular gas density, $\rho_{{\rm H}_2}(z)$, a key discriminant between competing models of galaxy quenching: ``equilibrium'' models attribute the decline of the global SFR density since $z \sim 2$ to the depletion of gas reservoirs, whereas alternatives invoke a decreasing star-formation efficiency \cite{Tacconi2020}. A volume-averaged inventory of the molecular gas can break this degeneracy.

The \cii{} line, one of the principal coolants of the ISM and typically the brightest emission line of star-forming galaxies \cite{Bernal:2022jap}, provides the complementary SFR side of the ledger. In the local Universe its luminosity correlates tightly with the total SFR, but its behavior at high redshift is debated \cite{Lagache2018}: observations of ultra-luminous infrared galaxies reveal a ``\cii{} deficit'', a marked drop in the \cii{}/FIR luminosity ratio \cite{Stacey2010}, whereas studies of typical main-sequence galaxies out to $z > 4$ find little to no evolution in the \cii{}--SFR relation \cite{Schaerer2020}. The discrepancy plausibly reflects selection: targeted observations preferentially probe extreme systems. At $z \sim 1$--$4$ the line (rest frequency $1900.5$\,GHz) is observed at ${\sim}380$--$950$\,GHz, well within the high-frequency coverage of the proposed spectrometers, and an intensity-mapping measurement is sensitive to the luminosity-weighted average of \textit{all} emitters in the cosmic volume. Such a measurement would directly establish whether the deficit is a ubiquitous feature of the galaxy population or a selection effect.

\subsubsection{Statistical measurements and synergies with galaxy surveys}
The statistics of the intensity fluctuations carry information beyond the mean. One-point statistics of the intensity field (variance, skewness) constrain the CO and \cii{} line-luminosity functions without resolving individual sources; the faint-end slope and the position of the knee encode the duty cycle of star formation and the feedback processes regulating galaxy growth \cite{Li2016}, effectively probing the dwarf-galaxy population that remains undetectable in targeted surveys. On large scales, the angular resolution of ${\sim}1$--$2^\circ$ is well matched to the linear regime where the cosmological clustering signal is robust, and the spectral resolution permits a tomographic analysis in redshift shells of thickness $\Delta z \sim 0.5$ across cosmic noon. A detection of the 3D power spectrum of the line emission yields the bias-weighted mean intensity, $b_{\rm line}\bar{I}_{\rm line}$; since the bias depends strongly on halo mass, this measurement ties the star-forming gas reservoirs to their host dark matter halos, connecting the baryon cycle to the hierarchical growth of structure \cite{Bernal:2022jap}.
 
The principal obstacle to all of these measurements is continuum foregrounds, chiefly Galactic dust emission and the CIB itself. The most robust route to the high-redshift line signal is cross-correlation with galaxy catalogs of known redshift \cite{Serra2016}: correlating a given frequency channel with galaxies at the matching redshift isolates the line emission originating from that shell, while foregrounds, which are uncorrelated with the galaxy positions, average away. Full-sky spectral surveys overlap by construction with the spectroscopic and photometric samples of upcoming wide-field surveys such as Euclid, DESI and the Rubin Observatory/LSST, which can thus be used to ``light up'' the gas and dust associated with known galaxy populations. Together with local benchmarks and the high-resolution interferometric snapshots from facilities like ALMA, an absolutely calibrated measurement of the CO and \cii{} backgrounds and of their large-scale clustering would assemble a coherent, volume-averaged history of the baryon cycle at the peak epoch of cosmic star formation.

\subsection{The Cosmic Infrared Background} \label{Sec:CIB}
The far-IR/sub-mm CIB is understood to originate predominantly from dust emission in galaxies powered by obscured star formation, with only a minor contribution from AGN-heated dust. It therefore represents the cumulative emission from dust-obscured star formation integrated over cosmic time and provides an integral constraint on the cosmic star-formation rate density (SFRD).
It contributes roughly half of the total extragalactic background light and directly traces the history of dusty star formation \cite{dole2006}.
Space-based spectrometers such as \FOSSIL aim to measure the absolute spectrum of the CIB i.e., the mean sky brightness as a function of wavelength after subtraction of major foreground components (zodiacal light, Galactic cirrus, CMB). 

\subsubsection{CIB monopole and dipole measurements \label{CIB_past_measurements}}

To date, direct measurements of the absolute CIB intensity have only been achieved using FIRAS and DIRBE  on board COBE. The first detection was reported by \cite{Puget1996} followed by improved analyses from \cite{fixsen1998,lagache1999}. These measurements required extremely careful component separation (\cite{lagache2000}, see also, \cite{auclair2024, lenz2019} for the anisotropies).

Despite their pioneering nature, the FIRAS signal-to-noise ratio (SNR) at low frequencies ($\nu\lesssim$400\,GHz) remains low ($\ll$1 at 150\,GHz, \cite{odegard2019}). This is precisely the regime where the contribution from high-redshift ($z\gtrsim$5) dusty galaxies is expected to peak. This limited SNR currently prevents a precise determination of the long-wavelength CIB slope, leaving large uncertainties in the high-redshift contribution.

An alternative, indirect estimate of the CIB intensity is obtained by integrating galaxy number counts from deep surveys with, e.g., Herschel/SPIRE (\cite{bethermin2012}), SCUBA-2 (\cite{zavala2017,gao2024}) or ALMA (\cite{Fujimoto2024}). Forward modeling efforts reconstruct the total background by extrapolating faint-end populations below detection thresholds.
While these approaches have greatly improved our understanding of dusty galaxy evolution, they rely on completeness corrections and model assumptions. They cannot replace a direct absolute measurement of the mean CIB intensity.

In addition to its mean intensity (the monopole), the CIB must exhibit a dipole anisotropy induced by our motion with respect to the cosmic rest frame  \cite{piat2002}, analogous to the dipole observed in the CMB.  A high-SNR measurement between $\sim$100-1500\,GHz would therefore not only determine the monopole more precisely but could also enable the first robust characterization of the CIB dipole spectrum. Additional cosmologically relevant signals appear from the scattering if the CIB by free electrons in clusters of galaxies \citep{Sabyr2022, Acharya2023CIBSZ}. This effect is similar to the SZ effect but with the CIB replacing the CMB as a backlight. The related CIB-distortion has a broad spectral shape that in principle could be used to test the properties of the CIB across cosmic history by studying the lowest multipoles of the CIB.

\subsubsection{Cosmological constraints from the CIB}
The absolute CIB intensity provides an integral constraint on the cosmic SFRD and the CIB spectrum strongly constrains the normalization and redshift evolution of dusty star formation \cite{gispert2000}.
However, one of the largest remaining uncertainties concerns the contribution of dusty star formation at z$\gtrsim $5. Accurately determining the long-wavelength CIB intensity, where the high-z tail contributes most strongly,  would quantify the contribution of obscured star formation in the first billion years of cosmic history. This is essential for: (i) Completing the census of early galaxy growth, (ii) Understanding dust production timescales, (iii) Constraining early metal enrichment, (iv) Testing models of rapid baryonic collapse in massive halos.

Just as the dipole of the CMB arises from our peculiar velocity relative to the CMB rest frame, the CIB dipole would be induced by Doppler boosting of the Cosmic Infrared Background. Measuring it would: (i) Test whether the CIB rest frame coincides with the CMB rest frame \cite{land-strykowsk2025}, (ii) Constrain bulk flows at low redshift, (iii)
Directly probe the spectral slope of the CIB, and thus constrain dust temperature evolution.

Finally, because of its large PSF, \FOSSIL is not well suited for anisotropy measurements, except on the very largest angular scales. Provided they can be separated from Galactic dust emission, these very large-scale CIB anisotropies measured by \FOSSIL at $\ell  \lesssim 60$ 
could be exploited for several cosmological applications.

In particular, they can serve as a tracer of the underlying matter distribution, enabling probes of the integrated Sachs–Wolfe (ISW) effect through cross-correlation with CMB temperature maps. Because the ISW signal arises from the late-time decay of gravitational potentials, it is most prominent on large angular scales, where \FOSSIL retains sensitivity despite its limited angular resolution. The broad redshift distribution of the CIB, which peaks at $z \sim 1$–3, makes it a particularly effective integrated tracer for ISW studies \citep{2019A&A...621A..32M}. \FOSSIL CIB maps could therefore provide an additional large-scale structure tracer to improve statistical constraints on the ISW signal, complementing existing galaxy-based analyses. However, this approach is extremely challenging in practice. As shown by \cite{2019A&A...621A..32M}, preserving the signal-to-noise ratio of ISW measurements within 10\% over 40\% of the sky requires Galactic dust contamination to be reduced to the level of $\sim 0.01\%$ in power. Achieving such a level of cleaning is highly demanding, and potentially unrealistic, yet it is a necessary condition for exploiting the CIB as a competitive cosmological tracer in this context.

They also provide a promising avenue for constraining primordial non-Gaussianity. As shown theoretically by \cite{2016MNRAS.463.2046T}, observations of the CIB at frequencies of a few hundred GHz have the potential to  constrain the nonlinear parameter $f_{\mathrm{NL}}$ through the scale-dependent bias on large angular scales. In particular, \cite{2016MNRAS.463.2046T} highlight that the inclusion of high-frequency channels in the range $\sim$800--1000\,GHz is crucial to mitigate contamination from Galactic dust, a regime that represents a key strength of \FOSSIL or \PIXIE. As an illustrative case, they consider eight frequency bands between 200 and 600\,GHz, assuming a sensitivity of 1\,Jy\,sr$^{-1}$ and an angular resolution of 1.6\,degrees, and forecast a statistical uncertainty of $\sigma(f_{\mathrm{NL}}) \simeq 0.08$ under the assumption of 1\% residual dust contamination. However, a more realistic assessment will require a more comprehensive treatment of the contaminants as well as modelling uncertainties in the CIB and bias prescriptions.

Further information could be extracted from cross-correlations between the large-scale CIB anisotropies and the CMB lensing potential \citep{2023PhRvD.108h3522M} which is expected to have reduced sensitivity to primordial non-Gaussianity, but improved robustness to Galactic dust contamination.

At smaller scales, CIB anisotropies have been measured with Planck \cite{planck_CIB_2011, planck_CIB_2014} and Herschel \cite{viero2013,cao2020} and indirectly (through models in likelihood) in ground-based experiments (current and future) at $\sim100-200$\,GHz. However, precise modeling of CIB anisotropies requires knowledge of the monopole \cite{maniyar2018}. The zero level sets the normalization of emissivity and halo occupation models. Currently, this anchor is uncertain (see Sect.\,\ref{CIB_past_measurements}).

\subsubsection{The \FOSSIL CIB legacy}

More than 25 years after COBE, the long-wavelength CIB intensity remains poorly constrained. The $\nu\lesssim$400\,GHz window is both the least precisely measured and the most critical for probing the high-redshift dusty Universe.
\FOSSIL’s capability to deliver a high-SNR, absolute measurement of the CIB intensity and the dipole in this regime would: (i) Provide the missing anchor for dusty SFRD at z$>$5, (ii) Fix the zero level required for anisotropy modeling, and thus anchor halo-model interpretations of CIB clustering measurements, (iii) Clarify the role of early dust-obscured star formation, (iv) Strengthen connections between dust emission and cosmic gas reservoirs, (v) Provide an independent test of our cosmic velocity, (vi) Test the isotropy of the Universe.

A definitive measurement of the CIB monopole and dipole at long wavelengths is a legacy-level contribution to our understanding of cosmic star formation history and large-scale matter distribution.

\section{The Local Universe}
The cosmological signals described in the preceding sections must all be observed through the emission of our own Galaxy and Solar System. For an absolutely calibrated spectrometer, however, these local foregrounds are simultaneously science targets, as the same survey that characterizes them for removal returns a detailed physical measurement of each. In Sec.~\ref{sec:galactic_dust}, we discuss thermal dust, which dominates the Galactic sky above $\sim 100\,$GHz. Absolute spectroscopy directly measures the monopole of the dust emission, a quantity out of reach for differential imagers such as \textit{Planck}, and can detect departures from the canonical modified blackbody ($T_{\rm d} \simeq 15$--$25\,$K, $\beta \simeq 1.5$--$2$) that discriminate between physical grain models. Moreover, full-sky maps of far-infrared cooling lines (\textsc{C\,ii}, \textsc{N\,ii}, and the CO rotational ladder), first delivered by FIRAS, would trace star formation and the thermodynamic state of the interstellar medium, providing physical priors for component separation in CMB $B$-mode searches.

Sec.~\ref{AME} addresses the anomalous microwave emission, which accounts for as much as $\sim 50\%$ of the Galactic emission at $30\,$GHz, yet whose sky-averaged amplitude is constrained only by ARCADE2 (a $0.4 \pm 0.1$ fraction of the Galactic plane emission at $23\,$GHz). Measurements above its $\sim 20$--$25\,$GHz peak, combined with low-frequency data from TMS, would enable a robust separation of this component and shed light on its debated physical origin. Finally, Sec.~\ref{sec:zodiacal} considers the zodiacal emission from interplanetary dust. Long-duration absolute monitoring would disentangle genuine temporal variability of the cloud from instrumental drifts, extend spectral constraints on its composition, and constrain the rate at which comets and asteroids replenish it. Beyond enabling the cosmological program, these measurements constitute a legacy characterization of the local microwave sky, from the interplanetary dust cloud to the diffuse interstellar medium.

\subsection{Galactic dust properties} \label{sec:galactic_dust}
By delivering a highly sensitive spectrometric survey of the full sky, missions targeting spectral distortions  will enable an unprecedented characterization of the spectral energy distribution (SED) of thermal dust emission, the dominant Galactic foreground at high frequencies ($\gtrsim 100$ GHz). Such precise determination of the SED is essential for robust component separation and the reliable extraction of faint spectral distortions. At the same time, it will yield critical insights into the microphysical properties of interstellar dust grains and the thermodynamical and radiative conditions of the interstellar medium (ISM) in our Galaxy.

While interstellar dust grains constitute only approximately $\sim 1\%$ of the mass of the ISM, they play a critical role on its dynamics, thermodynamics and chemistry \cite{Draine2011}. The characteristic sizes of the grains span $a \sim 0.3\,\mathrm{nm}$ [large molecules and Polycyclic Aromatic Hydrocarbons (PAHs)] to $a \sim 0.1{-}0.3\,\mathrm{\mu m}$. Composed primarily of amorphous silicates and carbonaceous materials, these grains absorb a substantial fraction (of order $\sim 30{-}50\%$) of the stellar radiation field in star-forming galaxies and re-emit it thermally at infrared to submillimeter wavelengths, with equilibrium temperatures typically $T_d \sim 15{-}25\,\mathrm{K}$ in the diffuse ISM \cite{Krugel2003}. Beyond their radiative impact, dust grains provide catalytic surfaces for the formation of molecular hydrogen, which constitutes the dominant H$_2$ production channel in cold neutral gas, with formation rates $R_{\mathrm{H}_2} \sim 3 \times 10^{-17}\,\mathrm{cm}^3\,\mathrm{s}^{-1}$ under Milky Way conditions \cite{CazauxTielens2004}. Through photoelectric emission, they also dominate the heating of the diffuse neutral ISM, converting a few percent of absorbed UV photon energy into gas heating \cite{BakesTielens1994}.

A fine characterization of the thermal dust SED in intensity encodes quantitative information on grain growth, shattering, coagulation, and destruction processes, thereby tracing the lifecycle of baryonic matter and the dynamical state of the ISM across Galactic environments. In the diffuse ISM, the canonical description of thermal dust emission is given by a modified blackbody (MBB),
\begin{equation}
I_\nu \;=\; \tau_{\nu_0} \left(\frac{\nu}{\nu_0}\right)^{\beta} B_\nu(T_d),
\end{equation}
where $\tau_{\nu_0}$ is the dust optical depth evaluated at a reference frequency $\nu_0$, $\beta$ the dust emissivity spectral index, related to grain composition, $T_d$ the effective dust temperature, and $B_\nu(T_d)$ the Planck function. For a recent review on dust emission and a discussion of the MBB law, see e.g. \cite{Shirley2026}. Observations of the Milky Way typically find $T_d \simeq 15{-}25\,\mathrm{K}$ and $\beta \simeq 1.5{-}2$ in the far-infrared and submillimeter \cite{Planck2014XI}.
The sensitive spectrometric coverage of the full sky of a spectral distortion mission is expected to constrain the mean dust optical depth, temperature and spectral index with unprecedented precision. Regarding optical depth, missions targeting spectral distortions have a major advantage over CMB missions targeting anisotropies. Indeed, while differential imagers such as \textit{Planck} only measure relative intensity variations on the
sky --- leaving the zero level (monopole) of the dust optical depth
$\tau_\nu$ undetermined, so that map offsets must be fixed a
posteriori through external priors such as the correlation with
\textsc{H\,i} column density \citep{Planck2014XI,Planck2018XII}, spectral distortion missions, as absolutely calibrated FTS
continuously comparing the sky to an internal reference blackbody,
will directly measure the absolute specific intensity
$I_\nu \simeq \tau_\nu B_\nu(T_d)$ and hence determine the zero point
of the dust optical depth in every frequency channel. 

More importantly, the exquisite measurements provided by spectral distortion missions will allow us to characterize the deviations of the SED away from this canonical law. Indeed, laboratory measurements, theoretical considerations and detailed dust models indicate that intrinsic deviations from a MBB description are to be expected. In particular, the optical properties of amorphous silicates exhibit temperature- and frequency-dependent emissivities, leading to departures from a simple power-law behavior at millimeter wavelengths \cite{Demyk2017,Demyk2022}. Such behavior arises naturally in physical models of amorphous solids, in which absorption by two-level systems (TLS) and disordered charge distributions produces a millimeter emissivity excess that increases with grain temperature \citep{Meny2007,Paradis2011}. These effects reflect the disordered internal structure of interstellar grains and imply that $\beta$ is neither strictly constant nor universal, and that $T_d$ and $\beta$ can be intrinsically (anti-)correlated. Whether such correlation, observed in Planck data \citep{Planck2014XI}, is physical or is an artifact of data analysis is one of the open questions of contemporary Galactic science and high precision absolutely calibrated data from a mission targeting spectral distortions over a large frequency range might help settling this long standing debate.  

Physically-motivated dust models further predict departures from the MBB approximation. The recent \emph{Astrodust} framework provides a unified description of interstellar dust consistent with extinction, emission, polarization, and elemental abundance constraints, predicting subtle but measurable curvature in the millimeter SED \citep{Draine2021,Hensley2023}. Similarly, the THEMIS framework \citep{Jones2013,Jones2017} describes interstellar dust as a population of amorphous carbonaceous grains and large amorphous silicates with carbonaceous mantles, whose properties evolve with the local radiation field and density. Its latest incarnation, THEMIS~2.0 \citep{Ysard2024}, is built on laboratory-measured optical constants of interstellar silicate analogues spanning $5\,\mathrm{\mu m}$ to $1\,\mathrm{mm}$ and self-consistently reproduces extinction, emission, and polarization constraints. Because the model emissivity depends on grain composition, mantle thickness, and temperature, THEMIS~2.0 predicts spectral indices varying over $\beta \simeq 1.4{-}1.9$ in the diffuse ISM together with frequency-dependent curvature of the millimeter SED, i.e., measurable departures from a single MBB at the sensitivity level targeted by spectral distortion missions. Discriminating between the distinct millimeter predictions of Astrodust and THEMIS-like models is precisely the kind of measurement that only an absolutely calibrated, full-sky spectrometric survey can deliver.

In addition, it is clear both from the point of view of the physical models discussed above and from Planck data in intensity that $\beta$ and $T_d$ must vary significantly over the whole sky \citep{Planck2014XI}. Averaging the signal over regions with spatial variations in $T_d$ and $\beta$ naturally produces effective SED distortions relative to a pure MBB. Such averaging occurs unavoidably both along the line-of-sight (in the 3D Galaxy) and in-between line of sights (within the beam of the instrument). As the MBB is not a linear law in $\beta$ and $T_d$, the average of multiple MBB associated with different parameters will be distorted. Such distortions can be described using a moment expansion around a reference MBB spectrum, leading to higher-order corrections that become relevant at high sensitivity \cite{Chluba2017Moments,Rotti2021}. Such moment-induced distortions are unavoidable in realistic Galactic environments and must be accounted for in precision component separation, also in the context of $B$-mode searches \citep[e.g.,][]{Remazeilles2020cMILC, Mangilli2021}. 

The level of precision and the frequency coverage of spectral distortion missions will allow statistically significant detection of intrinsic and line-of-sight–induced departures from the single-MBB paradigm, far more precise than one can be achieved with imagers \citep{Mangilli2021}. From these distortions, it will be possible to infer the statistics associated to the variations of dust temperature and spectral indices in different regions of the Galaxy. 

Precise measurements of the SED will ultimately allow to understand the physical origin of the observed fluctuations and distinguish between dust models in which the emission is driven by a single dust component whose properties fluctuate continuously in the Galaxy and models containing physically distinct and clearly separate co-existing dust components, as the four component model recently proposed in \cite{Cosmoglobe2026}. Ultimately, synergetic analysis exploiting the combination of spectral distortion data with CMB imager measurements in intensity and polarization as well as last stellar extinction data will undoubtedly allow to revolutionize our understanding of dust emission in the microwave and its underlying physics.

As the mis-characterization of the canonical dust SED and its distortions is one of the major obstacles faced by CMB imagers searching for the primordial $B$-modes, spectral distortion missions will give invaluable ancillary data and physical priors for the imagers to use in their component separation, providing that the complexity of dust in intensity is comparable to the one in polarisation\footnote{Additional complexity is present in polarization due to its geometrical nature arising from the  perpendicular alignment of the grains with the Galactic magnetic field \citep{Tassis2015,Vacher2023}. However, if the same grain population(s) are responsible for the intensity and the polarized signal, the fluctuations of $\beta$ and $T_d$ across the Galaxy should be the same for the two fields. Having measurements or priors on the amplitude of these fluctuations would already be invaluable for CMB imagers science.}.

One of FIRAS's long-ranging astrophysical achievements was mapping emission lines across the entire sky \citep{bennett1994}. In particular, they were able to make full-sky maps of \textsc{N\,ii} ($205\,\mathrm{\mu m}$) and \textsc{C\,ii} ($\mathrm{158\,\mu m}$). These bright emission lines directly give estimates of the cooling-rates in the local ISM. The existence of certain ions and molecules instead of others gives an estimate of the amount of shielding, local density, and photoionization rate in the Milky Way \citep{tielens1985}.

FIRAS also made direct measurements of a series of emission lines, including the CO transition ladder from $J=1-0$ to $J=8-7$, as well as two \textsc{C\,i} lines, one \textsc{C\,ii} line, two \textsc{N\,ii} lines, one \textsc{O\,i} line, and CH line \citep{fixsen1999}. Most of these lines were only detected along the Galactic plane. Future missions with higher signal-to-noise and improved FWHM will necessitate mapping of these emission lines across the entire Milky Way. This will give direct unbiased tracers of star formation regions across the sky, including ones that have not been targeted explicitly.

As demonstrated in \citep{gjerlow2026}, specific emission line maps can be used to better trace different populations of thermal dust, and better inform component separation efforts. FIRAS was somewhat limited in its angular and frequency resolution; this is a clear way that future spectral distortion missions will be able to help with modeling the 3D galaxy and its more complex structure.
\subsection{Anomalous Microwave Emission}\label{AME}
Anomalous Microwave Emission (AME) is a major component of the diffuse Galactic emission at microwave frequencies, with a characteristic SED peaking in the range between approximately 10 and 60\,GHz (see \cite{AMEreview} for a review). It was originally identified in CMB observations as an excess emission spatially correlated with far-infrared radiation from interstellar dust, but inconsistent with classical synchrotron or free–free emission mechanisms \cite{Kogut1996, Leith1997, Oliveira1998, Oliveira1999}. Subsequent observations have confirmed that this correlation persists down to arcminute angular scales, while the emission remains largely diffuse on degree scales \cite{Watson2005, Casassus2008, Tibbs2010, Dickinson2010, Battistelli2019, Arce2020}.

AME is now known to be ubiquitous across the Milky Way and can contribute up to roughly half of the total Galactic emission at 30\,GHz, as established by observations from Planck and QUIJOTE \cite{Planck2016_A25, Mateo2023}.
Spectral information in the $10$--$20$\,GHz range (for example, from the upcoming TMS experiment) is particularly important for obtaining unbiased estimates of the AME amplitude and peak frequency \cite{Fred2023, Hoerning2026}. As a consequence, AME represents one of the dominant astrophysical foregrounds for CMB intensity measurements. Although its polarization appears to be weak \cite{Ricardo2017, Gonzalez2025}, even small residual levels may impact future high-sensitivity CMB experiments, particularly those targeting primordial polarization signals \cite{Remazeilles2016}.

The spectral energy distribution of AME exhibits a characteristic broad peak around 20--25\,GHz \cite{Mateo2023, Roke2025}. While this behaviour is commonly parameterized by a simple phenomenological (log-Gaussian) model \cite{Fred2023} as
\begin{equation}
I_\nu \;=\; A_{\rm AME} \exp\left[-\frac{1}{2}\frac{\ln^2(\nu/\nu_{\rm AME})}{W_{\rm AME}^2}\right],
\end{equation}
the observed spectral width $W_{\rm AME}=0.56\pm0.10$ is significantly broader than predicted by current theoretical descriptions \cite{Mateo2023, Roke2025}, highlighting substantial gaps in our understanding of the underlying physical mechanisms using the {\tt SpyDust} framework \citep{Zhang2025SpyDust, Zhang2026}. Future observations of AME-emitting regions with higher angular resolution \cite{AME-SKA} will be essential for distinguishing between competing models and fully exploiting AME as a diagnostic of interstellar grain physics and the small-scale structure of the interstellar medium.

In connection to spectral distortions, and despite this recent significant progress in characterizing the spatially varying AME signal, its sky-averaged (monopole) contribution remains poorly constrained. To date, the only direct estimate comes from absolute measurements by the balloon-borne experiment ARCADE2 \cite{Kogut2011}, which suggest that AME contributes to $0.4\pm0.1$ of the total Galactic plane emission at 23\,GHz. This value is consistent with the $0.46 \pm 0.08$ measured for diffuse AME emission in the Galactic plane \cite{Mateo2023}. 

The relatively large uncertainty associated with this estimate currently limits the accuracy of foreground modelling in studies of CMB spectral distortions.
In this context, COSMO, BISOU and \FOSSIL will provide high-precision absolute measurements at frequencies above the AME peak,  enabling a robust separation of the AME monopole from other Galactic components by combining with existing (ARCADE2) and planned (TMS, \cite{TMS}) low frequency absolute measurements below the peak of the AME SED at $\sim 20$\,GHz.

\subsection{Zodiacal dust emission} \label{sec:zodiacal}

Interplanetary dust within the Solar System is the source of the highest-intensity diffuse emission in the infrared. Early optical observations of scattered optical emission brightest around the  Sun (see, e.g., \cite{Leinert1975} and references therein) had been used to infer the three-dimensional structure of the so-called zodiacal dust cloud, but the emission being modeled by a line of sight integral is poorly constrained problem, with many degenerate solutions. Data taken from the  \textit{IRAS} observatory were used to make detailed observations of the thermal component of the zodiacal dust emission, making meaningful constraints on the dominant interplanetary dust cloud shape, as well as detecting bands of excess emission coincident with asteroidal bands \citep{Good1986,Sykes1988,RowanRobinson1990}. 
Data from the DIRBE satellite has been used to make further extensions to the \textit{IRAS} models, by including features such as an Earth-trailing component caused by the wake of the Earth moving through the diffuse cloud, as well as a circumsolar ring  aligned with the orbit \citep{kelsall1998,Wright1998}.
These models are usually referred to as the K98 and Wright models after the authors -- other analyses often modify these models as a baseline for more advanced analysis \citep{Tsumura2010, Ootsubo2016, San2024, obrien2025,Crill2025}. In particular, \citep{Ootsubo2016} find that the high resolution of \textit{AKARI} does not permit an azimuthally symmetric asteroidal dust band, but uses the Wright model for removing the interplanetary dust cloud.

In addition to the movement of the Earth, intrinsic temporal variation in the zodiacal emission is expected and has been observed. For one concrete example of this, \citep{kelsall1998} measured a 27 day variation in the $4.9\,\mathrm{\mu m}$ emission at the 1\,\% level, which they hypothesized to be due to changes in UV flux due to the solar rotation period. Additionally, the AKARI IRC detected a variable component of diffuse emission at $9\,\mathrm{\mu m}$, which was suggested to be a small cloud associated with a coronal mass ejection that occurred between two observations at the same region one year apart \citep{ishihara2017}. Finally, we expect the angular momentum of Jupiter to interact with the interplanetary dust, with intrinsic realignment of the dust coinciding with Jupiter's 11.96 year period.

The knowledge of zodiacal emission derived from \textit{IRAS} was limited by its mission length (10 months), calibration uncertainty ($\sim5\%)$, and incomplete sky coverage ($\sim90\%$). Similarly, DIRBE's cold mission length of 285 days did not allow for a full re-observation of pixels at different phases of the Earth's orbit. These limitations make it difficult to determine whether the brightness of the zodiacal emission is changing due to intrinsic variation or instrumental changes. In this respect, it is noteworthy the FIRAS team did not attempt a new fit to the spatial structure of the zodiacal emission, but did use the K98 model as a template above 600\,GHz, finding no evidence for it below these frequencies, with a maximum amplitude of $0.01\,\mathrm{MJy\,sr^{-1}}$ at this cutoff \citep{fixsen1998}.
As a striking example of this, archival HST observations indicate that new isotropic components could exist that have not yet been incorporated into current models \citep{obrien2025}. At the same time, there have been studies showing variation over time of the K98 model comparing data taken over several decades, e.g., \citep{pyo2010}. It is very difficult to tell whether this is due to calibration or bandpass errors, or a genuine time variable change in emissivity.

Beyond the spatial structure and its spatial and temporal variations being constrained, the spectral dependence of the dust is poorly constrained. 
Typical measurements simply take emissivities as measured in individual bands rather than fitting for a phenomenological model such as a modified blackbody, or more ideally, detailed dust modeling.
Measurements from \textit{Planck} found unexpected changes in emissivity, implying the dust composition is highly non-uniform \cite{planck2014}.
The measurement of the spectrum of the various regions of the IPD cloud can also constrain emission line measurements, which have been expected and hinted at in \citep{Fixsen2002}.
This alone can give a new view into the chemical composition of the Solar System that would otherwise only be possible through in situ measurements. In addition, the total mass, the replenishment rate, and the size distribution can be probed through these measurements. The FIRAS analysis will be superseded by a new spectral distortions experiment, and differences between the two will put directly limits on the rate at which comets and asteroids contribute to the composition of the IPD cloud.

In a broader context, it is extremely difficult to measure a three-dimensional structure that we are embedded within. Solving this problem requires data from as many vantage points as possible so that we can fully understand the IPD structure. For full sky measurements, this process began in earnest with \textit{IRAS} in the 1980s, continued with COBE in the 1990s, AKARI in the 2000's, WISE in the 2010's, and SPHEREx in the 2020's.  
An all-sky CMB spectral distortion mission 
would be unique in its combined temporal, spatial, and frequency coverage, and perhaps most importantly its absolute calibration. The absolute calibration and length of such a mission will be critical for fully understanding Solar System, and in generating this absolutely calibrated IPD model, the astronomical community will be able to determine the true temporal variation of interplanetary dust.
\section{Dark Matter and Astroparticle Physics}
The gravitational evidence for dark matter is overwhelming, yet every attempt to detect its non-gravitational interactions, whether in the laboratory or in astrophysical environments, has so far proved fruitless. The CMB frequency spectrum offers a fundamentally different strategy. Any coupling active at $z \lesssim 10^6$ between the dark sector and the photon-baryon plasma, whether it injects energy, removes it, or exchanges photons directly, is encoded as a departure of the CMB from a blackbody. In this section, we examine what future spectral measurements would reveal about the particle nature of dark matter, moving from canonical (MeV--TeV mass range) interaction channels to light and macroscopic candidates.

In Sec.~\ref{sec:DM_interactions}, we consider dark matter's most direct couplings to Standard Model particles: scattering, annihilation, and decay. Elastic scattering with baryons or photons continuously cools the plasma, driving the chemical potential negative and probing interaction rates well beyond the reach of anisotropy data for sub-GeV masses. Whereas velocity-independent annihilation is already tightly capped by anisotropy bounds, velocity-suppressed ($p$-wave) channels can generate distortions as large as $\mu \simeq 3\times 10^{-8}$ for masses in the $10$--$50\,$MeV range. Decaying particles source signals across nearly thirty orders of magnitude in lifetime. Light ($m_\chi \lesssim 20\,$keV) states inject photons that survive as distinct non-thermal populations, while heavier particles trigger electromagnetic cascades whose relativistic distortion shapes encode the mass and decay channel. Notably, for lifetimes exceeding the age of the Universe the post-recombination signal can reach $\Delta I/I \sim 10^{-6}$, comparable to the expected late-time $y$-distortion. Annihilation around recombination further modifies the cosmological recombination lines at the tens-of-percent level.

Light (sub eV) candidates are the subject of Secs.~\ref{sec:axions} and \ref{sec:dark_gamma}. For axions and dark photons, the expansion of the Universe itself performs a mass scan: as the free-electron density falls, the photon plasma mass sweeps across roughly ten decades ($\sim 10^{-15}$--$10^{-5}\,$eV), and resonant conversions can occur when $m_{\gamma} = m_{\rm dm}$. The resulting distortions carry spectral templates distinct from $\mu$ and $y$, while plasma inhomogeneities generate patchy anisotropies that can be cross-correlated with galaxy surveys. Future measurements would tighten the dark photon kinetic mixing by two orders of magnitude beyond \COBEF and, for primordial magnetic fields near current limits, the axion-photon coupling by three to four.

Finally, Sec.~\ref{sec:alt_DM} confronts the possibility that the dark sector rivals the visible one in complexity. Atomic dark matter can undergo its own recombination at $10^{3} \lesssim z \lesssim 10^{6}$, triggering resonant conversions within the primordial distortion epoch. Composite states with internal structure absorb and emit photons through a dark analogue of the hydrogen spin-flip transition. Macroscopic candidates are also considered. Axion quark nuggets, whose antimatter component annihilates with ambient baryons at a rate tied to both number densities, would generate $\mu \simeq 5\times 10^{-8}$ in a $(10$--$1000)\,$g mass window unconstrained by existing probes. 

\subsection{Dark Matter-SM particle interactions, decays, and annihilations}
\label{sec:DM_interactions}
Future spectral measurements offer a way to use calorimetric properties of the CMB as a way to probe dark matter (DM) properties, including masses, lifetimes, annihilation cross sections, and scattering interactions with Standard Model (SM) particles, across parameter space inaccessible to terrestrial experiments and complementary to CMB anisotropy constraints.

\paragraph{Baryon--dark matter scattering.} 
If DM scatters elastically with photons, electrons, or nuclei, it acts as an additional heat sink for the photon--baryon plasma, analogous to the cooling imprinted by baryons themselves through Compton scattering \cite{Ali-Haimoud:2015pwa}. This process generates a \emph{negative} chemical potential $\mu$, with an amplitude that can be as large as a few times the DM-to-photon number ratio $n_\chi/n_\gamma$, and is therefore most pronounced for sub-GeV DM masses. For generic power-law cross sections $(\sigma \propto v^n)$, current COBE/FIRAS limits are capable of constraining DM--proton, electron, and photon scattering rates. For velocity independent cross-sections, DM--proton scattering is constrained at the level of $\sigma_{\rm DM\text{-}p} \lesssim 10^{-24}\,\mathrm{cm}^2\,({\rm keV}/m_\chi)^{1/2}$ \cite{Ali-Haimoud:2015pwa}. A more sophisticated treatment accounting for the gradual thermal decoupling of DM from the plasma, rather than the instantaneous-decoupling approximation, demonstrates that the spectral distortion signal is systematically \emph{larger} than originally estimated \cite{Ali-Haimoud:2021lka}. With \FOSSIL sensitivity of $|\mu| \sim 10^{-8}$, the accessible cross-section regime would extend orders of magnitude below current CMB-anisotropy upper limits, reaching DM masses up to $\sim 1\,$GeV \cite{Ali-Haimoud:2021lka}. Indeed, when the cross section has a velocity dependence, the scattering induced $\mu$ distortion remains one of the most promising ways to constrain direct $\chi p$, $\chi e$ and $\chi \gamma$ interactions \cite{Ali-Haimoud:2021lka}. If only a small fraction of the DM undergoes scattering, spectral distortion constraints (which scale linearly with the scattering fraction) are also more robust than bounds from the CMB anisotropies \cite{Slatyer:2018aqg}; CMB anisotropies lose constraining power even for very high scattering cross sections once the scattering component of DM is small enough, as it can effectively be absorbed into uncertainties on the baryon density \cite{Boddy2018b}. 
Importantly, the spectral distortion signal is complementary to constraints from CMB anisotropies and the 21\,cm absorption trough, which probe lower-redshift effects of the same interactions \cite{Slatyer:2018aqg}.

\paragraph{Annihilation: $s$-wave and $p$-wave channels.}
Residual DM annihilations at $z\lesssim 2\times 10^6$ will inject electromagnetic energy into the plasma, sourcing spectral distortions whose type depends on the epoch of energy deposition. For velocity-independent ($s$-wave) annihilation, the stringent limits from Planck on modifications to the ionization history at recombination severely restricts the allowed energy injection at earlier times, rendering any resulting spectral distortion unobservable even at the sensitivity of a \FOSSIL-type instrument \cite{Chluba2013fore,
Lucca:2019rxf,Li:2024xlr}, at least if the injected particles are energetic enough to ionize hydrogen. 

This can be seen by a simple calorimetric estimate: the power injected by $s$-wave annihilation per unit volume, over a Hubble time, scales as $(1+z)^6/H(z)$, and consequently the ratio of the injected energy density to the total energy density scales as $\delta \rho/\rho \propto (1+z)^2/H(z)$. During radiation domination, $H(z) \propto (1+z)^2$ and so $\delta \rho/\rho$ is roughly constant with respect to redshift, while $\delta \rho/\rho$ falls slowly ($\propto (1+z)^{1/2}$) once the universe enters matter domination. Thus up to $\mathcal{O}(1)$ factors, we should expect the spectral distortion from $s$-wave annihilation to be of the same order as $\delta \rho/\rho$ around the epoch of matter-radiation equality. Because matter-radiation equality is in close proximity to the epoch of last scattering, the overall spectral distortion should also be of the same order as $\delta \rho/\rho$ at recombination. However, soon after recombination, the amount of additional ionization is strongly constrained. If the energy is injected in the form of particles above the ionization threshold, then generically an $\mathcal{O}(1)$ fraction of the injected energy proceeds into ionization (e.g.~\cite{Slatyer2015}); the amount of energy needed to ionize all the hydrogen in the universe is only $\mathcal{O}(10)$ eV/baryon, and changes to the high-redshift ionized fraction at the level of $10^{-3}$ can be constrained by Planck. An injection of $0.01$ eV/baryon corresponds to $\sim 0.01 \eta = 6 \times 10^{-12}$ eV/photon (here $\eta$ is the photon-to-baryon ratio), or $\delta \rho/\rho \sim 3\times 10^{-11}$ for typical CMB photon energies at recombination. Consequently, while there are certainly multiple $\mathcal{O}(1)$ factors missing from this simple estimate, our expectation should be that the maximum $s$-wave annihilation signal will be around the $\delta \rho/\rho \sim 10^{-10}$ level for models that produce ionizing particles after recombination.

The situation changes qualitatively for velocity-dependent ($p$-wave) annihilation, where $\langle\sigma v\rangle \propto v^{2}$. Because thermal velocities are large in the $\mu$-era ($z \sim 10^{4}$--$10^{6}$) but small at recombination, $p$-wave models can produce significant $\mu$-distortions while remaining consistent with CMB anisotropy bounds \cite{Chluba2013fore,
Li:2024xlr}. Under current constraints from other cosmological observables (BBN, diffuse photon backgrounds, etc.) $\mu$-distortions as large as $\mu \simeq 3\times 10^{-8}$ could be realised for DM masses in the 10--50\,MeV range with kinetic decoupling temperatures around 1\,keV \cite{Li:2024xlr}. This lies squarely within \PIXIE or \FOSSIL's detection threshold. The non-detection of such a signal would strengthen upper bounds on the $p$-wave cross section by an order of magnitude beyond existing limits, while a detection would point directly to the DM mass scale and epoch of kinetic decoupling \cite{Li:2024xlr}.

\paragraph{Decaying particles: low-mass regime.}

Light particles ($m_{\chi} \lesssim 20\,$keV) that decay directly into photon pairs produce distortions with a rich spectral structure \cite{Chluba2013fore, Chluba2015,
Bolliet:2020ofj}. Unlike pure heating scenarios, photon injection at energies comparable to or below the CMB peak frequency $h\nu \lesssim k T_{\rm CMB}$ do not thermalise efficiently through Compton scattering; instead, the injected photons can survive as a distinct non-thermal population \cite{Bolliet:2020ofj}. Using \texttt{CosmoTherm}~\cite{Bolliet:2020ofj},  a comprehensive library of spectral distortion solutions covering injection energies $E_{\rm inj} \simeq (10^{-10}$--$10)\,$keV and lifetimes $\tau_X \simeq (10^{5}$--$10^{33})\,$s has been produced. Using this, it was shown that \COBEF constraints are already competitive with other bounds for the stimulated decay of axion-like particles with masses $m_{a} \gtrsim 27\,$eV. Future spectral distortion experiments have the potential to improve upon FIRAS sensitivity by several orders of magnitude across this parameter space, providing a powerful and independent probe of axions, dark photons, and other light relics \cite{Bolliet:2020ofj,Liu2023}.

\paragraph{High-mass decays and electromagnetic cascades.}
For DM masses with $m_{\chi} \gtrsim {\rm MeV}$, decay products initiate electromagnetic cascades through pair production, inverse Compton scattering, and photon--photon interactions in both the pre-and post-recombination plasma. The resulting spectral distortions are qualitatively different from the standard $\mu$- or $y$-type shapes: relativistic electrons in the cascade upscatter CMB photons to energies well above the thermal bath, imprinting a characteristic non-thermal relativistic (ntr-type) distortion whose shape depends on the injection energy, decay channel, and redshift of injection \cite{Acharya:2018iwh,Acharya:2019owx, Liu2023}. 

For sub-GeV DM (or metastable non-DM particles) decaying pre-recombination, the distortion shape is sensitive to both the DM mass and the specific decay channel to SM particles, while for masses above $\sim 10\,$GeV the shape becomes universal, depending only on the electromagnetic branching fraction~\cite{Acharya:2019owx}. The constraints from properly evolving the full particle cascade are stronger by up to an order of magnitude compared to the simplified assumption that all injected energy thermalises as heat \cite{Acharya:2019owx}. The presence of non-degenerate spectral shapes could also allow future telescopes to 
distinguish the underlying injection mechanism from low-redshift $y$-distortions produced during reionisation. This complementarity between spectral distortions, CMB anisotropies, and BBN constraints, spanning lifetimes from $10^{5}\,$s to $10^{25}\,$s, has been demonstrated within a unified framework \cite{Poulin:2016anj,Lucca:2019rxf}, confirming that future spectrometers could surpass BBN bounds by two to three orders of magnitude for decaying species.

The post-recombination contribution to spectral distortions from decaying DM has also been computed precisely, using \texttt{DarkHistory} to model the secondary particle cascade and track the atomic interactions that are relevant in the presence of neutral gas \cite{Liu2023}. Even prior to recombination, the presence of a small amount of neutral gas in the universe can significantly enhance the spectral distortion signal by opening up new pathways for photons to redistribute their energy via photoionization and excitation. For decays with lifetimes longer than the age of the universe, the post-recombination contribution generally dominates the signal \cite{Liu2023}. 

Again, this can be understood by a simple energetic argument like the one given above for annihilation: the energy injected per volume over a Hubble time scales as $(1+z)^3/H(z)$, so the contribution from that Hubble time to the distortion scales as $\delta\rho/\rho \propto 1/(H(z)(1+z))$, which increases at low $z$ with a scaling between $(1+z)^{-3}$ (in radiation domination) and $(1+z)^{-1}$ (in dark energy domination). This means the signal is dominated by low redshifts where the competing CMB anisotropy constraints on excess ionization (or similar bounds like Lyman-$\alpha$ constraints on excess heating, e.g.~\cite{Liu:2020wqz}) are much less severe, due in part to the presence of astrophysical reionization and heating. The decaying DM lifetimes constrained by such ionization and heating bounds are at the level of $10^9 \times$ the age of the universe, meaning the amount of power injected by decay in a Hubble time today is at the level of $10^{-9}\times 5$ GeV/baryon (here we have taken the DM mass per baryon to be $\sim 5$ GeV),  which corresponds to $\delta \rho/\rho \approx( 5 \cdot  \eta$ eV/photon)/($2\times 10^{-4}$ eV/photon) $\sim 10^{-6}$.

\begin{figure*}[t]
    \centering
    \includegraphics[width=0.8\linewidth]{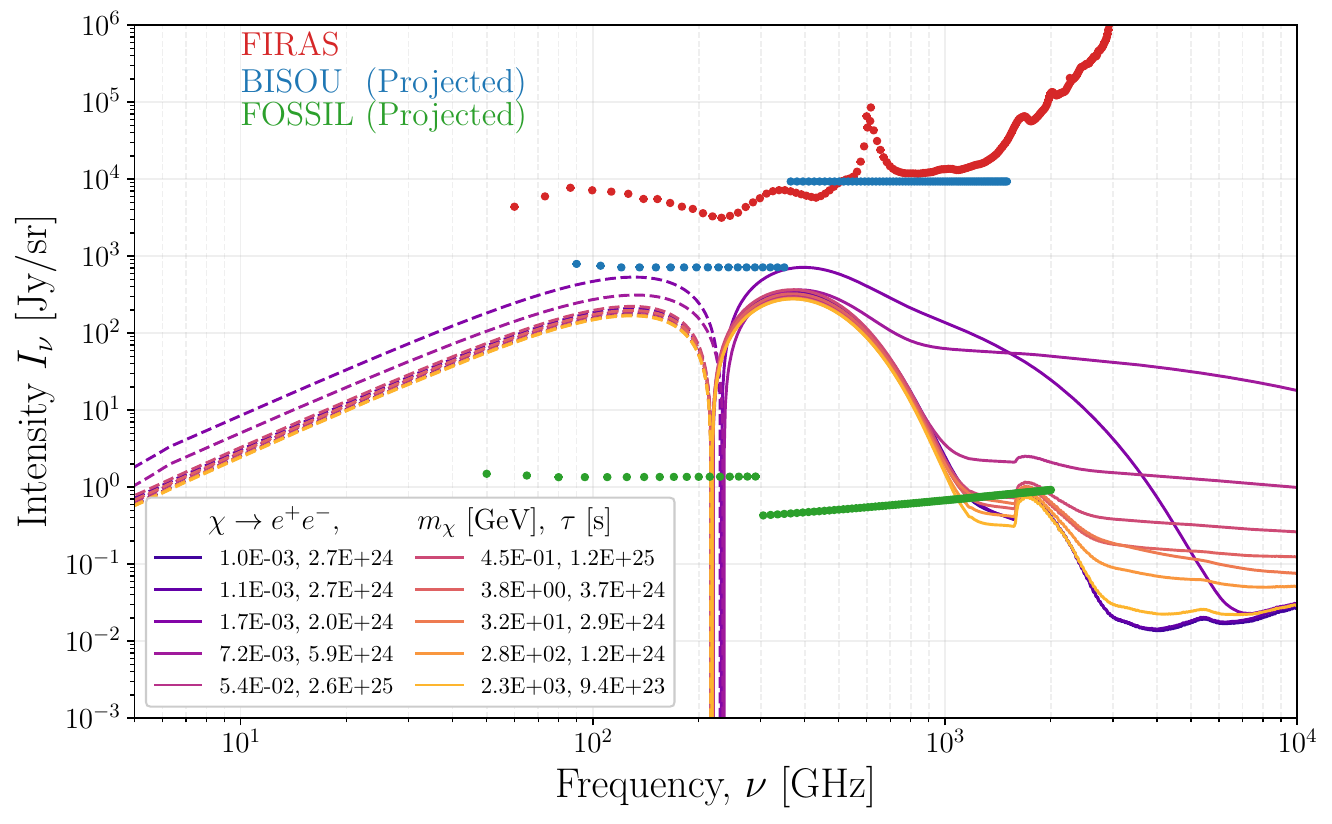}
    \caption{Spectral distortion signatures computed using \texttt{DarkHistory} for the $\chi \rightarrow e^+e^-$ process. The  variety of dark matter masses and lifetimes are chosen such that they saturate upper limits from CMB anisotropy studies \cite{Liu2023}. The sensitivities of \COBEF (red), BISOU (blue) and \FOSSIL (green) are also plotted. }
    \label{fig:DH_DM_decay}
\end{figure*}

It has been shown that indeed, when a detailed calculation is performed, the largest post-recombination signals arise from DM in the $\sim$\,MeV--GeV range decaying to $e^{+}e^{-}$, producing distortions up to $\Delta I/I\sim 10^{-6}$ of the CMB intensity at the current exclusion boundary from CMB anisotropies \cite{Liu2023}; this is similar in size to the expected signal from reionisation, providing yet another optimal target for future spectrometers.  
As in the pre-recombination case, the shape of these distortions can differ substantially from $\mu$ or $y$-distortions; in particular, high-frequency non-thermal tails are ubiquitous. A selection of these signals are plotted in Fig.~\ref{fig:DH_DM_decay}.

\paragraph{Recombination radiation as a probe of DM annihilation.}
Beyond the standard $\mu$/$y$ signals, DM annihilation during the recombination epoch ($z \sim 1000$) can modify the cosmological recombination spectrum itself, thereby opening the possibility to distinguish pre- from post-recombination $y$-distortions \cite{Chluba2009, Chluba2010}. Extra ionizations induced by DM annihilation products are followed by a recapture and subsequent radiative cascade, generating additional photons in the hydrogen and helium recombination lines, particularly the Balmer, Paschen, and Brackett series \cite{Chluba2009}. These DM-induced spectral features appear at the same frequencies as the standard recombination radiation but with amplitudes that depend on the branching of deposited energy into ionisations, excitations, and heating \cite{Chluba2009,Slatyer2009,Chluba2010}. For annihilation efficiencies near the current upper limits, changes to the H\,\textsc{i} recombination spectrum can reach $\sim$\,10--70\% in some spectral bands, while the He\,\textsc{i} spectrum is affected at the $\sim$\,10--20\% level \cite{Chluba2009}. The detailed spectral shape of these modifications is sensitive to the underlying DM model: different annihilation channels yield different ratios of ionisation to excitation energy, offering a unique avenue to characterise the nature of the annihilating particle \cite{Chluba2009,Slatyer2009}. A future spectrometer with broad frequency coverage would enable a search for these recombination-line signatures, providing an independent and highly complementary cross-check on DM annihilation constraints derived from the CMB power spectrum.

\subsection{Axions and ALPs} \label{sec:axions}
Axions and axion-like particles (ALPs) are among the most well-motivated extensions to the Standard Model, offering solutions to the strong CP problem and providing viable dark matter candidates across a vast range of masses, $m_{\rm a}$~\cite{Peccei1977,Weinberg1978,Wilczek1978,Marsh2016}. Through the Chern-Simons coupling $\mathcal{L} \supset -\frac{1}{4}g_{a\gamma\gamma}\, a\, F_{\mu\nu}\tilde{F}^{\mu\nu}$, photon-axion conversions in the presence of magnetic fields source spectral distortion signatures whose amplitude, spectral shape, and spatial morphology encode the axion mass, coupling strength, and properties of the intervening medium. The combination of absolute spectral calibration, broad frequency coverage, and high sensitivity to departures from a perfect blackbody that should be possible with future measurements of the CMB spectrum will open multiple new windows into axion physics, probing previously inaccessible regions of the $(m_a, g_{a\gamma\gamma})$ parameter space. The distortions from resonant and non-resonant conversion mechanisms have, in particular, different spectral shapes, distinguishable from each other as well as from the $y$ and $\mu$-type distortions. 

\paragraph{Homogeneous (sky-averaged) distortions from resonant and non-resonant conversions.}
In the presence of primordial magnetic fields ($B$), resonant $\gamma \leftrightarrow a$ conversions occur when the photon plasma mass equals the axion mass, $m_\gamma(z) = m_a$, at a redshift primarily governed by the evolving free electron density. For conversions at $z \lesssim 2 \times 10^6$, photon extractions/injections from the CMB produce $\mu$- and $y$-type distortions scaling as $g_{a\gamma\gamma} B$. Early constraints using COBE/FIRAS established limits of $g_{a\gamma\gamma} B \lesssim 10^{-13}\;\text{GeV}^{-1}\,\text{nG}$~\cite{Mirizzi2009}, with implications for the string axiverse~\cite{Tashiro2013}. It was later shown these bounds are weaker than claimed for $m_a \lesssim 5 \times 10^{-13}\;\text{eV}$~\cite{Mukherjee2018}. Updated constraints using \texttt{CosmoTherm} refined the photon plasma mass modeling (including helium recombination effects), determined the precise spectral template of the axion distortion (distinct from both $\mu$ and $y$), and treated the large-distortion regime for the first time~\cite{Cyr2024}. Crucially, this unique spectral shape means that constraints from spectral distortions will not merely improve raw sensitivity, but would also break degeneracies with other distortion sources by leveraging the full spectral information across hundreds of frequency channels. Under the assumption of primordial $B$-fields near current CMB limits ($\sim$nG), \FOSSIL would improve upon COBE/FIRAS constraints by $3$--$4$ orders of magnitude across $10^{-13}\;\text{eV} \lesssim m_a \lesssim 10^{-2}\;\text{eV}$, with the limits being most pronounced for masses where the distortion acquires its characteristic non-thermal shape.

\paragraph{Patchy screening.}
Resonant conversions within the magnetized plasma of large-scale structure produce anisotropic spectral distortions correlated with the galaxy distribution. This ``patchy screening'' effect was formalized within the halo model~\cite{Asgari2023,Mondino2024}, showing that the screening optical depth $\tau^a(\hat{n}) \propto \omega\, g_{a\gamma\gamma}^2$ has a characteristic linear frequency dependence that enables separation from other signals via ILC (Internal Linear Combination \cite{Tegmark2003}) techniques. The cross-power spectrum between the axion distortion and large scale structure ($C_\ell^{T^a g}$) typically provides the dominant sensitivity channel. The first observational constraints from this mechanism, obtained by cross-correlating foreground-cleaned \textit{Planck} maps with the \textit{unWISE} galaxy catalogue, yielded $g_{a\gamma\gamma} \lesssim 2 \times 10^{-12}\;\text{GeV}^{-1}$ (95\% c.l.) for $10^{-13}\;\text{eV} \lesssim m_a \lesssim 10^{-12}\;\text{eV}$~\cite{Goldstein2024}, competitive with the tightest astrophysical bounds. Future constraints on spectral distortions will improve this patchy program primarily through its ability to produce improved foreground models which will allow for a cleaner extraction of the axion-template signal. This, in conjunction with the influx of galaxy survey data from DESI, Euclid, and LSST, could lead to orders-of-magnitude improvement in the $g_{a\gamma\gamma}$ constraints. 

\paragraph{Conversions in voids, the IGM, and Milky Way.}
CMB photons can get converted into ALPs in the magnetic fields of local voids, the intergalactic medium, and our own Galaxy \cite{Mukherjee2018}. For $10^{-13}~ {\rm eV} \lesssim m_a \lesssim 10^{-11}~{\rm eV}$, resonant conversion can occur in the large scale coherent Galactic magnetic field. If this happens, it leaves a characteristic large scale anisotropic signature correlated with the Galactic magnetic field morphology in addition to the induced spectral distortion.  For $m_a \lesssim 10^{-14}~{\rm eV}$, the resonance condition ($m_{a} = m_{\gamma}$) is never satisfied. It is possible, however, for CMB photons to convert (non-resonantly) to ALPs in stochastic magnetic fields, yielding a large scale anisotropy in the spectral distortion correlated with the magnetic field and electron density distribution in our Galaxy and the surrounding IGM and voids. Since we have good understanding of magnetic fields in our own Galaxy, the degeneracy due to uncertainty in the magnetic fields is absent in the Galactic signals.

\paragraph{Additional distortion probes.}
Complementary constraints on heavier ALPs ($m_a \gtrsim 27\;\text{eV}$) arise from perturbative and stimulated $a \to \gamma\gamma$ decays, injecting photons directly into the CMB. Significant work has been done to perform a comprehensive \texttt{CosmoTherm} analysis combining COBE/FIRAS, EDGES, and \textit{Planck} data~\cite{Bolliet:2020ofj}. Future measurements of the CMB spectrum will extend these decay constraints by several orders of magnitude, probing regions relevant to the QCD axion band. Conversions in individual galaxy clusters also provide independent limits of $g_{a\gamma\gamma} \lesssim \mathcal{O}(10^{-11})\;\text{GeV}^{-1}$~\cite{Schlederer2016}, with projections for SO and CMB-S4 reaching $\sim 10^{-13}\;\text{GeV}^{-1}$~\cite{Mukherjee2020,Mehta2024a}. If such measurements are absolutely calibrated, they will complement these angular-resolution-driven searches by constraining the sky-averaged monopole distortion, similar to the patchy screening scenario.

\subsection{Dark photons}
\label{sec:dark_gamma}
The dark photon $A'$ is one of the simplest low-energy extensions of the Standard Model \cite{Holdom:1985ag,Fabbrichesi:2020wbt}, and has been proposed as a dark matter candidate or as a member of a larger dark sector (see~\cite{Fabbrichesi:2020wbt} for a review). 
$A'$ is the gauge boson of a dark $U(1)'$ symmetry, and has two model parameters: its mass $m_{A'}$ and the kinetic mixing parameter $\epsilon$, which controls the strength of its coupling to the Standard Model photon. The Lagrangian can be written as
\begin{equation}
	\mathcal{L} \supset -\frac{1}{4} F^2_{\mu\nu} - \frac{1}{4}\left(F_{\mu\nu}^\prime\right)^2 +\frac{1}{2}m_{A'}^2(A'_\mu)^2-\frac{\epsilon}{2} F_{\mu\nu}^{\prime}F^{\mu\nu}\, ,
\end{equation}
where $F_{\mu\nu}^{\prime}$ and $F^{\mu\nu}$ are the field strength tensors of the dark and Standard Model photons, respectively. 
Through this kinetic mixing, $\gamma \leftrightarrow A'$ oscillations can occur, which in turn lead to the injection or removal of SM photons, thereby causing distortions to the CMB blackbody spectrum~\cite{Chluba2013Green,Chluba2015}. 
The probability of conversion is generally extremely small, but Ref.~\cite{Mirizzi:2009iz} pointed out that it is resonantly enhanced when the effective photon mass in a plasma, $m_{\gamma}^2(z) \propto n_e(z)$ is equal to $m_{A'}^2$, where $n_e$ is the free electron number density. 
This resonant condition means that $\gamma \leftrightarrow A'$ conversions typically occur close to a characteristic redshift $z_*$ set by requiring that the mean plasma mass $\overline{m_\gamma^2}(z_\star) = m_{A'}^2$. 
Below, we first discuss $\gamma \to A'$ conversions, which occur as long as $\epsilon>0$ and do not rely on dark photons being the dark matter, leading to energy removal from the CMB; since the physics of these conversions differ significantly depending on whether they happen before or after recombination, we discuss these two regimes separately.
Here, we also discuss SD anisotropies induced by $\gamma \to A'$ conversions. 
Then, we consider $A' \to \gamma$ conversions for the case where the $A'$ is an ultralight dark matter candidate, first discussed in Ref.~\cite{Arias:2012az}.  

\paragraph{$\gamma \to A'$ pre-recombination conversions.} 
For dark photons with $\qty{e-10}{\eV}\lesssim m_{A'} \lesssim \qty{5e-5}{\eV}$, resonant conversions occur before recombination~\cite{McDermott:2019lch,Chluba2024,Arsenadze2024}. 
Because $\gamma \to A'$ conversions remove photons from the CMB while Compton scattering is efficient, sophisticated modeling is needed to track how the CMB spectrum evolves with time in order to understand the full distortion created by dark photons~\cite{Chluba2015}. 
Early work on the subject \cite{Mirizzi:2009iz,McDermott:2019lch} provided some estimates of the spectral distortion, but more recent work then modeled the full photon removal effects using both a computational \cite{Chluba2024} and a Green's function approach \cite{Chluba2015, Chluba2024, Arsenadze2024}. 
Both works found that these effects can flip the sign and increase the amplitude of the distortion relative to earlier energy-injection-based estimates, setting limits on the kinetic mixing parameter of $\epsilon \lesssim 10^{-7}$ over most of the relevant mass range. 
Furthermore, Ref.~\cite{Hook:2023smg} also demonstrated that spectral distortion measurements are able to constrain a more complicated dark sector containing a dark photon and an axion. 
The mixed coupling to photons and axions induces photon-to-dark-sector conversions with model-dependent spectral signatures, including Doppler and free-streaming distortions with shapes that differ qualitatively from standard $\mu$/$y$ signals. 
Currently, COBE/FIRAS measurements set the strongest constraints on dark photons over much of this mass range, and forecasts show that space-based spectral distortion experiments would be sensitive to kinetic mixing $\epsilon$ two orders of magnitude smaller than the FIRAS bounds ~\cite{Chluba2024,Arsenadze2024}.

Additionally, measuring SD anisotropies could provide further constraining power in the search for dark photons beyond the monopole signal~\cite{Evangelista2026}. 
These anisotropies can be generated by fluctuations in local quantities (e.g. baryon density and blackbody temperature) and variations in local clocks, typically parametrized by the potential $\Psi^{(1)}$, that can alter the redshifts at which $\gamma \to A'$ conversions occur and the probability of conversion. 
Notably, also purely isotropic conversions can give rise to SD anisotropies, since the background monopole enters as a source term in the perturbed frequency hierarchy through the collision terms in the Boltzmann equations (see Ref.~\cite{kite_spectro-spatial_2023-III} and Sec.~\ref{sec:CMB-sds-anisotropies} for a more complete discussion). 
Recent work computed constraints on $\epsilon$ using the $\mu \times T$ cross-power spectrum and \textit{Planck} data, finding them just marginally weaker than those obtained with \textit{COBE/FIRAS} for the monopole \cite{Evangelista2026}. Moreover, interesting patterns resembling iso-curvature-type perturbations have been seen as a correction to the temperature power spectrum, even for early $\gamma \to A'$ conversions during epochs where other SDs are not present~\cite{Evangelista2026}. 
	
\paragraph{$\gamma \to A'$ conversions in the free-streaming era.} 
SD experiments are also sensitive to lighter dark photons with $\qty{e-15}{\eV} \lesssim m_{A'} \lesssim \qty{e-10}{\eV}$, which undergo resonant conversions during the free-streaming era.
At this point, CMB photons are unable to redistribute themselves efficiently, leaving any imprint of $\gamma \to A'$ conversions frozen in place in the CMB spectrum \cite{Arsenadze2024}.
Initial work focused on resonant $\gamma \to A'$ conversions assuming a homogeneous value for $m_{\gamma}^2(z)$ \cite{Mirizzi:2009iz}.
Later work expanded on this simplified treatment by considering inhomogeneities in the plasma mass, which means that multiple resonant conversions can occur along a single line of sight, significantly modifying the predicted magnitude of the signal~\cite{Bondarenko:2020moh,Garcia:2020qrp,Caputo2020,Caputo:2020rnx}.
These distortions to the CMB spectrum measured by FIRAS remain one of the strongest limits on the kinetic mixing parameter $\epsilon$, which a space-based SD experiment could again exceed by up to two orders of magnitude \cite{Kunze:2015noa,Caputo2020}.
Beyond distortions to the sky-averaged CMB blackbody spectrum, inhomogeneities in the plasma mass also create patchy spectral distortions.
Inhomogeneities in $m_\gamma^2$ mean that CMB photons traveling along two different lines of sight have different probabilities of conversions, creating new CMB anisotropies.
Searching for these anisotropies and their cross-correlation with large-scale structure set a limit of $\epsilon\lesssim 4\times 10^{-8}$ for $\qty{2e-13}{\eV}\lesssim m_{A'} \lesssim \qty{2e-12}{\eV}$  \cite{Pirvu:2023lch,Aramburo-Garcia:2024cbz,McCarthy2024}. 
Future cross-correlations of large-scale structure with radio measurements of this patchy screening effect are forecasted to improve upon this sensitivity by a factor of $\sim 4$~\cite{Baker2025a, Baker2025b}.
These probes require precise modeling of astrophysical foregrounds in order to separate out the dark photon signal.
SD experiments promise to significantly improve our understanding of these foregrounds, which could greatly improve our ability to extract a faint dark photon signal from maps of CMB anisotropies. 

\begin{figure*}[t]
    \centering
    \includegraphics[width=0.7\linewidth]{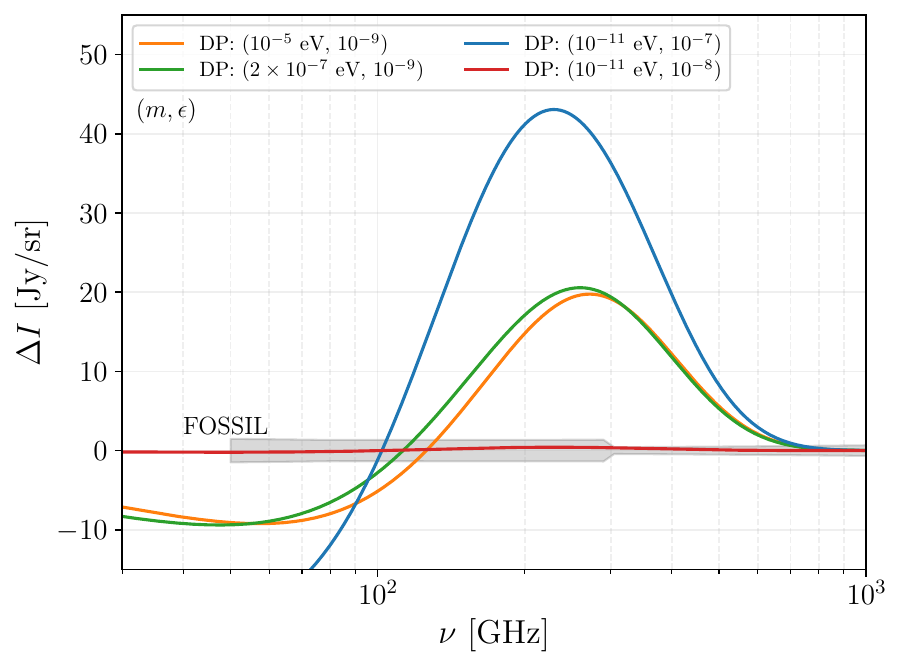}
    \caption{A variety of dark photon spectral distortions ($\gamma \rightarrow A'$). For resonant conversions occurring in the $\mu$-epoch (orange), during the $\mu$-y transition (green), and post-recombination (blue/red), we show parameter space points near the detection threshold of \FOSSIL. The ordered pairs are $(m,\epsilon)$, and each distortion is computed in the homogeneous limit.}
    \label{fig:axion_DP_distortion}
\end{figure*}

\paragraph{$A' \to \gamma$ conversions for $A'$ dark matter.}
If dark photons comprise a substantial fraction of the dark matter, then $A' \to \gamma$ conversions can also occur.
These conversions inject additional photons and energy into baryons.
By energy conservation, these photons have energy equal to $m_{A'}$, which is $\lesssim \qty{5e-5}{\eV}$ for the cases that we consider here. 
These photons are then rapidly absorbed via inverse-Bremsstrahlung, causing substantial heating effects. 
If $A' \to \gamma$ conversions occur before recombination, this heating effect can create significant spectral distortions which constrain $\epsilon \lesssim 10^{-12}$ for dark photon dark matter \cite{Arias:2012az,McDermott:2019lch}.
Ref.~\cite{Witte:2020rvb} also showed that a space-based SD experiment could be sensitive to spectral distortions caused by baryon heating from $A' \to \gamma$ conversions at late times as well. 
However, it is important to note that recent work has suggested that the heating effects from $A' \to \gamma$ conversions could be much smaller than previously thought, which would significantly weaken these constraints \cite{Hook:2025pbn}. Representative spectral distortion signatures for dark photons are shown in Fig.~\ref{fig:axion_DP_distortion}.

\subsection{Alternative dark matter paradigms} \label{sec:alt_DM}
The non-detection of dark matter in direct and indirect experiments motivates us to consider that the dark sector may be as complex as the visible sector. A wide class of dark matter (DM) candidates beyond the canonical WIMP or axion paradigms predict distinctive imprints on the CMB frequency spectrum.  These range from dark sectors with their own gauge interactions, to composite or macroscopic objects formed at early cosmological phase transitions.  The signatures of dark matter in these cases on cosmological observables may be subtle and quite different from what the standard searches are designed for. In any viable dark matter model, the interactions with the visible sector must die down to negligible levels by the time of recombination in order to satisfy CMB anisotropy and low redshift/post recombination constraints. However, dark matter interactions are allowed to be significant before recombination, constrained only by the  COBE-FIRAS experiment in the 1990s. The next CMB spectral distortion experiment will therefore have a substantial window of discovery for these very interesting dark matter candidates. 

\paragraph{Millicharged particles and atomic dark sectors.}
As already discussed in the previous sections, dark photons are among the simplest extensions to the Standard Model of particle physics. If in addition to the dark photon, there exist fermions charged under the new $U(1)_{D}$, they generate an effective plasma mass for the dark photon, enabling resonant $\gamma \leftrightarrow A'$ oscillations that distort the CMB spectrum~\cite{Berlin:2022hmt}, thus allowing constraints to be set on scenarios where the vacuum mass of the dark photon is $m_{A'}^{\rm vac}=0$.  Even as a small subcomponent of the total DM density ($f_\mathrm{DM} \lesssim 10^{-2}$), millicharged particles (mCPs) with masses in the range $m_\chi \sim (1$--$100)\;\mathrm{GeV}$ and effective charges $q_\chi \gtrsim 10^{-7} \times (m_\chi/\mathrm{GeV})^{1/2}$ produce signals accessible to current COBE/FIRAS bounds \cite{Berlin:2022hmt}, complementary to stellar cooling and direct detection limits.  For millicharged dark radiation produced through freeze-in, existing FIRAS data already probes energy densities well below the $\Delta N_\mathrm{eff}$ threshold, and a \FOSSIL-class mission would reach the irreducible freeze-in abundance itself.

These scenarios become particularly rich when bound states form in the dark sector.  In atomic dark matter (aDM) models, dark protons and electrons can form dark hydrogen~\cite{Kaplan:2009de} when the dark sector radiation ($A'$) is sufficiently cold.  For a wide range of parameter space, the dark plasma undergoes its own recombination at redshifts $10^3 \lesssim z \lesssim 10^6$, causing the dark photon thermal mass to decrease exponentially~\cite{Adams:2026xxx}.  This allows the resonant conversion condition to be satisfied in the primordial distortion epoch ($z \gtrsim 10^3$), generating characteristic $\mu$- and $y$-type patterns.  It was recently shown that in the positronium limit ($m_{p_{\rm D}} = m _{e_{\rm D}}$), COBE/FIRAS already constrains the dark milli-charge to $q_\mathrm{D} \lesssim 10^{-7}$--$10^{-6}$ across a broad range of dark fermion masses $m_{e_\mathrm{D}} \lesssim 9\;\mathrm{MeV}$, with \FOSSIL projected to improve these bounds by more than two orders of magnitude~\cite{Adams:2026xxx}. Generally speaking, data from \FOSSIL would provide us with leading constraints in regions of parameter space traditionally difficult to probe using direct detection experiments, $\Delta N_\mathrm{eff}$, and supernovae.

\paragraph{Composite dark matter with electromagnetic moments.}
Even electrically neutral DM can couple to the radiation field through higher-order electromagnetic moments (charge radius, polarizability, or magnetic dipole) if it possesses internal structure~\cite{pospelov2000}.  Such composite bound states would invariably have many internal levels and transitions between these levels would be possible through absorption and emission of photons. The lowest states in such systems would be of most interest in general, as they would be most easily excited. For example, we may have a transition similar to the spin flip transition of hydrogen for a composite state made up of two fermions. Such systems may be modeled phenomenologically as two-level systems separated by a transition energy $\Delta E$ that can range from radio frequencies  to gamma rays \cite{Ganguly.2024}. For cosmology and CMB spectral distortions, we can parameterize such systems in terms of the transition energy, mass of the composite state, and the strength of transition. These parameters can then easily be mapped to specific particle physics models.  Electromagnetic transitions between these internal levels of dark matter absorb or inject photons at specific frequencies. The resulting spectral distortions can have distinctive non-thermal spectral features~\cite{Ganguly:2025}, in addition to the thermal distortions from cooling or heating of the CMB. The actual shape of the CMB spectral distortion depends sensitively on the parameters of the model (e.g. the mass of dark matter, its internal transition energies, and the strength of transition) and thus is rich in information about the intrinsic properties of dark matter. An important aspect of spectral distortions in these dark matter models is that they depend not just on the interaction of dark matter with the visible sector but also on dark matter self interactions. For $\mathrm{MeV}$-mass DM with $\Delta E \sim 1$--$10\;\mathrm{eV}$, it was found that COBE/FIRAS constraints on the electromagnetic coupling are already three orders of magnitude stronger than terrestrial direct detection limits.  The next CMB spectral distortion experiment would considerably extend this reach further. In  particular, for transition energies in the $0.1$--$10\;\mathrm{eV}$ window and for light dark matter with DM masses up to $m_{\chi} \simeq 100$ MeV, the non-thermal distortion shapes provide an additional diagnostic handle allowing these spectral distortions to be distinguished from thermal as well as non-thermal spectral distortions resulting from energy injection in the early Universe. 

\paragraph{Macroscopic dark matter.}
Another class of non-particle dark matter candidates consists of macroscopic composite objects. Macroscopic composites of quarks or baryons (``Macros'') can be assembled during phase transitions in the early Universe, and represent a qualitatively different DM candidate~\cite{Witten:1984rs,Kumar2019}.  Their hot, nuclear-density interiors act as a heat reservoir, cooling slowly compared to the ambient plasma, emitting photons from the surface and neutrinos from the bulk throughout the distortion epoch.  The resulting $\mu$ and $y$ signals are mass-independent for fixed internal density, depending instead on the equation of state and surface composition~\cite{Kumar2019}.  For nuclear-density Macros comprising the entirety of the DM, the predicted distortions would be detectable by the next generation CMB spectral distortion experiment such as \FOSSIL or \PIXIE.  The photon emission from the Macros may persist well beyond recombination as a distinct late-time background. One such realization of macroscopic composite dark matter is provided by axion quark nuggets, as discussed below.

\paragraph{Axion quark nuggets.}
Axion quark nuggets (AQNs), formed during the QCD phase transition through axion-induced charge separation, offer a unified explanation for the observed similarity between the dark and baryonic matter abundances, $\Omega_\mathrm{DM} \sim \Omega_\mathrm{{\rm b}}$ ~\cite{Zhitnitsky:2002qa, Zhitnitsky_2021,vanwaerbeke2026qcddrivendarkmatteraqns}. Unlike conventional particle dark matter candidates, AQNs are macroscopic composite objects composed of stable quark matter with masses ranging from grams to kilograms while retaining microscopic (sub-micron) radii. In this framework, the dark matter population consists of both matter and antimatter nuggets, with the latter generating observable signatures through annihilation with ambient baryons.

AQNs could be matter or antimatter objects. Antimatter AQNs annihilate with ambient baryons, injecting energy at a rate proportional to both the baryon and dark matter number densities. Although AQNs behave as cold dark matter on cosmological scales, interactions between antimatter nuggets and the surrounding baryonic plasma continuously inject energy into the primordial medium ~\cite{Majidi_2024}. The annihilation rate depends on the evolving physical conditions of the plasma, including the ionization fraction, gas temperature, and the relative baryon--AQN velocity. This scaling, distinct from conventional scenarios of DM annihilation or decay, produces a unique redshift dependence for the energy injection, allowing model discrimination ~\cite{Majidi2025}. Unlike conventional particle dark matter annihilation, this energy injection history cannot be captured by a single effective annihilation parameter; instead, it requires following the evolving properties of both the plasma and the AQN population. 

AQN-baryon annihilation may also inject relativistic particles into the plasma, triggering electromagnetic particle cascades. The resulting non-thermal spectral distortions have a very different shape from the standard thermal distortions \cite{Acharya:2018iwh,Acharya:2019owx}. The shape of the distortions depends on the annihilation channels as well as how the rate of annihilation evolves with time \cite{Acharya:2018iwh,Acharya2019}.  Consequently, this distinctive energy injection history, along with the non-thermal nature of the spectral distortions, makes CMB spectral distortions a powerful probe of the AQN scenario and provides an observational signature that differs from conventional dark matter models.

Recent calculations using a modified version of the CLASS Boltzmann code ~\cite{Lucca:2019rxf,Majidi2025} have shown that AQNs in the currently unconstrained mass range of $(10$--$1000)\;\mathrm{g}$ produce both $\mu-$ and $\mathrm{y}-$type spectral distortions while leaving the CMB anisotropy power spectra and the inferred cosmological parameters largely unaffected. Unlike anisotropies, which are primarily sensitive to perturbations in the matter and radiation fields, spectral distortions directly probe the integrated thermal history of the Universe. Since AQN annihilation continuously injects energy into the primordial plasma over an extended range of redshifts, its cumulative effect is encoded much more efficiently in the CMB frequency spectrum than in the anisotropy power spectra. As a result, the predicted signal is dominated by the $\mu$-distortion component, reflecting the greater sensitivity of the CMB spectrum to energy injection at higher redshifts. The predicted distortion reaches $\mu\simeq 5\times 10^{-8}$, making it easily within reach of the \FOSSIL sensitivity. In addition, the predicted $\mathrm{y}-$distortion remains below current observational limits while falling within the projected sensitivity of the future CMB spectral distortion experiments. Spectral distortions therefore provide information complementary to CMB anisotropy measurements by probing the integrated thermal history of the Universe rather than the evolution of cosmological perturbations. 

Future CMB spectral distortion missions, in particular \FOSSIL, will therefore provide a unique opportunity to test macroscopic dark matter candidates that remain largely unconstrained by existing observations. Together, these results highlight the potential of CMB spectral distortions as a powerful probe of macroscopic composite dark matter models.

\section{Distortions Beyond the Monopole}
\label{sec:CMB-sds-anisotropies}
Most studies of CMB SDs have concentrated on sky-averaged monopole signals. This is a natural starting point: the distortion monopole records the total energy and photon production throughout the thermal history of the Universe and can be measured only through absolute spectroscopy. However, the physical processes responsible for SDs need not be spatially uniform. Perturbations in the matter density, radiation field and gravitational potentials modulate both the sources of distortions and the efficiency with which they thermalise and propagate through the medium. The resulting SD signal therefore varies across the sky, providing information that goes far beyond the average spectrum \citep{Chluba:2012gq, Chluba:2022efq, kite_spectro-spatial_2023-III}.

SD anisotropies combine spectral and spatial information. Their frequency dependence can distinguish changes in the temperature from genuine departures away from a blackbody, while their angular distribution traces the spatial fluctuations of the underlying source in an epoch-dependent manner. Cross-correlations with the primary CMB temperature and polarisation anisotropies are especially valuable because they isolate signals associated with the primordial perturbations and are less susceptible to uncorrelated foreground contamination. They can also reveal processes that produce only a weak monopole distortion, or none at all, but leave a spatially-varying signature through their coupling to the inhomogeneous Universe.

The thermal Sunyaev-Zeldovich effect provides the best-known example of a SD anisotropy: inverse Compton scattering by hot electrons in groups and clusters gives rise to a spatially varying $y$-type distortion that traces the distribution and thermodynamic state of ionised gas \citep{Zeldovich1969, Sunyaev1980ARAA, Rephaeli1995ARAA, Carlstrom2002, Mroczkowski2012}. This can be simply thought of as a late-time anisotropic heating process caused by the structure formation process which today is routinely used to detect clusters (see Sec.~\ref{sec:SZ_effect} for broader perspectives). However, before recombination, analogous anisotropies can be generated by the dissipation of primordial perturbations, spatially varying energy or photon injection, particle decay and photon conversion. These early signals probe physical conditions over $10^{3}\lesssim z\lesssim 2\times10^{6}$, connecting the frequency-dependent CMB sky to epochs that are largely inaccessible through conventional temperature and polarisation anisotropies.

Calculating these signals requires the simultaneous evolution of their spatial, angular and frequency dependence. A direct solution of the frequency-dependent photon Boltzmann equation for every Fourier mode is computationally prohibitive. This difficulty was overcome through the development of the frequency hierarchy (FH) method \citep{Chluba:2022xsd,
Chluba:2022efq, kite_spectro-spatial_2023-III, Chluba2026}. The key simplification realized in those works was to decompose the photon spectrum into a compact basis containing the temperature-shift spectrum, $G(x)$, the standard $y$- and $\mu$-distortion shapes, $Y(x)$ and $M(x)$, and a sequence of boosted $y$-type shapes $Y_i(x)$, as introduced in Eq.~\eqref{eq:basis}. The latter describe the intermediate distortions generated during the transition between the $\mu$- and $y$-eras \citep{Chluba:2022xsd}. The spectral amplitudes associated with this basis are then evolved through a generalised Boltzmann hierarchy, reducing the dimensionality of the problem by more than two orders of magnitude while retaining an accurate description of the relevant spectral evolution \citep{Chluba:2022efq, kite_spectro-spatial_2023-III, Evangelista2025}.

The FH formalism incorporates the effects of gravitational and Doppler driving, Compton scattering, double Compton emission, Bremsstrahlung, perturbed thermalisation and spatially varying energy release. It therefore allows distortion transfer functions and angular power spectra to be calculated in much the same way as conventional CMB temperature and polarisation spectra. Recent improvements ensure that the correct equilibrium solutions are built directly into the evolution equations and include stimulated-scattering effects, kinematic corrections, corrections to the standard perturbation variables and explicit photon source terms \citep{Chluba2026}. These additions are particularly important when distinguishing genuine distortions from apparent spectral components generated by changes in the reference temperature, and when considering processes that add or remove photons rather than simply heating the plasma.

This framework opens several distinct scientific opportunities. Four particularly relevant examples are discussed below: primordial non-Gaussianity, decaying dark matter, photon–dark-photon conversion and departures from the standard CMB temperature–redshift relation. Importantly, SD anisotropies can also be extracted with existing, ongoing and planned CMB imaging experiments, thus defining a  wide range of synergistic targets for future exploration in conjunction with CMB spectroscopy.

\begin{figure}
    \centering
    \includegraphics[width=0.50\linewidth]{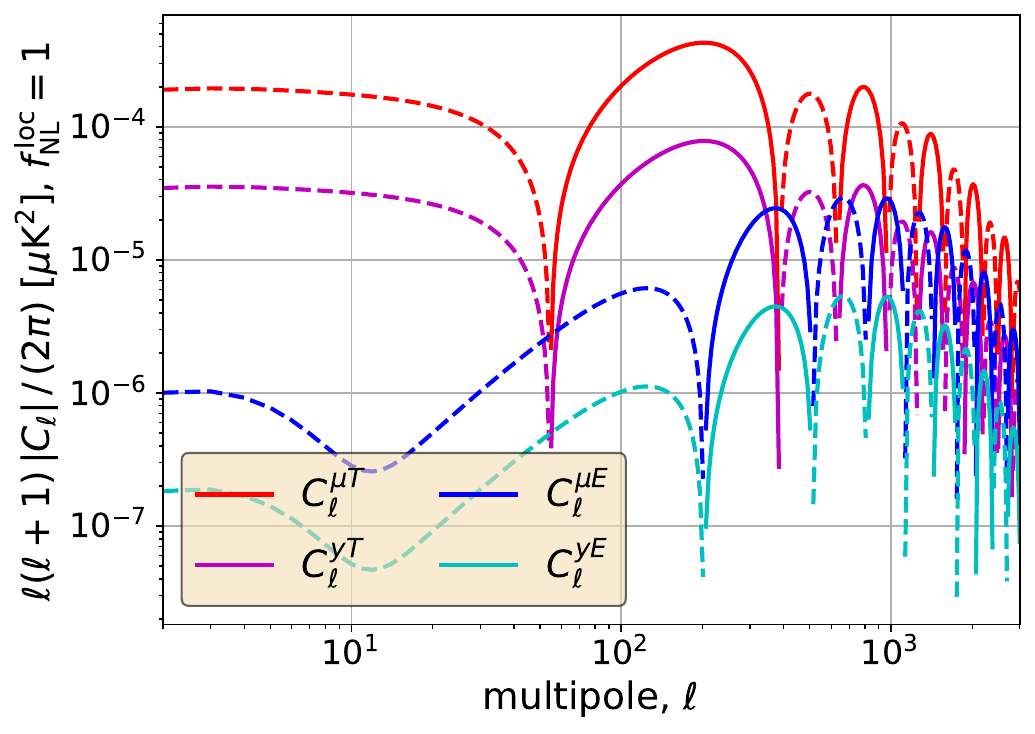}
    \includegraphics[width=0.47\columnwidth]{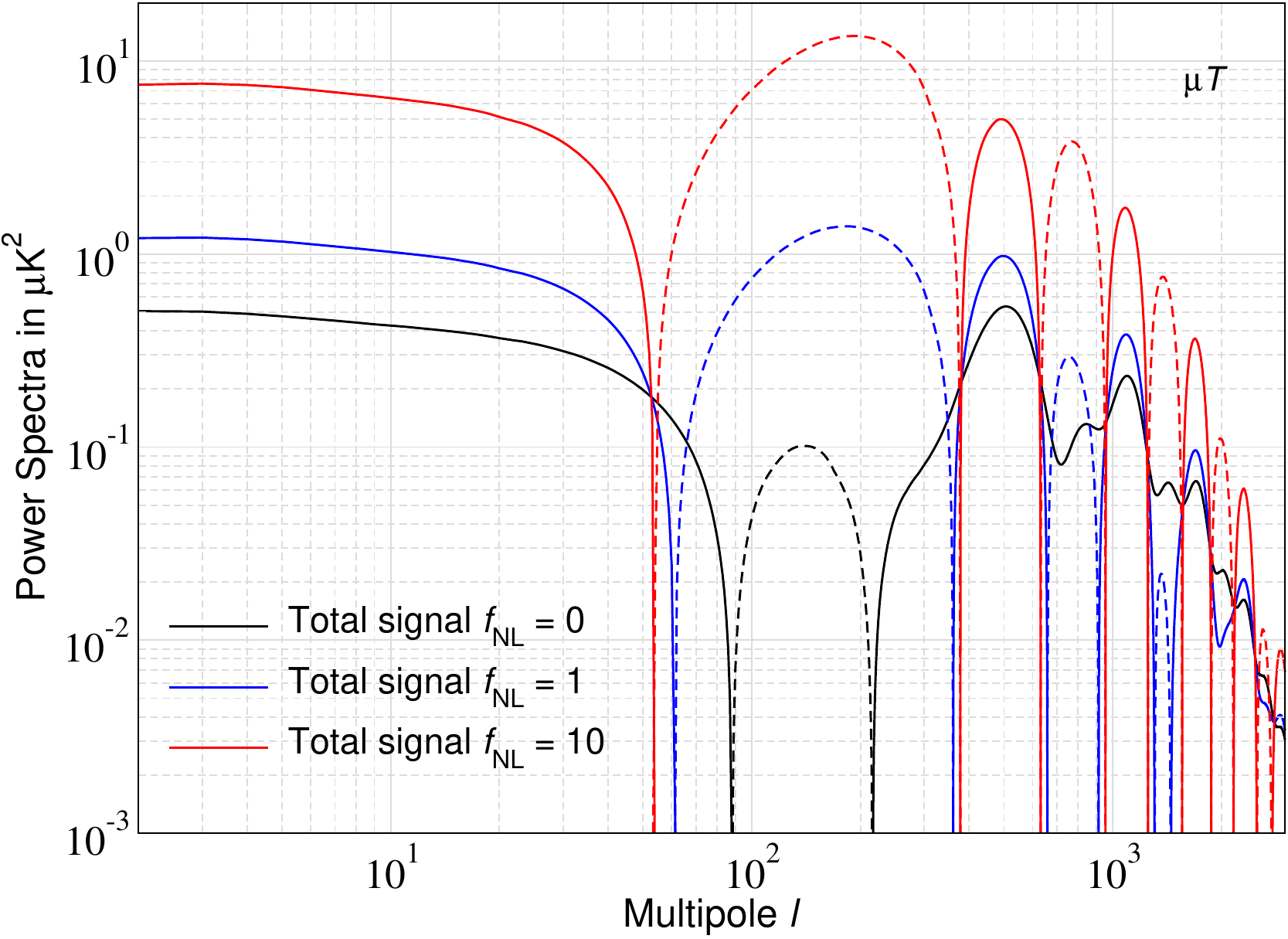}
    \caption{Left: Absolute values of the angular cross-power spectra between primordial spectral distortion anisotropies and primary CMB anisotropies for $\langle\mu\rangle =2\times 10^{-8}$, $\langle y\rangle =4\times 10^{-9}$, and $f_{\rm NL}^{\rm loc}=1$: $C_\ell^{\mu T}$ (red), $C_\ell^{yT}$ (purple), $C_\ell^{\mu E}$ (blue), and $C_\ell^{yE}$ (cyan). Solid (dashed) lines indicate positive (negative) values of the cross-spectra.  The theoretical predictions are based on the calculations presented in \cite{Ravenni2017} and were recently refined in \citep{Chluba2026muT}. Right: Illustration of the $\mu T$ signals for a scenario with enhanced small-scale power (a factor of $\simeq 10^3$ in amplitude around $k_{\rm p}\simeq 10^2\,{\rm Mpc}^{-1}$). We then illustrate the $\ell$ dependence of the total signal for varying values of $f_{\rm NL}$, demonstrating how the spectro-spatial information can in principle be used to distinguish different models and break degeneracies. The figure is taken from \citep{Chluba2026muT}.}
    \label{fig:sd_crossspectra}
\end{figure}

\subsection{Primordial non-Gaussianity}
SD anisotropies provide a unique probe of primordial non-Gaussianity on scales far smaller than those accessible through conventional measurements of CMB anisotropies or large-scale structure. As discussed in Section~\ref{sec:small-scale-structure}, the dissipation of primordial density perturbations on small scales in the early Universe inevitably generates an average (monopole) $\mu$- and $y$-type spectral distortion of the CMB blackbody radiation. In the presence of primordial non-Gaussianity (PNG), however, couplings between long- and short-wavelength perturbations modulate the amplitude of small-scale power across the sky. This spatial variation in the dissipation rate leads to anisotropic spectral distortions that are correlated with large-scale CMB temperature and polarization anisotropies \citep{Pajer2012, Ganc2012, Biagetti2013, Ota2015aniso_iso, Emami:2015xqa, Khatri2015aniso, Chluba2017, Ota:2016mqd, Ravenni2017, Cabass:2018jgj}.

A wide class of well-motivated inflationary scenarios predict such signatures. These include multi-field inflation models, scenarios with non-standard (non-Bunch-Davies) initial conditions, and other mechanisms that generate squeezed primordial bispectra and higher-order correlations \citep[e.g.,][]{Ganc2012, Dimastrogiovanni2016, Bartolo2016trispec, Shiraishi2016}. In general, non-Gaussian information is encoded in higher-order correlation functions of the primordial curvature perturbations, notably the bispectrum and trispectrum. Measurements of the CMB temperature and polarization anisotropies by \Planck\ have already placed stringent constraints on these quantities on large scales \citep{Planck2013ng, Akrami:2019izv}. Spectral distortion anisotropies extend these tests to a completely different regime, probing primordial fluctuations at wavenumbers up to $k \simeq 10^3\,{\rm Mpc}^{-1}$, many orders of magnitude beyond the reach of conventional CMB and galaxy surveys.

Particularly powerful are cross-correlations between spectral distortion anisotropies and large-scale CMB temperature ($T$) or polarization ($E$) anisotropies. These observables are sensitive to squeezed configurations of the primordial bispectrum and trispectrum, in which one mode has a wavelength much larger than the others \citep{Ganc2012, Biagetti2013, Ota2015aniso_iso, Bartolo:2015fqz}. A canonical example is the local-type bispectrum, parameterized by the amplitude $f_{\rm NL}^{\rm loc}$. Because single-field inflation predicts an extremely small signal in this limit, any statistically significant detection of a $\mu T$, $\mu E$, $y T$, or $y E$ correlation would rule out all standard single-field inflation models \citep{Cabass:2018jgj}. Conversely, increasingly stringent upper limits would constrain broad classes of multi-field inflation models \citep{Dimastrogiovanni2016} and non-standard initial-condition scenarios \citep{Ganc2012}.

Spectral distortion anisotropies can be measured with both absolute CMB spectrometers and differential CMB imaging experiments. The $\mu T$ angular cross-power spectrum can be written approximately as \citep[e.g.,][]{Ganc2012, Chluba2017}
\begin{equation}
	C_\ell^{\mu T} 
	\simeq
	- 12 \,\langle\mu\rangle \, f_{\rm NL}^{\rm loc} \, 
	\int \! {\rm d} k \,\frac{2}{\pi}\, k^2 \,
	\mathcal{T}_\ell^T \! (k)
	\,\frac{j_{\ell}(k \, r_{\rm ls})}{5}\,
	P(k) \, {\rm e}^{-\frac{15k^2}{8k_{{\rm rec}}^2}}\,,
\label{eq:ClXi-X_squeezed}
\end{equation}
where $\langle \mu \rangle$ is the average $\mu$-distortion generated by dissipation of small-scale perturbations, $\mathcal{T}_\ell^T$ is the temperature transfer function, $r_{\rm ls}$ is the comoving distance to last scattering, and $k_{\rm rec}$ denotes the diffusion-damping scale at recombination (see left panel of Figure~\ref{fig:sd_crossspectra}). The amplitude of the cross-correlation is directly proportional to both the average distortion and the level of PNG. Consequently, in this scenario an absolute measurement of the distortion monopole is required to break the degeneracy and fully interpret any observed anisotropic signal, highlighting the importance of combining spectroscopic and imaging CMB observations \citep{Chluba2017, Chluba2026muT}. Additional opportunities for constraining and distinguishing various models arise in scenarios with enhanced small scale power (see right panel of Fig.~\ref{fig:sd_crossspectra} for illustration). As recently demonstrated \cite{Chluba2026muT}, in these cases two distinct contributions to the $\mu T$ signal arise, one from the propagation of the average distortions through the perturbed medium and the other from anisotropic dissipation. Together these give a rich spectro-spatial phenomenology that can be explored with CMB imaging and spectroscopy.

Detecting spectral distortion anisotropies is observationally challenging, because the signal is extremely faint and must be separated from bright Galactic and extragalactic foregrounds. Cross-correlation with CMB temperature and $E$-mode polarization maps greatly enhances detectability and robustness against astrophysical foregrounds, and has therefore been identified as a promising new observable for future CMB imaging experiments \citep{Remazeilles2018:mu, Remazeilles2022:mu, Zegeye2023, Zegeye2025}, with preliminary upper bounds already obtained from current data \citep{Rotti2022, Bianchini:2022dqh}.

\begin{figure}
    \centering
    \includegraphics[width=0.5\linewidth]{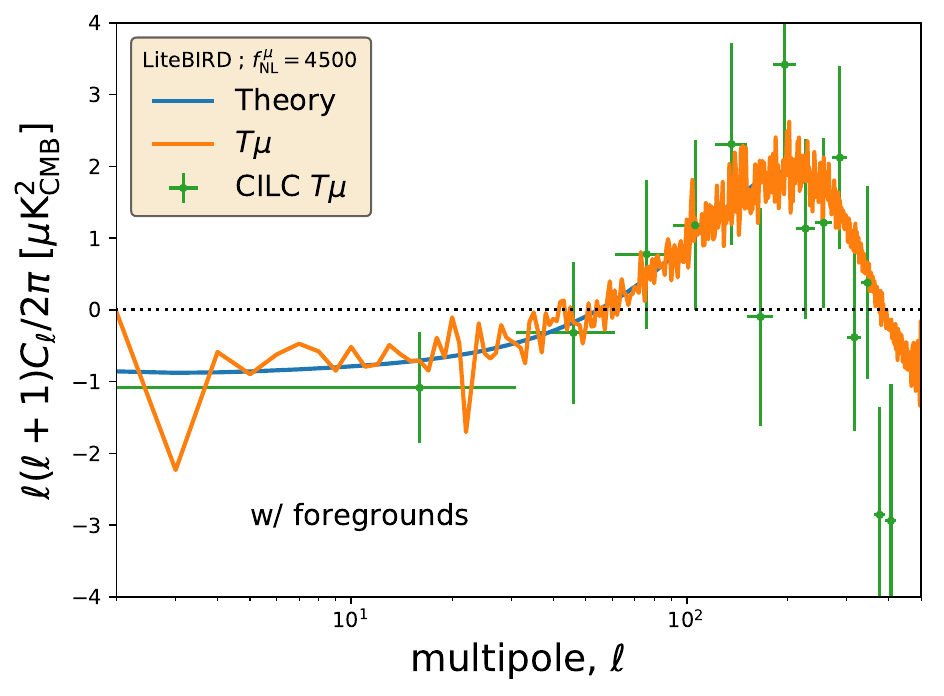}~
    \includegraphics[width=0.5\linewidth]{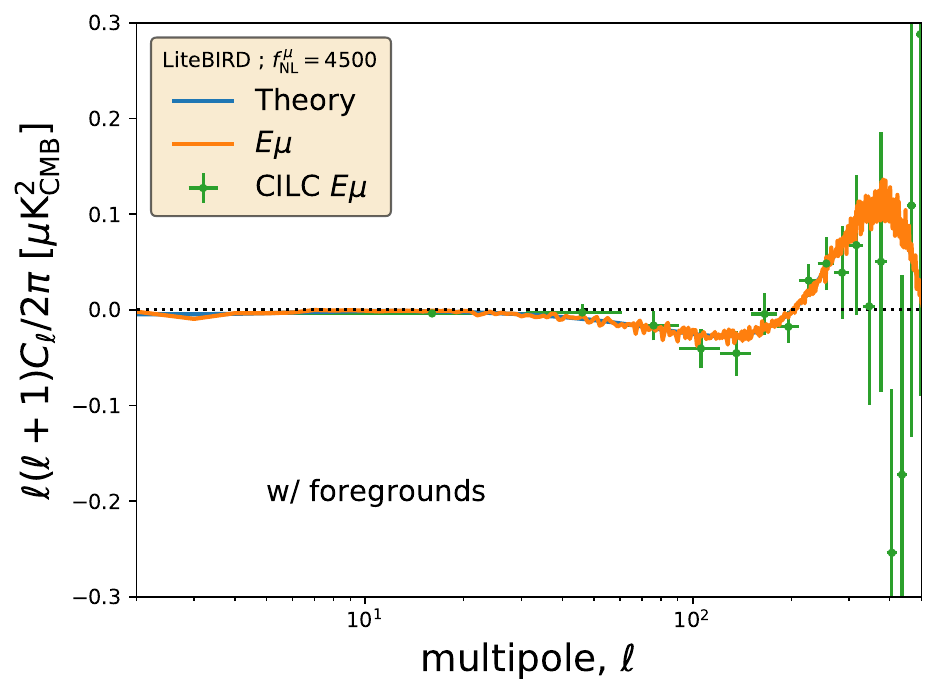}
    \caption{Reconstructed $\mu T$ (left) and $\mu E$ (right) cross-power spectra for a LiteBIRD-like mission, assuming $\langle\mu\rangle = 2\times10^{-8}$ and $f_{\rm NL}^{\rm loc}(k\simeq740\,{\rm Mpc}^{-1})=4500$ \citep{Remazeilles2022:mu}. After foreground mitigation and component separation using the Constrained ILC (CILC) method, the $\mu T$ and $\mu E$ correlations are recovered without significant bias over the multipole range $2\leq \ell \leq 500$. Combining the recovered modes over the multipoles, the joint $\mu T$ and $\mu E$ measurements yield a $5\sigma$ detection of $f_{\rm NL}^{\rm loc}=4500$ on scales $k\simeq740\,{\rm Mpc}^{-1}$ inaccessible to conventional CMB anisotropy or large-scale structure observables.}
     \label{fig:sd_anisotropies_litebird}
\end{figure}
Forecasts for future CMB imaging missions indicate that joint measurements of $\mu T$ and $\mu E$ cross-correlations can place the first direct constraints on local-type non-Gaussianity at scales around $k \simeq 740\,{\rm Mpc}^{-1}$, thereby probing possible scale dependence of the non-Gaussianity parameters \citep{Biagetti2013,Emami:2015xqa}. For example, the LiteBIRD mission is expected to reach $\left|f_{\rm NL}^{\rm loc}(k \simeq 740,{\rm Mpc}^{-1})\right| < 900$ (68\% CL), while the combination of LiteBIRD with SKAO could improve this to $\left|f_{\rm NL}^{\rm loc}(k \simeq 740,{\rm Mpc}^{-1})\right| < 90$ (68\% CL) \citep{Remazeilles2022:mu, Zegeye2025}. Figure~\ref{fig:sd_anisotropies_litebird} shows the reconstructed $\mu T$ cross-power spectrum obtained from simulated LiteBIRD sky observations, after foreground cleaning and deprojection of residual CMB temperature anisotropies in the reconstructed $\mu$ map using the Constrained-ILC technique \citep{Remazeilles2011a}, thereby eliminating spurious residual $TT$ correlations in the $\mu T$ cross-power spectrum \citep{Remazeilles2018:mu,
Rotti:2022gcl} that hampered early constraints \citep{Khatri2015aniso}. For a fiducial signal with $\langle\mu\rangle = 2\times10^{-8}$ and $f_{\rm NL}^{\rm loc}(k\simeq740 \,{\rm Mpc}^{-1})=4500$, the correlation can be recovered without bias over the multipole range $2\leq \ell \leq 500$, yielding an overall  $5\sigma$ detection significance of $f_{\rm NL}^{\rm loc}(k\simeq740,{\rm Mpc}^{-1})=4500$ for LiteBIRD \citep{Remazeilles2022:mu}. Such measurements would place constraints on even mildly scale-dependent non-Gaussianity, corresponding to $n_{\rm NL}\lesssim 0.7$ for $f_{\rm NL}^{\rm loc}(k)=f_{\rm NL}^{\rm loc}(k_0)(k/k_0)^{n_{\rm NL}}$, $k_0=0.05\,{\rm Mpc}^{-1}$, and $f_{\rm NL}^{\rm loc}(k_0)\simeq 5$. They therefore provide a powerful complement to constraints from \Planck\ \citep{Akrami:2019izv} and present and future large-scale structure surveys such as \SphereX\ and SKAO \citep{Dore:2014cca, Camera2015}, which probe substantially larger scales. Even if the resulting limits on $f_{\rm NL}^{\rm loc}$ are weaker than those obtained at CMB scales $k_0 = 0.05\,{\rm Mpc}^{-1}$, they would test whether PNG remains constant across more than four orders of magnitude in scale, providing valuable information about the physics of inflation.

For PNG scenarios, most of the signal in $\mu$-distortion anisotropy correlations is concentrated on large angular scales ($\ell \lesssim 500$), favouring full-sky observations with moderate angular resolution and broad spectral coverage. A mission such as \FOSSIL or \PIXIE is ideally suited to this science case. Its dense frequency sampling across a wide spectral range enables the separation of the distinctive $\mu$- and $y$-distortion anisotropies from foreground emission while simultaneously suppressing contamination from primary CMB anisotropies. Studies have shown that broad frequency coverage, particularly extending both below $\sim40\,{\rm GHz}$ and above several hundred GHz, provides greater leverage for reconstructing distortion anisotropies than simply increasing sensitivity within a narrower band \citep{Remazeilles2018:mu}. Since the bulk of the $\mu$ distortion resides in the low-frequency tail of the CMB blackbody, where Galactic synchrotron and free-free emission constitute major foreground contaminants, additional synergy with low-frequency surveys, including QUIJOTE \citep{MFI2}, TMS \citep{TMS_BB_2024}, and SKAO \citep{Camera2015}, can further enhance component separation performance and improve the signal-to-noise ratio of reconstructed anisotropic distortion maps \citep{Zegeye2025}.

The scientific return is maximized when a spectroscopic mission is combined with a next-generation CMB imager. Experiments such as LiteBIRD \citep{LiteBIRD2023} will deliver cosmic-variance-limited measurements of large-scale temperature and $E$-mode polarization anisotropies, while \FOSSIL can provide the complementary measurement of the average spectral distortion and its anisotropies. Together, these datasets break key degeneracies and enable a robust interpretation of spectral-distortion correlations in terms of PNG and early-Universe physics.

\begin{figure}
    \centering
    \includegraphics[width=0.47\linewidth]{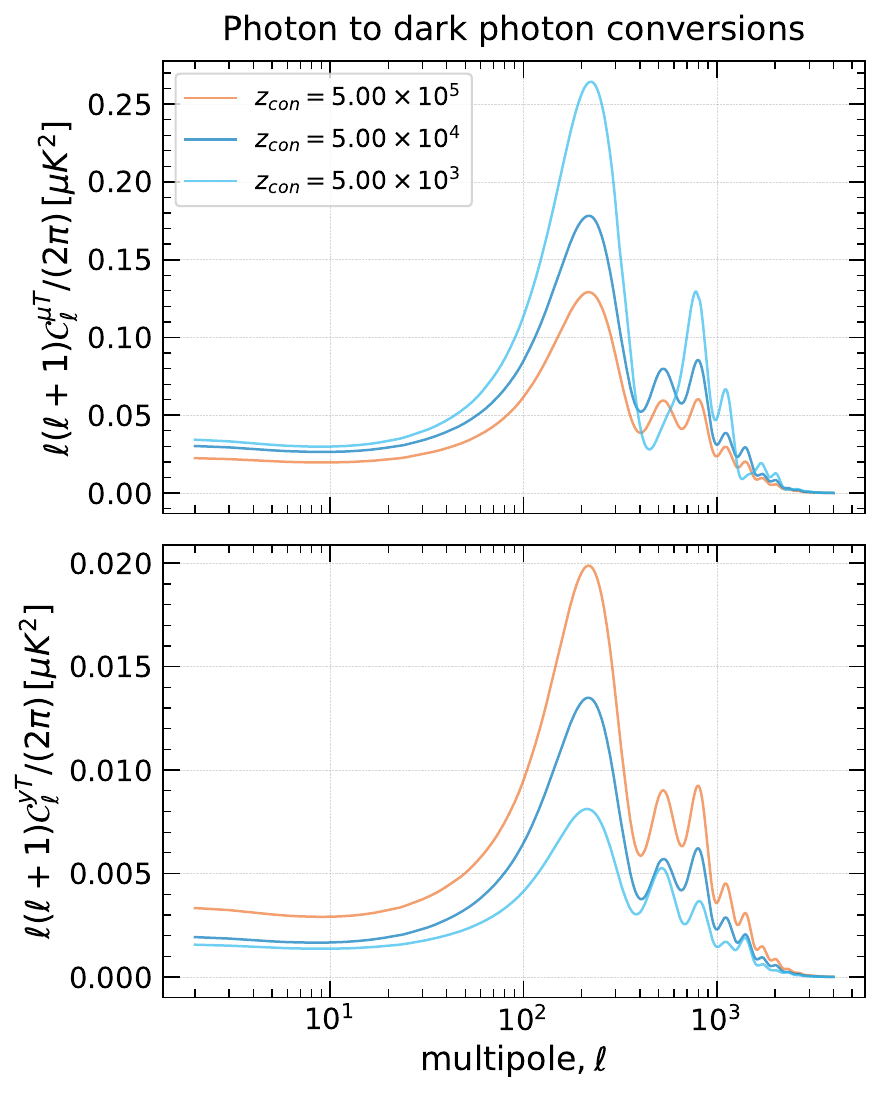}\hspace{2mm}
    \includegraphics[width=0.47\linewidth]{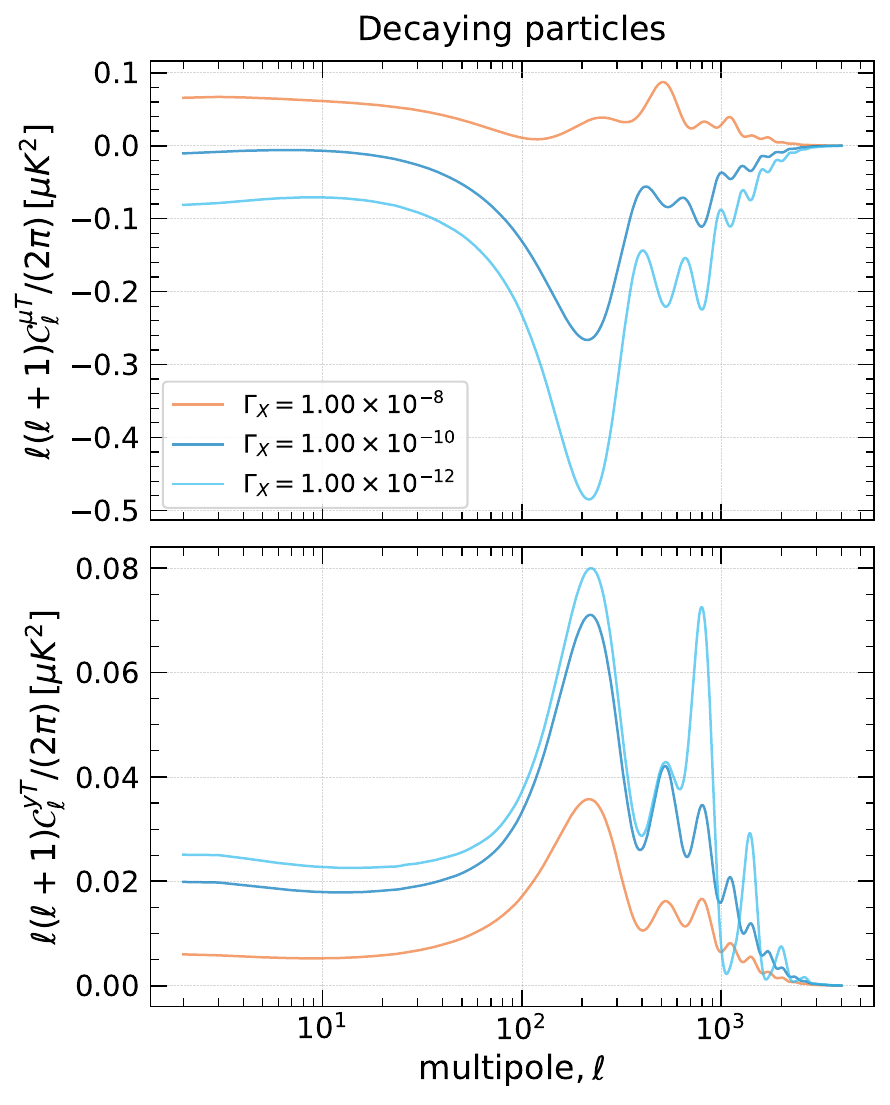}
    \caption{Cross correlation power spectra of the CMB temperature with the standard distortions $\mu$ and $y$. Two different generation mechanisms for SD anisotropies have been considered: on the left, photons converting into dark photons varying the number density, on the right, particles decaying and injecting heating into the CMB. For both scenarios, we considered three different values for the conversion redshift  ($z_{\rm con}$) and decay rates ($\Gamma_X$ in ${\rm sec}^{-1}$), respectively, spanning both the $\mu$- and $y$-epoch before recombination.}
     \label{fig:sd_anisotropies}
\end{figure}

\subsection{Decaying dark matter}
Particles that decay electromagnetically before recombination can inject energy or photons into the primordial plasma, as also described in Sec.~\ref{sec:DM_interactions}. Depending on the decay products, lifetime and injection energy, this can heat the electrons, ionise the medium and/or directly modify the CMB photon distribution \citep[e.g.,][]{Hu1993,
ChlubaSunyaev:2012,Chluba2013c,Bolliet:2020ofj, Acharya:2019owx}. The sky-averaged distortion constrains the total energy released, while distortion anisotropies encode how that release is coupled to the primordial density and gravitational perturbations \citep{kite_spectro-spatial_2023-III}.

Even when the decay rate is homogeneous in the background, the corresponding distortion need not be spatially uniform. Perturbations alter the local density of the decaying species, while gravitational effects cause a fixed stage of the decay history to be reached at slightly different times in different regions. The background distortion also enters the perturbed photon collision terms and is redistributed by gravitational effects, Thomson scattering and thermalisation. Consequently, an average energy release sources correlated $\mu$-, $y$- and residual-distortion anisotropies.

The relative amplitudes and angular structure of these components directly depend on the particle lifetime \citep{kite_spectro-spatial_2023-III}. Energy release at $z\gtrsim 5\times10^{4}$ is efficiently Comptonised and predominantly produces $\mu$-type signatures, whereas later injection increasingly appears as a $y$-type or residual distortion. The $\mu T$ and $y T$ spectra therefore contain epoch-dependent information that is partly lost when the signal is compressed into a single monopole amplitude. Representative spectra for several particle lifetimes are shown in the right-hand panels of Figure~\ref{fig:sd_anisotropies}.

Calculations using the FH demonstrate that these cross-correlations can reach observable amplitudes without violating the existing COBE/FIRAS monopole limits \citep{kite_spectro-spatial_2023-III}. They consequently provide an independent method of constraining early energy release with CMB imaging data. To reiterate, however, combining the anisotropies with a measurement of the absolute spectrum from e.g. \FOSSIL would provide the ideal scenario: the monopole fixes the total distortion, while the anisotropic signal tests its spatial origin and redshift evolution. This combination could help distinguish decaying dark matter scenarios from other mechanisms that produce similar average $\mu$- or $y$-type spectra.

\subsection{Dark-photon induced anisotropies}
Dark photons provide an interesting case study in which distortions are produced by changing the photon number rather than by conventional heating. If ordinary photons kinetically mix with a massive dark photon, resonant conversion can occur when the effective plasma mass of the photon matches the dark-photon mass (see Sec.~\ref{sec:dark_gamma} for more details). Conversion of CMB photons into dark photons removes photons from the observable spectrum and generates a frequency-dependent decrement whose form depends on the conversion redshift and subsequent thermalisation \citep{McDermott:2019lch, Chluba2024, Arsenadze2024,
Evangelista2026}.

The conversion probability and resonance condition both depend on local quantities, such as the free-electron density, baryon density and CMB temperature \citep{Mirizzi:2009iz}. Their perturbations therefore cause the conversion surface and conversion efficiency to vary across the sky. Gravitational perturbations provide an additional modulation by changing the relation between coordinate time and the local physical conditions at which resonance occurs. Moreover, even a conversion process that is isotropic at the background level produces distortion anisotropies because the evolving monopole spectrum appears as a source in the perturbed Boltzmann hierarchy \citep{kite_spectro-spatial_2023-III}.

These effects generate correlated temperature, $\mu$- and $y$-type signals, as illustrated in the left-hand panels of Figure~\ref{fig:sd_anisotropies}. The balance between the different spectral components depends on when the conversion occurs. At high redshifts, thermalisation drives the photon deficit towards a nearly $\mu$-type spectrum, while later conversions will cause the intrinsic dark photon distortion shape to be retained, even until today. Corrections to the ordinary temperature power spectrum may also arise and can resemble an isocurvature contribution, providing information that is not contained in the SD monopole alone  \citep{Evangelista2026}.

A consistent treatment requires the separation of genuine photon-removal distortions from temperature shifts and the inclusion of explicit photon source or sink terms. These are among the principal extensions introduced in the improved FH formulation \citep{Chluba2026}. Recent calculations indicate that constraints derived from the $\mu T$ correlation using existing Planck data can already approach those obtained from the COBE/FIRAS monopole \citep{Evangelista2026}. \FOSSIL would strengthen this test by measuring the absolute photon deficit and its detailed spectral shape while providing the multifrequency information needed to reconstruct the associated anisotropies. Joint analysis of the monopole, distortion auto-spectra and cross-correlations with the primary CMB could therefore test both the dark-photon mixing strength and the redshift dependence of conversion. 

\subsection{Changes to the temperature–redshift relation}
In $\Lambda$CDM, adiabatic expansion and photon-number conservation imply
$T_{\rm CMB}(z)=T_{0}(1+z)$. Noticeable departures from this relation can arise in models involving photon production or absorption, decaying vacuum energy, photon conversion or interactions with additional relativistic species \citep[e.g.,][]{Lima1996}. A commonly used phenomenological parametrisation is
\begin{equation}
T_{\rm CMB}(z)=T_{0}(1+z)^{1-\beta}
\end{equation}
where $\beta=0$ recovers the standard scaling. Although photon production can be arranged so that the background spectrum remains approximately blackbody, maintaining exact adiabaticity is difficult without generating spectral distortions \citep{Chluba:2014wda}. Nevertheless, several constraints have been obtained on the value of $\beta$ using a wide range of methods at late \citep{Songaila1994, Battistelli:2002ie, Luzzi:2009ae, Noterdaeme2011, Hurier:2013ona} and early times \citep{Planck2015params, Ivanov:2020mfr}.

The improved FH treatment makes it possible to follow both the change in the background temperature and the associated perturbations consistently \citep{Chluba2026}. This is important because a spectral component resembling a $y$-distortion can arise merely from expressing a shifted blackbody relative to an inappropriate reference temperature. The revised FH formalism separates this apparent signal from genuine non-equilibrium SDs and ensures that the solution relaxes towards the correct local blackbody state.

A changing temperature–redshift relation can nevertheless generate observable anisotropic signatures even when {\it no} monopole distortion is produced by construction. Different regions experience the photon source at slightly different local times, and their temperature perturbations do not respond instantaneously. Modes that remain tightly coupled approach the new equilibrium rapidly, whereas larger-scale modes can retain a phase shift and amplitude offset that record when the departure from the standard relation began. If the change starts sufficiently early, incomplete Comptonisation also produces correlated $\mu$- and $y$-type components. Their relative amplitudes can therefore help determine both the magnitude of the departure and its onset redshift \citep{Chluba2026}.

This example illustrates a broader advantage of distortion anisotropies: they can remain observable when the corresponding monopole signal is weak or vanishes. \FOSSIL would provide the absolute determination of $T_{0}$ and the background spectrum needed to anchor such a test, while its multifrequency maps and cross-correlations with external CMB data would search for the spatially varying residual. The result would be a direct consistency test linking the present CMB temperature, its redshift evolution and the frequency-dependent anisotropy field. This science case is complementary to the dedicated discussion of $T_{0}$ and the temperature–redshift in relation Sec.~\ref{sec:T0}.

\subsection{Additional sources of SD anisotropies}
Beyond the aforementioned important examples, SD anisotropies open a broader observational window onto new physics. Essentially {\it all} mechanisms leading to average distortion signals are inevitably expected to cause distortion anisotropies. To give a few additional examples, SD anisotropies can probe anisotropic heating generated by annihilating particles \citep{kite_spectro-spatial_2023-III}, signatures associated with clustered primordial black holes \citep[e.g.,][]{Ozsoy2021}, axion-photon conversion processes \citep[e.g.,][]{Cyr2024} and primordial magnetic fields, although these examples have not been exhaustively treated thus far. Again, since the information resides predominantly in large-scale spatial correlations, the complementarity between future CMB absolute spectrometer experiments and CMB imaging missions cannot be overstated.

These applications show that SD anisotropies are not merely a secondary extension of monopole spectroscopy. They provide an independent observable that connects the thermal history of the CMB to the spatial structure of the Universe. By combining absolute spectroscopy, broad frequency coverage and angular cross-correlations, \FOSSIL and other space-based endeavors can distinguish sources with similar average spectra, test processes that produce little or no monopole distortion, and extend precision cosmology into physical regimes that cannot be reached using the conventional CMB temperature and polarisation fields alone.

\section{Upcoming and Proposed Experiments} \label{sec:missions}
It has been more than 30 years since the last CMB spectrometer mission. Therefore, the science case developed in the preceding sections rests on measurements that do not yet exist. In this section, we describe an ensemble of instrumental efforts that is rejuvenating the field and which aims at transforming CMB SD measurements into a precision probe for cosmology and particle physics, thereby writing an entirely new chapter in CMB cosmology. Today, this these activities extend from proposed concepts, e.g. SPECTER, to pathfinder experiments under study or being developed, such as ASPERa, TMS, COSMO and BISOU, through to proposed space missions such as \PIXIE and \FOSSIL that aim at actual measurements of the spectral distortions.

Each tier reduces the risk of the next while delivering highly impactful science results along the way. Together, these efforts trace a concrete path from the \COBEF era to the regime where the $\Lambda$CDM prediction of $\mu \simeq 2\times 10^{-8}$ shifts from forecast to measurement. \FOSSIL represents the present realisation of this path forward and constitutes a {\it major} scientific mission with a broad array of science that can be achieved, as described in detail above. Assuredly, it will open a new  window in our Universe that will inspire even more ambitious missions in the long term. 

\subsection{Ground-based} \label{sec:ground}
On the ground, APSERa ($2$--$6\,$GHz), TMS ($10$--$20\,$GHz), and COSMO (two atmospheric windows spanning $130$--$300\,$GHz, from Dome C, Antarctica) target the frequency ranges accessible from Earth. Beyond their primary goals, which include a measurement of the post-recombination Comptonization ($y_{\rm pr} \simeq 2\times 10^{-6}$), these experiments will characterize the low-frequency foregrounds (synchrotron, AME, and the radio synchrotron background) that limit any distortion measurement.

\subsubsection{TMS}

TMS (Tenerife Microwave Spectrometer\footnote{TMS: \url{https://research.iac.es/proyecto/tms/}}) \cite{TMS} is a ground-based spectrometer to be deployed at the Teide Observatory (Tenerife, Spain). It will perform precise measurements of the absolute spectral distortions of the sky in the 10--20\,GHz frequency range, down to levels of 10--20\,Jy/sr with an angular resolution of about 2 degrees on sky \cite{tmsIMO}. TMS is based on a pseudo-correlation architecture, comprising two radiometer chains that use both orthogonal linear polarizations. The optical system includes two corrugated feedhorns, one facing an internal calibrator that provides a cold reference signal at approximately 6\,K, and the other pointing to the sky through the cryostat window via an offset-fed reflector \cite{TMS_optics}. This optical configuration ensures excellent cross-polarization performance ($\leq -30 $\,dB) and symmetric beams with minimal frequency dependence.

With this target sensitivity, TMS will provide key measurements to characterise the spectral properties of Galactic synchrotron and anomalous microwave emission (AME), as well as to constrain the properties of extragalactic emission in the 10--20\,GHz range. In particular, TMS aims to provide the most precise measurement to date of the frequency dependence of the radio synchrotron background (RSB) in this frequency range.
Moreover, TMS will provide an absolute calibration reference for the existing QUIJOTE-MFI data \cite{2023MNRAS.519.3383R} and future MFI2 maps, as well as an accurate relative calibration scale at the sub-percent level for all these datasets.
\begin{figure*}[h]
\centering
\includegraphics[width=0.9\linewidth]{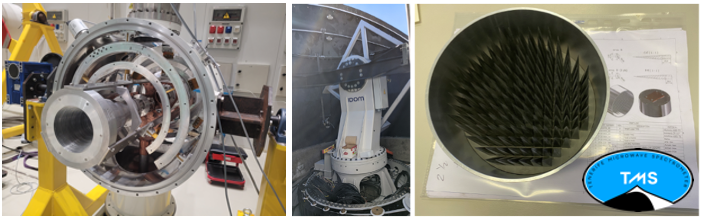}
\caption{Left: Assembly of the TMS internal structure in the IAC integration clean room (June 2026). Center: TMS platform and supporting structure, installed at the Teide Observatory inside the TMS dome (November 2022). Right: TMS cold load (black-body internal calibrator), fabricated by INAF-OAS (Bologna, Italy).}
\label{fig:tms1}
\end{figure*}

The project is now in its final implementation phase (see Figure~\ref{fig:tms1}). All the hardware subsystems of the instrument are already fabricated. The TMS platform is already installed at the Teide Observatory. The cryostat can is complete, while its internal cold structure is in the final stages of integration. The full-octave horns \cite{2022JInst..17P6041D} have been manufactured and measured, and the OMTs and hybrids are currently being tested in the laboratory. The internal blackbody calibrator is fabricated and tested \cite{TMS_BB_2024}. The digital acquisition system (DAS), built around SoC FPGA technology, is currently being tested on the QUIJOTE MFI2 instrument \cite{MFI2}. Full system integration is expected by mid 2027.

\subsubsection{COSMO}
The Cosmic Monopole Observer\footnote{COSMO:  \url{https://cosmo.roma1.infn.it}} is a pioneering ground-based experiment aiming to detect the largest isotropic spectral distortions of the Cosmic Microwave Background (CMB) \cite{Masi:2021azs} —specifically, the Comptonization due to post-recombination ionized matter along the line of sight and the high-frequency extension of the so-called ARCADE radio excess — while validating an innovative method for atmospheric emission removal. The target spectral distortion signals are extremely small compared to the emission from the instrument, the atmosphere, and Galactic foregrounds. COSMO has been designed to minimize, or accurately measure and remove, these instrumental and atmospheric foregrounds. Additionally, the CMB blackbody background will be canceled out by the differential nature of the spectrometer, which compares the sky to a highly accurate internal blackbody load at 2.725 K. 

\begin{figure*}[h]
    \centering
    \includegraphics[width=0.9\linewidth]{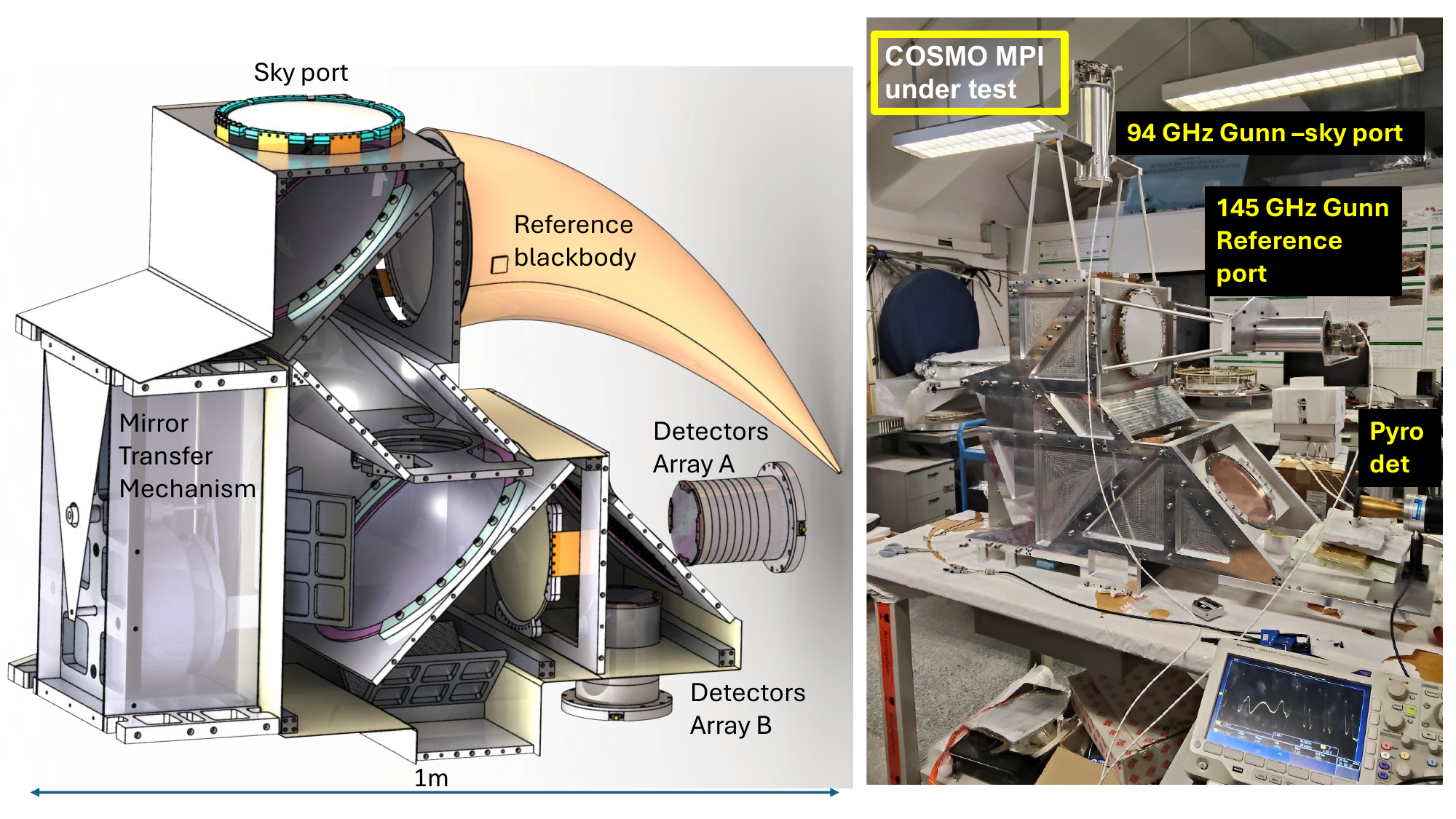}
    \caption{Left: rendering of the COSMO DFTS. All parts shown are inserted in a pulse-tube cooled cryostat at temperatures below 3K. Right: the COSMO DFTS in the laboratory during room temperature validation tests (June 2026).}
    \label{fig:cosmo1}
\end{figure*}

To mitigate atmospheric emissions, COSMO employs a cryogenic Differential Fourier Transform Spectrometer (DFTS) (see Figure \ref{fig:cosmo1}) and operates from the Concordia Station at Dome C, Antarctica. 

Fast sky dips effectively counter atmospheric fluctuations, enabled by the use of rapid Kinetic Inductance Detectors (KIDs). Window emission is accounted for by precisely monitoring its temperature and calibrating its emissivity, while precise temperature control of other instrumental components further mitigates any residual imbalanced emissions. The scan strategy covers diverse Galactic latitudes and maps interstellar dust emission. Over two years of continuous integration covering approximately 1,800 square degrees of the sky, the experiment is expected to reach a target sensitivity for the average spectral brightness on the order of $300$ Jy/sr per spectral bin across ten 10-GHz-wide bins within the 200–300 GHz atmospheric window. It will also reach roughly $10$ Jy/sr per spectral bin across five 10-GHz-wide bins within the 130–170 GHz window [see \cite{COSMO2022} for details. The separation of spectral distortions, Galactic dust emission, and the CIB will be achieved by exploiting the lack of covariance between their respective spectra. These sensitivities will allow us to estimate the average Comptonization parameter $y_{pr}$ from post-recombination ionized matter (which is on the order of $2 \times 10^{-6}$) with a 1$\sigma$ error on the order of $2 \times 10^{-7}$. At this target sensitivity, COSMO will also shed light on the physical origin of the ARCADE excess, delivering crucial new high-frequency data. 

In addition to measuring the CMB spectrum via sky dips, COSMO will monitor the emission of the Antarctic atmosphere within two millimeter-wave windows. The strong molecular lines bracketing these two windows originate from O$_2$ at 118.75 GHz, H$_2$O at 183.31 GHz, and H$_2$O at 325.15 GHz; furthermore, narrow O$_3$ lines are present in the 200–300 GHz window. The continuum above 150 GHz is dominated by the tails of high-frequency H$_2$O lines. The year-round, precise, medium-resolution atmospheric measurements provided by COSMO will be compared to synthetic spectra to monitor atmospheric temperature and humidity profiles, the mixing ratios of H$_2$O, O$_2$, and O$_3$, and their variability under the extreme conditions of the Antarctic climate. These measurements will establish a comprehensive reference database for atmospheric physics and Antarctic climatology studies.

\subsubsection{APSERa}
\begin{figure}[h]
    \centering
    \includegraphics[width=0.8\linewidth]{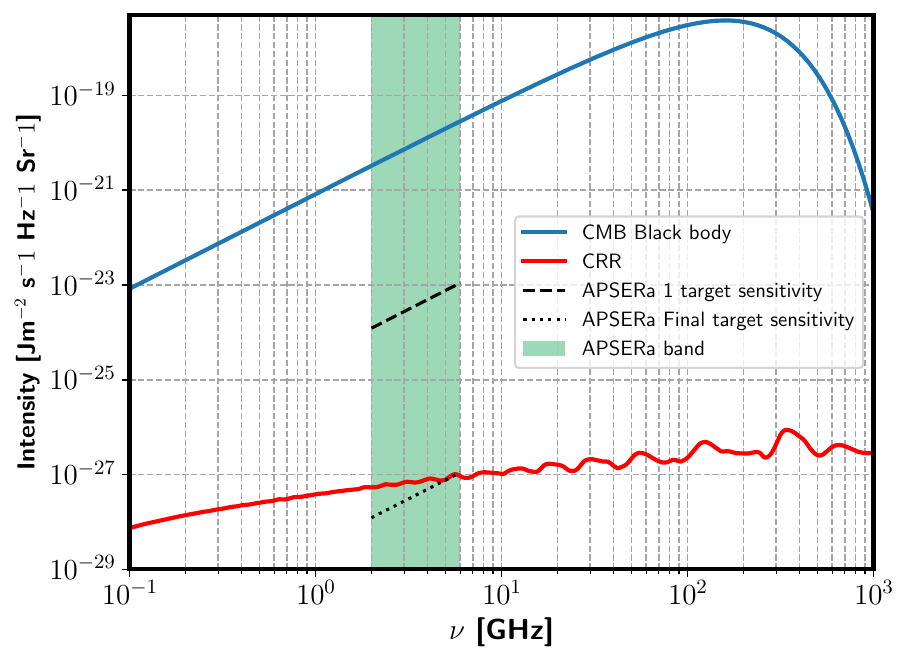}
    \caption{The Cosmological Recombintaion Radiation is 9 orders of magnitude fainter than the CMB blackbody spectrum. Though faint, it has a well predicted and rich spectral signature which is advantageous for signal detection. The APSERa band is highlighted in the shaded region. APSERa in its final phase will be an array of 128 cooled radiometers deployed in a high altitude radio quiet site, with a primary science goal of detecting the CRR. CRR signature from COSMOSPEC \cite{Chluba2016CosmoSpec}.}
    \label{fig:CRR_dyn_range}
\end{figure}

APSERa - Array of Precision Spectrometers for the Epoch of RecombinAtion -- is an upcoming ground-based experiment operating in an octave band over 2-6 GHz (Visit the \href{https://cmbdistortionlab.wixsite.com/distortion-lab-rri/apsera}{APSERa website} here). The primary science goal of APSERa is to detect the cosmological recombination radiation (CRR) which is predicted to appear as an additive distortion to the CMB spectrum. The CRR is an inevitable prediction of $\Lambda$CDM cosmology \cite{sunyaev2009signals}, deviations from predicted signatures is a sign of new physics, and the nature of Dark Matter \cite{Chluba2010}. The CRR is a unique signature that can directly probe prestellar helium abundance ($Y_p$) \cite{chluba2010cosmological} and time evolution of the thermal and ionization history of the Universe probing into the surface of last scattering \cite{sunyaev2009signals}. The CRR signature is sensitive to early Dark Energy and (time varying) fundamental constants \cite{Hart:2022agu}.  APSERa in its first phase will be operational in the S-band, over $2.5- 4$ GHz. The CRR signal is inherently faint. Its amplitude relative to the underlying CMB blackbody spectrum is readily informed by the ratio of number of baryons ($N_b \propto \Omega_bh^2$)  to photons $N_{\gamma}$.  Whereas the signature of the CRR is predicted to span from $\sim 100~\textrm{MHz}$ through to $\sim 1~\textrm{THz}$, the choice of band for APSERa is determined by two criteria, namely maximizing signal to noise ratio for a ground-based detection, and having multiple spectral signatures in an octave band for effective foreground separation (See Figure \ref{fig:CRR_dyn_range}). 
The APSERa feasibility study \cite{sathyanarayana2015detection}, assuming ideal antennas, suggest a 90\% confidence detection observing the sky for one antenna year with an array of 128 cooled receivers. Recent simplistic beam-model simulations suggest non-ideal antenna behavior can be tolerated in detection experiments due to clear and rich spectral signature prediction that effectively enables smooth foreground separation \cite{krishna2025detecting}.

The planned deployment site for APSERa is in the high altitude terrains of the Ladakh plateau in India. This region is favorable on two fronts, namely low local radio frequency interference, which in this band is dominated by satellite communications that are globally present \cite{sathish2024rfi} as well low atmospheric water vapor \cite{singh2026site} which can contribute $O(1)$ Kelvin to the system temperature in the S-band. Due to the frequency of operation, APSERa can leverage all techniques used in radio-astronomy such as coherent detection, cooled HEMT amplifiers, and direct digitization using latest generation digital hardware \cite{srivani2024digital}. APSERa will be deployed in phases. The first phase will be a single element Dicke switching radiometer the first with a wide-beam antenna \cite{sathish2024antenna, 11426669} that has spectrally smooth operation over $2.5-4$~GHz targeting a CMB absolute temperature measurement with mK sensitivity. Due to bright synchrotron foregrounds and measurement challenges, the error bars on the CMB temperature measurement in the Rayleigh-Jeans tail remain large, typically $20\%$ or greater. Interesting physics can lie in these large error bars and is important to constrain \cite{Pospelov:2018kdh}. This will be the science goal of APSERa in phase one, in addition to studies of the synchrotron background itself with important implications for probing the excess radiation reported by the ARCADE-2 experiment. This first phase is expected to be commissioned in a high altitude radio quiet location over the summer of 2028 with expected science results in 2029. Future phases will see expansion to antenna arrays with beam forming options, and increased on-sky sensitivity, with the final array comprising 128 antenna elements with cooled receivers and precision calibration for detecting the CRR, with a target sensitivity of O(10nK), the timeline subject to funding for expansion to a 100 element array.

\subsection{Balloon Experiments} \label{sec:balloons}
From the stratosphere, only one experiment is presently under study. BISOU ($90\,$GHz--$1.5\,$THz) is designed to measure the $y$ monopole at $5$--$6\sigma$, surpassing \COBEF by more than an order of magnitude in sensitivity, while maturing the technologies required for a space mission and characterizing the various foreground emissions present at high frequencies.

\subsubsection{BISOU}
The \textit{Balloon Interferometer for Spectral Observations of the primordial Universe} (BISOU) \cite{Maffei:2024SPIE} is a balloon-borne spectrometer dedicated to measurements of CMB spectral distortions. Its primary scientific objective is to achieve the first detection of the monopole Compton $y$-distortion, while simultaneously improving measurements of the Cosmic Infrared Background (CIB), refining the determination of the absolute CMB temperature, and observing Galactic fine-structure lines such as C\,{\sc i}, C\,{\sc ii}, N\,{\sc ii}, and O\,{\sc i}. Operating from the stratosphere at an altitude of roughly 40\,km allows for a much larger spectral coverage (approximately 90\,GHz to 1.5\,THz) necessary for an accurate foreground subtraction, and provides a substantial atmospheric emission reduction compared to ground-based facilities.

BISOU builds upon the instrument concepts originally developed for the \textit{PIXIE} and \textit{FOSSIL} mission proposals. As such, in addition to its scientific objectives, it will act as a technological and scientific pathfinder for future missions. It will offer a lower-cost and lower-risk platform to validate state of the art spectrometer technologies, calibration strategies, and data analysis methods that will be required for future space mission. 

The instrument design is based on a cryogenic Martin--Puplett differential Fourier Transform Spectrometer performing differential measurements between the sky and an internal blackbody reference load, maintained close to the CMB temperature (2.725\,K). This architecture largely removes the CMB blackbody monopole and suppresses common instrumental emission, allowing the detectors to directly measure small spectral differences between the sky and the internal reference blackbody. 

\begin{figure}[t]
\centerline{
\includegraphics[width=0.8\linewidth]{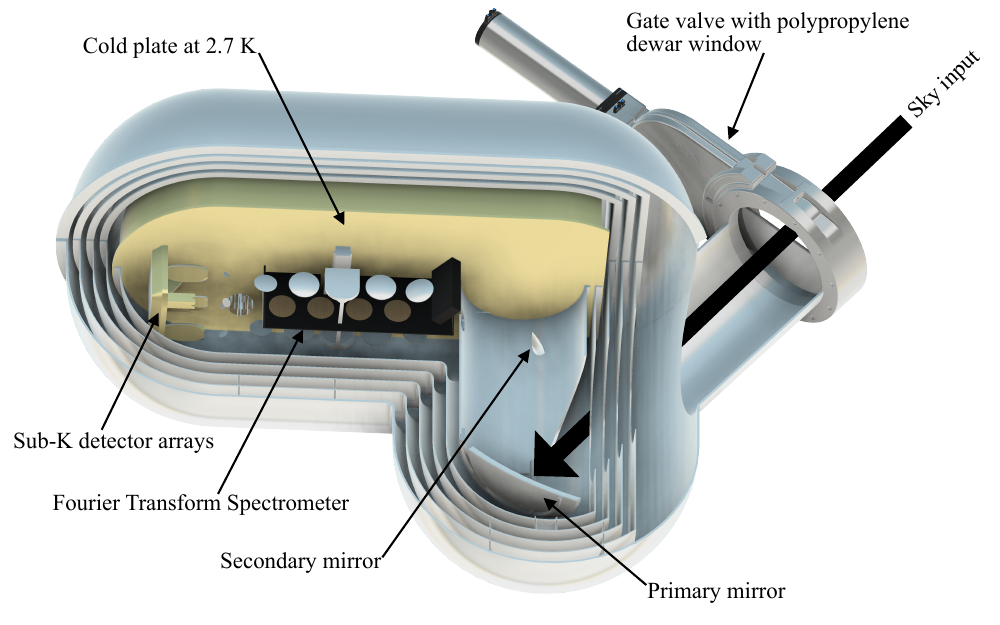}}
\caption{A rending view of the BISOU instrument: the various sub-systems are mounted on the cold plate (gold) and inside the L-shaped dewar. The whole instrument is cooled down to 2.7 K using liquid helium, while the detectors are at about 100 mK. Image courtesy V. Sauvage.}
\label{fig:BISOU}
\end{figure}

While a balloon platform introduces additional constraints in the form of power, mass and dimensional restrictions for instance, with respect to a ground-based or space project, it also has several advantages. Unlike the original \textit{PIXIE} concept, BISOU has been specifically redesigned to measure distortions and uses a single sky-facing telescope, while the second input port permanently observes the internal blackbody reference. This simplified optical configuration reduces the overall payload mass and complexity, making it compatible with the CNES CARMEN balloon gondola while preserving the differential measurement scheme required for absolute spectroscopy. The whole instrument is hosted within a liquid-helium cooled cryostat (see Fig.~\ref{fig:BISOU}), which takes advantage of the lower atmospheric pressure at high altitudes (3\,mbar typically). With this, the spectrometer and all of its optical elements (including the telescope) can be maintained at 2.7\,K, which helps to reduce instrument systematics. Only the focal planes will need to be cooled to sub-K temperatures in order to reach the required photon-noise-limited sensitivity that will be achieved by the use of a few multi-moded feedhorn-coupled detectors. The association of the telescope, with a primary mirror projected aperture diameter of 30\,cm, and multi-mode optics, provides a nearly frequency-independent equivalent Gaussian beamwidth of $\sim 2.5-3^{\circ}$ across the full spectral range.

An end-to-end spectro-photometric model, including the optical loading from the instrument together with astrophysical foregrounds and the residual atmospheric emission at balloon altitude, has been developed to optimize the instrument design and assess its scientific performance. Its results, together with the dimensions constraints, led to a spectral coverage ranging from 90\,GHz to 1.5\,THz, divided in two sub-bands using a dichroic beam splitter: a low-frequency channel optimized for measurements of the CMB spectral distortions and a high-frequency channel dedicated primarily to Galactic foreground and CIB characterization. This configuration significantly reduces the photon noise contamination of the low-frequency band by the highest frequencies while preserving the information required for accurate component separation. 
The nominal spectral resolution is 15\,GHz, sufficient to recover the broad spectral signatures of the expected distortions while keeping the interferometer stroke below 1\,cm. The optical design has also been optimized to minimize instrumental systematics, including stray light, polarization leakage and beam distortions over the full frequency range.

Assuming a five-day transatlantic balloon flight with a single detector and a 75\% observing efficiency, BISOU is expected to detect the CMB $y$-distortion monopole with a signal-to-noise ratio of approximately 5--6, corresponding to an improvement of more than one order of magnitude over the \textit{COBE/FIRAS} constraint. The instrument will also improve measurements of the CIB monopole and provide valuable broadband observations of Galactic emission required for foreground modelling.

BISOU entered a two-year Phase~A programme in 2024, due to end by mid 2027. Provided it is selected for development, a first BISOU test (short) flight is planned for summer 2030, followed by a longer transatlantic science flight between Kiruna, Sweden and Canada by June 2032. 
In parallel, a cryogenic breadboard model is being developed to validate the key technologies, including the optical chain, detectors, calibration system, and atmospheric mitigation strategy under representative operating conditions. These developments will increase the technological readiness of the critical sub-systems, and provide the scientific and technological foundation required for the construction and calibration of the flight instrument. They will also allow for a systematic effects analysis, together with the development of a comprehensive end-to-end instrument model.
%

\subsection{Space-based} \label{sec:space}
In space, several FTS-based concepts were proposed either to ESA or NASA such as \PIXIE (NASA MIDEX 2011, 2016), PRISM (ESA L-mission), PRISTINE (ESA, F1), \FOSSIL (ESA M7). The most mature \FOSSIL (ESA M8) and \PIXIE (NASA MIDEX) instrument concepts described below would deliver full-sky, absolutely calibrated spectra across $50\,$GHz--$2\,$THz and $28\,$GHz--$6\,$THz respectively, improving on \COBEF by three orders of magnitude. A new radiometer-based, alternative to the FTS concepts, SPECTER, would cover the range $1$—$2000\,$GHz.

\subsubsection{\FOSSIL}
\begin{figure}[t]
\centerline{
\includegraphics[width=0.8\linewidth]{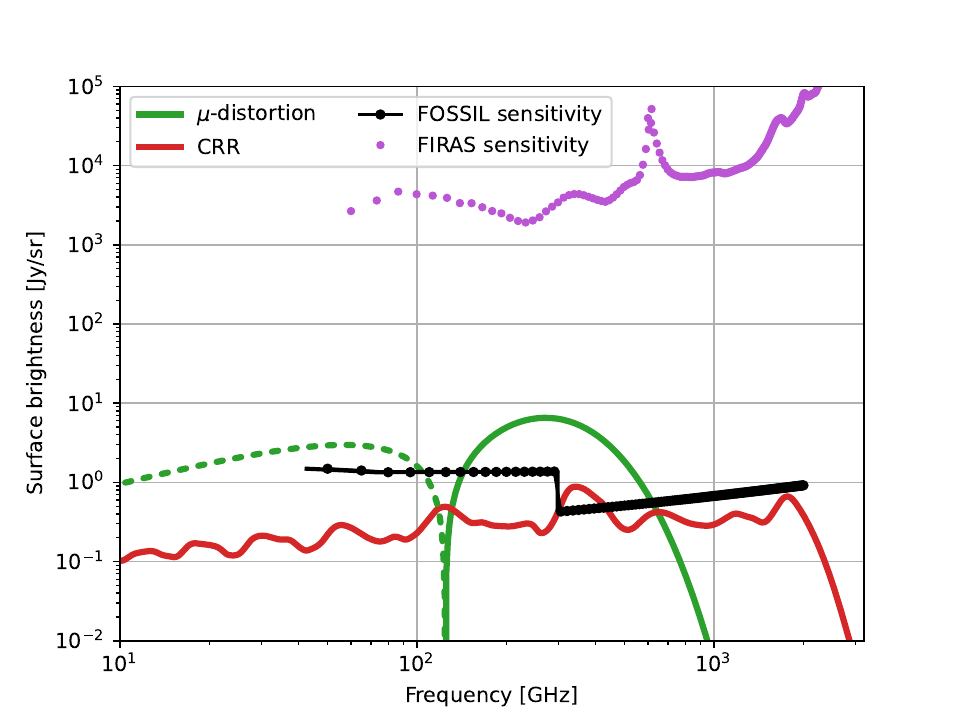}}
\caption{Sensitivity of the \FOSSIL mission (black) compared to the \COBEF sensitivity (purple). The $\mu$-distortion (green) and the CRR (red) spectra are also shown for reference. Solid curves indicate positive values, while dashed curves indicate negative values.}
\label{fig:FOSSIL_sensitivity}
\end{figure}

Proposed as the ESA M8 mission, to be launched in the 2040s, {\it {F}TS f{O}r CMB {S}pectral di{S}tort{I}on exp{L}oration} (\FOSSIL) \cite{FOSSILPlaceholder} presently represents the sole unique opportunity to achieve a major step forward in observational cosmology by measuring the frequency spectrum of the CMB at an unprecedented level of precision (see Fig.~ \ref{fig:FOSSIL_sensitivity}). 

\FOSSIL will address key open questions in cosmology about the nature of dark matter, the physics of inflation, and the growth of structure. These science goals can be achieved through \FOSSIL’s unique capability to perform an absolute measurement of the sky brightness over the entire celestial sphere at a resolution of 1--2$^\circ$ from 50\,GHz to 2\,THz with a forecasted improvement of three orders of magnitude over the \COBEF satellite. 

\FOSSIL will map the sky through a continuous scanning strategy, spinning around its axis while performing spectroscopic measurements with a two-input--two-output absolute Fourier-transform spectrometer (FTS), similarly to \PIXIE. Beam-switch mechanisms, with holes and mirrors, enable \FOSSIL to switch the two FTS inputs, so that the each of the two arms can either view the sky or a blackbody internal reference load cooled to close to the CMB temperature ($\sim$2.7~K). The requirements set by the primary science goals are reached by using four multi-moded feedhorn-coupled detectors, instead of the thousands of detectors typically used for CMB imagers. 

%
Splitting the spectral band into two sub-bands further reduces the photon noise, most notably at low-frequencies, optimising the measurement of the $\mu$-type spectral distortion. The two FTS outputs result in four detector units cooled to 50~mK and mounted within a focal plane assembly. Kinetic Inductance Detectors are chosen as the baseline. The instrument and surrounding baffles are actively cooled to 4.5~K by a 4~K cryocooler, which also provides supplemental cooling at 25~K to a shield, in addition of three passively cooled V-grooves. Temperatures below 4.5~K are provided by a sub-K multi-ADR (adiabatic demagnetisation refrigerator) system developed for NewAthena/X-IFU and LiteBIRD.

\FOSSIL will deliver full-sky, absolutely-calibrated maps of the sky emission across 130 frequencies. This exceptional data set will form a long-lasting reference. It will provide an absolute calibration anchor for existing and future multi-frequency sky surveys, significantly enhancing their scientific return and allow for synergistic opportunities with all major space and ground-based facilities.

\subsubsection{\PIXIE}
The Primordial Inflation Explorer (\PIXIE)
\cite{Kogut:2024vbi}
is a mission concept 
developed for NASA's MIDEX Explorer program
to map the absolute intensity
and linear polarization
of the CMB and diffuse foregrounds.
A single cryogenic Fourier transform spectrometer
compares the sky to an external blackbody calibration target,
measuring the Stokes $I, Q, U$ parameters
to levels $\sim$200~Jy/sr
in each 2.65$^\circ$ diameter beam
over the full sky,
in each of 300 frequency channels
from 28~GHz to 6~THz.
Over a baseline 2-year mission
from the Sun-Earth L2 point,
\PIXIE would achieve sensitivity
$\delta y = 9 \times 10^{-9}$
for the mean Comptonization parameter
and
$\delta \mu = 5 \times 10^{-8}$
for the chemical potential.
With sensitivity over 1000 times greater than COBE/FIRAS,
\PIXIE opens a broad discovery space
for the origin, contents, and evolution of the Universe.

Cosmological signals are small compared to the
instantaneous instrument noise,
requiring strict control of instrumental signals.
\PIXIE's single instrument consists of a polarizing
Fourier transform spectrometer 
which compares the sky to an external blackbody calibrator.
Thermometers embedded in the absorbing structure
monitor temperatures to 0.1~mK absolute accuracy
and few-nK relative precision
within the calibrator
\cite{2020JCAP...05..041K}.
To minimize instrumental artifacts,
the entire optical path
and surrounding walls
are maintained at 2.725~K,
forming an isothermal enclosure
matching the CMB monopole.
The instrument design provides
multiple levels of null operation,
signal modulation,
and signal differences,
with only few-percent systematic error suppression
required at each level.
Jackknife tests
based on discrete instrument symmetries
provide an independent means to
identify, model, and remove remaining instrumental signals
\cite{2023JCAP...07..057K}.

%
\begin{figure}[t]
\centerline{
\includegraphics[width=0.8\linewidth]{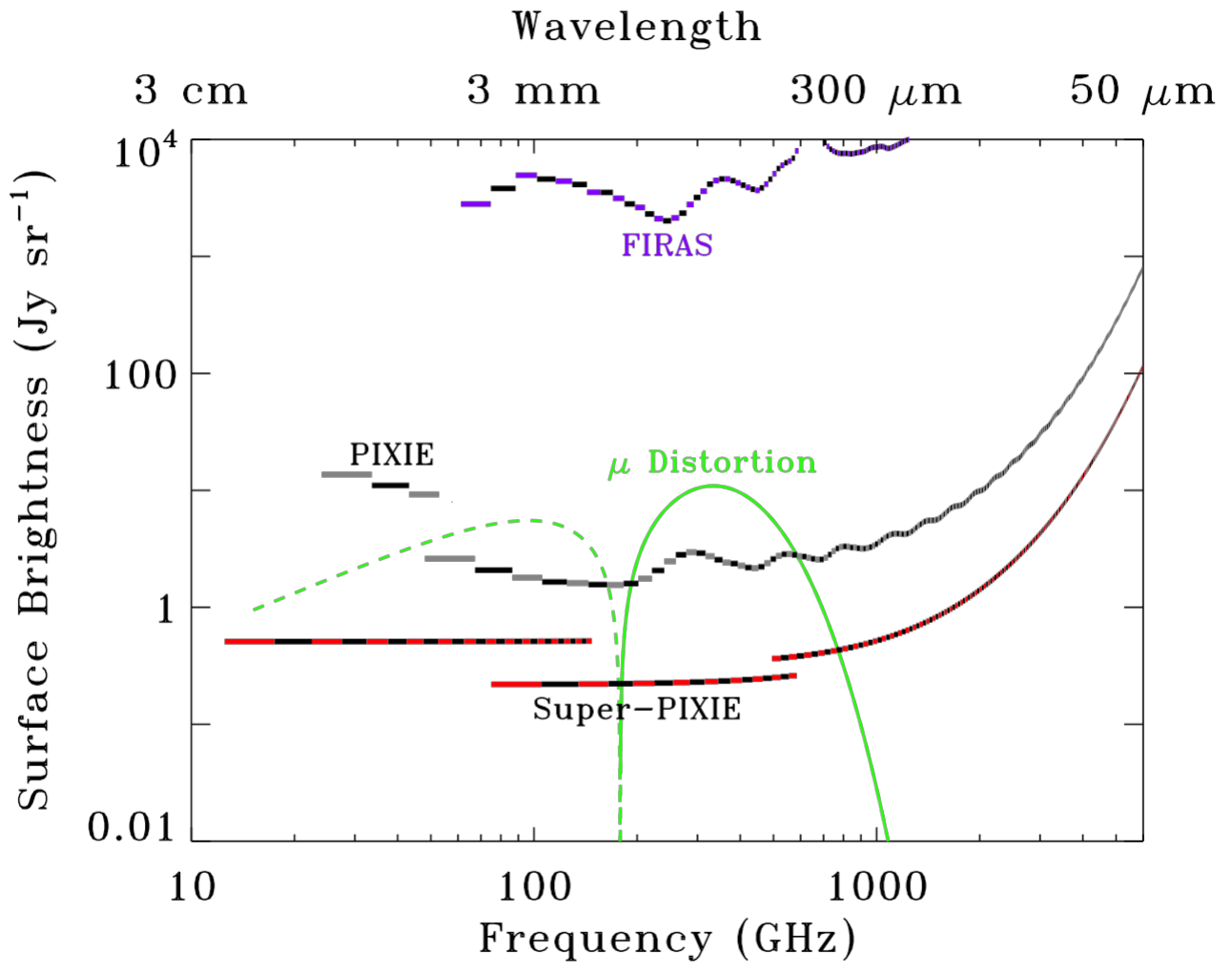}}
\caption{
A Super-\PIXIE mission
with 3 FTS modules
would tune the channel width 
and optical passband of each module
to optimize the combined sensitivity to 
foregrounds and CMB spectral distortions.
\label{pixie_vs_phoenix}}
\end{figure}

\PIXIE would operate at the background limit of photon noise
from the CMB and astrophysical foregrounds.
The forecasted sensitivity to CMB spectral distortions
is limited primarily by confusion from low-frequency foregrounds
(synchrotron, free-free, and anomalous microwave emission).
Improved foreground component separation
requires additional channels at lower frequencies,
without sacrificing sensitivity at higher frequencies.
This cannot be accomplished with a single FTS:
lengthening the optical phase delay to reduce the channel width
(thereby squeezing in additional low-frequency channels)
degrades the sensitivity to continuum signals
(which do not benefit from the higher spectral resolution).
A more ambitious mission
(either through NASA's Astrophysics Probe line
or ESA's L-class mission)
would allow an additional order of magnitude improvement in sensitivity.
Such a ``Super-\PIXIE'' 
would employ 3 (or more) cryogenic FTS modules,
each using the basic \PIXIE design,
but with channel width
and optical passband (hence photon NEP)
tuned to optimize the combined sensitivity
to selected CMB signals.
A low-frequency module
would employ 5~GHz channel width
over a passband 15--150~GHz
to characterize the low-frequency foregrounds
while avoiding photon noise from the CMB monopole
at higher frequencies.
A mid-frequency module
would use wider 30~GHz channels 
over a passband 90--600~GHz
to cover the cosmological signals.
A high-frequency module
would use 30~GHz channel width
over the passband 450--6000~GHz
to characterize the high-frequency foregrounds,
setting the lowest frequency at 450~GHz
to minimize the photon noise contribution from the CMB monopole.
Figure \ref{pixie_vs_phoenix} 
compares the resulting sensitivity 
to both \PIXIE and FIRAS.
Such a Super-\PIXIE mission
could provide a $3\sigma$ or greater detection
of the minimal chemical potential distortion
from dissipation of primordial density perturbations
within the $\Lambda$CDM model.

\subsubsection{SPECTER}
The Spectral Photometry Experiment for Cosmic ThErmal distoRtions (\textit{SPECTER}) is a recently proposed concept for a satellite mission with the main objective of detecting the $\mu$-distortion at $\gtrsim 5\sigma$, even after marginalizing over a broad set of foreground templates and absorbing the associated penalty in the final $\mu$ error bar~\cite{specter}. The key characteristic of the \textit{SPECTER} concept is the use of total-power radiometers with absolute temperature calibration, different from previous and currently proposed FTS-based spectral distortion instruments like \textit{COBE/FIRAS}.
Such design allows optimization of \textit{SPECTER}'s sensitivity towards detecting the $\mu$-distortion in the presence of astrophysical foregrounds via the precise choice of the frequency bands and the number of detectors per band, as determined via detailed optimization using state-of-the-art sky emission models.
\begin{figure*}[h]
    \centering
    \includegraphics[width=0.8\linewidth]{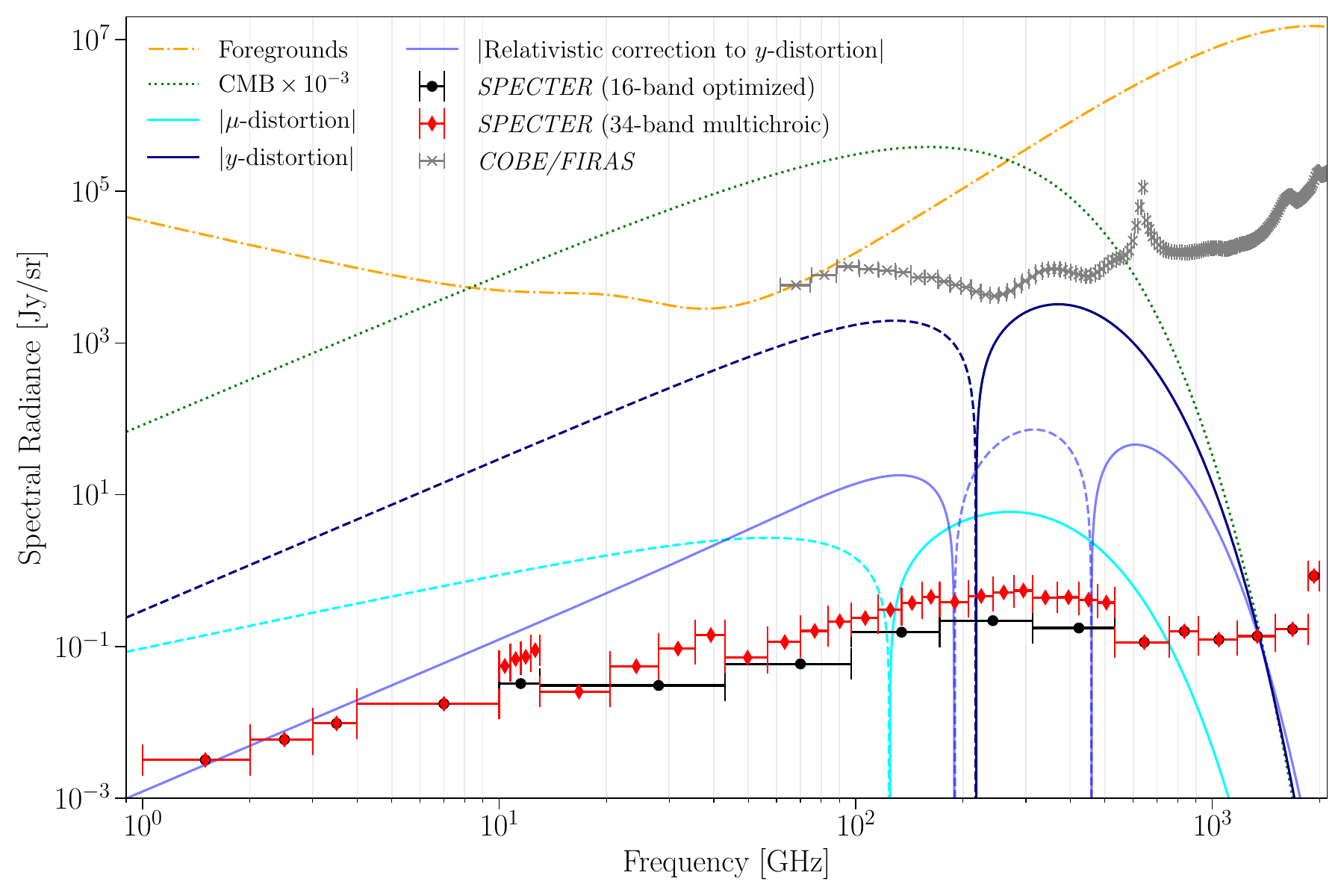}
    \caption{Sensitivities for the \textit{SPECTER} 16-band optimized (black circles) and the 34-band multichroic configurations (red rombs) assuming a full sky observation and one year of spectral distortion integration time. Total astrophysical foregrounds (dot-dashed orange), $\mu$-distortion (cyan), and $y$-distortion (dark blue) and its relativistic correction (light blue) are also shown. Note that the absolute values of the distortions are plotted with dashed lines corresponding to the negative values. \textit{COBE/FIRAS} sensitivity (grey crosses) and the scaled CMB blackbody emission (dotted green) are plotted for reference. Figure adapted from Ref.~\cite{specter}.}
    \label{fig:specter}
\end{figure*}

The \textit{SPECTER} concept comprises several instruments, each with absolute temperature calibration systems, spanning a frequency range between $1-2000$ GHz: two Low-Frequency Foregrounds Imagers (LFFI, $1.5-12$ GHz), one Primary CMB Imager (PCI, $30-420$ GHz), and one High-Frequency Foregrounds Imager (HFFI, $650-2000$ GHz). The instrument includes 1100 detectors in total, with HEMT amplifiers used at low frequencies ($\lesssim10$ GHz) and bolometric detectors at high frequencies. Each imager has its own optical system consisting of a crossed-Dragone telescope, a detector array, two absolute temperature calibrators, and a load-selecting mirror, which allows switching between the sky and the onboard calibrators. The sensitivities across channels range between $\sim10^{-3}-1$ Jy/sr assuming one year of spectral distortion integration time ($\sim$few$-10^{10}\mu$K$_{\rm CMB}$-arcmin in map-domain noise). The optimized configuration includes 16 radiometric bands, with higher spectral resolution set-ups possible via multichroic detector technology at minimal extra cost, which provide robustness against uncertainties in the current foreground sky emission models at small sensitivity penalty (see 34-band configuration results in Ref.~\cite{specter}). In its current design, \textit{SPECTER} has a roughly $1\deg$ angular resolution at 150 GHz. 

With this configuration and assuming a total mission time of 2 (8) years, \textit{SPECTER} would be able to detect the $\mu$-distortion at $\sim5\sigma$ $(10\sigma)$, and to provide sub-percent and percent-level constraints on the $y$-distortion and its relativistic correction, respectively. We show the \textit{SPECTER} sensitivity compared to CMB spectral distortions, foregrounds, \textit{COBE/FIRAS} noise, and the CMB blackbody in Figure \ref{fig:specter}. The figure is adapted from Ref.~\cite{specter}, which uses the sky model from Ref.~\cite{Abitbol2017}.

\textit{SPECTER} is in the initial design phase with further development required to move this instrument concept forward. In particular, further development is needed in two main directions: (1) the size of the LFFIs is a limiting factor at present, necessitating further revision of the LFFI design either through development of more effective amplifiers,  use of external data to constrain low-frequency foregrounds, or other alternate solutions; (2) the absolute temperature calibration through the use of the load-selecting mirror and cross-channel band calibration needs to be tested via, for example, a smaller \textit{SPECTER} prototype experiment. While additional research and development is needed for any type of spectral distortion instrument in order to make a high significance measurement of the faint $\mu$ signal, we highlight the specific areas that need improvement and the important follow-up actions for \textit{SPECTER.}

In summary, \textit{SPECTER} is an instrument concept which presents a new and additional approach to measuring spectral distortions beyond the standard spectrometer-based designs. It offers a way to employ mature detector
technology used in CMB anisotropy experiments and to optimize the design of an instrument to target CMB spectral
distortions in the presence of much brighter foreground contamination.

\subsection{Towards a space-based mission}
The long dormancy between CMB spectrometer experiments necessitates a staged program to rejuvenate the field and transform CMB SD measurements into a precision probe for cosmology and particle physics, as described above. Successful measurements of distortions would allow us to write an entirely new chapter in CMB cosmology. One option proposed in the ESA Voyage 2050 white paper on CMB spectral distortions \citep{Chluba2021} envisioned an M-class mission such as \FOSSIL providing transformational data to the field. The \FOSSIL mission is designed to {\it revolutionize} CMB spectroscopy by replacing several long-standing upper limits with actual detections, exploring a broad range of new physics scenarios, and producing an absolutely-calibrated multifrequency view of the sky with immense legacy value for the broad astrophysics community. The \FOSSIL mission would firmly establish the instrumental, observational and computational foundations necessary to utilize space-based spectrometer technology, while growing a strong community to continue pushing forward the frontiers of this rich scientific target. 

\begin{figure}[h]
    \centering
    \includegraphics[width=1\columnwidth]{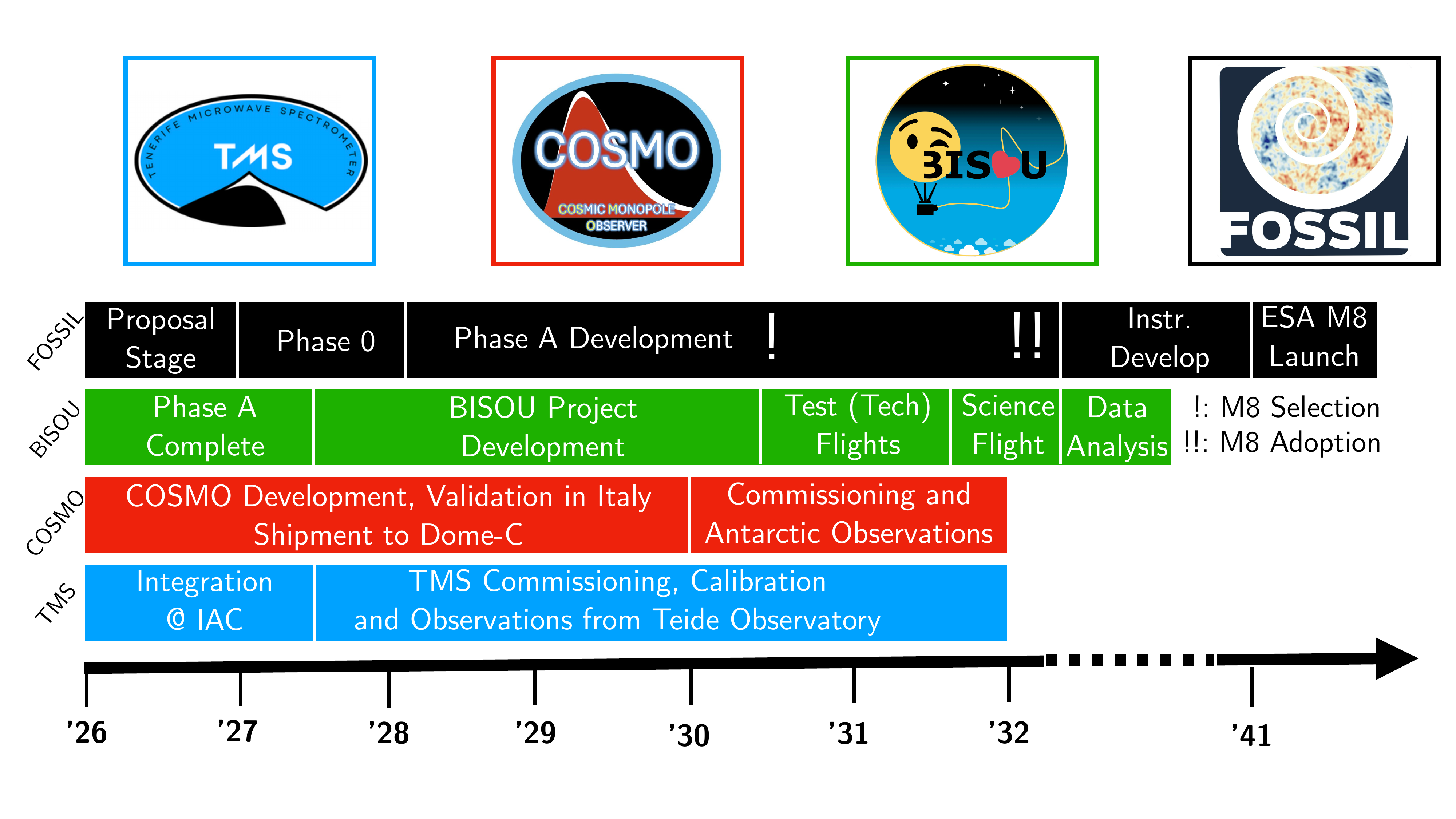}
    \caption{A heuristic timeline of operations for the more mature TMS, COSMO, BISOU, and \FOSSIL experiments. }
    \label{fig:roadmap}
\end{figure}

A subset of the more mature observational efforts discussed above (TMS, COSMO and BISOU) can naturally be thought of as a sequence of pathfinder activities, whose heuristic timelines are shown in Fig.~\ref{fig:roadmap}. Together they address crucial preparatory questions from sub-orbital platforms, while promising detections of the average $y$-distortions and shedding new light on the origin of the RSB excess within the next 5-10 years. 
Their role is therefore complementary to \FOSSIL: they will advance and develop the experimental techniques, calibration strategies, foreground knowledge and analysis methods needed for absolute spectroscopy before the proposed launch of the M-class mission around 2041.
In many ways, TMS, COSMO and BISOU will thus play the role of COBE/DMR for CMB anisotropies, with \FOSSIL performing the first detection of truly primordial physics through the diffusion damping $\mu$-distortion. Beyond the standard model, \FOSSIL will constrain vasts swaths of parameter space for contemporary models of dark matter (e.g. axions, ALPs, WIMPs, and dark photons), and completely rule out canonical PBH formation scenarios for any cosmological fraction of supermassive black holes. 

The resulting roadmap extends from the pathfinding measurements of TMS, COSMO and BISOU, through to \FOSSIL as a major milestone of CMB spectroscopy in space. Each stage has an independent major scientific return, but its broader value lies in progressively opening the SD window and establishing the knowledge needed to realise the next stage, while building the community to carry the field into the future. Assuredly, the measurements made by \FOSSIL will open a new window into our Universe that will inspire even more ambitious missions in the long term.

\section{Conclusions} \label{sec:conclusions}
More than three decades after \COBEF, the frequency spectrum of the CMB is now the least explored of its primary observables, with current limits of $|\langle \mu \rangle| \leq 4.7\times 10^{-5}$ and $\langle y \rangle \leq 5.2\times 10^{-6}$, roughly a factor of a few above the largest distortions guaranteed within $\Lambda$CDM. Once photon number-changing processes freeze out at $z \simeq 2\times 10^{6}$, perturbations away from a blackbody can no longer be fully thermalized, and any subsequent exchange of energy or entropy leaves a permanent imprint whose spectral shape records the epoch and amplitude of release. The CMB spectrum is therefore an exquisitely sensitive calorimeter for energy release from $z \simeq 10^{6}$ to the present day.

The successive freeze-out of number-changing ($z \simeq 2\times 10^{6}$) and energy-redistributing ($z \simeq 5\times 10^{4}$) interactions partitions the distortion history into the $T$, $\mu$, residual, and $y$ eras (Sec.~\ref{sec:Intro}), each a distinct limit of the photon Boltzmann equation. The residual era in particular carries timing information that helps discriminate between injection scenarios. Full numerical solutions are obtained routinely with \CT and \texttt{DarkHistory}, which are complementary in their treatment of low- and high-energy injections, while Green's function methods and the frequency hierarchy provide highly accurate approximations to the distortion amplitudes.

The science case for spectral distortions is extraordinarily diverse. In the early Universe, the dissipation of small-scale acoustic modes yields a robust $\Lambda$CDM prediction of $\mu \simeq 2\times 10^{-8}$. This signal is an integrated and essentially astrophysics-free measure of the primordial power spectrum over $1\,{\rm Mpc^{-1}} \lesssim k \lesssim 10^{4}\,{\rm Mpc^{-1}}$, roughly ten $e$-folds of inflationary evolution beyond the reach of the anisotropies (Sec.~\ref{sec:small-scale-structure}). A detection at this level would support the simplest plateau models of inflation, while the same measurement sharply constrains any scenario with enhanced small-scale power. As discussed in Sec.~\ref{sec:BH-formation}, \FOSSIL could exclude primordial black holes as seeds of the earliest supermassive black holes even in the most extreme non-Gaussian scenarios up to $k \simeq 6\times 10^{4}\,{\rm Mpc^{-1}}$, along with the not-quite-primordial channel. A null result would leave direct collapse as one of the few viable heavy seed mechanisms. Distortions also probe cosmological gravitational wave backgrounds (Sec.~\ref{sec:GWBGs}), directly through tensor dissipation in the otherwise inaccessible picohertz window down to $\Omega_{\rm GW} \sim 10^{-8}$, and indirectly through the enhanced fluctuations that source scalar-induced backgrounds relevant to pulsar timing arrays and to possible $B$-mode contamination.

Spectral distortions also offer a way to test extensions beyond the standard model (Sec.~\ref{sec:BSM}), for example through probing topological defect networks with symmetry breaking scales up to $\sim 10^{13}\,{\rm GeV}$, evaporating or accreting primordial black holes, and early-Universe resolutions of the Hubble tension. Superconducting strings source negative $\mu$-distortions due to abundant photon production, evaporating black holes probe the primordial power spectrum at $k \sim 10^{14}$--$10^{17}\,{\rm Mpc^{-1}}$, and many of these scenarios produce non-standard spectral shapes that allow for powerful model discrimination. In addition, distortions can help probe magnetogenesis scenarios through the dissipation of primordial magnetic fields in the early Universe. 

Another underexplored feature of the CMB spectrum is the cosmological recombination radiation (Sec.~\ref{CRR}), which offers a direct spectroscopic view of the three recombination epochs and responds sensitively to non-standard recombination histories. Excitingly, \FOSSIL forecasts indicate that the Balmer-$\alpha$ line from $z \approx 1300$ may be in reach. Finally, an order-of-magnitude improvement on $T_0$ (Sec.~\ref{sec:T0}) would sharpen BBN predictions and remove an uncertainty that degrades constraints on $\omega_{\rm b}$ by up to $50\%$ for upcoming anisotropy experiments.

After recombination, the thermal SZ effect provides a guaranteed distortion at $\langle y \rangle \simeq 10^{-6}$, which is a central target for the ground and balloon-borne efforts of TMS \cite{TMS}, COSMO \cite{COSMO2022}, and BISOU \cite{bisou2024}. From space, \FOSSIL \cite{FOSSILPlaceholder} would measure this at $200\sigma$ while detecting relativistic corrections to the signal at $\simeq 20\sigma$ (Sec.~\ref{sec:SZ_effect}). A measurement at this level would provide percent-level constraints on supernova and AGN feedback, and few-percent constraints on the baryonic suppression of the matter power spectrum. Beneath this lies the $y_{\rm reion} \sim 10^{-7}$ contribution of the reionized intergalactic medium (Sec.~\ref{sec:CD_and_reionization}). In the Rayleigh-Jeans tail, low-frequency distortions couple to the 21cm signal at Cosmic Dawn through soft photon heating, and an absolute measurement near $50\,$GHz would discriminate between the proposed origins of the unexplained radio synchrotron background (Sec.~\ref{sec:RSB}). A by-product of an absolute spectrometer in space is that it can act simultaneously as a full-sky, absolutely calibrated spectral survey. Its coverage of the CO rotational ladder and the \cii{} line enables tomographic line-intensity mapping from the epoch of reionization to the present (Sec.~\ref{sec:LIM}) and a volume-averaged census of molecular gas and star formation at cosmic noon (Sec.~\ref{sec:cosmic_noon}) (albeit with a wide beam angle). It would also deliver a high-significance measurement of the CIB monopole and dipole at $\nu \lesssim 400\,$GHz, where dust-obscured star formation at $z \gtrsim 5$ contributes most strongly (Sec.~\ref{Sec:CIB}).

Within our own Galaxy and Solar System, the foregrounds that must be removed to reach these signals are themselves science targets. An absolute spectrometer measures the zero level of the Galactic dust optical depth, which is inaccessible to differential imagers, and can detect departures from a modified blackbody that would allow us to discriminate between physical grain models (Sec.~\ref{sec:galactic_dust}). Together with full-sky maps of the far-infrared cooling lines, this data would provide physical priors for $B$-mode component separation. Combined with low-frequency data from TMS, such a survey would determine the poorly constrained monopole of the anomalous microwave emission (Sec.~\ref{AME}), and long-duration absolute monitoring would separate true temporal variability of the zodiacal cloud from instrumental drifts (Sec.~\ref{sec:zodiacal}).

The dark sector offers perhaps the richest discovery space. Velocity dependent dark matter-baryon scattering can produce a negative $\mu$ distortion, and with \FOSSIL-level sensitivity, one could probe cross sections orders of magnitude below anisotropy bounds for sub-GeV masses, with a similar story for $p$-wave annihilations in the dark sector (Sec.~\ref{sec:DM_interactions}). Decays spanning nearly thirty orders of magnitude in lifetime produce non-thermal signals that can be searched for. In the case of cosmologically long-lived particles, a distortion at the level of $\Delta I/I \sim 10^{-6}$ can be reached when saturating anisotropy bounds. For axions and dark photons (Secs.~\ref{sec:axions} and \ref{sec:dark_gamma}), the monotonically decreasing plasma frequency of the standard model photon performs a natural scan across roughly ten decades in  mass, with resonant conversions imprinting templates distinct from $\mu$ and $y$. Beyond the global signals, plasma inhomogeneities in these resonant conversion scenarios produce patchy anisotropies that can searched for in our local Universe. Future measurements would tighten limits on the kinetic mixing parameter by two orders of magnitude beyond \COBEF and, for primordial magnetic fields near current limits, on the axion-photon coupling by three to four. While it is not feasible to discuss every possible distortion induced by the hidden sector, recent work on atomic and millicharged dark matter, composite states with electromagnetic moments, and macroscopic candidates such as axion quark nuggets have also been shown to produce strong distortions in wide regions of their respective parameter spaces (Sec.~\ref{sec:alt_DM}).
\begin{figure}[H]
    \centering
    \includegraphics[width=1\columnwidth]{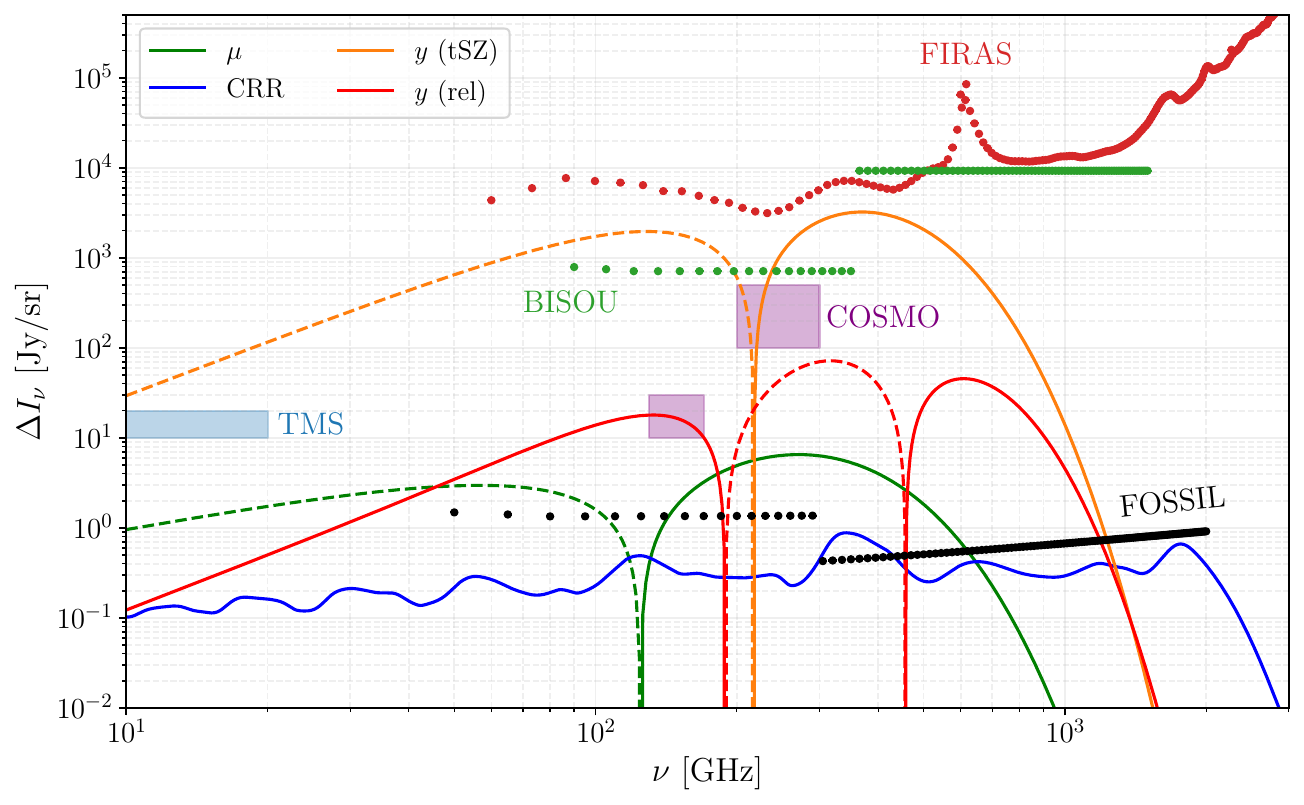}
    \caption{A summary of various $\Lambda$CDM distortions and forecasted sensitivities from the most advanced experimental designs. The $\mu$ distortion is sourced by the damping of small scale power under the assumption of a nearly scale invariant spectrum of perturbations down to $k \simeq 10^{5} \, {\rm Mpc}^{-1}$. The $y$ distortion signatures are from the thermal (orange) Sunyaev-Zel'dovich effect and its relativistic corrections (red). Finally, the spectrum of cosmological recombination radiation (CRR) as computed by \texttt{CosmoSpec} is shown in blue.}
    \label{fig:dist_overview}
\end{figure}

Distortion anisotropies extend this program beyond the sky-averaged monopole (Sec.~\ref{sec:CMB-sds-anisotropies}). They trace the spatial structure of the sources, and the frequency hierarchy formalism now allows their power spectra to be computed in a similar manner to the temperature and polarization spectra. Cross-correlations such as $\mu T$ and $\mu E$ provide a pathway to probe primordial non-Gaussianity on very small scales, where a significant detection would rule out standard single-field inflation. Distortion anisotropies also encode epoch-dependent information on decaying particles, resonant photon conversions, and the temperature-redshift relation that the monopole alone cannot provide. Because the amplitude of these correlations scales with the average distortion, an absolute spectrometer such as \FOSSIL is required to interpret them, ideally alongside a next-generation imager such as LiteBIRD.

Realizing this science requires measurements that do not yet exist, and in Sec.~\ref{sec:missions} an overview of the experimental landscape was presented. Hardware is being integrated and tested today across three complementary tiers. From the ground, TMS ($10$--$20\,$GHz) and COSMO ($130$--$170\,$ and $200$--$300$ GHz) will characterize the low-frequency foregrounds and target the global tSZ effect. From the stratosphere, BISOU ($90\,$GHz--$1.5\,$THz) aims to detect this $y$ monopole at $5$--$6\sigma$, surpassing the \COBEF sensitivity by more than an order of magnitude at low frequencies. Each of these efforts delivers standalone science within the next $5$--$10$ years while maturing the technologies and analysis methods required for space.

From space, the \FOSSIL satellite has been proposed as part of the ESA M8 call. If successful, it would launch in the 2040s and is a cornerstone of the spectral distortion program. By delivering full-sky, absolutely calibrated spectra in $130$ channels from $50\,$GHz to $2\,$THz at $1$--$2^{\circ}$ resolution, it would improve on \COBEF by three orders of magnitude, replace several long-standing upper limits with detections, and bring the $\Lambda$CDM $\mu$-distortion within reach. \FOSSIL would provide the first detection of truly primordial physics within the CMB spectrum while establishing a comprehensive understanding of the foregrounds and systematics that forecasts alone cannot provide. Its maps would moreover serve as an absolute calibration anchor for existing and future multi-frequency surveys. Complementary concepts such as \PIXIE and SPECTER offer alternative paths to comparable sensitivity. A summary of the main science goals and projected sensitivities of the most advanced experimental designs is presented in Fig.~\ref{fig:dist_overview}.

A single absolute measurement of the CMB frequency spectrum simultaneously tests the nature of small scale primordial perturbations, the various potential properties of dark matter, the thermal history of cosmic structure, and the composition of our own Galaxy and Solar System. With pathfinders under construction and \FOSSIL on the horizon, the community is well positioned to open this window, peering beyond the veil of the surface of last scattering, and deep into the primordial Universe.

\section*{Acknowledgments}
BC would like to acknowledge support from an NSERC Banting Fellowship, as well as the Simons Foundation (Grant Number 929255). The COSMO experiment has been funded by the Italian National Antarctic Research Program (PNRA), and by the Italian "Fondo Italiano per la Scienza" (FIS2). A validation campaign at the INAF observatory in Campo Imperatore (2200 m osl) is planned for 2027, before shipping to Antarctica. The work on the BISOU programme is funded by the Centre national d’études spatiales (CNES), France (ROR: https://ror.org/04h1h0y33) and by r\'egion \^Ile-de-France IDF-DIM-ORIGINES-2023-4-07 and IDF-DIM-ORIGINES-2024-1-06. The work on \FOSSIL has received funding from CNES.

\pagebreak

\appendix

\section{Author Affiliations}
\printauthoraffiliations

\bibliographystyle{JHEP}
\bibliography{master_refs}

\end{document}